\documentclass[a4paper,fleqn]{cas-sc}

\usepackage{siunitx} 
\usepackage[numbers,sort&compress]{natbib}
\usepackage{etoolbox} 
\AtBeginEnvironment{thebibliography}{\small\setlength{\bibsep}{0pt}}
\usepackage{doi}
\usepackage{tikz} 
\usetikzlibrary{positioning, fit}
\usetikzlibrary{arrows.meta} 
\usetikzlibrary{calc}
\usepackage{framed}
\usepackage{bm}
\usepackage[utf8]{inputenc}
\usepackage{xifthen}
\usepackage{float}
\usepackage{morefloats}
\usepackage{amsthm}
\usepackage{mathtools}
\usepackage[normalem]{ulem}         
\usepackage{tabularx} 

\hypersetup{hypertexnames=false}

\newtheorem{Remark}{Remark}

\def\epow{\mathrm{e}}               
\def\cphase{c_{\mathrm{ph}}}        
\def\cgroup{c_{\mathrm g}}          
\def\cpress{c_\mathrm{p}}           
\def\cshear{c_\mathrm{s}}           
\def\cplateflex{c_\mathrm{fl}}      
\def\cplateax{c_\mathrm{ax}}        
\def\wins{n_\mathrm{s}}             
\def\winl{n_\mathrm{l}}             
\def\wine{n_\mathrm{e}}             
\def\sta{\bar{S}_{E}^{\mathrm{STA}}}      
\def\lta{\bar{S}_{E}^{\mathrm{LTA}}}      
\def\statolta{\bar{r}}              
\def\er{\bar{S}_\mathrm{er}}        
\def\mer{\bar{S}_\mathrm{mer}}      
\def\var{\mathrm{var}}              
\def\aic{\chi}                      
\def\ansys{ANSYS\textsuperscript{\textregistered}}                  
\def\ethrthr{e_{33}}                
\def\ethrone{e_{31}}                
\def\eonefiv{e_{15}}                
\def\epsoneone{\epsilon_{11}}       
\def\epsthrthr{\epsilon_{33}}       
\def\psvfull{Polytec Scanning Vibrometer 400 3D M~} 
\def\dt{\Delta t}                   
\def\dx{\Delta x}                   
\def\dy{\Delta y}                   
\def\tdom{\tau_\mathrm{dom}}        
\def\Ra{R_a}                        
\def\tfirstlb{t_{\mathrm{lb}1}}     
\def\tfirstub{t_{\mathrm{ub}1}}     
\def\tmax{t_\mathrm{max}}           
\def\tam{t_\mathrm{am}}             
\def\tfe{t_\mathrm{fe}}             
\def\tfb{t_\mathrm{fb}}             
\def\tfa{t_\mathrm{fa}}             
\def\sla{SLA}                       
\def\Thetav{\boldsymbol \Theta}
\def\ev{\bm {\mathrm e}}
\def\etap{\eta_{\mathrm p}}
\def\etas{\eta_{\mathrm s}}
\def\energy{S_{\mathrm{E} }}        
\def\snr{\mathrm{SNR}}
\def\mathnoise{\mathcal{M}}
\def\xb{\mathbf{x}}
\def\sensor{\mathrm{S}}
\def\tso{t_{S_0}}                   
\def\tao{t_{A_0}}                   
\def\coi{COI}

\def\cso{c_{S_0}}                   
\def\cao{c_{A_0}}                   
\def\wavnr{k}
\def\iop{\mathcal{S}}
\def\iob{\mathcal{I}}
\def\allen{\beta}
\newcommand{\mhyphen}{\mbox{-}}
\def\aicglob{AIC-GM}
\def\aicloc{AIC-LM}

\definecolor{darkpastelgreen}{rgb}{0.01, 0.75, 0.24}
\definecolor{ayechblue}{rgb}{0.09, 0.768, 1}
\definecolor{orchid}{rgb}{0.855, 0.439, 0.839}
\definecolor{OliveGreen}{rgb}{0,0.6,0}

\newcommand\ayechsout{\bgroup\markoverwith{\textcolor{ayechblue}{\rule[0.5ex]{2pt}{0.8pt}}}\ULon} 

\newcommand{\dnachdx}[3][]{\ifthenelse{\isempty{#1}}{\dfrac{\partial {#2}}{\partial {#3}}} {\dfrac{\partial^{#1} {#2}}{\partial {#3}^{#1}}}}
\newcommand{\ton}[1][]{\ifthenelse{\isempty{#1}}{t_{\mathrm{on}}}{t_{\mathrm{on}{,#1}}}}
\newcommand{\relerrt}[1][]{\ifthenelse{\isempty{#1}}{\epsilon_{t}}{\epsilon_{t_{#1}}}}
\newcommand{\abserrt}[1][]{\ifthenelse{\isempty{#1}}{\delta_{t}}{\delta_{t_{#1}}}}
\newcommand{\test}[1][]{\ifthenelse{\isempty{#1}}{\hat{t}}{\hat{t}_{#1}}}
\newcommand{\ttrue}[1][]{\ifthenelse{\isempty{#1}}{\bar{t}}{\bar{t}_{#1}}}

\newcommand{\imp}[1]{I_{#1}}
\renewcommand{\comment}[1]{}

\newcommand{\sclogr}[3]{\mathcal {S_W} \{ #1 \}(#2,#3)}

\DeclareMathOperator{\curl}{curl}

\makeatletter
\def\hlinewd#1{%
\noalign{\ifnum0=`}\fi\hrule \@height #1 \futurelet
    \reserved@a\@xhline}
\makeatother

\DeclareMathOperator{\diverg}{div}
\DeclareMathOperator{\grad}{grad}
\DeclareMathOperator{\trace}{tr}
\DeclareMathOperator{\rms}{rms}
\DeclareMathOperator*{\argmax}{argmax}
\DeclareMathOperator*{\argmin}{argmin}
\DeclareMathOperator{\sym}{sym}

\newcommand{\stress}{\mathbf T}
\newcommand{\strain}{\boldsymbol \varepsilon}

\newcommand{\I}{\mathbf I}

\newcommand{\fv}{\bm f}
\newcommand{\uv}{\bm u}
\newcommand{\ToA}{TOA}

\newcommand{\sth}{s_{\rm th}}

\newcommand{\fourier}[1]{\mathcal F \{ #1 \}(\omega)}
\newcommand{\stft}[3]{\mathcal F_g \{ #1 \}(#2,#3)}
\newcommand{\cwt}[3]{\mathcal W_\psi \{ #1 \}(#2,#3)}

\begin{document}
\let\WriteBookmarks\relax
\def\floatpagepagefraction{1}
\def\textpagefraction{.001}

\shorttitle{Assessment of Time-of-Arrival Estimation Methods}
\shortauthors{L. Grasboeck et~al.}

\title[mode=title]{Assessment of Time-of-Arrival Estimation Methods for Impact Detection in Isotropic Plates using Piezoceramic Sensors}

\author[1,2]{Lukas Grasboeck}[orcid=0009-0007-0662-4913]
\cormark[1]
\ead{lukas.grasboeck@lcm.at}
\credit{Conceptualization, Methodology, Software, Investigation, Writing - Original Draft, Visualization}

\author[2]{Alexander Humer}[orcid=0000-0001-8510-5423]
\credit{Supervision, Conceptualization, Methodology, Software, Validation, Writing - Review \& Editing}

\author[3,4]{Ayech Benjeddou}[orcid=0000-0002-4760-4800]
\credit{Co-Supervision, Validation, Resources, Writing - Review \& Editing}

\affiliation[1]{organization={Linz Center of Mechatronics GmbH},
                city={Linz},
                country={Austria}}

\affiliation[2]{organization={Institute of Technical Mechanics, Johannes Kepler University Linz},
                city={Linz},
                country={Austria}}

\affiliation[3]{organization={Institut Supérieur de Mécanique de Paris},
                city={Saint Ouen},
                country={France}}

\affiliation[4]{organization={Université de Technologie de Compiègne, Roberval (Mechanics, Energy \& Electricity)},
                city={Compiègne},
                country={France}}

\cortext[cor1]{Corresponding author}

\begin{abstract}
This work describes and assesses different methods for estimating the time-of-arrival (\ToA{}) of impact-induced waves in isotropic plate-like structures.
The methods considered include threshold crossing (TC), continuous wavelet transform (CWT), short/long term average (\sla), modified energy ratio (MER), and the Akaike information criterion (AIC).
Their advantages, limitations, and sensitivities to method-specific parameters are systematically investigated.
The assessment is based on synthetic data from transient finite element simulations that are experimentally calibrated with respect to excitation and dispersion characteristics.
Wave propagation is monitored using piezoceramic patch sensors bonded to the plate surface, and robustness is evaluated for impacts of varying positions and force profiles, including noise-contaminated sensor signals in order to account for practically relevant measurement conditions.
The results show that the methods are capable of detecting the fundamental Lamb wave modes, with nearly all capturing both the symmetric and anti-symmetric mode arrivals under noise-free conditions.
In particular, noise primarily impairs the detection of the earliest symmetric-mode arrivals, while meaningful anti-symmetric-mode \ToA{}-estimates can still be obtained by suitable preprocessing or time-frequency analysis.
Besides, new contributions to the assessed \ToA{}-estimation methods include a frequency-domain threshold crossing within the CWT framework that improves both robustness and accuracy of \ToA{}-estimation, and the consideration of local minima in the AIC that proves effective for detecting the \ToA{} of the fundamental symmetric mode.
Beyond these findings, the research provides practical guidelines and insights into the specific characteristics of each assessed method, supporting accurate and reliable \ToA{}-estimation for applications such as impact localization.
\end{abstract}


\maketitle

\section{Introduction}
Detection, localization and quantification of impacts on load-bearing structures are crucial tasks in the field of \emph{structural health monitoring} (SHM).
Knowledge on impact events -- specifically, whether an impact has occurred, where it occurred, and what kind of impact it was -- is essential for effective maintenance and damage assessment.
In SHM, detection involves identifying the occurrence of an impact, localization pinpoints the exact location of the impact on the structure, and quantification evaluates the severity and potential consequences of the impact.
An impact induces elastic waves that propagate in a structure and are supposed to be detected by appropriate sensors and sensor networks.
Therefore, detection and localization of impacts can be regarded as \emph{passive processes} since no external actuation is required. 
In thin-walled structures, waves primarily propagate as \emph{guided waves}, i.e., along preferred directions of structures, see, e.g., \cite{Lamb_1917, Giurgiutiu_2014,Achenbach_1999}.

Localization of impacts, which represent acoustic sources, is an inverse problem: Given sensor signals related to waves induced by an impact, the source's point of origin is to be reconstructed.
There are numerous methods for localization as, e.g., triangulation \cite{Kundu_2014}, beamforming \cite{McLaskey_2010, He_2012}, strain rosettes \cite{LanzaDiScalea_2007}, optimization-based techniques (deducted from triangulation approach) \cite{Ciampa_2010, Ciampa_2012b, Staszewski_2009, Jang_2015, Merlo_2017}, elastic Poynting vector method \cite{Guyomar_2009, Guyomar_2011b} and data-based approaches such as time reversal \cite{Ciampa_2012, Park_2012, Qiu_2011} or machine learning techniques \cite{Jang_2015, Tabian_2019b, Haywood_2005, LeClerc_2007, Seno_2019}.
These methods differ, among other things, in their applicability in terms of material properties, e.g., whether a material (or structure) is isotropic or anisotropic, and also in terms of the complexity of the structure's geometry, which may range from simple planar panels to fuselage components of aircrafts \cite{Kundu_2014}.

Irrespective of the actual localization method, however, accurate detection of impacts, i.e., the determination of the time at which waves induced by impacts are captured by a sensor---the so-called \emph{time-of-arrival} (\ToA{})---is a fundamental prerequisite.
Note that we distinguish between the common terms \emph{time-of-flight} (TOF) and the \ToA{}.
The former defines the timespan for the wave to propagate from the point of impact to a specific sensor; \ToA{}, however, only relates to the moment when the wave reaches the sensor.
In other words, \ToA{} refers to a specific point in time, whereas TOF denotes a time interval.
Many methods have their roots in the field of seismology, where, e.g., localization of the hypocenter of a seismic activity \cite{Akram_2016, Han_2009, Trnkoczy_} requires \ToA{}-estimation.
We refer to Akram and Eaton \cite{Akram_2016} for an overview on \ToA{}-estimation methods in seismology, which includes hybrid approaches that combine aspects of several methods.
Common methods for \ToA{}-estimation include, e.g., threshold crossing (TC) method \cite{Kundu_2009, Sanchez_2016, SharifKhodaei_2012b, Mallardo_2013}, continuous wavelet transform (CWT) \cite{Gaul_1998, Hamstad_2005, Mallat_2009, Dris_2020, Meo_2005}, short/long-term average (SLA) algorithm \cite{Akram_2016, Han_2009}, modified energy ratio (MER) algorithm \cite{Han_2009, J.Wong_2009}, cross-correlation techniques \cite{vandecar1990determination, DeMeersman1998, Molyneux_1999}, higher order statistics \cite{Saragiotis_2004, Lokajicek_2006, Liu_2021}, Akaike information criterion (AIC) \cite{Allen_1982, Sedlak_2013, Simone_2017, Hou_2025, Barile_2025} and neural network approaches \cite{McCormack_1993, Dai_1995, Zonzini_2022}. 

The goal of the present assessment is to evaluate selected \ToA{}-estimation methods, which are well-established in their respective fields of application, in the context of impact detection of thin-walled structures.
For this purpose, the following methods are investigated:
Firstly, we start with the TC (in time-domain) method, which defines the \ToA{} as that point in time when a sensor's signal level exceeds a pre-defined threshold.
Secondly, CWT provides information on the spectral content over time by convolution of the sensor signal with a set of scaled wavelet functions.
With this additional information, we aim to bypass the complexity introduced by the dispersive nature of guided wave propagation.
Thirdly, creating averages over small portions of the sensor signal and relating them to each other is the approach of both \sla{} and MER methods.
By sliding these averaging windows over time-signals, \sla{} adopts windows of different widths, while MER uses windows of the same width but relates energy-based quantities instead of averages.
Lastly, we make use of a method from information theory for statistical model selection \cite{Stoica_2004}, namely the AIC method.
It is based on the assumption that a signal can be divided into two parts: A \emph{non-informative} part (noise) and an \emph{informative} part (signal), which are separated by the \ToA{} \cite{Simone_2017, Kitagawa_1978}.

We aim to explore the benefits, limitations and provide guidelines for applications of the individual methods.
Propagation of guided waves in thin-walled three-dimensional (3D) structures is undoubtedly a very complex phenomenon in many respects.
For our assessment, we rely on finite-element (FE) models to simulate the propagation of waves in elastic media. 
To represent impacts on the structure as realistically as possible, the 3D model of the thin-walled panel is calibrated to experiments. 

Following this introduction, Sec. \ref{sec::theory_wav_prop} outlines the fundamentals of wave propagation in 3D continua, as well as the prerequisite for guided wave propagation in thin-walled structures.
Section \ref{sec::FE_exp_model} describes the model for the transient analysis of Lamb waves that propagate in a thin-walled panel.
We also describe the experimental setup, which is used to calibrate the FE model of the thin-walled panel.
In Sec. \ref{sec::toa}, we explain the fundamentals of the individual \ToA{} estimation methods and define the \ToA{} for each one.
Afterwards, in Sec. \ref{sec::results}, the \ToA{}-estimation methodologies are comprehensively applied to the 3D FE model data, with the subsequent presentation and interpretation of the corresponding results.
Finally, we summarize the results that emerged from this work and highlight the advantages and disadvantages of the methods,  and provide guidance for successful \ToA{}-estimation in the context of impact-induced guided wave propagation in thin-walled structures.
\section{Lamb Wave Propagation in Elastic Media}
\label{sec::theory_wav_prop}
To begin with, we recall the fundamentals of wave propagation in 3D continua, from which the theory of Lamb waves is obtained as a special case. 
The (local) balance of linear momentum, see, e.g., \cite[p. 149]{Altenbach_2015}, provides the foundation for both theoretical considerations and the numerical realization of models for wave propagation by means of the FE method:
\begin{equation}\label{eq::strong_form}
    \rho \ddot \uv = \diverg \stress + \fv.
\end{equation}
In the above relation, $\uv$ is the displacement field, $\rho$ denotes the mass density (per unit current volume) and $\stress$ is Cauchy's stress tensor; $\fv$ are volume forces imposed on the material body under consideration. 
Material time-derivatives are indicated by superimposed dots.

We stay within the framework of \emph{linear elasticity}, i.e., deformations are sufficiently small to be described by the linearized strain tensor $\strain$, which is defined as the symmetric part of the displacement gradient:
\begin{equation}\label{eq::linearized_strain_tensor}
    \strain = \frac 1 2 \left\{ \grad \uv + \left( \grad \uv \right)^T \right\} = \sym (\grad \uv). 
\end{equation}
Further, we assume the stress response to be linear and isotropic, 
\begin{equation}\label{eq::cauchy_stress_tensor}
    \stress = \lambda \left( \trace \strain \right) \I + 2 \mu \strain ,
\end{equation}
where $\lambda$ and $\mu$ denote Lamé's first parameter and the shear modulus, respectively.
Substituting the definition of strains \eqref{eq::linearized_strain_tensor} and Hooke's law \eqref{eq::cauchy_stress_tensor} into the balance of linear momentum \eqref{eq::strong_form}, we obtain the \emph{Navier-Cauchy} equations
\begin{equation}
    \label{eq::navier_cauchy}
    \begin{aligned}
        \rho \ddot \uv 
        &= \diverg \left\{ 2 \mu \sym (\grad \uv) + \lambda \left(\diverg \uv \right) \I \right\} + \fv \\
        &= \left(\lambda + \mu \right) \grad \left(\diverg \uv \right) + \mu \diverg \left(\grad \uv \right) + \fv ,
    \end{aligned}
\end{equation}
i.e., the equations of motion expressed in terms of displacement field \cite[p. 307]{Altenbach_2015}.
The \emph{Navier-Cauchy} equations govern the propagation of \emph{bulk waves} in unbounded elastic continua, see, e.g., \cite[p. 276]{Giurgiutiu_2014}.
The most fundamental form of bulk waves in solids are \emph{longitudinal} and \emph{transverse} waves.
In longitudinal waves, particle motion is parallel to the direction of propagation; they are often referred to as \emph{pressure waves} or \emph{P-waves}.
Transverse waves are characterized by particle motion perpendicular to the direction of propagation; hence, they are commonly denoted as \emph{shear waves} or \emph{S-waves}, see, e.g., \cite[p. 123]{Achenbach_1999}.

Not least from a practical point of view, we are primarily interested in waves in thin-walled structures, which are characterized by the fact that so-called \emph{guided waves} are the main mode of propagation.
For the case of isotropic homogeneous plates, guided waves are referred to as \emph{Lamb waves}, cf. \cite{Lamb_1917}. 
As the term \emph{guided} implies, waves are guided along a plate's or shell's preferred dimensions between its top and bottom faces \cite[p. 309]{Giurgiutiu_2014}. 
Owing to their thin-walled nature, bulk waves excited, e.g., by an impact on the structure, are more or less immediately reflected at a panel's faces. 
Upon repeated reflections, longitudinal and transverse waves interfere and are eventually converted into Lamb waves, which propagate \emph{in-plane}, i.e., waves are guided along the preferred dimensions of thin-walled structures.
Lamb waves materialize as symmetric ($ S_i $) and antisymmetric ($ A_i $) modes of order $ i $ and are highly dispersive, i.e., waves of different frequencies propagate at different velocities.
Besides material properties, the dispersion depends on the product of frequency $ f $ and the plate thickness $ 2d $.
Dispersion curves, i.e., the relation between phase velocity $ \cphase $ and frequency-half-thickness product $ fd $, are obtained by solving the so-called \emph{Rayleigh-Lamb} equations.

\begin{Remark}
    In the context of dispersive wave propagation, one must inevitably deal with the concepts of \emph{phase} and \emph{group velocities}.
    In literature, these two terms are usually explained within the framework of one-dimensional wave propagation, either in terms of \emph{wave packets} or in terms of two waves with slightly different frequencies, see, e.g., \cite{Giurgiutiu_2014, Tipler_2012, Crawford_1968, LIGHTHILL_1965}. 
    We consider a wave packet, which is also referred to as \emph{tone burst}.
    The latter is composed from a monochromatic signal of a certain angular frequency, say $ \omega_0 $, which is modulated in amplitude by a window function (e.g., Gaussian, Hanning, Blackmann, etc.) to obtain a wave packet-like shape \cite[p. 234]{Giurgiutiu_2014}.
    Due to the modulation, the wave packet is no longer monochromatic.
    The individual harmonic waves which the wave packet is composed of in the sense of a Fourier series propagate with the phase velocity $ \cphase $, whereas the envelope of the wave packet travels at the speed of the group velocity $ \cgroup $.
    The fact that different wavelengths may propagate at different velocities is mathematically described by the dependency of the phase velocity on the wavelength $\lambda$:
    \begin{equation}
        \cphase = \cphase (\lambda) .
    \end{equation}
    Using the relation among wave speed, wavelength and (angular) frequency, $\lambda = \cphase / f = 2 \pi \cphase / \omega$, we rewrite $\cphase(\lambda)$ in terms of the wavelength-dependent angular frequency $\omega (\lambda)$ as
    \begin{equation}
        \cphase (\lambda) = \lambda f(\lambda) = \frac{\lambda}{2 \pi} \omega(\lambda) .
    \end{equation}
    The function $\omega (\lambda)$ is the \emph{dispersion relation}, which is commonly expressed in terms of the wave number $\wavnr = 2 \pi / \lambda$:
    \begin{equation}
        \omega(k) = k \, \cphase (k) .
        \label{eq:dispersion_relation}
    \end{equation}
    The group velocity $\cgroup$ is defined by the phase velocity's derivative with respect to the wave number:  
    \begin{equation}\label{eq::def_group_velocity}
        \cgroup (\wavnr) = \frac{d\omega (k)}{d\wavnr} = \cphase(\wavnr) + \wavnr \frac{d \cphase(k)}{d \wavnr},
    \end{equation}
    see, e.g., \cite{Tipler_2012} and \cite{Crawford_1968}.
    In \emph{non-dispersive media}, $\omega (k)$ is linear in the wave number $k$, see Eq.~\eqref{eq:dispersion_relation}, which implies that the phase velocity is constant, i.e., $d \cphase / d\wavnr = 0$, and identical to the group velocity.
    The shape of a wave packet is preserved as all frequency-components propagate at the same speed.
    In \emph{dispersive media}, $\omega(k)$ is a non-linear function of the wave number $k$ and group and phase velocities no longer coincide. 
    As individual frequencies of a wave packet propagate at different velocities, the packet’s shape is not preserved and becomes dispersed during propagation.
\end{Remark}

\begin{figure}[pos=!htbp]
    \centering
    \begin{tikzpicture}				
        \node[anchor=south west,inner sep=0] (image1) at (0, 0) {\includegraphics[width=\textwidth, trim=0 0 0 0, clip, ]{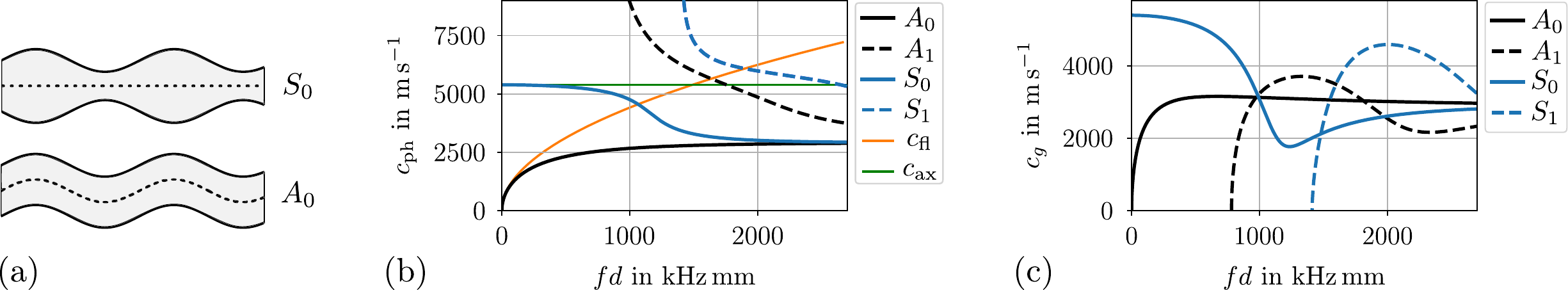}};        
        \begin{scope}[x={(image1.south east)},y={(image1.north west)}]
        \end{scope} 
    \end{tikzpicture}
    \caption{(a) Schematic illustration of the fundamental mode shapes $ S_0 $ and $ A_0 $; (b) solutions of Rayleigh-Lamb equation (expressed in terms of phase velocities $\cphase$) for symmetric ($ S_0 $ and $ S_1 $) and antisymmetric ($ A_0 $ and $ A_1 $) cases.
    Additionally, the phase velocities of flexural ($ \cplateflex $) and axial ($ \cplateax $) plate waves according to Kirchhoff-Love plate theory are illustrated.
    (c) Corresponding group velocities $\cgroup$ for the respective Lamb wave modes.}
    \label{fig::lamb_wave_disp_curve}					
\end{figure}

To obtain the dispersion relations of Lamb waves, we use the fundamental theorem of vector calculus, i.e., any vector field can be expressed as the sum of the gradient of a \emph{scalar potential} $ \Phi $ and the curl of a \emph{vector potential} $ \Thetav = \Theta_x \ev_x + \Theta_y \ev_y + \Theta_z \ev_z $ \cite[p. 280]{Giurgiutiu_2014}:\footnote{
    The fundamental theorem of vector calculus is also referred to as Helmholtz decomposition. Depending on the literature, the theorem can also be found in the form $ \uv = -\grad \Phi + \curl \Thetav $.}
\begin{equation}\label{eq::wave_potential}
    \uv = \grad \Phi + \curl \Thetav.
\end{equation}
Substituting the above decomposition into the Navier-Cauchy equations~\eqref{eq::navier_cauchy}, we obtain, after omitting external volume forces, the following relation:
\begin{equation}\label{eq::wave_potential_in_navier_cauchy_1}
    \grad \left\{ \left( \lambda + 2\mu \right) \diverg \left( \grad \Phi \right) - \rho \ddot \Phi \right\} + \curl \left\{ \mu \diverg \left( \grad \Thetav \right) - \rho \ddot \Thetav \right\} = \bm 0 .
\end{equation} 
Any non-trivial solution of the earlier equation requires the arguments of both the gradient and the curl to vanish, which gives a set of uncoupled partial differential equations (PDEs) for the scalar potential $\Phi$ and the vector potential $\Thetav$, respectively:
\begin{equation}\label{eq::wave_potential_in_navier_cauchy_2}
    \left( \lambda + 2\mu \right) \diverg \left( \grad \Phi \right) - \rho \ddot \Phi = 0 , \qquad
    \mu \diverg \left( \grad \Thetav \right) - \rho \ddot \Thetav = \bm 0 .
\end{equation}
Lamb waves follow as a special case from the Navier-Cauchy equations \eqref{eq::navier_cauchy}.
We assume a thin plate-like structure under plane strain conditions, which is aligned with the $x$-axis of a global Cartesian frame; the $z$-axis is perpendicular to the plate. 
Therefore, the displacement in $y$-direction vanishes identically as do all related components of the strain tensor:
\begin{equation}\label{eq::navier_cauchy_y_invariant}
    u_y = 0 , \qquad  \varepsilon_{yy} = \varepsilon_{xy} = \varepsilon_{yz} = 0 .  
\end{equation}
We further assume that only pressure waves and transverse (vertical, $xz$-plane) shear waves occur, which, along with plane-strain conditions, imply that only the $y$-component of the vector potential is non-trivial.
Equations~\eqref{eq::wave_potential_in_navier_cauchy_2} then reduce to
\begin{equation}
    \label{eq::wave_potential_in_navier_cauchy_4}
    \cpress^2 \diverg \left( \grad \Phi \right) - \ddot \Phi = 0 , \qquad
    \cshear^2 \diverg \left( \grad \Theta_y \right) - \ddot \Theta_y =  0,
\end{equation}
where $ \cpress $ and $ \cshear $ are phase velocities at which pressure (longitudinal) and shear (transverse) waves propagate:
\begin{equation}
    \label{eq::cpress_cshear}
    \cpress = \sqrt{\frac{\lambda+2\mu}{\rho}} , \qquad 
    \cshear = \sqrt{\frac{\mu}{\rho}}.
\end{equation}
We seek solutions for waves that propagate in $x$-direction. 
Accordingly, the following ansatz is used for $ \Phi $ and $ \Theta_y $, see, e.g., \cite[p. 310]{Giurgiutiu_2014}
\begin{equation}\label{eq::wave_potential_ansatz}
    \Phi = \phi(z)\epow^{j(\wavnr_x x - \omega t)} , \qquad 
    \Theta_y = j\theta(z)\epow^{j(\wavnr_x x - \omega t)},
\end{equation}
where $ \wavnr_x $ denotes the wave number in $x$-direction.
Upon the separation of variables \eqref{eq::wave_potential_ansatz}, Eqs.~\eqref{eq::wave_potential_in_navier_cauchy_4} give the following set of (uncoupled) ordinary differential equations (ODEs)
\begin{equation}\label{eq::wave_potential_ansatz_in_navier_cauchy}
    \frac{d^2\phi(z)}{dz^2} + \etap^2\phi(z) = 0 , \qquad 
    \frac{d^2\theta(z)}{dz^2} + \etas^2\theta(z) = 0,
\end{equation}
where
\begin{equation}\label{eq::def_etap_and_etas}
    \etap = \sqrt{\frac{\omega^2}{\cpress^2} - \wavnr_x^2} , \qquad 
    \etas = \sqrt{\frac{\omega^2}{\cshear^2} - \wavnr_x^2},
\end{equation}
have been introduced for the sake of brevity.
The linear, second-order ODEs~\eqref{eq::wave_potential_ansatz_in_navier_cauchy} governing $\phi(z)$ and $\theta(z)$ have the following general solutions:
\begin{equation}\label{eq::lamb_wave_solution_ansatz_1}
    \phi(z) = C_0 \sin \etap z + C_1 \cos \etap z , \qquad
    \theta(z) = C_2 \sin \etas z + C_3 \cos \etas z.
\end{equation}
Constants $ C_i$, $i=0,\ldots,3$ are determined by the boundary conditions at the top and bottom faces of the plate. 
Assuming both faces to be free, normal and shear stresses must vanish \cite[p. 287]{Giurgiutiu_2014}, i.e.,
\begin{equation}\label{eq::z_invariant_bc}
    T_{zz}|_{z=\pm d} = 0 , \qquad  T_{xz}|_{z=\pm d} = 0 ,
\end{equation}
where $d = h/2$ is half of the plate thickness. 
To determine $ C_0, C_1, C_2 $ and $ C_3 $, the stresses in Eq.~\eqref{eq::z_invariant_bc} must be expressed in terms of the potentials $\Phi$ and $\Thetav$ and, subsequently, in terms of $ \phi $ and $ \theta $, see Eqs.~\eqref{eq::lamb_wave_solution_ansatz_1}:
\begin{equation}
    \begin{aligned}\label{eq::stresses_with_ansatz}
        T_{zz} &= - \left[ \left( \lambda\wavnr_x^2 + (\lambda + 2\mu) \etap^2 \right) \phi(z) - 2\mu\wavnr_x \dnachdx{\theta(z)}{z} \right]\epow^{j(\wavnr_x x - \omega t)}, \\
        T_{xz} &= j\mu \left[ 2 \dnachdx{\phi(z)}{z} \wavnr_x + (\wavnr_x^2 - \etas^2) \theta(z) \right] \epow^{j(\wavnr_x x - \omega t)}.
    \end{aligned}
\end{equation}
Substituting the above relations for stresses \eqref{eq::stresses_with_ansatz} into the boundary conditions \eqref{eq::z_invariant_bc}, we obtain the following system of algebraic equations for $C_i$, $i=0,\ldots,3$:
{\small\begin{equation}\label{eq::z_invariant_bc_substituted}
    \begin{bmatrix}
        (\wavnr_x^2 - \etas^2) \sin \etap d &  (\wavnr_x^2 - \etas^2) \cos \etap d & 2 \wavnr_x \etas \cos \etas d & -2 \wavnr_x \etas \sin \etas d \\
        - (\wavnr_x^2 - \etas^2) \sin \etap d &  (\wavnr_x^2 - \etas^2) \cos \etap d & 2 \wavnr_x \etas \cos \etas d & 2 \wavnr_x \etas \sin \etas d \\
        2 \wavnr_x \etap \cos \etap d & -2 \wavnr_x \etap \sin \etap d & (\wavnr_x^2 - \etas^2) \sin \etas d & (\wavnr_x^2 - \etas^2) \cos \etas d \\
        2 \wavnr_x \etap \cos \etap d & 2 \wavnr_x \etap \sin \etap d & -(\wavnr_x^2 - \etas^2) \sin \etas d & (\wavnr_x^2 - \etas^2) \cos \etas d
    \end{bmatrix} 
    \begin{bmatrix}
        C_0 \\
        C_1 \\
        C_2 \\
        C_3
    \end{bmatrix} = 
    \begin{bmatrix}
        0 \\
        0 \\
        0 \\
        0
    \end{bmatrix} .
\end{equation}}
Adding and subtracting the first two and last two equations, respectively, we obtain two uncoupled systems of two equations each.
For each system, the determinant of the coefficient matrix must vanish to admit non-trivial solutions, which gives the so-called \emph{Rayleigh-Lamb} equations, cf., e.g., \cite[p. 310]{Giurgiutiu_2014}:
\begin{equation}\label{eq::rayleigh_lamb_S0}
    \frac{\tan \etap d}{\tan \etas d} = - \left( \frac{(\wavnr_x^2 - \etas^2)^2}{4\wavnr_x^2\etap\etas} \right)^{\pm 1} ,
\end{equation}
where positive and negative exponents correspond to symmetric and anti-symmetric \emph{wave modes}, respectively.
The \emph{Rayleigh-Lamb} equations are transcendental equations with an infinite number of roots, each of which corresponds to a particular wave mode.
We find eigenfunction solutions \emph{symmetric} about the center-plane $S_i$ and denote the \emph{anti-symmetric} ones as $A_i$.
Figure~\ref{fig::lamb_wave_disp_curve}~({a}) illustrates the fundamental symmetric ($S_0$) and anti-symmetric ($A_0$) modes.
Corresponding dispersion curves for the phase velocities are shown in Fig.~\ref{fig::lamb_wave_disp_curve}~({b}) and for the group velocities in Fig.~\ref{fig::lamb_wave_disp_curve}~({c}).
At this point, we note that the mode shapes also depend on the frequency-half-thickness product $ fd $.
For small values of $ fd $, the wave speeds of the $ A_0 $ mode and the flexural plate waves coincide.
Similarly, the wave speeds of axial plate waves and the $ S_0 $ mode are very close to each other for small values of $ fd $.
We also note that the first order modes ($S_1$ and $A_1$) appear only above certain values of $ fd $.

\section{Finite Element Model and Experimental Calibration}
\label{sec::FE_exp_model}
In what follows, we describe the 3D simulation model used in the present assessment and motivate the excitation representing impacts.
Both the structural model and the excitation are derived from experimental setups, against which simulations are calibrated.

\subsection{Free Aluminum Plate with Four Piezoceramic Patches}
\label{ssec:3D_fe_model}
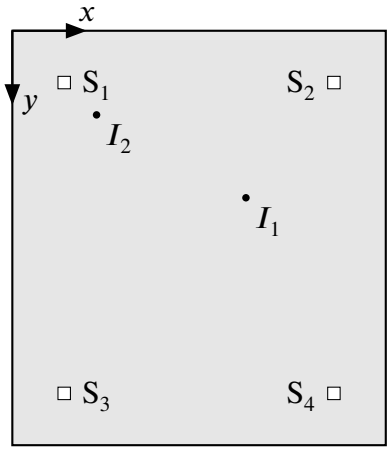
\begin{figure}[pos=!htbp]
    \centering
    \def\plateSensorLayoutWidth{0.30\textwidth}
    \providecommand{\plateSensorLayoutWidth}{0.65\textwidth}

\begin{tikzpicture}[
    x=\dimexpr\plateSensorLayoutWidth/900\relax,
    y=\dimexpr-\plateSensorLayoutWidth/900\relax,
    font=\large
]

\def\PlateW{900}
\def\PlateH{1000}
\def\HalfSensor{15}

\coordinate (S1) at (125,125);
\coordinate (S2) at (775,125);
\coordinate (S3) at (125,875);
\coordinate (S4) at (775,875);

\coordinate (I1) at (563,403);
\coordinate (I2) at (203,203);

\fill[gray!20] (0,0) rectangle (\PlateW,\PlateH);
\draw[thick] (0,0) rectangle (\PlateW,\PlateH);

\draw[-{Triangle[length=2.8mm,width=2.0mm]}, thick] (0,0) -- (180,0) node[above] {$x$};
\draw[-{Triangle[length=2.8mm,width=2.0mm]}, thick] (0,0) -- (0,180) node[right] {$y$};

\foreach \S/\idx/\anchor/\dx/\dy in {
    S1/1/west/22/6,
    S2/2/east/-22/6,
    S3/3/west/22/6,
    S4/4/east/-22/6
}{
    \draw[fill=white]
        ($(\S)+(-\HalfSensor,-\HalfSensor)$)
        rectangle
        ($(\S)+(\HalfSensor,\HalfSensor)$);
    \node[anchor=\anchor] at ($(\S)+(\dx,\dy)$) {$\sensor_{\idx}$};
}

\fill (I1) circle[radius=9];
\node[anchor=north west] at ($(I1)+(0,0)$) {$I_1$};

\fill (I2) circle[radius=9];
\node[anchor=north west] at ($(I2)+(0,0)$) {$I_2$};

\end{tikzpicture}
    \caption{Positions of sensors ($\sensor_i;~i=1\mhyphen4$) and impacts ($\imp{i};~i=1,2$) on aluminum plate.}
    \label{fig::3d_al_plate_sketch}
\end{figure}

We study a rectangular aluminum panel ($L_x \times L_y \times h = \qtyproduct{900 x 1000 x 2}{\milli\meter}$), which is equipped with four piezoelectric patches to detect waves induced by impacts, see Fig.~\ref{fig::3d_al_plate_sketch}.
The boundaries of the panel are assumed as free. 
We use piezoelectric transducers by PI Ceramic~\cite{pi_ceramic} made from PIC255 material, for which the parameters are listed in Appendix \ref{app:a}.
The positions $ \xb_{s_i}$ of sensors $ \sensor_i $, $i=1,\ldots,4$, i.e., the respective centers of the patches, and locations  of the impacts $ \xb_{\imp{i}}$, ${i=1,2}$ are given in Tab. \ref{tab::3d_fe_model_sens_imp_pos}.
\begin{table}[pos=!htbp]
    \centering
    \small
    \caption{Positions of sensors and impacts, and their distances from each other in the 3D FE simulation model of the aluminum plate.}
    \label{tab::3d_fe_model_sens_imp_pos}
    \begin{tabular}{cccc}
        \toprule
         & $ \left( x, y \right) / \si{\milli\meter}$   & $ \Vert \xb_{s_i} - \xb_{\imp{1}} \Vert / \si{\milli\meter} $ & $ \Vert \xb_{s_i} - \xb_{\imp{2}} \Vert / \si{\milli\meter} $ \\
        \midrule
        $ \sensor_1 $     & $ (125, 125) $ & $ 518.8 $ & $ 110.3 $\\
        $ \sensor_2 $     & $ (775, 125) $ & $ 349.6 $ & $ 577.3 $\\
        $ \sensor_3 $     & $ (125, 875) $ & $ 643.9 $ & $ 676.5 $\\
        $ \sensor_4 $     & $ (775, 875) $ & $ 517.4 $ & $ 882.5 $\\
        $ \imp{1} $        & $ (563, 403) $ &     --     &    --   \\
        $ \imp{2} $        & $ (203, 203) $ &     --     &    --   \\      
        \bottomrule            
    \end{tabular}
\end{table}
In our investigations, two different impact locations $ \imp{1} $, $ \imp{2} $ are studied.
While the first is near the center of the plate, the second impact occurs comparatively close to sensor $ \sensor_1 $.

The 3D FE model of the aluminum plate is implemented in the commercial FE software \ansys{} 2022 R2 \cite{ansys_help}.
Both the aluminium plate and the sensors are discretized by means of second-order hexahedral elements (\num{20} nodes). 
For the plate, conventional SOLID186 elements are used; the piezoelectric patches (\qtyproduct{30 x 30 x 0.5}{\milli\meter}) are discretized by means of SOLID226 elements, in which the electric potential serves as additional nodal degree of freedom to account for the electro-mechanical coupling.
Equipotential conditions are imposed on the patches, which are the voltages we measure.
Perfect bonding is assumed; meshes of the plate and the sensors are conforming.
Convergence was analyzed by means of the time-history and the frequency-content (up to \SI{200}{\kilo\hertz}) of the voltage signals of the piezoelectric patches.
In-plane, a global element size of $\SI{2}{\milli\metre}$, which is reduced to $\SI{1.5}{\milli\metre}$ at the patches and the impacted area, has proven sufficient.
To accurately capture flexural deformation induced by anti-symmetric wave modes, both the plate and the patches are discretized using three layers over the thickness; in total, the model has \num{3.6e6} DOF.
The patches are polarized in positive $z$-direction.
The impact is realized as a uniformly distributed surface load in positive $z$-direction over an area of \qtyproduct{6 x 6}{\milli\meter}.
The actual force signals are determined experimentally as described in Sec.~\ref{ssec::exp_setup}.
For the transient analysis of the wave propagation, the \emph{trapezoidal rule} is used, which is a second-order accurate, implicit special case of the family of Newmark-Beta methods~\cite{newmark1959method}, where the Newmark coefficients are set to $\beta=1/4$ and $\gamma=1/2$, respectively.
For the time-discretization, a constant time-step of $\dt = \SI{0.2}{\micro\second}$ is used.

\subsection{Experimental Setup and Measurements}\label{ssec::exp_setup}
In our assessment of \ToA{}-estimation methods, we primarily rely on numerical analysis to generate representative sensor signals, not least for the sake of reproducibility.
Nonetheless, we aim at being as close as possible to real-world problems.
For this reason, we compare and calibrate our numerical model against a laboratory setup.
In particular, we focus on the following two aspects: Firstly, the dispersion characteristics of the numerical model should accurately match those of the actual aluminium plate. 
Secondly, representative impact loads, which are imposed in the FE model, are to be derived from experiments. 

To determine the dispersion characteristics of the plate, transient waves need to be measured as they propagate through the structure, which, undoubtedly, is a challenging problem.
Capturing wave patterns requires both high temporal and spatial resolutions.
To meet these requirements, we use Laser-Doppler vibrometry (Polytec PSV-400 3D M, see Appendix \ref{app:b}) in our experimental setup.
A common way to obtain the dispersion relation of Lamb waves, i.e., $\omega(k)$, see Eq.~\eqref{eq:dispersion_relation}, is to measure the out-of-plane velocity along a line, i.e., to perform a so-called \emph{B-scan} \cite{Hellier_2001}.
To excite vibrations, a piezoelectric transducer (PI Ceramic P876.SP1~\cite{pi_ceramic}) is applied at the center of the plate using X60 adhesive by HBM \cite{hbm}.
A $ \SI{50}{\micro\second} $ rectangular (rect) pulse created by an external function generator is used as excitation.
The transducer is driven by an amplifier (LE 200/070 EBW by Piezomechanik GmbH \cite{piezomech}).
The actual voltage applied to the transducer and its frequency-spectrum are shown in Figs.~\ref{fig::al_plate_exp_piezo_rect_signal} (a) and (b).
The distortion of the rectangular pulse is due to the characteristics of the amplifier.
The experimental setup for the B-scan is schematically shown in Fig. \ref{fig::al_plate_exp_piezo_rect_signal} (c).
The first measurement point is located at a distance of $ \SI{50}{\milli\metre} $ from the excitation to minimize the influence of the piezoelectric transducer on the dispersion curves.
The spatial resolution along the line, i.e., the spacing of the measurement points, is approximately $ \SI{676.5}{\micro\metre} $.
The scanning head is placed at a distance of $ \SI{507}{\milli\meter} $ from the plate and the velocity decoder with the highest measurement resolution is used (VD-09 5 mm/s/V).
Applying a two-dimensional (2D) Fourier Transform (FT) (one in time and one in space) to the B-Scan, yields a frequency-wavenumber representation of the measurement, i.e., the dispersion characteristics of the aluminum plate, see Fig. \ref{fig::al_plate_exp_piezo_0_disp_curve}.

To obtain representative impact loads that are used as excitation in the FE model, we proceed as follows.
In recent works, the impact was conducted by pencil lead break tests \cite{Cheng_2021}, impact hammer \cite{Coverley_2003}, object drop \cite{Merlo_2017} or self-made manufactured impact stands \cite{Tabian_2019b}, to give a few examples.
Our focus lies on impacts a structure may experience during operation, that is why pencil lead break tests are not relevant in the present context.
We want to keep the measurement setup as simple as possible; so, we opted for impact excitation by hitting the structure with an electromagnetic shaker.
For this purpose, a shaker from TIRA Schwingtechnik (model TV $ 50009 $) is used together with the corresponding amplifier (DA 200).
The control of the shaker is performed with the internal function generator of the laser scanning vibrometer.
A $ \SI{100}{\hertz} $ half-sine is used to impact the plate every second.
A force sensor from PCB Electronics (model 208C01) is mounted on the shaker tip to record the force during the impact.
The shaker is placed in such a way that just before impact, the control of the shaker is terminated and the shaker tip hits the plate only with its own inertia.
More details of the measurement setup for impacting the plate with the shaker are given in Appendix \ref{app:b}.

\subsection{Calibration of the 3D FE Simulation Model}
\begin{figure}[pos=!htbp]
    \centering
    \begin{tikzpicture}				
        \node[anchor=south west,inner sep=0] (image1) at (0, 0) {\includegraphics[width=0.7\textwidth, trim=0 0 0 0, clip, ]{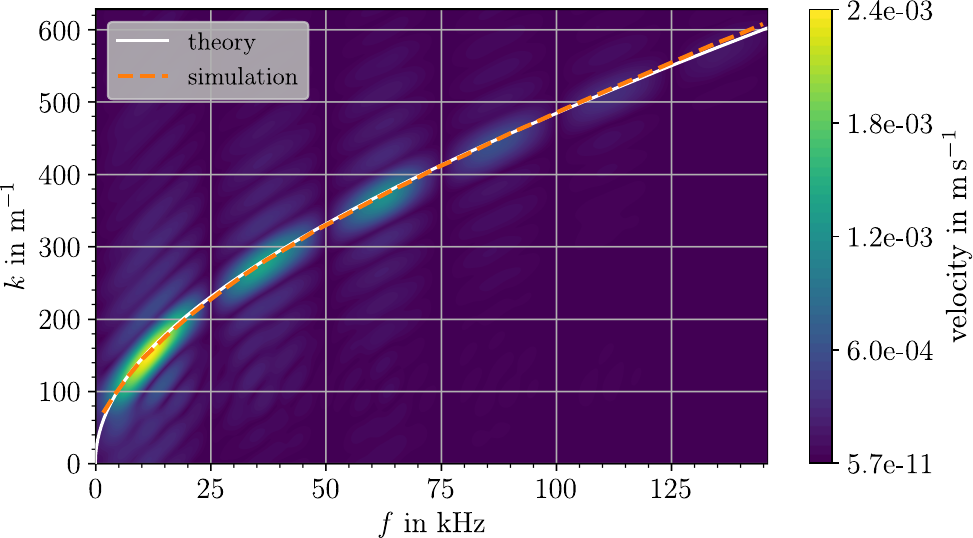}};        
        \begin{scope}[x={(image1.south east)},y={(image1.north west)}]
        \end{scope} 
    \end{tikzpicture}
    \caption{
    Dispersion characteristics of the aluminum plate: The contour plot shows the experimental results, the orange dashed line shows the results obtained from the simulation (line of local maxima in the wavenumber-domain as a function of the frequency) and the solid white line is the dispersion curve of the $ A_0 $ mode from Rayleigh-Lamb equations \eqref{eq::rayleigh_lamb_S0}.}
    \label{fig::al_plate_exp_piezo_0_disp_curve}						
\end{figure}

The FE simulation model is here calibrated with experimental results on the basis of the dispersion characteristics.
The basis of the calibration is the solution of the Rayleigh-Lamb equation for the first antisymmetric mode $ A_0 $.
A complete validation between simulation and measurement is beyond the scope of the present paper.
Indeed, it would require more than matching the dispersion properties, e.g., material damping, wave radiation from plate to ambient air, influence of the support (foam mat), material and geometric imperfections as, e.g., a non-uniform thickness of the plate, to name a few.
The experimental results are illustrated in Fig. \ref{fig::al_plate_exp_piezo_0_disp_curve}, which shows the measured out-of-plane velocity in the wavenumber-frequency domain.
\begin{figure}[pos=!htbp]    \centering
    \begin{tikzpicture}				
        \node[anchor=south west,inner sep=0] (image1) at (0, 0) {\includegraphics[width=0.9\textwidth, trim=0 0 0 0, clip, ]{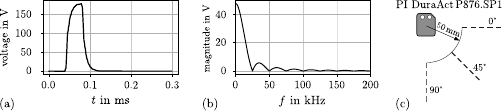}};             
    \end{tikzpicture}
    \caption{Driving voltage of piezo to obtain the dispersion characteristics of the aluminum plate (a) time-history and (b) frequency-spectrum. Subfigure (c) shows schematically how B-scans are performed for different orientations upon excitation with a piezoelectric transducer. }
    \label{fig::al_plate_exp_piezo_rect_signal}						
\end{figure}
Note that the frequencies that are a multiple of \SI{25}{\kilo\hertz} are not excited by the piezoelectric actuator, see Fig.~\ref{fig::al_plate_exp_piezo_rect_signal}.
Local maxima of the simulated results are illustrated by the orange dashed line as a function of the frequency in the wavenumber-domain.
The solid white line indicates the dispersion curve of the $ A_0 $ Lamb wave mode.
Geometry and mass of the aluminum plate translate into a density of $\rho = \SI{2660}{\kg\per\metre\cubed}$.
According to the manufacturer's data sheet, Poisson's ratio is set to a value of $\nu = \SI{0.33}{}$.
By choosing Young's modulus as $E = \SI{69}{\giga\pascal}$, the dispersion properties of the $A_0$ mode in measurements and simulations can be matched.

\begin{figure}[pos=!htbp]
    \centering
    \begin{tikzpicture}
        \node[anchor=south west,inner sep=0] (image1) at (0, 0) {\includegraphics[width=\textwidth, trim=0 0 0 0, clip, ]{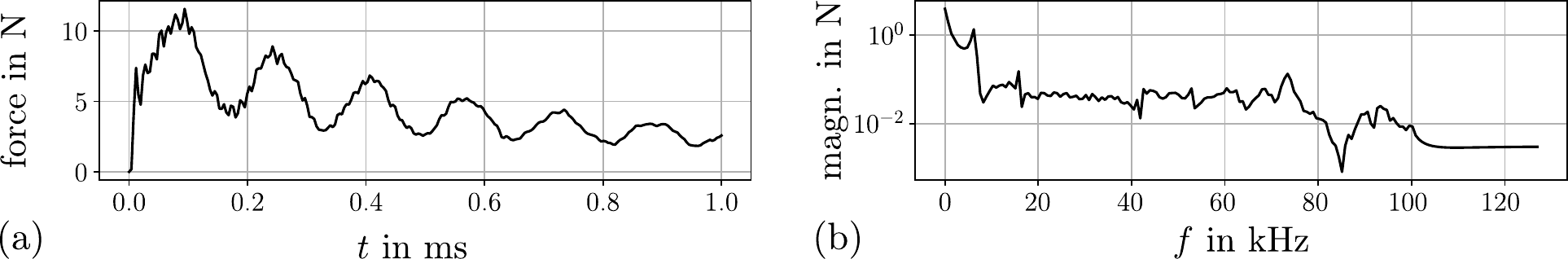}};      
    \end{tikzpicture}
    \caption{Recorded force during an impact on the aluminum plate: (a) time-signal and (b) the corresponding frequency-spectrum.}
    \label{fig::3d_shaker_signal}					
\end{figure}

In what follows, we discuss the properties of the impacts which are applied to the aluminum plate in our experimental investigation.
The impact of the shaker on the plate represents a transient load with certain temporal and spectral properties, see Figs.~\ref{fig::3d_shaker_signal} (a) and (b), respectively.
In Fig.~\ref{fig::3d_shaker_signal} (a), a steep increase of the force can be observed in the first $ \SI{20}{\micro\second} $, followed by comparatively low-frequency oscillations (superimposed by high-frequency components) for the rest of the time-domain considered.
The spectrum is relatively broadband and has frequency-components up to $ \SI{100}{\kilo\hertz} $.
The low-frequency oscillations mentioned above occur at a frequency of $ \SI{6}{\kilo\hertz} $, and their origin is not immediately clear.
To gain further insight, we determined how long the tip of the shaker is in contact with the plate by attaching a switch that closes an electric circuit to the shaker.
As a matter of fact, we found that the shaker and plate stay in contact for the entire period considered, and the dynamics of the shaker affects our force.
As similar imperfections can be expected to occur also in real-world impacts, we deliberately decided to still use this particular type of impact in our investigations. We refer to it as \emph{experiment-based impact} and corresponding results are indicated by a superscript ``~$\iop$~''.

\begin{figure}[pos=!htbp]
    \centering
    \begin{tikzpicture}
        \node[anchor=south west,inner sep=0] (image1) at (0, 0) {\includegraphics[width=\textwidth, trim=0 0 0 0, clip, ]{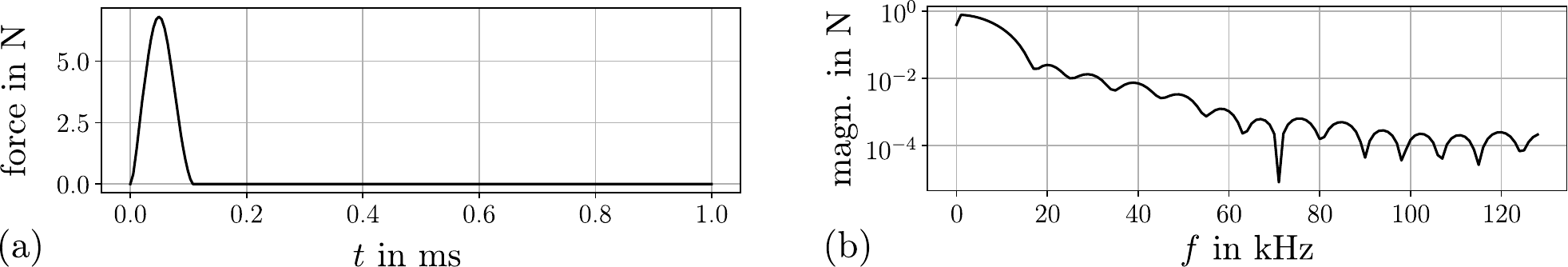}};      
    \end{tikzpicture}
    \caption{Recorded force during an impact on a massive aluminum block: (a) time-signal and (b) the corresponding frequency-spectrum.}
    \label{fig::3d_hammer_signal}					
\end{figure}

As a second impact, we chose to generate an idealized force signal by impacting a massive aluminum block with an impact hammer.
In this case, the time-signal of the force sensor truly resembles an impulse-like excitation, see Fig.~\ref{fig::3d_hammer_signal} (a).
Due to the stiffness of the impacted block, the time-of-contact with the hammer is much shorter as compared to the experiment-based impact on the thin plate.
The corresponding frequency-spectrum is shown in Fig. \ref{fig::3d_hammer_signal} (b), where, similarly to the impact on the thin plate, the maximum occurs at a low frequency and the magnitude decreases with the increase of frequency.
For the remainder of this work, we refer to the impact visualized in Fig. \ref{fig::3d_hammer_signal} as \emph{idealized impact}; quantities referring to the idealized impact are indicated by a superscript ``~$\iob$~''.
Both types of impacts, i.e., the idealized impact and the experiment-based impact, are realized in our 3D FE model of the plate.
The impact-induced energy can be calculated by determining the mechanical power on the structure through FE simulation. 
To achieve this, the velocities of the nodes in the impact area are multiplied by the applied force and then summed over the respective area.
Integrating the mechanical power over the entire time-interval considered provides the mechanical work done on the structure.
Specifically, for the idealized impact, the calculated energy is $\SI{0.015}{\milli\joule}$, while for the experiment-based impact, it amounts to $\SI{0.203}{\milli\joule}$.
These energies are insufficient to damage the plate.
It is important to clarify that in this context, an impact is not synonymous with damage.
Unlike the common understanding in many SHM applications, where an impact signifies structural damage or the formation of a crack, our definition refers to an impact as an event, when something hits the structure.
This can be debris, a dropped tool, or manually applied force (such as the idealized and experiment-based impacts in this paper).

\section{Time-of-Arrival Estimation Methods}\label{sec::toa}
Before we introduce different methods to determine the \ToA{}, we need to address the notion of \ToA{} in the present problem:
Clearly, an impact constitutes a relatively broadband excitation of the structure, which distinguishes the present problem from SHM-applications in literature, in which the excitation is not only ``well-defined'', but typically also narrow-banded \cite{Giurgiutiu_2014}.
For thin-walled structures, in which Lamb waves develop and propagate, we also face the challenge of a ``response'' spectrum that differs considerably from the spectrum of the impact load, cf. spectra in Figs. \ref{fig::3d_shaker_signal} and \ref{fig::3d_hammer_signal}.
The frequency-range of waves excited by impacts plays a pivotal role in view of the dispersive nature of waves in 3D continua: Lamb waves with different frequencies propagate at different speeds.
The immediate implications on \ToA{}-estimation may seem rather unspectacular at first glance: 
There is no such thing as a single \ToA{}, much rather we need to be specific about particular frequencies for \ToA{}-estimation.
In other words, waves of different frequencies do not arrive simultaneously, but at different times.
Note that we are not interested in \ToA{} \emph{per se}. 
We want to apply \ToA{}-estimation methods as essential means in impact detection, localization and, eventually, quantification. 
In this context, the dispersive nature of Lamb waves may intuitively appear as additional complication. 
If the dispersion properties of a structure are known, however, we may be able to exploit that knowledge for our purposes.

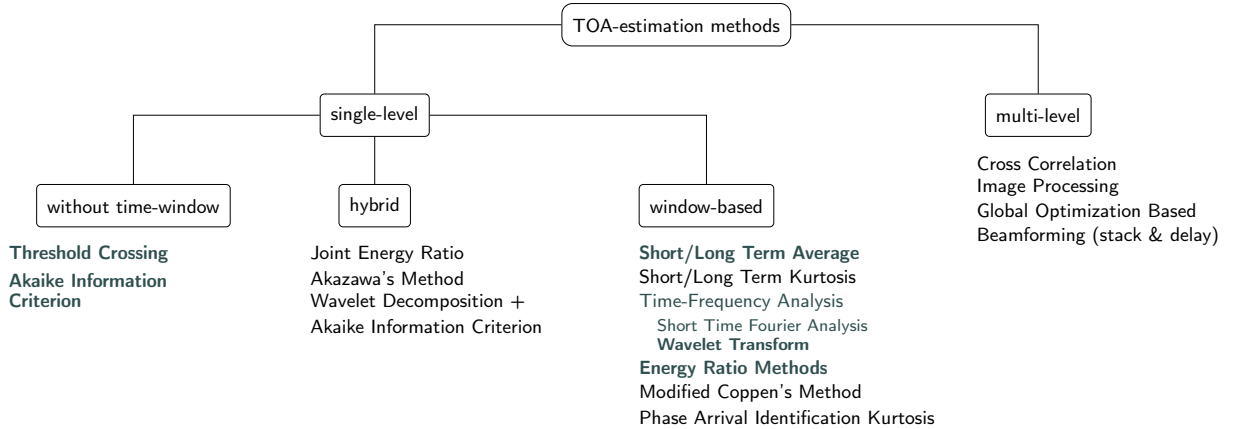
\begin{figure}[pos=!htbp]
    \centering
    \resizebox{\textwidth}{!}{
    \begin{tikzpicture}				
        \node [rectangle, solid, draw=black, inner sep=5pt, minimum height=20pt,rounded corners=5] (box1) at (0, 0) {\ToA{}-estimation methods};
        \node [rectangle, solid, draw=black, inner sep=5pt, minimum height=20pt,rounded corners=2] (box2) at (-5, -1.5){single-level};
        \node [rectangle, solid, draw=black, inner sep=5pt, minimum height=20pt,rounded corners=2] (box21) at (-9, -3){without time-window};
        \node [anchor=west, inner sep=5pt] (box211)  at  (-11.25, -3.8){\small \textbf{\hyperref[ssec::tc]{Threshold Crossing}}};
        \node [anchor=west, inner sep=5pt] (box211)  at  (-11.25, -4.4){\small \textbf{\hyperref[subsec::aic]{\shortstack[l]{Akaike Information\\Criterion}}}};
        \node [rectangle, solid, draw=black, inner sep=5pt, minimum height=20pt,rounded corners=2] (box22) at (-5, -3){hybrid};
        \node [anchor=west, inner sep=5pt] (box221)  at  (-6.25, -3.8) {\small Joint Energy Ratio};
        \node [anchor=west, inner sep=5pt] (box222)  at  (-6.25, -4.2) {\small Akazawa's Method};
        \node [anchor=west, inner sep=5pt] (box223)  at  (-6.25, -4.6) {\small Wavelet Decomposition +};
        \node [anchor=west, inner sep=5pt] (box224)  at  (-6.25, -5.0) {\small Akaike Information Criterion};
        \node [rectangle, solid, draw=black, inner sep=5pt, minimum height=20pt, rounded corners=2] (box23) at (0.5, -3){window-based};
        \node [anchor=west, inner sep=5pt] (box231)  at  (-0.80, -3.8) {\small\textbf{\hyperref[ssec::1D_SLA]{Short/Long Term Average}}};
        \node [anchor=west, inner sep=5pt] (box232)  at  (-0.80, -4.2) {\small Short/Long Term Kurtosis};
        \node [anchor=west, inner sep=5pt] (box233)  at  (-0.80, -4.6) {\small \hyperref[sec::TFA]{Time-Frequency Analysis}};
        \node [anchor=west, inner sep=5pt] (box2331) at  (-0.50, -5.0) {\footnotesize \hyperref[ssec::stft]{Short Time Fourier Analysis}};
        \node [anchor=west, inner sep=5pt] (box2332) at  (-0.50, -5.3) {\footnotesize \textbf{\hyperref[ssec::cwt]{Wavelet Transform}}};
        \node [anchor=west, inner sep=5pt] (box234)  at  (-0.80, -5.7) {\small \textbf{\hyperref[ssec::1D_MER]{Energy Ratio Methods}}};
        \node [anchor=west, inner sep=5pt] (box235)  at  (-0.80, -6.1) {\small Modified Coppen's Method};
        \node [anchor=west, inner sep=5pt] (box236)  at  (-0.80, -6.5) {\small Phase Arrival Identification Kurtosis};
        \node [rectangle, solid, draw=black, inner sep=5pt, minimum height=20pt,rounded corners=2] (box3) at (6, -1.5){multi-level};
        \node [anchor=west, inner sep=5pt] (box31)   at  (4.80, -2.3)  {\small Cross Correlation};
        \node [anchor=west, inner sep=5pt] (box32)   at  (4.80, -2.7)  {\small Image Processing};
        \node [anchor=west, inner sep=5pt] (box33)   at  (4.80, -3.1)  {\small Global Optimization Based};
        \node [anchor=west, inner sep=5pt] (box34)   at  (4.80, -3.5)  {\small Beamforming (stack \& delay)};
        \draw (box1.west) -| (box2.north);
        \draw (box1.east) -| (box3.north);
        \draw (box2.west) -| (box21.north);
        \draw (box2.south) -| (box22.north);
        \draw (box2.east) -| (box23.north);
    \end{tikzpicture}
    }
    \caption{Schematic overview of selected \ToA{}-estimation methods based on \cite{Akram_2016}. The methods in bold are investigated hereafter.}
    \label{fig::overview_TOA_methods_tikz}			
\end{figure}

In the present section, we introduce several methods for \ToA{}-estimation, which can be categorized from different points of view.
Depending on the number of signals that are being simultaneously used to make a single estimate, we distinguish \emph{single-level} and \emph{multi-level} methods \cite{Akram_2016}.
Single-level methods use the signal of a single sensor at some specific positions.
Multi-level algorithms, on the other hand, rely on signals from several sensors located at different positions.
Methods that apply different kinds of time-windows to sensor signals are referred to as \emph{window-based}, see, e.g., \cite{Akram_2016}.
We can further distinguish methods solely operating in the \emph{time-domain} from so-called \emph{time-frequency analysis} (TFA) methods, which also include the frequency-domain of sensor signals.
Information on a signal's frequency-content allows us to exploit a-priori knowledge of dispersion properties for \ToA{}-estimation.
A large variety of \emph{hybrid methods} that combine individual aspects of different \ToA{}-estimation methods have been proposed \cite{Akram_2016}.
Figure \ref{fig::overview_TOA_methods_tikz} is meant to give an overview of various techniques for estimating the \ToA{} and to highlight those retained here.
\begin{Remark}
    Some of the methods assessed in what follows originate in the field of seismology, which is characterized not only by a different frequency-range, but also by different `modes' of wave propagation.
    In seismology, longitudinal $P$-waves and transverse $S$-waves propagate in the interior of a material body, e.g., soil.
    During propagation, these waves are reflected at interfaces, where mode conversion occurs, leading to interference between the waves. 
    The interference generates waves that are guided along the surface, known as  \emph{Rayleigh} and \emph{Love} waves \cite[p. 215]{Shearer_2009}.
    Typical seismographs first show a sensor output due to the arrival of $P$- and $S$-waves, followed by a response related to surface waves.
    Hence, many \ToA-methods in seismic problems focus on the initial arrival of $P$- or $S$-waves \cite{Akram_2016}.
    As opposed to seismic problems, waves in thin-walled structures are characterized by the fact that, though impacts initially trigger longitudinal and transverse waves, these are almost instantaneously converted into Lamb waves due to the close distance of top and bottom faces of structures.
    In seismology, signal and noise differ in frequency-contents, which allows effective filtering.
    Problems of guided waves in plates, however, are characterized by signal and noise sharing the same frequency-range, see, e.g., \cite{Akram_2016}.
\end{Remark}    

In subsequent sections, we describe several methods to estimate the \ToA{} of waves in thin-walled structures from sensor signals.
In our description, we make repeated use of fundamental concepts of signal processing, which are to be briefly outlined to begin with.  
Starting with notation, time-continuous sensor signals are denoted by $ s(t) $ throughout this work.
Practically, sensors signals are recorded over finite time-intervals, which we assume as $[0, T]$, i.e., $T$ denotes both the end-time and the duration of the signal. 
Irrespective of whether physical measurements or numerical simulations are performed, we generally deal with time-discrete signals, i.e., sequences of data $\{ s[i] = s(t_i) \}$ obtained by sampling time-continous sensor signals at times $t_i \in \mathcal T = \{ t_i, i = 0, \ldots, N-1 \}$, where $N$ denotes the number of samples.
Throughout our assessment, we assume a constant sampling rate, i.e., time-increments have a uniform length of $\Delta t = T / (N-1)$ and the $i$-th sample corresponds to the time $t_i = i \Delta t$.
In signal theory, the \emph{energy} of a general complex-valued signal is defined as
\begin{equation}\label{eq::energy_of_signal}
    \energy = \int_{-\infty}^{\infty} |s(t)|^2 dt , \qquad
    \energy = \sum_{i = -\infty}^\infty |s[i]|^2 ,
\end{equation}
see, e.g., \cite{Cohen_1995}.
For finite time-intervals, we correspondingly have
\begin{equation}\label{eq::energy_of_signal_fin_t}
    \energy = \int_0^{T} |s(t)|^2 dt , \qquad
    \energy = \sum_{i = 0}^{N-1} |s[i]|^2 .
\end{equation}
The integrand $ |s(t)|^2 $ accordingly represents a signal's (instantaneous) \emph{power}.
Taking absolute values can be omitted for real-valued signals.
\emph{Variance} is another important property of a signal, which, in case of uniformly distributed samples, is defined by \cite{Billingsley_1995}
\begin{equation}
    \label{eq::definition_variance}
    \var\left\{ s(t) \right\} = \frac{1}{T}\int_0^{T}\left( s\left(t\right) - s_\mu \right)^2 dt, \qquad
    \var\left\{ s \right\} = \frac{1}{N} \sum_{i=0}^{N-1}\left( s[i] - s_\mu \right)^2,
\end{equation}
where $ s_\mu $ is the signal's average value, i.e., its arithmetic mean:
\begin{equation}\label{eq::definition_mean}
    s_\mu = \frac{1}{T} \int_0^T s(t) dt, \qquad
    s_\mu = \frac{1}{N} \sum_{i=0}^{N-1} s[i].
\end{equation}
Both measured and synthetic sensor signals are generally subject to noise. 
The former is usually affected by, e.g., quantization noise while the latter is, in most cases, only subjected to computational noise \cite{More_2011}.
The \emph{signal-to-noise ratio} (SNR), which is defined as the ratio between the power of the signal and the power of noise, serves as a measure of the intensity of the noise $ \mathnoise $ \cite{Bennett_1948}:
\begin{equation}\label{eq::snr}
    \snr = \frac{\rms\left\{s(t)\right\}}{\mathrm{rms}\left\{\mathnoise\right\}}, 
\end{equation}
where $ \rms\{\cdot\} $ denotes to root mean square value and is defined as
\begin{equation}\label{eq::definition_rms}
    \rms\left\{ s(t) \right\} = \sqrt{\frac{1}{T} \int_{0}^{T} s(t)^2 dt} , \qquad 
    \rms\left\{ s \right\} = \sqrt{\frac{1}{N} \sum_{i=0}^{N-1} s[i]^2}.
\end{equation}
In the field of signal processing, the \emph{Fourier transform} (FT) of signals  plays an unparalleled role in frequency-domain analysis.
The FT of the time-continuous signal $s(t)$ is an integral transform, for which we adopt a functional notation $\fourier{s}$, defined by
\begin{equation}
    \fourier{s} = \frac{1}{\sqrt{2\pi}}\int_{-\infty}^\infty s(t) e^{-j \omega t} dt ,
    \label{eq::fourier}
\end{equation}
see, e.g., \cite[p. 6]{Cohen_1995}.

\subsection{Threshold-Crossing in Time-Domain}\label{ssec::tc}
TC is the most intuitive method for determining the \ToA{} of waves excited by an impact.
The \ToA{} is \emph{defined} by the time at which a sensor's signal level exceeds some pre-defined threshold value $\sth$, 
\begin{equation}\label{eq::def_threshold}
    \test = \min_{t_i \in \mathcal T} \{t_i \, : \, \vert s(t_i) \vert > \sth \} ,
\end{equation}
see, e.g., \cite{Kundu_2009, Sanchez_2016, SharifKhodaei_2012b, Mallardo_2013}.
Clearly, the proper choice of the threshold value is pivotal when applying TC.
Two fundamental considerations constrain the admissible range for threshold value.
First of all, the value needs to be chosen sufficiently greater than the noise level of the sensor signal.
Secondly, the threshold value needs to be small enough to indicate ``relevant'' impacts.
In the context of SHM, impacts are often characterized by the kinetic energy of the object that impacts the structure of interest.
The respective application also plays a crucial role when choosing threshold values.
It makes a difference whether impacts are only to be detected, or whether the \ToA{} of waves excited by the impact is to be estimated.
For impact detection, threshold values are less critical as compared to \ToA{}-estimation (and subsequent impact localization), in which thresholds need to be adapted to amplitudes of sensor signals related to particular wave modes.
In any case, thresholds are typically chosen based on heuristic considerations, in which problem-specific experience plays an import role.

\subsection{Time-Frequency Analysis}
\label{sec::TFA}
The fact that wave speed in dispersive media is frequency-dependent allows us to infer information on the distance traveled from the (temporal) evolution of a sensor signal's frequency-spectrum.  
For this reason, we can use TFA methods, which describe the evolution of the spectral content, in \ToA{}-estimation, see, e.g., \cite[p. 70]{Cohen_1995}.
In the following two sub-sections, we discuss two of the most important methods in TFA, i.e., \emph{short-time Fourier transform} (STFT) and CWT.
In both methods, a trade-off between resolution in the time-domain and resolution in the frequency-domain needs to be made, see, e.g., \cite{Cohen_1995} and \cite{Mallat_2009}.
Beyond \ToA{}-estimation, recent approaches have also addressed the separation of overlapping guided-wave modes and boundary reflections, highlighting the broader relevance of TFA for interpreting multimodal Lamb-wave signals \cite{Jana_2026}.

\subsubsection{Short-Time Fourier Transform}
\label{ssec::stft} 
To determine how a signal's frequency-spectrum evolves over time, i.e., to localize spectral components in the time-domain, FT is applied to a series of consecutive, comparatively short time-intervals taken from the signal.
For this purpose, the signal $s(t)$ is multiplied by a (fixed-length) window function $g(t)$, which is shifted in time to extract individual time-intervals of the signal. %
Formally, STFT is defined as the FT of the product of the signal and the window function:
\begin{equation}\label{eq::STFT}
    \stft{s}{\omega}{b} = \dfrac{1}{\sqrt{2\pi}}\int_{-\infty}^{\infty} s(t)g(t-b)\epow^{-j\omega t}dt, \qquad
    \stft{s}{\omega}{b} = \sum_{k = -\infty}^\infty s_k g_{k-b} e^{-j \omega k} ,
\end{equation}
where  $ b $ denotes the (temporal) center position of the time-window, see, e.g., \cite{Gaul_1998}.
A large variety of window functions is being used in the context of TFA.
Exemplarily, we mention \emph{rectangular}, \emph{Hamming}, \emph{Gaussian}, \emph{Hanning} and \emph{Blackmann} windows. 
The result of the FT depends on the properties of the window function in both time and frequency domains. 
We refer to \cite[pp. 98-101]{Mallat_2009} for a discussion of advantages and disadvantages of the respective window functions. 
Temporal and frequency resolutions are subject to Gabor's limit, which is regarded as an uncertainty principle, see, e.g., \cite{Gabor_1946}, i.e., the resolutions are inversely proportional to each other.
Wide windows allow us to accurately resolve frequencies, whereas narrow windows allow for a good temporal resolution. 

\subsubsection{Continuous Wavelet Transform}
\label{ssec::cwt}
The main limitation of STFT is the fact that, by selecting a time-window function with certain spectrum and temporal support, the resolutions in the time-domain and the frequency-domain are determined for the entire time-frequency window of interest.
As a matter of fact, we typically do not require the same resolutions throughout the time-frequency window.
This idea is reflected in CWT, which, as opposed to STFT, does not rely on a single window-like function that is shifted along the time-axis, but the signal $s(t)$ is projected onto a family or ``dictionary''~\cite{Mallat_2009} of wavelets.

From a general perspective, the eponymous \emph{wavelet functions} can be described as wave-like oscillations with a finite support (in time).
A family of wavelets is obtained by shifting and dilating a generating function $\psi(t)$, i.e., the so-called \emph{mother wavelet}.
Let $a > 0$ denote the scaling parameter that governs the dilation of the mother wavelet and $b$ its shift in time, the family of wavelets derived from the mother wavelet $\psi(t)$ is given by
\begin{equation}\label{eq::cwt_shifted_scaled_wavelet}
    \psi_{a,b}(t) = \dfrac{1}{\sqrt{a}}\psi\left(\dfrac{t-b}{a}\right) ,
\end{equation}
see, e.g., \cite[p. 102]{Mallat_2009}.
In the context of CWT, functions with zero mean and unit square norm are used as mother wavelets,
\begin{equation} \label{eq:cwt_zero_mean_unit_norm}
    \int_{-\infty}^{\infty} \psi(t) dt = 0 ,
    \qquad
    \int_{-\infty}^{\infty} \left\vert \psi(t) \right\vert^2 dt = 1 , 
\end{equation} 
which imply that $\psi(t)$ must be both integrable and square integrable.
For an inverse transform to exist, wavelets used in CWT need to further satisfy the so-called admissibility condition, i.e., 
\begin{equation}
    \label{eq::cwt_admiss_cond}
    \int_{-\infty}^{\infty}\dfrac{|\Psi(\omega)|^2}{\omega}d\omega < \infty,
\end{equation}
where $ \Psi(\omega) = \fourier{\psi}$ is the Fourier transform of $\psi(t)$, see, e.g., \cite[p. 825]{Giurgiutiu_2014}.
The CWT of a signal $ s(t) $ is defined as the scalar product (projection)
\begin{equation}
    \label{eq::CWT}
    \cwt{s}{a}{b} = \int_{-\infty}^{\infty} \psi_{a,b}^* s(t) dt, 
\end{equation}
where $ \psi_{a,b}^* $ denotes the complex conjugate of the wavelet $\psi_{a,b}$.\footnote{
    For the definition and implementation of CWT applied to time-discrete signals we refer to, e.g., \cite{Arts_2022}.
}
The set of wavelet coefficients $\cwt{s}{a}{b}$ constitutes a highly redundant time-frequency representation (``image'') of the signal $s(t)$ \cite[p. 17]{Mallat_2009}.
The shift $b$ defines the position of the wavelet $\psi_{a,b}$ in the time-domain.
Assuming that $\psi$ is centered at $t=0$, this means that $\psi_{a,b}$ is centered at $t=b$. 
The scaling factor $a$, on the other hand, defines the position of the wavelet $\psi_{a,b}$ 
in the frequency-domain.
Small time-scales $a$ translate into a compression of the mother wavelet in the time-domain, which emphasizes the high-frequency content of $s(t)$ in the projection, see Eq.~\eqref{eq::CWT}, whereas large scales stretch the mother wavelet such that the low frequency-content of $s(t)$ contributes most to the projection. 
Let $\omega_c$ denote the center (angular) frequency of the mother wavelet $\psi(t)$, which---making use of the fact that we require wavelets to have a unit square norm, cf. Eq.~\eqref{eq:cwt_zero_mean_unit_norm}---is given by 
\begin{equation}
    \omega_c = \frac{\int_{-\infty}^\infty \omega \left\vert \Psi (\omega) \right\vert^2 d\omega}{\int_{-\infty}^\infty \left\vert \Psi (\omega) \right\vert^2 d\omega} 
    = \int_{-\infty}^\infty \omega \left\vert \Psi (\omega) \right\vert^2 d\omega ,
\end{equation}
see, e.g., \cite{Mallat_2009}.
The center angular frequency of the dilated wavelet $\psi_{a,b}$ follows as 
\begin{equation}\label{eq::scale_to_frequency}
    \omega_{c,a} = \frac{\omega_c}{a} , 
\end{equation}
and is also referred to as \emph{pseudo}- or \emph{instantaneous} (angular) frequency see, e.g., \cite{Rioul_1991} and \cite[p. 116]{Mallat_2009}. 

\begin{figure}[pos=!htbp]
    \centering
    \begin{tikzpicture}				
        \node[anchor=south west,inner sep=0] (image) at (0, 0) {\includegraphics[width=0.8\textwidth, trim=0 0 0 0, clip, ]{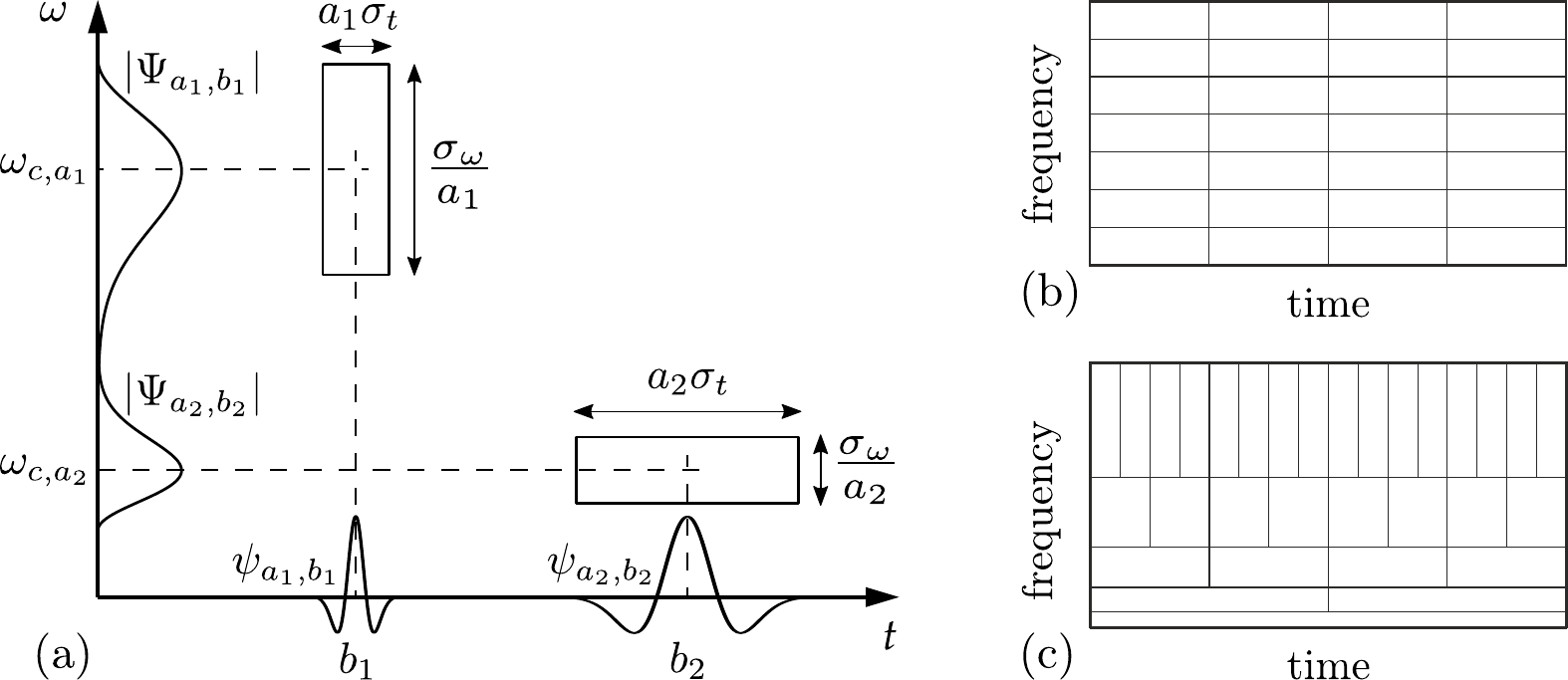}};   
        \begin{scope}[x={(image.south east)},y={(image.north west)}]				
        \end{scope}     																
    \end{tikzpicture}
    \caption{Characteristics of resolution in time and frequency domains: (a) Time-frequency boxes of two wavelets $ \psi_{a_1,b_1} $ and $ \psi_{a_2,b_2} $, adapted from \cite[p. 18]{Mallat_2009}. 
    Sub-figures (b) and (c) show the tiling of the time-frequency plane in time-frequency boxes when applying TFA on a signal. 
    In case of STFT (b) the dimensions of boxes are constant, which correspond to a constant time-frequency resolution in the entire time-frequency plane.
    The CWT (c) divides the time and frequency axes in separate intervals of varying sizes, which gives CWT the property of multi-resolution.}
    \label{fig::cwt_time_frequency_boxes}			
\end{figure}

Owing to the scaling of wavelets, CWT is a \emph{multi-resolution} method, i.e., the resolution can be varied in time and frequency. 
A comparison to STFT reveals what is meant by multi-resolution property.
For this purpose, let $ \sigma_t $ denote the support of the mother wavelet $ \psi(t) $ in the time-domain and $ \sigma_\omega $ the support of its Fourier transform $\Psi(\omega)$ in the frequency-domain, see, e.g., \cite{Rioul_1991, Cohen_1995, Mallat_2009} for their formal definitions. 
We consider two wavelets $ \psi_{a_i,b_i} $, $i=1,2$ for which $a_2 > a_1$ and $b_2 > b_1$ hold, see Fig. \ref{fig::cwt_time_frequency_boxes} (a).
The wavelets are centered at $ b_i $ in the time-domain and their temporal supports are related to the support of the mother wavelet by $\sigma_t a_i $, $i=1,2$.
The greater dilation $ a_2 > a_1 $, results in a greater temporal support of $ \psi_{a_2,b_2} $ as compared to $ \psi_{a_1,b_1} $.
The support in the frequency-domain, however, is inversely proportional to the dilation, i.e., $\sigma_\omega / a_i$, $i = 1,2$.
As $ a_2 > a_1 $, the support of the Fourier transform $\Psi_{a_2, b_2}$ in the frequency-domain is compressed (narrower) as compared to $\Psi_{a_1, b_1}$.
From Eq.~\eqref{eq::scale_to_frequency}, we find $\omega_{c,a_2} < \omega_{c,a_1}$, i.e., the spectrum of $\Psi_{a_2, b_2}$ is shifted to lower frequencies as compared to $\Psi_{a_1, b_1}$.
Hence, in the case of $ a_1<a_2 $, the temporal resolution increases (reduced time-support) and the frequency-resolution decreases (dilated frequency-support), as illustrated in Fig. \ref{fig::cwt_time_frequency_boxes} (a).
This property of time-frequency resolution can be illustrated as a box in the time-frequency plane, where smaller box widths correspond to a higher resolution in the time-domain and smaller box heights represent a higher frequency-resolution.
Regardless of how the mother wavelet is scaled, the product of the supports in the time and frequency-domains remains constant, i.e., the area of the boxes in Fig.~\ref{fig::cwt_time_frequency_boxes} is preserved.
In fact, the product of the supports in the time and frequency-domains, i.e., $\sigma_t~\sigma_\omega$, cannot be less than $1/2$, see, e.g., \cite[p. 44]{Mallat_2009}. 
This is a consequence of the trade-off between time and frequency resolutions and we refer for a detailed discussion to, e.g., \cite{Cohen_1995, Mallat_2009}.
In Figs. \ref{fig::cwt_time_frequency_boxes} (b) and (c), the resolution characteristics of STFT and CWT are illustrated.
In case of the STFT, all boxes have the same dimensions due to the constant window width.
Considering the tiling of the time-frequency plane obtained by CWT, the aspect ratios of the rectangles change, but the area remains constant for a particular mother wavelet $ \psi $ \footnote{
    The adjective ``continuous'' in CWT suggests that a ``discrete'' kind of wavelet transform (DWT) exists, too.
    Both CWT and DWT can be applied to continuous and discrete signals.
    The term continuous or discrete does not refer to the property of the signal, but rather to how the scale parameter is discretized.
    While the scaling parameters for CWT can be selected as finely as possible but in principle as desired, those for DWT may only be selected according to a specific rule.
    In DWT, the signal is decomposed into a set of wavelets orthogonal to each other, which is the major difference compared to CWT \cite[p. 830]{Giurgiutiu_2014}.
    Furthermore, the wavelets are constructed from a scaling function which needs to be orthogonal to its discrete translations; so, the shift parameter is always proportional to the scale parameter.
    For a detailed description of CWT and DWT we refer to relevant literature, e.g., \cite{Giurgiutiu_2014, Mallat_2009, Arts_2022, Rioul_1991}.
}.
There are numerous wavelet types, such as \emph{Daubechies}, \emph{Gabor}, \emph{Morlet}, \emph{Gaussian} and \emph{Haar}, where each wavelet has its specific properties with different advantages and disadvantages \cite{Li_2009}.
Complex analytic wavelets as, e.g., the complex Morlet wavelet, are mainly used to decouple vibration modes and to track the temporal evolution of frequency-components \cite{Haase_2003}.
We use the implementation and definition of the \emph{Morlet} wavelet provided by the Python package \emph{Scipy} \cite{2020SciPy-NMeth}:
\begin{equation}\label{eq::def_morlet_wavelet}
    \psi_\text{morl}(t) = \frac{1}{\sqrt[4]{\pi}}~\epow^{j\omega_c t}~\epow^{-\frac{t^2}{2}}.
\end{equation}
The definition of the Morlet wavelet \eqref{eq::def_morlet_wavelet} may vary, depending on the literature source.
In most cases, the definitions differ in terms of the normalization constant which can affect the properties of the wavelet. 
However, they all have in common that it is defined as a sinusoidal wave modulated by a Gaussian envelope \cite{Mallat_2009}.

According to Li et al. \cite{Li_2009}, the normalized Shannon entropy of the wavelet coefficients can be used as a metric for choosing an appropriate mother wavelet for a given signal.
Their reasoning is based on the idea that it is essential for the mother wavelet to be as similar as possible to the main component of the signal $ s(t) $, i.e., where the highest energy of the signal concentrates.

It has been proven that the square modulus of the wavelet coefficients $ \cwt{s}{a}{b} $, i.e., the \emph{scalogram}, gives information about the energy density in the time-frequency domain at ($t{=}b$, $ f_{c,a}{=}f_c / a$) \cite[p. 109]{Mallat_2009} and is given by
\begin{equation}\label{eq::scalogram}
    \sclogr{s}{t}{f_{c,a}} = |\cwt{s}{a}{b}|^2=  |\cwt{s}{a}{b}|~|\cwt{s}{a}{b}^*|.
\end{equation}
In the literature, the \ToA{} was determined by the global maximum of the scalogram under use of some prior knowledge of the sensor signal, see e.g., Gaul et al. \cite{Gaul_1998} and Hamstad and O'Gallagher \cite{Hamstad_2005}.
However, in our case of Lamb wave propagation in the aluminum plate, we discovered that both global and local maxima of the scalogram do not yield consistent results among all sensor signals under different impact positions.
Therefore, we propose the concept of \emph{threshold crossing} in the \emph{frequency-domain}, which provides effective means to consistently estimate the \ToA{}, see Sec. \ref{ssec::3d_results_CWT}.

With the information on the energy density of the signal in the time-frequency domain provided by the scalogram $\sclogr{s}{t}{f_{c,a}}$, we can estimate the wave speed of the wave propagation using the dispersion characteristics of the material under test.
Knowing the speed at which the wave propagates at a given instantaneous frequency helps to localize impacts.
Similar to the FFT, boundary effects are present in the CWT due to the consideration of finite time-intervals.
For an admissible wavelet function $ \psi(t) $, the wavelet is scaled and shifted when a wavelet transform is performed. Wavelets which are located near the boundary of the considered time-interval cause boundary effects as known from the FT \cite{Torrence_1998}.
Various techniques exist to mitigate or account for those effects.
The \emph{cone of influence} (\coi{}) is the domain in the wavelet spectrum which marks regions that are affected by boundary effects.
For a scaled and shifted wavelet $ \psi_{a,b}(t) $, the \coi{} is the set of all $ t $, which are inside the support of the wavelet:
\begin{equation}\label{eq::def_coi}
    | t - b | \leq Ca,
\end{equation}
where the time-interval $ [-C, C] $ denotes the compact support of the mother wavelet \cite[p. 215]{Mallat_2009}.
For each dilation $a$, Eq. \eqref{eq::def_coi} allows us to determine which time-intervals lie inside the \coi{}, i.e., for which ``truncation'' occurs in time-integral defining CWT, cf. Eq.~\eqref{eq::CWT}.
The \coi{} supports interpretation of scalograms, since wavelet coefficients affected by boundary effects should be interpreted with caution.
In the scalograms presented later in this work, the \coi{} is visualized using white dash-dotted lines.
The region outside these lines corresponds to the \emph{interior} of the \coi{}, as defined by Eq. \eqref{eq::def_coi}.
In figures showing the full signal duration, the \coi{} appears symmetrically on both sides of the time-axis, e.g., Fig. \ref{fig:I1_wavelet_full_IOB}.
In many other figures (e.g., Figs. \ref{fig:I1_wavelet_S0_IOB}, \ref{fig:I1_wavelet_S0_IOP}, \ref{fig:I2_wavelet_S0_IOB}, etc.), however, only one side of the \coi{} -- typically on the left -- is visible, due to the plots being zoomed into early portions of the signal (e.g., around the \ToA{}-estimates).
In such cases, we refer to time-frequency points that fall \emph{outside} the white dash-dotted line (i.e., within the COI) as lying ``to the left of the \coi{} boundary'', and note that these regions are potentially affected by edge effects. This terminology is used consistently in the results section to avoid ambiguity.

\subsection{Short/Long Term Average}\label{ssec::1D_SLA}
The \sla{} method has its origins in the context of seismic data, i.e., earthquakes and localization of hypocenters \cite{Akram_2016} from accelerometer signals.
The main idea of \sla{} is to apply two (rectangular) time-windows of different lengths to the signal $s(t)$, i.e., a \emph{short-term} window and a \emph{long-term} window.
Given a specific point in time $t$ or, in the discrete case, a time-sample $i$, we compute the averages (mean values) over the respective time-windows 
\begin{equation}\label{eq::def_sta_lta}
    \bar{S}_{E}(t) = \frac{1}{t_a} \int_{t- t_a}^{t} s(\tau)^2~d\tau, \qquad \qquad \bar{S}_{E}(t_i) = \frac{1}{n} \sum_{j=i-n}^{i}s(t_j)^2,
\end{equation}
where $ t_a $ and $n$ determine the lengths of the averaging windows.
For the time-windows, preceeding $t_j$ are used and for values $ j < 0 $, a padding scheme is applied to account for those signal values, i.e., $s(t_j) = (s(t_0) + s(t_1))/2$ if $j<0$, see \cite{J.Wong_2009}.
We apply the average to the squared signal, i.e., $ s(t)^2 $, which represents the energy density in case of real signals Eq. \eqref{eq::energy_of_signal}.
Let $\wins$ and $\winl$ denote the number of time-samples in the short-term and  long-term windows, respectively, with $\wins < \winl$. 
By substituting $n$ in Eq. \eqref{eq::def_sta_lta} by $\wins$ and $\winl$, respectively, the short-term average (STA) $\sta(t_i)$ and the long-term average (LTA) $\lta(t_i)$ are obtained.
In SLA, the ratio between STA and LTA, i.e.,
\begin{equation}
    \label{eq::sta_to_lta}
    \statolta(t_i) = \frac{\sta(t_i)}{\lta(t_i)}
\end{equation}
is considered.
The \sla{} ratio $ \statolta(t_i) $ is an indicator similar to SNR, where STA should be sensitive to fast variations in signal amplitude and LTA should provide information about noise.
The \ToA{} is defined as the time-sample associated with the maximum of the forward finite-difference approximation of the time-derivative of $ \statolta(t_i) $:
\begin{equation}
    \label{eq::def_ton_stalta}
    \test = \argmax_{t_i \in \mathcal T}~\dot{\statolta}\left( t_i \right).
\end{equation}
The choices of window lengths have great influence on the result of the \ToA{}-estimation.
The window lengths $ \wins $ and $ \winl $ should be selected according to the frequency-characteristics of the sensor signal.
If $ \wins $ is too small, inappropriate averaging will result in fluctuations on the $ \statolta $-curve, leading to false \ToA{} estimates.
Too large values of $ \winl $, on the other hand, may cause the SLA-method to miss important events that are closely spaced in time.
Akram and Eaton \cite{Akram_2016} specifically suggest that the STA window should cover few periods of typical frequencies of the sensor signal and an optimal LTA window should exceed few periods of noise fluctuations.
Trnkoczy \cite{Trnkoczy_} discussed how to chose STA and LTA windows in context of seismic events. 
He described the quantity $ \sta $ as a signal filter and the smaller the window $ \wins $ is selected, the higher is the sensitivity to high-frequency signal components.
Longer LTA windows result in a slowly increasing $ \lta $ value.
This allows the quantity $ \statolta(t_i) $ to respond quickly to the seismic events, for which the STA window was designed to \cite{Trnkoczy_}. 
A short LTA window, however, could be advantageous when dealing with events that are not part of the wave propagation caused by the impact, but are present in the structure under investigation, e.g., vibrations under operating conditions of the respective component.
The influence of different choices of STA and LTA windows is discussed in Sec. \ref{sec::results}.
\subsection{Modified Energy Ratio}
\label{ssec::1D_MER}
The MER algorithm is similar to \sla-ratio, but it uses two time-windows of equal length that precede and succeed a given time $t$ or $t_i$.
It was proposed by Han et al. \cite{Han_2009} on the basis of the energy ratio, defined by
\begin{equation}\label{eq::def_er}
    \er(t) =  \frac{ \int_{t}^{t+t_a}s(\tau)^2~d\tau }{ \int_{t-t_a}^{t} s(\tau)^2~d\tau  }, \qquad \qquad \er(t_i) =  \frac{\sum_{j=i}^{i+\wine}s(t_j)^2}{ \sum_{j=i-\wine}^{i} s(t_j)^2  }.
\end{equation}
In the above equation, $t_a$ and $ \wine $ determine the lengths of the time-windows. 
Again, we need to deal with samples outside the time-range, for which Wong et al. \cite{J.Wong_2009} proposed the following padding scheme:
\begin{equation}\label{eq::MER_rule}
    s(t_j) = 
    \begin{cases}
        \frac{s(t_0) + s(t_1)}{2},    & \text{if } j<0\\[8pt]
        \frac{s(t_{n-2}) + s(t_{n-1})}{2},  & \text{if } j>n-1
    \end{cases}.
\end{equation}
To estimate the \ToA{}, Han et al. \cite{Han_2009} did not use the plain energy ratio itself.
By modifying the energy ratio, they improved accuracy for problems with low SNR signals.
They tested several modifications of the energy ratio on representative sensor signals and concluded that using the third power of its product with the absolute value of the signal $\left|s(t_i)\right|$ enhanced accuracy for low SNR problems:
\begin{equation}\label{eq::def_mer}
    \mer(t_i) = \left(\left|s(t_i)\right|\er(t_i)\right)^3.
\end{equation}
The \ToA{} is defined as the time associated with the maximum value of $ \mer(t_i) $, i.e.,
\begin{equation}\label{eq::def_ton_mer}
    \test = \argmax_{t_i \in \mathcal T} \mer \left( t_i \right) .
\end{equation}
As for the \sla{} method, the size of time-windows $ \wine $ of the MER method must be selected appropriately.
According to Akram et al. \cite{Akram_2016}, $ \wine $ should be chosen the same way as the STA window, i.e., the window should cover few periods of  typical frequencies of the sensor signal.
For poor signal quality (low SNR values), longer windows should be chosen for a better \ToA{}-estimation, since shorter windows are more sensitive to noise fluctuations \cite{Akram_2016}.

\subsection{Akaike Information Criterion}
\label{subsec::aic}
\begin{figure}[pos=!htbp]
    \centering
    \begin{tikzpicture}				
        \node[anchor=south west,inner sep=0] (image1) at (0, 0) {\includegraphics[width=\textwidth, trim=0 0 0 0, clip, ]{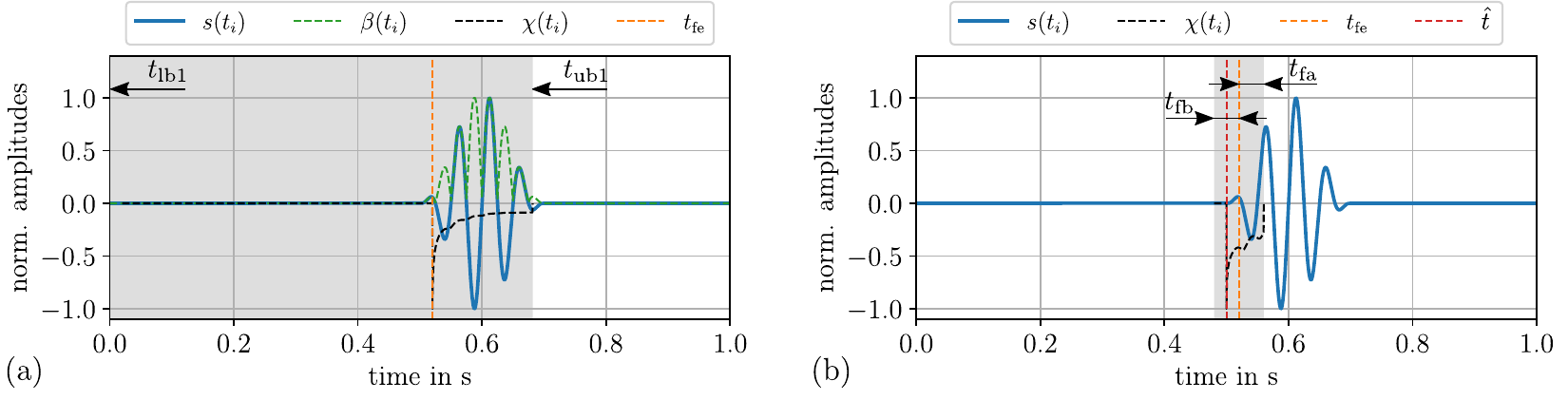}};        
        \begin{scope}[x={(image1.south east)},y={(image1.north west)}]
        \end{scope} 
    \end{tikzpicture}
    \caption{
        Illustration of the two-step AIC algorithm applied to a generic tone-burst signal.
        (a) In the first step, the sensor signal $s(t_i)$ (blue) and its characteristic function $\beta(t_i)$ (green) are shown.
        The AIC function $\aic \left( t_i \right)$ (black) is evaluated in the first AIC window $[\tfirstlb, \tfirstub $].
        The minimum of $\aic \left( t_i \right)$ in this window yields the first \ToA{}-estimate $\tfe$ (orange).
        (b) In the second AIC step, $\aic \left( t_i \right)$ is re-evaluated in the second AIC window $ [\tfe-\tfb, \tfe+\tfa] $ in the vicinity of the first estimate $\tfe$.
        The minimum of $\aic \left( t_i \right)$ in the second AIC window is the final \ToA{}-estimate $\test$ (red).
        }
    \label{fig::aic_1d_wav_pkg_demo}					
\end{figure}

The AIC originates in the field of information theory, where it was proposed as a measure to quantify the quality of stochastic models.
The AIC is defined as \cite{Akaike_1974}
\begin{equation}
    \mathrm{AIC} = 2P - 2\ln(\hat{L}),
\end{equation}
where $P$ is the number of parameters of the model and $\hat{L}$ is the maximum likelihood of the model given the observed data.
AIC describes the relative loss of information of models for the process that generates the data.
The idea behind it is to choose a model that provides the best trade-off between ``goodness-of-fit'' (through the likelihood) and model complexity (through the number of parameters).
Adding parameters can improve fit but risks overfitting, so AIC penalizes complexity \cite{Stoica_2004}.
Typically, a time-series or observed data can be described by a parametrized model, often an autorgressive (AR) model in context of seismic onset detection or \ToA{}-estimation.
An AR process of order $M$, models a time series $\{s(t_i)\}$ as a linear combination of its past $M$ values plus noise \cite{Sleeman_1999}
\begin{align}
    s(t_i) = \sum_{m=1}^{M} a_{m} s(t_{i-m}) + e(t_i),
\end{align}
where $a_m$ are the AR coefficients and $e(t_i)$ is typically white Gaussian noise.
The idea is that current values depend on recent past values, allowing the model to capture temporal structure and predict future values from past observations.
When dealing with non-stationary processes, the assumption that they can be modeled as a sequence of \emph{locally} stationary processes, which, in terms, are represented as AR processes is adopted \cite{Sedlak_2013, Kitagawa_1978, Kobayashi_2012}.
The AR-AIC picker, see e.g., \cite{Leonard_2000}, tries different split points in the time series, using AIC to find the split point that best explains the data.
Determining the correct order $M$ and then estimating the AR coefficients $\{a_m\}$ can be time-consuming and tedious.
Maeda \cite{Maeda_1985} proposed a simplification where no AR modeling is required.
Instead of fitting AR processes to each segment, he considered the time series $\{s(t_i)\}$ of length $N$ and a potential split point $k$ that divides the series into two segments:
\begin{itemize}
    \item Segment 1: $\{s(t_1),\dots,s(t_k)\}$
    \item Segment 2: $\{s(t_{k+1}),\dots,s(t_N)\}$
\end{itemize}
Instead of assuming each segment follows an AR model, Maeda's approach assumes a simpler statistical model: each segment is considered to be drawn from a (possibly different) stationary process with its own variance \cite{Maeda_1985}.
One can think of the time-series as switching from one variance regime to another, and this transition point minimizes the AIC.
The \ToA{} is chosen as the $k$ that minimizes this AIC value, reflecting a point where the variance structure changes most significantly.
Maeda's \cite{Maeda_1985} widely used implementation of AIC for \ToA{}-estimation of wave propagation is provided by
\begin{equation}\label{eq::def_aic}
    \aic(t_i) = i\ln\left(\var\left\{s(t_j)|_{j=0\dots i}\right\}\right) + (N-i-1)\ln\left(\var\left\{s(t_j)|_{j=i+1\dots N}\right\}\right)~,
\end{equation}  
where $ i $ is the index through all time-steps of the signal $ s(t_i) $, $ \var\{\cdot\} $ denotes the variance and $ N $ is the signal length.
It is worth noting that Eq. \eqref{eq::def_aic} should only be applied to a part of the signal, in order to work properly \cite{Simone_2017}.
Accordingly, recent AIC-based variants have focused on improving the selection of this signal portion and at increasing robustness against noise, e.g., by using characteristic-function-based, entropy-based or wavelet-denoising approaches \cite{Barile_2025, Hou_2025}.
Which portion of the signal should be considered is determined by so-called \emph{AIC pickers}.
In this work, we use an AIC picker based on a characteristic function in order to effectively separate the non-informative part from the informative part of the signal \cite{Simone_2017}.
There are various functions to use, e.g., the signal absolute value, envelope calculated by Hilbert transform, squared value or the envelope of squared value.
However, to enhance the sensitivity of periodical signal changes, the seismogram threshold picker of Allen \cite{Allen_1982}, which is effective for bulk specimens, is used.
Sedlak et al. \cite{Sedlak_2013} derived a modified version of Allen's formula for thin plates so that the AIC picker function $\allen(t_i)$ reads
\begin{equation}\label{eq::allens_formula}
    \allen(t_i) = |s(t_i)| + \Ra|s(t_i)-s(t_{i-1})|,
\end{equation}
where $ \Ra $ is a constant which value is determined by trial and error.
According to Sedlak et al. \cite{Sedlak_2013}, the constant $ \Ra $ was investigated in context of experimental results and should be close to $ \Ra=1 $ for artificial acoustic emission (AE) events (e.g. pencil-lead break test) and close to $ \Ra=4 $ for real AE sources with low SNR.
In order to obtain an accurate estimate of \ToA{}, it is crucial to carefully select the part of the signal where Maeda’s equation \eqref{eq::def_aic} is applied.
Therefore, we apply a window to the signal.
\ToA{}-estimation with AIC follows a two-step approach, as proposed by Sedlak et al \cite{Sedlak_2009}.
Each AIC step consists of two operations: (i) defining a time-window with the AIC picker and (ii) calculating the global minimum of the AIC \eqref{eq::def_aic} in the specified time-window \cite{Sedlak_2013}.
The first AIC step calculates a first estimate $\tfe$: 
\begin{equation}\label{eq::def_ton_aic}
    \tfe = \argmin_{t_i \in [\tfirstlb, \tfirstub]}  \aic \left( t_i \right), \qquad\qquad \tfirstub = \tmax + \tam , \qquad\qquad \tmax = \argmax_{t_i \in \mathcal T} \beta(t_i).
\end{equation}
The window of the first AIC step is defined between the lower and upper bounds $ \tfirstlb $ and $ \tfirstub $.
A time-delay, referred to as $ \tam $, is added to $ \tmax $, usually defined by trial and error.
In literature, a value of $ \tam=\SI{20}{\micro\second} $ is recommended for thin plate specimens \cite{Sedlak_2013}.
The second AIC step focuses on the immediate vicinity of the first estimate $ \tfe $ to increase the accuracy of the AIC algorithm \cite{Simone_2017}
\begin{equation}\label{eq::def_toa_aic}    
    \test = \argmin_{t_i \in [\tfe-\tfb, \tfe+\tfa]}  \aic \left( t_i \right).
\end{equation}
The second AIC window comprises two parts of length $ \tfb $ and $ \tfa $, respectively, which describe the lengths of the window parts before and after the first estimate $ \tfe $.
The width of the second AIC window, i.e. $ \tfb + \tfa $, was determined in recent literature with the values of $ \tfb=\SI{10}{\micro\second} $ and $ \tfa=\SI{30}{\micro\second} $ \cite{Sedlak_2013}.
In Fig. \ref{fig::aic_1d_wav_pkg_demo}, the two-step procedure of the AIC algorithm is illustrated, based on a generic tone burst signal.

\section{\ToA{} Methods Assessment}\label{sec::results}
In this section, we apply the present \ToA{}-estimation methods to the sensor signals of our FE model.
For this purpose, method-specific parameters are varied in order to perform a parametric study. 
The results serve as basis for the derivation of guidelines to accurately capture the \ToA{} of impact-induced waves by means of the methods assessed in the present paper.

\begin{figure}[pos=!htbp]
    \centering
    \begin{tikzpicture}
        \node[anchor=south west,inner sep=0] (image1) at (0, 0) {\includegraphics[width=\textwidth, trim=0 0 0 0, clip, ]{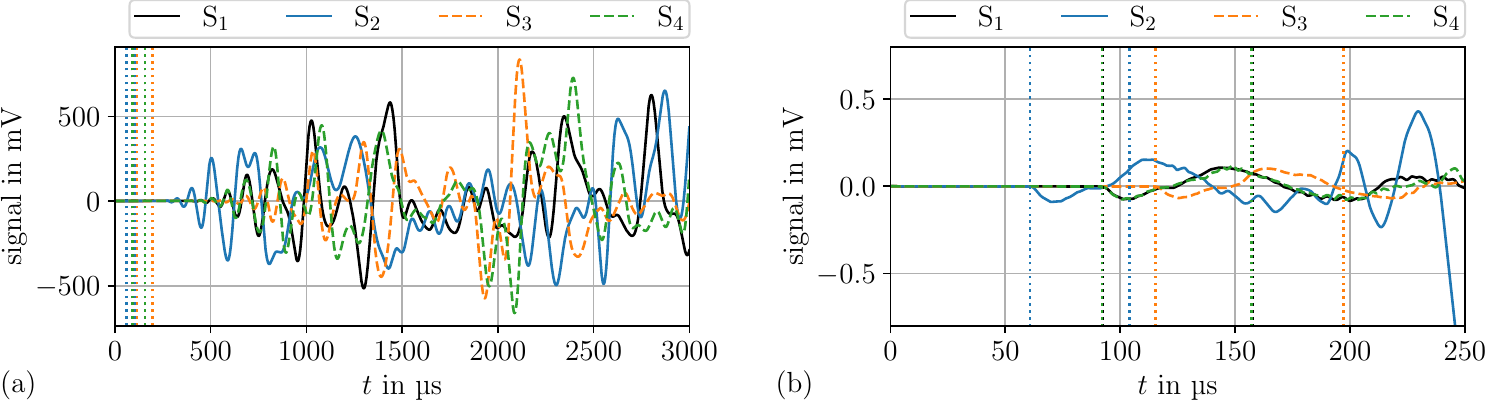}};      
    \end{tikzpicture}
    \caption{Illustration of the sensor signals for $ \imp{1}^{\iob} $. Time-history of the sensor signals over (a) the entire time-range considered and (b) a detailed view at the onset of the first events.}
    \label{fig::3d_sens_sig_imp_1_hammer+detail}					
\end{figure}

\begin{figure}[pos=!htbp]
    \centering
    \begin{tikzpicture}
        \node[anchor=south west,inner sep=0] (image1) at (0, 0) {\includegraphics[width=\textwidth, trim=0 0 0 0, clip, ]{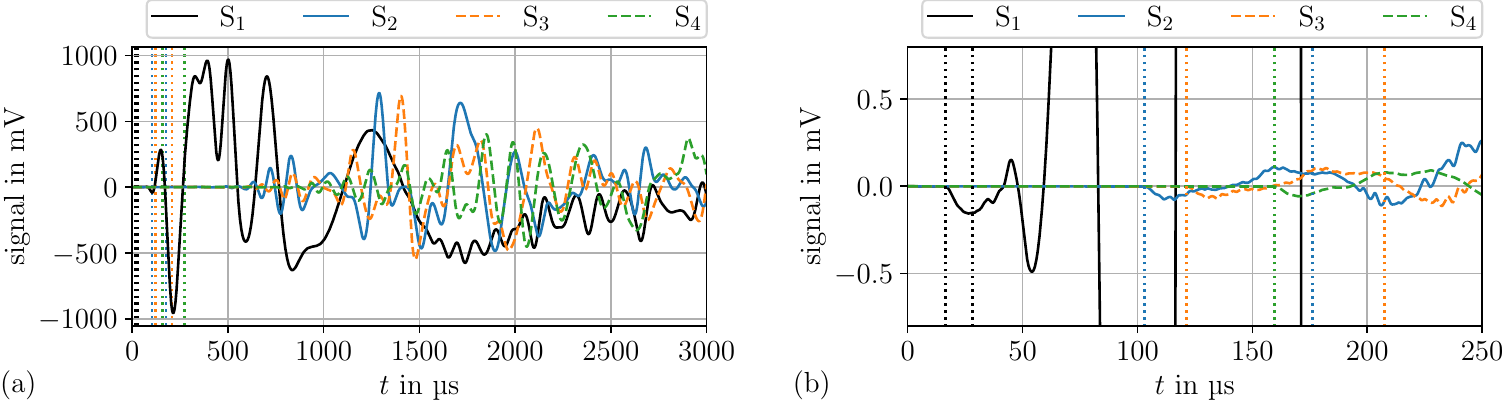}};      
    \end{tikzpicture}
    \caption{Illustration of the sensor signals for $ \imp{2}^{\iob} $. Time-history of the sensor signals over (a) the entire time-range considered and (b) a detailed view at the onset of the first events.}
    \label{fig::3d_sens_sig_imp_2_hammer+detail}					
\end{figure}

\begin{figure}[pos=!htbp]
    \centering
    \begin{tikzpicture}
        \node[anchor=south west,inner sep=0] (image1) at (0, 0) {\includegraphics[width=\textwidth, trim=0 0 0 0, clip, ]{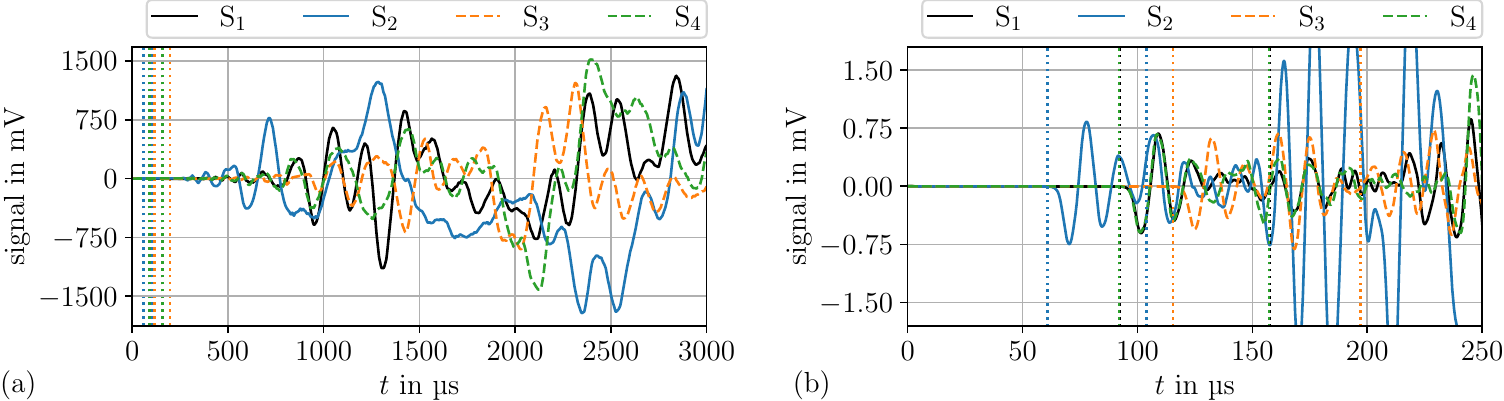}};      
    \end{tikzpicture}
    \caption{Illustration of the sensor signals at $ \imp{1}^{\iop} $. Time-history of the sensor signals over (a) the entire time-range considered and (b) a detailed view at the onset of the first events.}
    \label{fig::3d_sens_sig_imp_1_shaker+detail}					
\end{figure}

\begin{figure}[pos=!htbp]
    \centering
    \begin{tikzpicture}
        \node[anchor=south west,inner sep=0] (image1) at (0, 0) {\includegraphics[width=\textwidth, trim=0 0 0 0, clip, ]{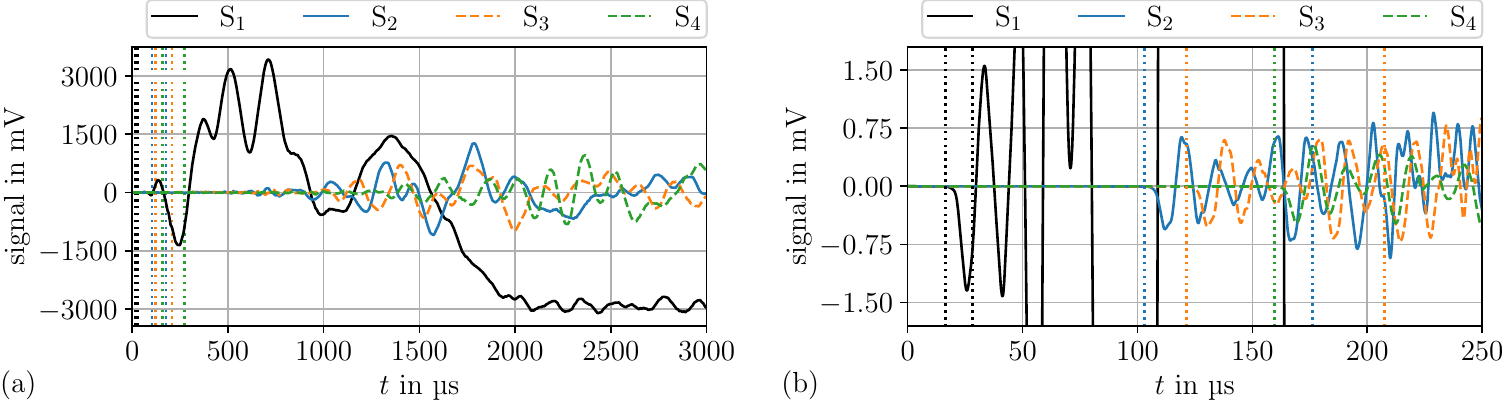}};      
    \end{tikzpicture}
    \caption{Illustration of the sensor signals at $ \imp{2}^{\iop} $. Time-history of the sensor signals over (a) the entire time-range considered and (b) a detailed view at the onset of the first events.}
    \label{fig::3d_sens_sig_imp_2_shaker+detail}					
\end{figure}

In what follows, we discuss the application of the \ToA{}-methods presented in the earlier section to waves excited upon impacts in plate-like structures.
For this purpose, we use the simulation model described in Sec. \ref{ssec:3D_fe_model}.
The simulated time is $ t_{\text{end}} = \SI{3}{\milli\second} $, where a constant time-increment of $ \dt = \SI{0.2}{\micro\second} $ is used, which amounts to \num{15000} time-steps.
The voltages measured at the electrodes of the four piezoelectric patches serve as  sensor signals.
We apply all investigated \ToA{}-estimation methods on sensor signals obtained from both idealized ($\imp{}^{\iob}$) and experiment-based ($\imp{}^{\iop}$) impacts.
The time-histories of the four sensor signals are illustrated in Figs. \ref{fig::3d_sens_sig_imp_1_hammer+detail} - \ref{fig::3d_sens_sig_imp_2_hammer+detail} for the idealized impact and in Figs. \ref{fig::3d_sens_sig_imp_1_shaker+detail} - \ref{fig::3d_sens_sig_imp_2_shaker+detail} for the experiment-based impact at both impact locations.
Irrespective of the latter, a first observation is that the signals differ significantly in amplitude within the time-interval considered. 
Intuitively, we assume sensor signals to be first measured at those sensors, that are closest to the impacts.
In spite of the complexity of interference patterns, we also expect amplitudes to be correlated to distances between sensors and the location of impact.
To help interpret the time scales, two key markers are defined and shown in the result plots: $ \tso $ and $ \tao $.
The former represents the \ToA{} of the fastest $S_0$ mode, associated with the minimum frequency-thickness product $f d$ (see Fig. \ref{fig::lamb_wave_disp_curve}), giving a wave speed of $ \cso = \SI{5392}{\metre\per\second} $ in our case.
Assuming that all longitudinal and transverse waves transform into Lamb waves, $ \tso $ marks the earliest possible time for an event to occur.
Similarly, $ \tao $ marks the \ToA{} of the fastest $A_0$ mode, which travels with $ \cao = \SI{3156}{\metre\per\second} $ at the frequency $ f_{A_0,\mathrm{max}} = \SI{674}{\kilo\hertz}$ for the investigated aluminum plate. see Fig. \ref{fig::lamb_wave_disp_curve}.
These time markers appear as vertical lines in Figs. \ref{fig::3d_sens_sig_imp_1_hammer+detail} - \ref{fig::3d_sens_sig_imp_2_shaker+detail}, with each line color-matched to its respective sensor signal trace.
To avoid crowding, we skip labeling the markers as they can be identified by the condition $\tso < \tao$.
Note that $ \tso $ and $ \tao $ are different for each sensor.
The corresponding values are listed in Tab. \ref{tab:ts0_and_ta0_markers}.
\begin{table}[pos=!htbp]
    \small
    \centering
    \begin{tabular}[]{r|rrrr|rrrr}
        \toprule        
        & \multicolumn{4}{c|}{$\tso$ in $\SI{}{\micro\second}$}
        & \multicolumn{4}{c}{$\tao$ in $\SI{}{\micro\second}$}
        \\
        & $\sensor_1$ 
        & $\sensor_2$ 
        & $\sensor_3$ 
        & $\sensor_4$ 
        & $\sensor_1$ 
        & $\sensor_2$ 
        & $\sensor_3$ 
        & $\sensor_4$ 
        \\        
        \midrule
        $\imp{1}$
        & \num{92.2}
        & \num{60.8}
        & \num{115.4}
        & \num{92.0}
        & \num{157.6}
        & \num{104.0}
        & \num{197.2}
        & \num{157.2}
        \\     
        $\imp{2}$
        & \num{16.4}
        & \num{103.0}
        & \num{121.4}
        & \num{159.6}
        & \num{28.2}
        & \num{176.2}
        & \num{207.6}
        & \num{272.8}
        \\        
        \bottomrule
    \end{tabular}
    \caption{
        Numerical values of time-markers $\tso$ and $\tao$ for each sensor and impact.
    }
    \label{tab:ts0_and_ta0_markers}
\end{table}

For impact location $ \imp{1} $, there is a strong similarity in the signals recorded by sensors $ S_1 $ and $ S_4 $ for both types of impacts, as shown in Figs. \ref{fig::3d_sens_sig_imp_1_hammer+detail} and \ref{fig::3d_sens_sig_imp_1_shaker+detail}.
This similarity is expected, given that these sensors are positioned at approximately the same distance from the impact location.
The sensor signals for the idealized impact \mbox{(Figs. \ref{fig::3d_sens_sig_imp_1_hammer+detail} and \ref{fig::3d_sens_sig_imp_2_hammer+detail})} generally show lower amplitudes compared to those from the experiment-based impact (Figs. \ref{fig::3d_sens_sig_imp_1_shaker+detail} and \ref{fig::3d_sens_sig_imp_2_shaker+detail}), due to the lower excitation force used in the idealized case.
The frequency spectra of the two impact types differ, as seen in Figs. \ref{fig::3d_shaker_signal} and \ref{fig::3d_hammer_signal}; resulting in distinct excitation frequencies for each case.
Despite the perpendicular impact direction relative to the plate surface, both the $ A_0 $ and $ S_0 $ wave modes are observed in the signals.
Literature confirms that simulations of guided waves under similar conditions typically produce both modes.
However, in experiments where impacts are perpendicular to the plate plane, only the $ A_0 $ mode is detected, with the $ S_0 $ mode either absent or too weak to be captured in the measurement data (see, e.g., \cite{Ciampa_2010, Merlo_2017, Kundu_2009, Zhu_2017, Grasboeck_smart2023}).

\begin{figure}[pos=!htbp]
    \centering
    \begin{tikzpicture}
        \node[anchor=south west,inner sep=0] (image1) at (0, 0) {\includegraphics[width=\textwidth, trim=0 0 0 0, clip, ]{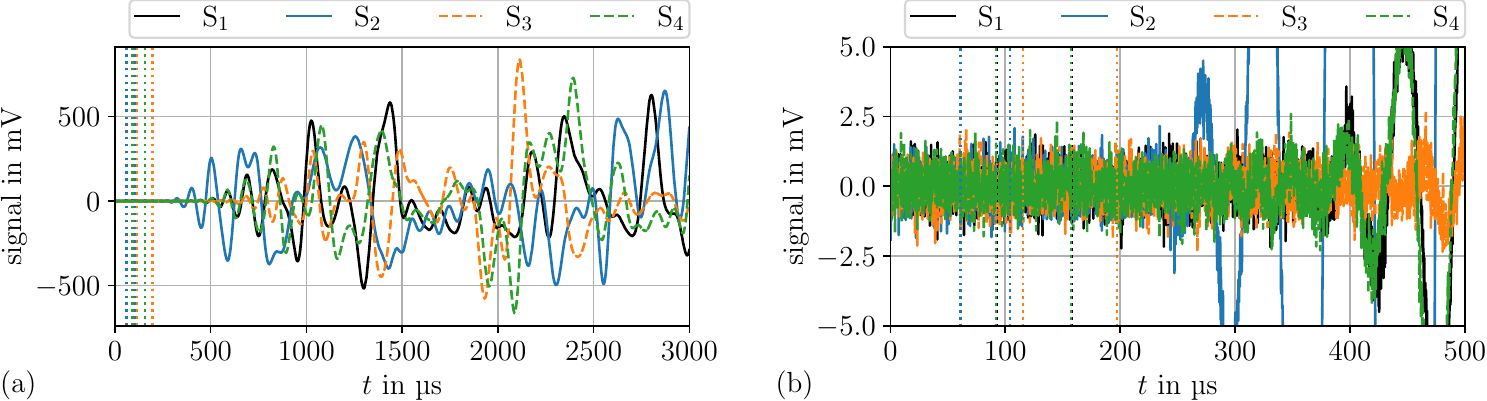}};      
    \end{tikzpicture}
    \caption{Illustration of the sensor signals at $ \imp{1}^{\iob} $. Time-history of the noise-contaminated sensor signals over (a) the entire time-range considered and (b) a detailed view at the onset of the first events.}
    \label{fig::3d_sens_sig_imp_1_hammer+detail_snr50db}
\end{figure}

\begin{figure}[pos=!htbp]
    \centering
    \begin{tikzpicture}
        \node[anchor=south west,inner sep=0] (image1) at (0, 0) {\includegraphics[width=\textwidth, trim=0 0 0 0, clip, ]{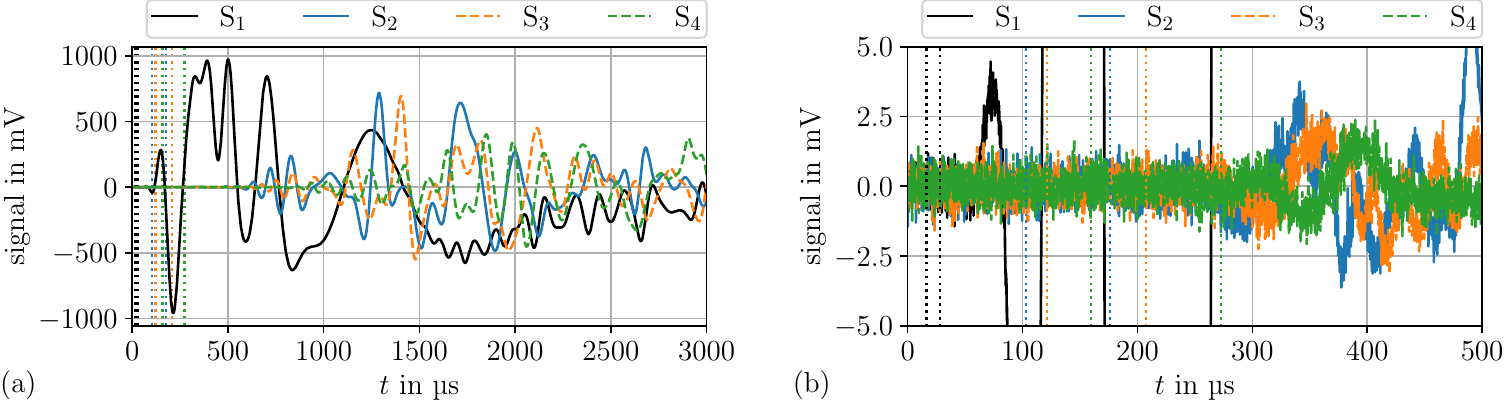}};      
    \end{tikzpicture}
    \caption{Illustration of the sensor signals at $ \imp{2}^{\iob} $. Time-history of the noise-contaminated sensor signals over (a) the entire time-range considered and (b) a detailed view at the onset of the first events.}
    \label{fig::3d_sens_sig_imp_2_hammer+detail_snr50db}
\end{figure}

\begin{figure}[pos=!htbp]
    \centering
    \begin{tikzpicture}
        \node[anchor=south west,inner sep=0] (image1) at (0, 0) {\includegraphics[width=\textwidth, trim=0 0 0 0, clip, ]{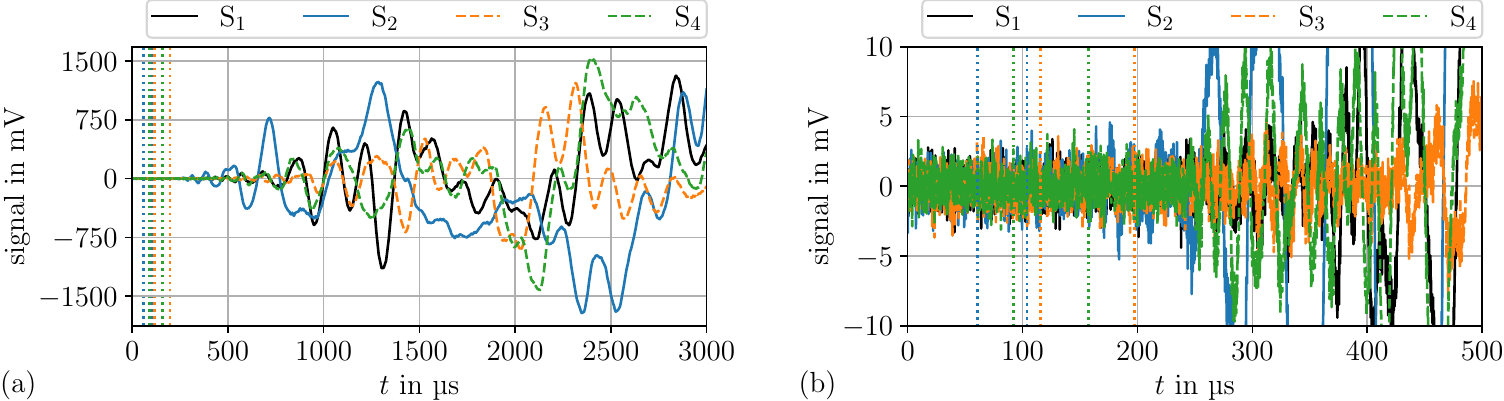}};      
    \end{tikzpicture}
    \caption{Illustration of the sensor signals at $ \imp{1}^{\iop} $. Time-history of the noise-contaminated sensor signals over (a) the entire time-range considered and (b) a detailed view at the onset of the first events.}
    \label{fig::3d_sens_sig_imp_1_shaker+detail_snr50db}
\end{figure}

\begin{figure}[pos=!htbp]
    \centering
    \begin{tikzpicture}
        \node[anchor=south west,inner sep=0] (image1) at (0, 0) {\includegraphics[width=\textwidth, trim=0 0 0 0, clip, ]{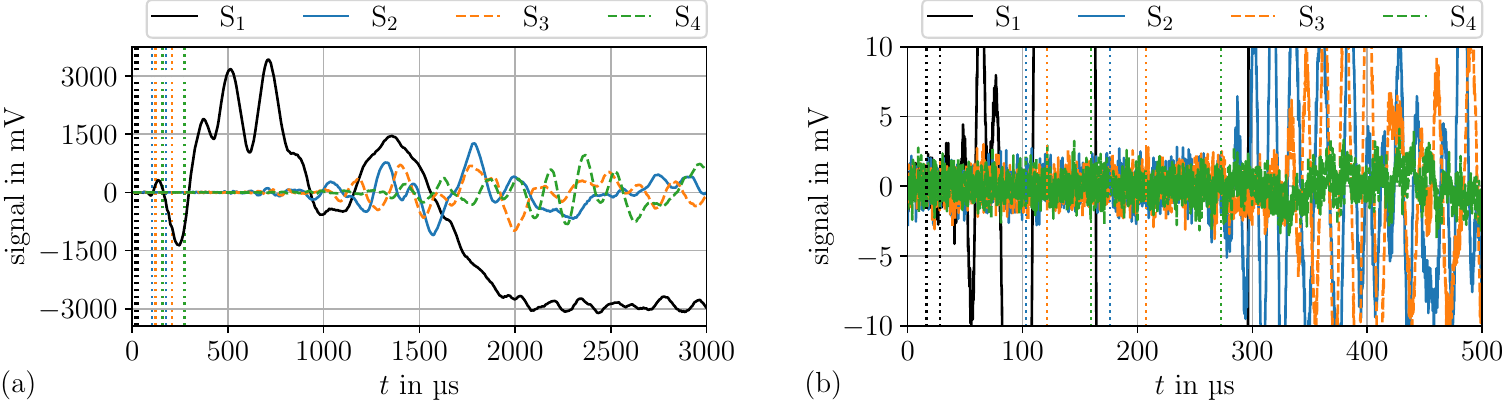}};      
    \end{tikzpicture}
    \caption{Illustration of the sensor signals at $ \imp{2}^{\iop} $. Time-history of the noise-contaminated sensor signals over (a) the entire time-range considered and (b) a detailed view at the onset of the first events.}
    \label{fig::3d_sens_sig_imp_2_shaker+detail_snr50db}
\end{figure}

To assess the robustness of the \ToA{}-estimation methods under more realistic conditions, artificial noise is added to the ideal sensor signals obtained from the FE-simulations.
The noise level is derived from experimental measurements reported in \cite{humer2025}, where an average SNR of \qty{45}{\decibel}, see Eq. \eqref{eq::snr}, was observed for comparable piezoceramic sensor setups.
Accordingly, additive white Gaussian noise is superimposed on the sensor signals obtained from the simulation model and scaled such that the resulting SNR equals \qty{45}{\decibel}.
The time-histories of the noise-contaminated sensor signals for the idealized impact at both locations are shown in Figs. \ref{fig::3d_sens_sig_imp_1_hammer+detail_snr50db} and \ref{fig::3d_sens_sig_imp_2_hammer+detail_snr50db}.
When considering the entire simulated time-interval, the noise-contaminated signals appear visually similar to the corresponding noise-free signals (c.f. Figs. \ref{fig::3d_sens_sig_imp_1_hammer+detail} and \ref{fig::3d_sens_sig_imp_2_hammer+detail}).
However, detailed views around the onset of the first wave arrivals reveal pronounced differences, with the initial arrivals being partially obscured by noise.
The same observiations apply to the experiment-based impact at both locations, as illustrated in Figs. \ref{fig::3d_sens_sig_imp_1_shaker+detail_snr50db} and \ref{fig::3d_sens_sig_imp_2_shaker+detail_snr50db} in comparison to the corresponding noise-free signals shown in Figs. \ref{fig::3d_sens_sig_imp_1_shaker+detail} and \ref{fig::3d_sens_sig_imp_2_shaker+detail}.



In the following, both ideal (noise-free) and noise-contaminated sensor signals are used to apply the investigated \ToA{}-estimation methods.
This allows a systematic assessment of the robustness of each method with respect to impact type and location ($ \imp{1}^{\iob} $, $ \imp{1}^{\iop} $, $ \imp{2}^{\iob} $ and $ \imp{2}^{\iop} $), as well as to the presence of noise.
\subsection{Threshold Crossing in Time-Domain}
\begin{figure}[pos=!htbp]
    \centering
    \begin{tikzpicture}				
        \node[anchor=south west,inner sep=0] (image1) at (0, 0) {\includegraphics[width=.8\textwidth, trim=0 0 0 0, clip, ]{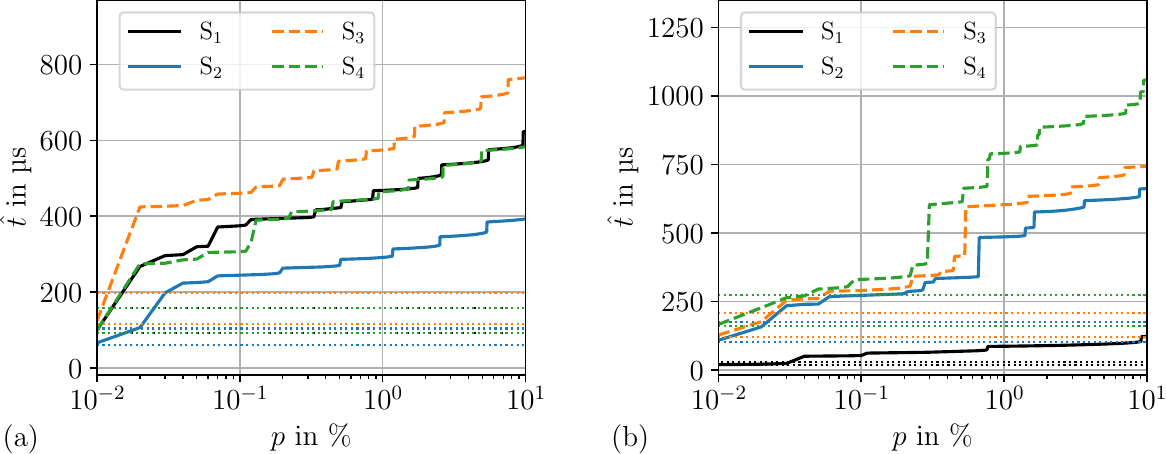}};        
        \begin{scope}[x={(image1.south east)},y={(image1.north west)}]
        \end{scope} 
    \end{tikzpicture}
    \caption{Results of \ToA{}-estimation with TC method applied on noise-free signals: Evolution of $ \test $ for each sensor signal for (a) $ \imp{1}^{\iob} $ and (b) $ \imp{2}^{\iob} $, when a common threshold is used for all sensor signals. The base value for the threshold determination, i.e., the smallest maximum of the sensor signals is $ \SI{135.4}{\milli\volt} $ for impact $ \imp{1}^{\iob} $ and $ \SI{40.6}{\milli\volt} $ for impact $ \imp{2}^{\iob} $.
    The dotted lines indicate the first arrivals of the fundamental Lamb wave modes $\tso$ and $\tao$, with each line color-matched to its respective sensor signal trace.}
    \label{fig::toa_lvl_iob_both_loc_cmn_thr_lvl}					
\end{figure}

\begin{figure}[pos=!htbp]
    \centering
    \begin{tikzpicture}				
        \node[anchor=south west,inner sep=0] (image1) at (0, 0) {\includegraphics[width=.8\textwidth, trim=0 0 0 0, clip, ]{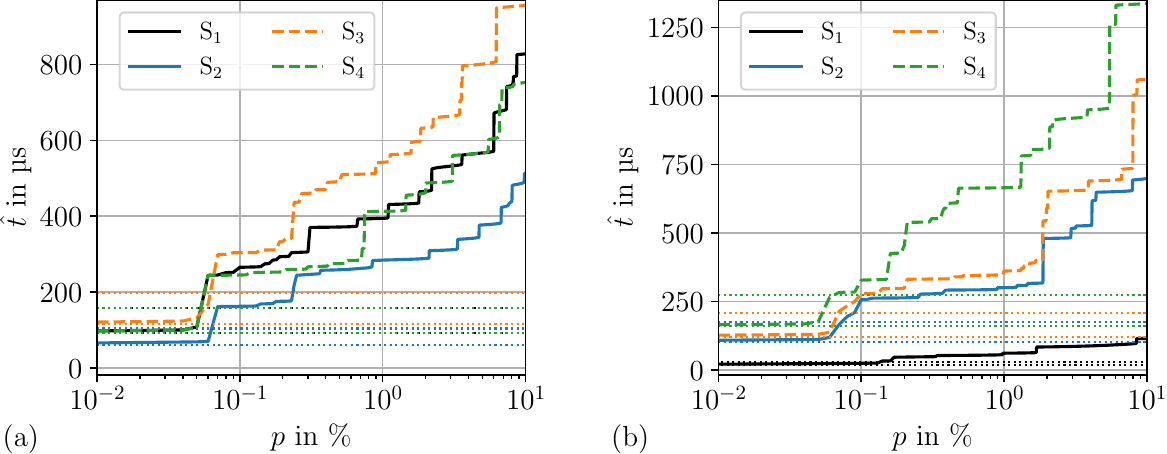}};        
        \begin{scope}[x={(image1.south east)},y={(image1.north west)}]
        \end{scope} 
    \end{tikzpicture}
    \caption{Same representation as Fig. \ref{fig::toa_lvl_iob_both_loc_cmn_thr_lvl}, but for the experiment-based impact positions (a) $ \imp{1}^{\iop} $ and (b) $ \imp{2}^{\iop} $.
    The base value for the threshold determination is $ \SI{180.4}{\milli\volt} $ for impact $ \imp{1}^{\iop} $ and $ \SI{53}{\milli\volt} $ for impact $ \imp{2}^{\iop} $.}
    \label{fig::toa_lvl_iop_both_loc_cmn_thr_lvl}					
\end{figure}

We start with the most intuitive method and examine the results of the TC method.
The choice of an appropriate threshold depends on the respective application. 
To derive a reasonable guideline, it makes sense to relate the threshold to characteristic properties of the given problem.
Noise level and maximum expected amplitudes are only two examples of reference values that can be used for this purpose.
As with every other method, signals below the noise level cannot be detected.
Too low values for the threshold increases the chance of ``false positives'', i.e., TC mistakes noise for impacts.
In case of high thresholds, on the other hand, initial waves related to impacts on the structure may be missed. 
To investigate how sensitive the \ToA{}-estimation for our sensor signals is with respect to the threshold level, we perform a parametric study in which the threshold is varied.
In an application scenario, rather than setting different thresholds for multiple sensors - Eq. \eqref{eq::def_threshold} - one would set a single threshold for all sensors.
In order to ensure that every sensor signal exceeds the predefined threshold value, we define a threshold value, which is, for a specific $ p $, the same for all sensor signals:
\begin{equation}\label{eq::def_threshold_lvl_const}   
    \sth = p~\min_j\left\{ \max_i \left|s_j(t_i)\right|  \right\} \quad \text{with} \quad j = 1\dots r,
\end{equation} 
where $ r $ is the number of sensors.
The smallest maximum value of all sensors is used as the base value for calculating the threshold.
This ensures that every sensor signal exceeds the threshold.
For the parametric study, we vary $ p $ from \numrange{1e-4}{1e-1}, which corresponds to \qtyrange{0.01}{10}{\percent} of the smallest maximum among all sensor signals.
The result of the parameter variation for a common threshold value for all noise-free sensors signals is illustrated for the idealized impact in Fig. \ref{fig::toa_lvl_iob_both_loc_cmn_thr_lvl} and for the experiment-based impact in Fig. \ref{fig::toa_lvl_iop_both_loc_cmn_thr_lvl}.
As the threshold level increases, the estimated \ToA{} shifts to later times, which is hardly surprising.
Larger steps of \ToA{}-estimates are related to local maxima in sensor signals, which are inherent to transient wave phenomena.
For signals with very high SNR, e.g., signals obtained by simulation where only numerical noise is present, any change in signal amplitude is higher than the maximum amplitude of numerical noise.
Hence, the TC method has the compelling advantage to set the threshold level particularly low to detect the very first event of the signal.
In our scenario, we assume that only the $0$th-order Lamb wave modes propagate, cf. Fig. \ref{fig::lamb_wave_disp_curve}.
This allows us to confidently assert that the $ S_0 $ mode, which has the highest group velocity within the relevant frequency range, is detected first.
Such information is very useful when determining the speed of a detected event is necessary, for example, when localizing an impact through triangulation.
Ideally, we aim to identify a range for the parameter $p$ where the \ToA{} for all sensor signals remains independent of the threshold and, preferably, also independent of the impact position.
While no such universally valid range can be identified in this case, a narrow threshold interval can be observed for the experiment-based impact, in which the estimated \ToA{} reliably coincides with $\tso$, c.f. Fig. \ref{fig::toa_lvl_iop_both_loc_cmn_thr_lvl}.
This behavior can be attributed to the more pronounced onset of the sensor signal during the experiment-based impact, see Fig. \ref{fig::3d_sens_sig_imp_1_shaker+detail}, whereas the idealized impact exhibits a more gradual amplitude build-up, see Fig. \ref{fig::3d_sens_sig_imp_1_hammer+detail}, causing the estimated \ToA{} to shift immediately as the threshold is increased.
For impact location $ \imp{1} $, we expect the \ToA{} curves for sensors  $ \sensor_1 $ and $ \sensor_4 $ to closely align, given that their distances to the impact location are nearly identical.
This close agreement is evident for the idealized impact across the entire parameter range, see Fig. \ref{fig::toa_lvl_iob_both_loc_cmn_thr_lvl} (a).
However, for the experiment-based impact, the curves do not align as closely as they do for the idealized impact, see Fig. \ref{fig::toa_lvl_iop_both_loc_cmn_thr_lvl} (a).
Additionally, we observe that for impact positions $ \imp{1} $ and $ \imp{2} $, the conditions $ \test[3]>\test[1]\approx \test[4]>\test[2] $ and $ \test[4]>\test[3]>\test[2]>\test[1] $, respectively, hold true for all values of $p$.
These orders correspond precisely to the order of the sensors' distances from the impact locations, see Tab. \ref{tab::3d_fe_model_sens_imp_pos}.

\begin{figure}[pos=!htbp]
    \centering
    \begin{tikzpicture}				
        \node[anchor=south west,inner sep=0] (image1) at (0, 0) {\includegraphics[width=.8\textwidth, trim=0 0 0 0, clip, ]{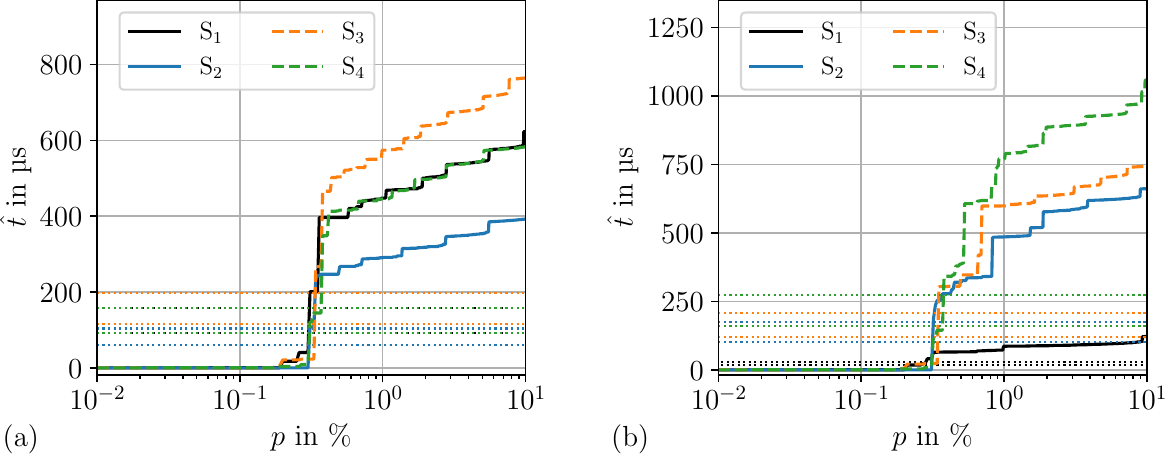}};        
        \begin{scope}[x={(image1.south east)},y={(image1.north west)}]
        \end{scope} 
    \end{tikzpicture}
    \caption{Same representation as Fig. \ref{fig::toa_lvl_iob_both_loc_cmn_thr_lvl}, but for noise-contaminated sensor signals corresponding to the idealized impact positions (a) $ \imp{1}^{\iob} $ and (b) $ \imp{2}^{\iob} $.
    The base value for the threshold determination is $ \SI{135.4}{\milli\volt} $ for impact $ \imp{1}^{\iob} $ and $ \SI{40.6}{\milli\volt} $ for impact $ \imp{2}^{\iob} $.}
    \label{fig::toa_lvl_iob_both_loc_cmn_thr_lvl_noise50dB}					
\end{figure}

\begin{figure}[pos=!htbp]
    \centering
    \begin{tikzpicture}				
        \node[anchor=south west,inner sep=0] (image1) at (0, 0) {\includegraphics[width=.8\textwidth, trim=0 0 0 0, clip, ]{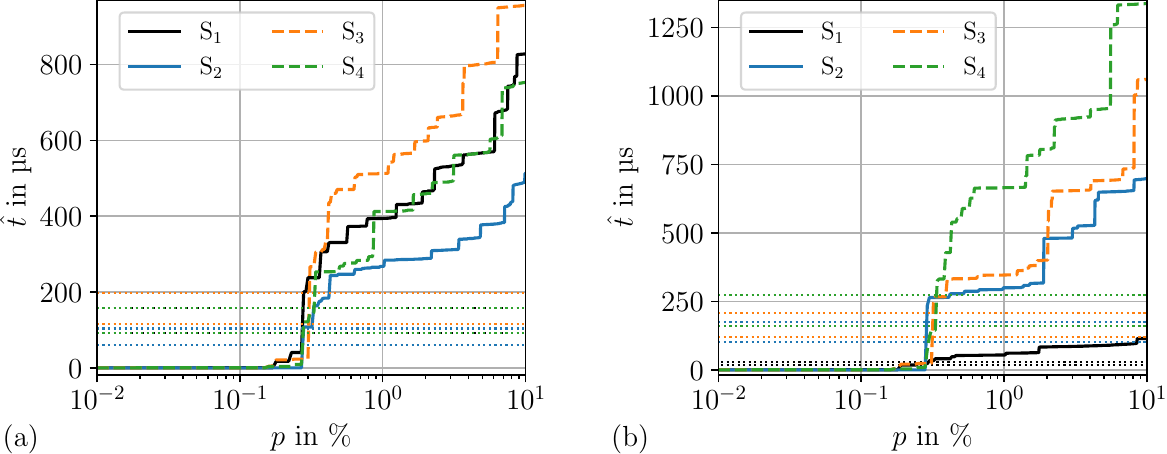}};        
        \begin{scope}[x={(image1.south east)},y={(image1.north west)}]
        \end{scope} 
    \end{tikzpicture}
    \caption{Same representation as Fig. \ref{fig::toa_lvl_iop_both_loc_cmn_thr_lvl}, but for noise-contaminated sensor signals corresponding to the experiment-based impact positions (a) $ \imp{1}^{\iop} $ and (b) $ \imp{2}^{\iop} $.
    The base value for the threshold determination is $ \SI{180.4}{\milli\volt} $ for impact $ \imp{1}^{\iop} $ and $ \SI{53}{\milli\volt} $ for impact $ \imp{2}^{\iop} $.}
    \label{fig::toa_lvl_iop_both_loc_cmn_thr_lvl_noise50dB}					
\end{figure}

The results for the noise-contaminated signals, shown in Figs. \ref{fig::toa_lvl_iob_both_loc_cmn_thr_lvl_noise50dB} and \ref{fig::toa_lvl_iop_both_loc_cmn_thr_lvl_noise50dB}, reveal a slightly different behavior of the TC-method compared to the noise-free case.
In particular, the advantage of selecting very small threshold levels to consistently detect the earliest $S_0$ mode arrivals is lost.
This is consistent with the time-history of the sensor signals, where the first arrival of the $S_0$ mode is masked by noise (see Figs. \ref{fig::3d_sens_sig_imp_1_hammer+detail_snr50db} - \ref{fig::3d_sens_sig_imp_2_shaker+detail_snr50db}), explaining why $\tso$ is no longer estimated in the TC results of Figs. \ref{fig::toa_lvl_iob_both_loc_cmn_thr_lvl_noise50dB} and \ref{fig::toa_lvl_iop_both_loc_cmn_thr_lvl_noise50dB}.
For sufficiently low values of the threshold parameter $p$, the noise floor alone exceeds the threshold, leading to identical \ToA{}-estimates for all sensors at $\test[i]=0$.
Only once the threshold exceeds the noise level do sensor-specific \ToA{}-estimates emerge, which then follow the same qualitative trends observed for the noise-free signals, i.e., a monotonic shift of $\test[i]$ to later times with increasing threshold.

The TC method becomes even more sensitive to the choice of the threshold in the presence of noise, which limits its robustness and practical applicability for reliable detection of early wave arrivals under realistic measurement conditions.
\subsection{Continuous Wavelet Transform} \label{ssec::3d_results_CWT}
\begin{figure}[pos=!htbp]
    \centering
    \includegraphics[width=0.9\linewidth]{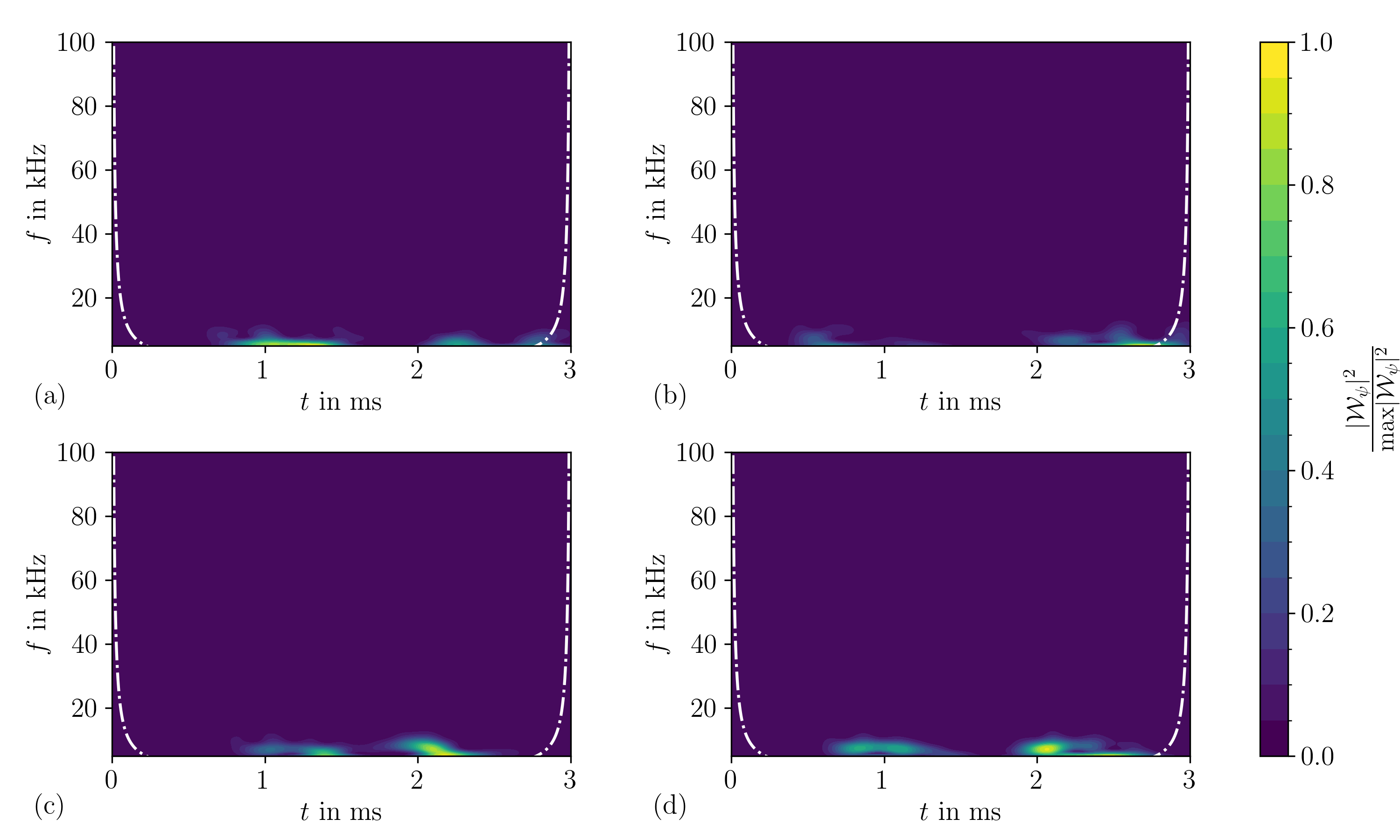}
    \caption{
        Scalograms of noise-free sensor signals $\sensor_1$--$\sensor_4$ for the \emph{idealized impact} $\imp{1}^{\iob}$: The normalized squares of the CWTs' absolute values are illustrated for the entire time-span (\SI{3}{\milli \second}) in the frequency-range of \SIrange{5}{100}{\kilo \hertz}.
        The \coi{} is indicated by white dash-dotted lines.
    } 
    \label{fig:I1_wavelet_full_IOB}
\end{figure}

To begin with, we consider noise-free sensor signals at the first impact location $\imp{1}$.
When performing the wavelet transform, we need to choose both the time-interval and the frequency-range of interest. 
As with the other \ToA{}-methods, we want to put as little prior knowledge as possible into our assessment, i.e., we assume not to know when the impact has occurred and which frequencies have been excited.
Although we have the time-histories of the excitation and of all sensor signals available, which give us a hint at which time-interval to look at when looking for inciding $S_0$ and $A_0$ waves, we deliberately consider the entire recorded time-interval of \SI{3}{\milli\second} and a frequency-range from \SIrange{5}{100}{\kilo\hertz} to begin with.
For this time-frequency window, Fig.~\ref{fig:I1_wavelet_full_IOB} shows the scalograms of the four sensor signals, i.e., the square of the absolute values of the respective CWTs, see Eq.~\eqref{eq::scalogram}, where a frequency-discretization of \SI{100}{\hertz} is used.
Note that we intentionally normalize each scalogram by its respective maximum value since the amplitudes of the signals vary considerably.
Such normalization allows us to illustrate details which are invisible if a common color scale is used. 

\begin{figure}[pos=!htbp]
    \centering
    \includegraphics[width=0.9\linewidth]{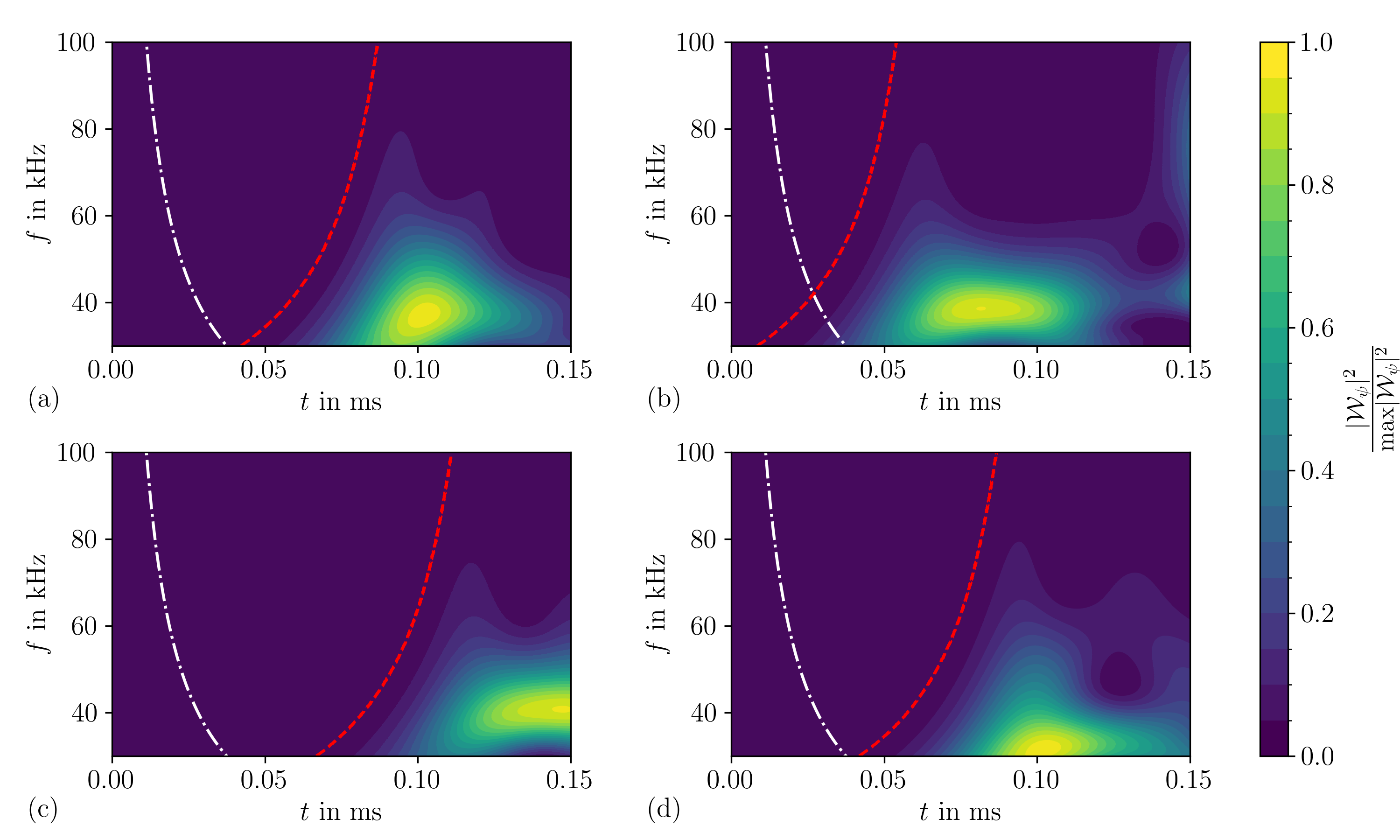}
    \caption{ 
        Scalograms of noise-free sensor signals $\sensor_1$--$\sensor_4$ for the \emph{idealized impact} $\imp{1}^{\iob}$: The normalized square of the CWT's absolute value is illustrated in time-spans of \SI{0.15}{\milli\second} and a frequency-range of \SIrange{30}{100}{\kilo \hertz}.
        Dashed red lines illustrate \ToA{}-estimates as a function of the frequency, i.e., times at which a (non-dimensional) threshold of $\sth = \num{1e-2}$ is reached.
        Dash-dotted white lines represent the \coi{}.   
    } 
    \label{fig:I1_wavelet_S0_IOB}
\end{figure}

For the four sensors $\sensor_1$--$\sensor_4$, the respective maxima of the scalograms occur after \SI{1}{\milli \second} well below \SI{20}{\kilo\hertz}. 
Nonetheless, we observe waves that arrive earlier, allowing to consider a smaller time-frequency window.
In other words, we adopt a zoomed view on the scalograms to identify further details. 
For this purpose, we consider a time-interval of \SI{0.15}{\milli\second} and a frequency-range of \SIrange{30}{100}{\kilo\hertz}, see Fig.~\ref{fig:I1_wavelet_S0_IOB}. 
The zoomed scalograms reveal local maxima that remain hidden in the global view of Fig.~\ref{fig:I1_wavelet_full_IOB}.
The dispersion curves, see Fig.~\ref{fig::lamb_wave_disp_curve}, suggest that the local maxima correspond to the symmetric $S_0$ mode, which, in the given frequency-range, propagates faster than the antisymmetric $A_0$ mode.
When it comes to identifying the \ToA{}, we note that local maxima of the scalograms occur at different frequencies.
For sensor $\sensor_4$, Fig.~\ref{fig:I1_wavelet_S0_IOB}~(d), the local maximum is clearly located at a lower frequency as compared to sensors $\sensor_1$--$\sensor_3$, Figs.~\ref{fig:I1_wavelet_S0_IOB}~(a)--(c).
Owing to the dispersive nature of lamb waves, i.e., different frequencies propagate at different velocities, we refrain from defining the \ToA{} in terms of temporal positions of local maxima in scalograms.
Instead, we propose a \emph{threshold-crossing} in the \emph{frequency-domain}, as previously introduced in \cite{Grasboeck_2025}.
In analogy to the time-domain, see Sec.~\ref{ssec::tc}, threshold-crossing in the frequency-domain means that we look for those times at which the scalogram exceeds a \emph{threshold value}.
For this purpose, we introduce the frequency as a parameter in our definition of the \ToA{} of sensor signal $s_i$, $i=1,\ldots,4$.  
{\small\begin{equation} \label{eq:cwt_tc}
    \test_i (f_{c,a}) = \min_{t_i \in \mathcal T} \left\{ t_i : \sclogr{s_i}{t}{f_{c,a}} / \bar{\mathcal W}^2 > \sth \right\} , \quad 
    \bar{\mathcal W} = \min_i \max_{t , f_{c,a}} \vert \mathcal W_\psi \{ s_i \} \vert , \quad
    i=1,\ldots,4.
\end{equation}}
As opposed to the time-domain, threshold-crossing in the frequency-domain does not give a single value for each sensor, but the \ToA{} is defined as a function of the frequency. 
Note that the threshold $\sth$ is a non-dimensional value since the scalograms are normalized by $\bar{\mathcal W}^2$, i.e., the smallest of the respective maximum values of the scalograms obtained from the four sensor signals in the considered time-frequency window.
The scaling eliminates the dependency on the amplitude of the excitation. 
The dashed red lines in Fig.~\ref{fig:I1_wavelet_S0_IOB} illustrate the \ToA{} functions for the four sensors $\sensor_1$--$\sensor_4$ for a threshold value of $\sth = \num{1e-2}$. 
Clearly, we see a non-linear dependency of $\test_i$ on the frequency, which may be related to the excitation on the one hand. 
On the other hand, it also reflects the fact that the $S_0$ mode propagates faster for lower frequencies, see Fig.~\ref{fig::lamb_wave_disp_curve}.

\begin{figure}[pos=!htbp]
    \centering
    \includegraphics[width=0.8\linewidth]{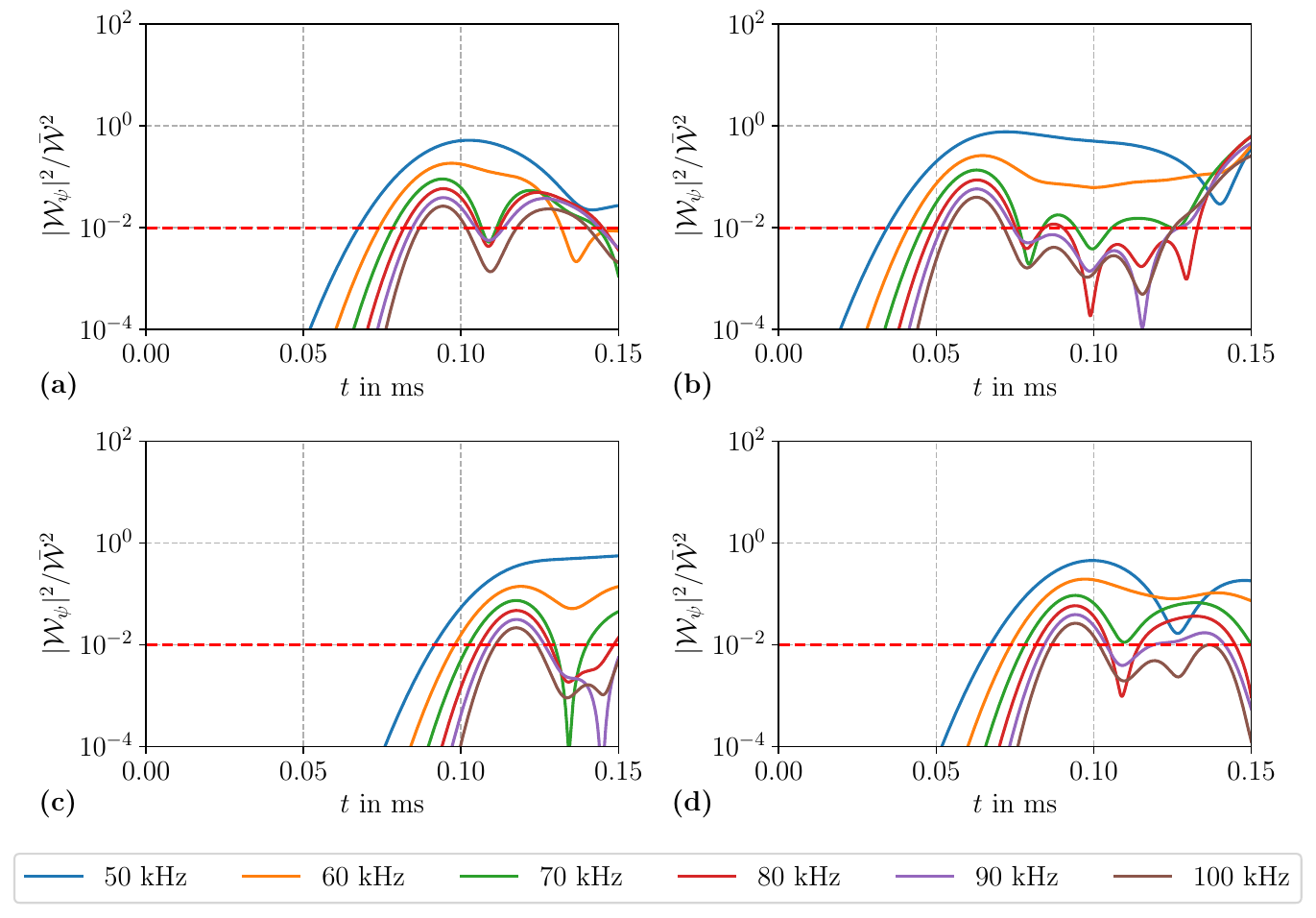}
    \caption{
        Sections of scalograms for noise-free sensor signals at \SIlist{50;60;70;80;90;100}{\kilo\hertz} for impact $\imp{1}^{\iob}$.
        The threshold $\sth = \num{1e-2}$ is indicated by dashed red lines; their (first) intersections with the scalograms' sections determine the \ToA{} for the respective frequencies.
    } 
    \label{fig:I1_wavelet_freq_S0_lines_IOB}
\end{figure}

To gain further insight, we consider sections along the temporal axis through the (normalized) scalograms for selected frequencies from \SIrange{50}{100}{\kilo\hertz} in steps of \SI{10}{\kilo\hertz}, see Fig.~\ref{fig:I1_wavelet_freq_S0_lines_IOB}, where the threshold is again indicated by dashed red lines. 
The first (time-wise) intersection of the threshold lines with the sections of the scalograms represent the \ToA{} for the proposed threshold crossing in the frequency-domain.
The corresponding numerical values for the \ToA{} at the sensors are listed in the first four columns of Tab.~\ref{tab:I1_wavelet_freq_lines}.
\begin{table}[pos=!htbp]
    \small
    \centering
    \begin{tabular}[]{r|cccc|cccc}
        \toprule
        Frequency in \SI{}{\kilo\hertz} 
        & \multicolumn{4}{l|}{Threshold crossing in \si{\milli\second} – $\imp{1}^{\iob}$} 
        & \multicolumn{4}{l}{Threshold crossing in \si{\milli\second} – $\imp{1}^{\iop}$} \\
        & $\sensor_1$ &  $\sensor_2$ &  $\sensor_3$ &  $\sensor_4$ 
        & $\sensor_1$ &  $\sensor_2$ &  $\sensor_3$ &  $\sensor_4$ 
        \\
        \midrule
        \num{50} & 
        \num{0.0676} & \num{0.0346} & \num{0.0916} & \num{0.0672} &
        \num{0.0726} & \num{0.0396} & \num{0.0966} & \num{0.0722}
        \\
        \num{60} & 
        \num{0.0740} & \num{0.0412} & \num{0.0982} & \num{0.0738} &  
        \num{0.0774} & \num{0.0446} & \num{0.1014} & \num{0.0772}
        \\
        \num{70} & 
        \num{0.0784} & \num{0.0456} & \num{0.1026} & \num{0.0782} & 
        \num{0.0812} & \num{0.0488} & \num{0.1052} & \num{0.0810}
        \\
        \num{80} & 
        \num{0.0820} & \num{0.0492} & \num{0.1060} & \num{0.0818} & 
        \num{0.0848} & \num{0.0522} & \num{0.1086} & \num{0.0846}
        \\
        \num{90} & 
        \num{0.0846} & \num{0.0518} & \num{0.1088} & \num{0.0844} & 
        \num{0.0876} & \num{0.0550} & \num{0.1114} & \num{0.0874}
        \\
        \num{100} & 
        \num{0.0870} & \num{0.0540} & \num{0.1110} & \num{0.0868} &
        \num{0.0902} & \num{0.0576} & \num{0.1140} & \num{0.0900}
        \\
        \bottomrule
    \end{tabular}
    \caption{
        Estimated \ToA{} at frequencies \SIlist{50;60;70;80;90;100}{\kilo\hertz} for a threshold of $\sth = \num{1e-2}$: Idealized impact $\imp{1}^{\iob}$ and experiment-based impact $\imp{1}^{\iop}$ for noise-free sensor signals.
    }
    \label{tab:I1_wavelet_freq_lines}
\end{table}%
To reduce the dependency of the \ToA{} on the frequency, we consider \emph{relative times}, i.e., time-differences $\test_i - \test_2$, $i=1,3,4$, relative to \ToA{} of sensor $\sensor_2$, which is located closest to the impact $\imp{1}$.
The results, which are given in the first three colums of Tab.~\ref{tab:I1_wavelet_freq_lines_diff} for the idealized impact $\imp{1}^{\iob}$ are remarkable:
The relative times coincide up to the sampling interval of \SI{0.2}{\micro\second} for all frequencies considered, which range from \SI{30}{\kilo \hertz} to \SI{100}{\kilo \hertz}.
In other words, the frequency-dependency of the \ToA{} is completely eliminated when considering time-differences rather than absolute times.
\begin{table}[pos=!htbp]
    \small
    \centering
    \begin{tabular}[]{r|ccc|ccc}
        \toprule
        Frequency in \SI{}{\kilo\hertz} 
        & \multicolumn{3}{l|}{Relative times in \si{\milli\second} – $\imp{1}^{\iob}$} 
        & \multicolumn{3}{l}{Relative times in \si{\milli\second} – $\imp{1}^{\iop}$} \\
        & ~~$\test_1 - \test_2$ ~~& ~~ $\test_3 - \test_2$ ~~& ~~ $\test_4 - \test_2$~~ 
        & ~~$\test_1 - \test_2$ ~~& ~~ $\test_3 - \test_2$ ~~ &~~  $\test_4 - \test_2$ ~~
        \\
        \midrule
        \num{50} & 
        \num{0.0330} & \num{0.0570} & \num{0.0326} &
        \num{0.0330} & \num{0.0570} & \num{0.0326} 
        \\
        \num{60} & 
        \num{0.0328} & \num{0.0570} & \num{0.0326} &
        \num{0.0328} & \num{0.0568} & \num{0.0326} 
        \\
        \num{70} & 
        \num{0.0328} & \num{0.0570} & \num{0.0326} &
        \num{0.0324} & \num{0.0564} & \num{0.0322} 
        \\
        \num{80} & 
        \num{0.0328} & \num{0.0568} & \num{0.0326} & 
        \num{0.0326} & \num{0.0564} & \num{0.0324} 
        \\
        \num{90} & 
        \num{0.0328} & \num{0.0570} & \num{0.0326} &
        \num{0.0326} & \num{0.0564} & \num{0.0324}  
        \\
        \num{100} & 
        \num{0.0330} & \num{0.0570} & \num{0.0328} & 
        \num{0.0326} & \num{0.0564} & \num{0.0324}  
        \\
        \bottomrule
    \end{tabular}
    \caption{
        Time-differences of \ToA{}-estimates for noise-free sensor signals at frequencies \SIlist{50;60;70;80;90;100}{\kilo\hertz}, relative to the closest sensor to the impact, for both the idealized impact $\imp{1}^{\iob}$ and the experiment-based impact $\imp{1}^{\iop}$.
    }
    \label{tab:I1_wavelet_freq_lines_diff}
\end{table}

\begin{figure}[pos=!htbp]
    \centering
    \includegraphics[width=0.9\linewidth]{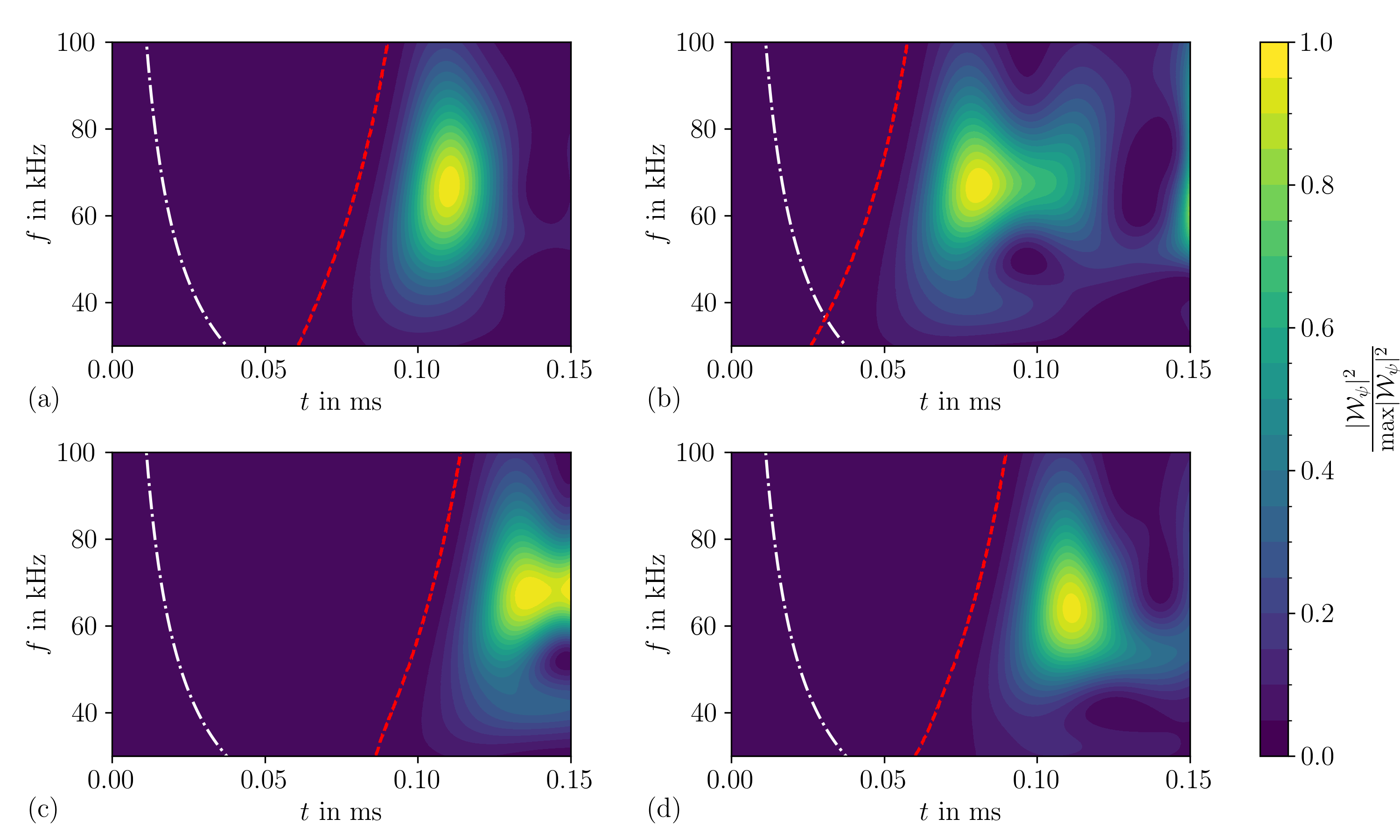}
    \caption{
        Scalograms of noise-free sensor signals $\sensor_1$--$\sensor_4$ for the \emph{experiment-based impact} $\imp{1}^{\iop}$: The normalized square of the CWT's absolute value is illustrated in time-spans of \SI{0.15}{\milli\second} and a frequency-range of \SIrange{30}{100}{\kilo \hertz}.
        Dashed red lines illustrate \ToA{}-estimates as functions of the frequency, i.e., times at which a (non-dimensional) threshold of $\sth = \num{1e-2}$ is reached. Dash-dotted white lines represent the \coi{}.   
    } 
    \label{fig:I1_wavelet_S0_IOP}
\end{figure}

\begin{figure}[pos=!htbp]
    \centering
    \includegraphics[width=0.8\linewidth]{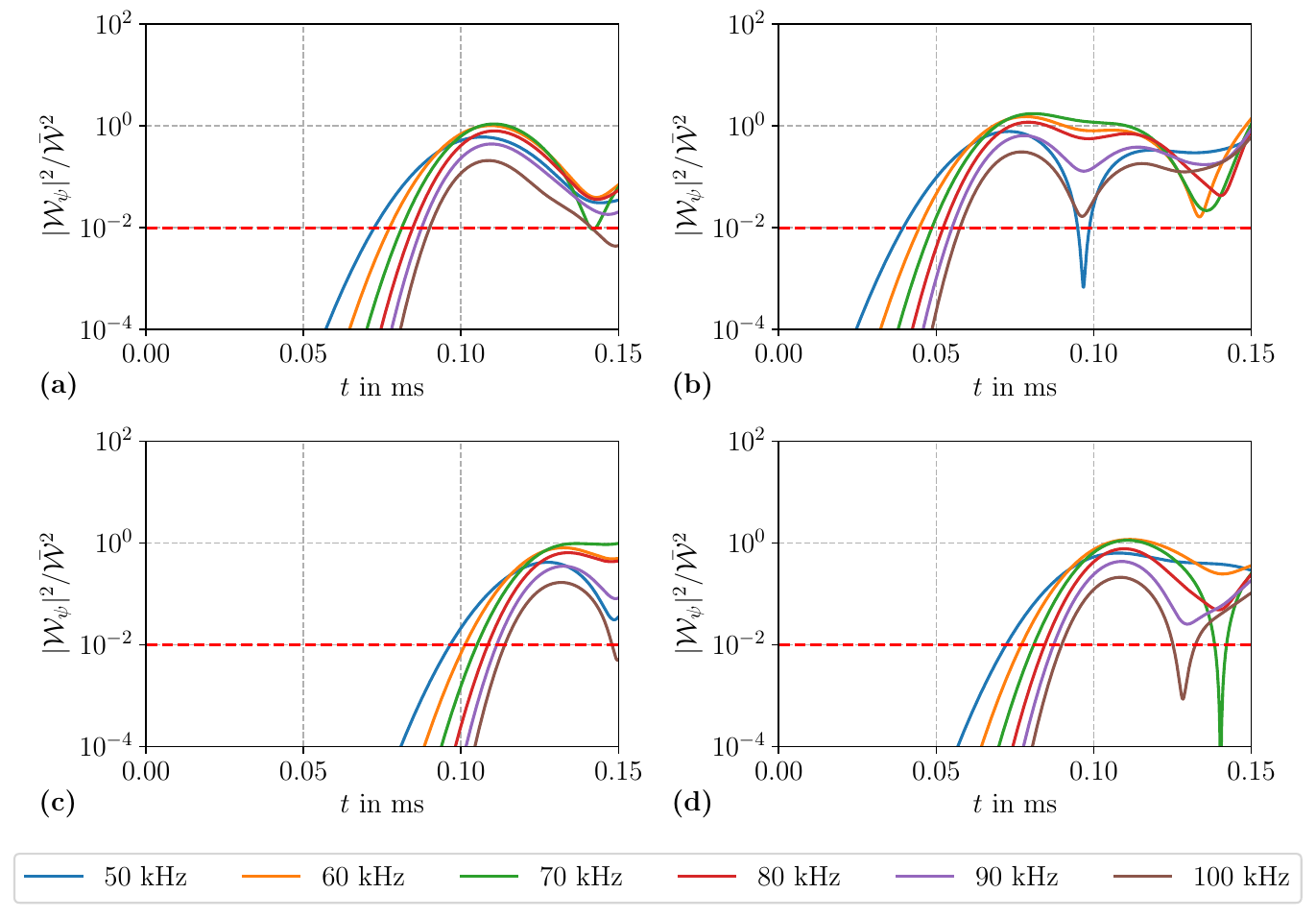}
    \caption{
        Sections of scalograms for noise-free sensor signals at \SIlist{50;60;70;80;90;100}{\kilo\hertz} for impact $\imp{1}^{\iop}$.
        The threshold $\sth = \num{1e-2}$ is indicated by dashed red lines; their (first) intersections with the scalograms' sections determine the \ToA{} for the respective frequencies. 
    } 
    \label{fig:I1_wavelet_freq_lines_IOP}
\end{figure}

\begin{figure}[pos=!htbp]
    \centering
    \includegraphics[width=0.9\linewidth]{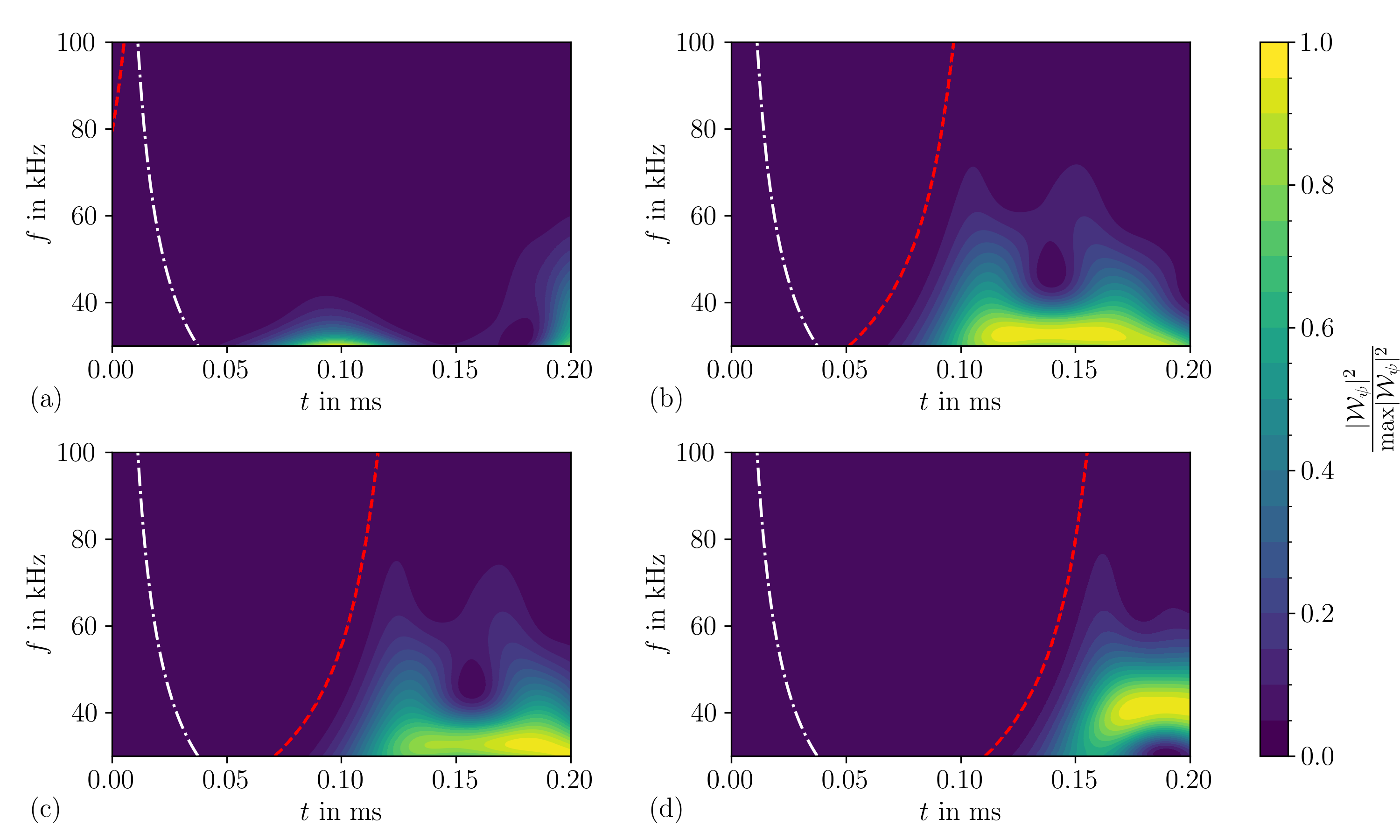}
    \caption{ 
        Scalograms of noise-free sensor signals $\sensor_1$--$\sensor_4$ for \emph{idealized impact} $\imp{2}^{\iob}$: The normalized square of the CWT's absolute value is illustrated for a frequency-range of \SIrange{30}{100}{\kilo \hertz} and time-windows of \SI{0.2}{\milli\second} each. 
        Dashed red lines illustrate \ToA{}-estimates as functions of the frequency, i.e., times at which a (non-dimensional) threshold of $\sth = \num{1e-2}$ is reached. Dash-dotted white lines represent the \coi{}.
    } 
    \label{fig:I2_wavelet_S0_IOB}
\end{figure}

We next consider the second kind of impact $\imp{1}^\iop$, which is derived from an experimental setup of a shaker impacting a plate.
Figure~\ref{fig:I1_wavelet_S0_IOP} shows the scalograms of the sensor signals for the same time-frequency window as for the idealized impact.
We observe that the local maxima of the scalograms are shifted to higher frequencies, which can be attributed to the modified excitation. 
Using the same non-dimensional threshold of $\sth = \num{1e-2}$, we also note  that the dependency of \ToA{} on the frequency is visibly different from that of the idealized impact.
As for the latter, however, the lower frequency-content reaches the threshold value earlier than high frequencies. 
In Fig.~\ref{fig:I1_wavelet_freq_lines_IOP}, sections through the scalogram at frequencies from \SI{50}{\kilo\hertz} to \SI{100}{\kilo\hertz} in steps of \SI{10}{\kilo\hertz} are illustrated together with the treshold value (dashed lines).
The numerical values of corresponding \ToA{} for these frequencies are given in the last four columns of Tab.~\ref{tab:I1_wavelet_freq_lines}.
Across the frequencies considered, the \ToA{} is slightly delayed as compared to the idealized impact $\imp{1}^{\iob}$.
Considering relative times, i.e., differences among estimated \ToA{} w.r.t. the second sensor $\sensor_2$, we obtain an almost perfect agreement with the results of the idealized impact, see Tab.~\ref{tab:I1_wavelet_freq_lines_diff}.
The relative times differ by only three sampling times, i.e., \SI{0.6}{\micro\second}, at most.
The consistency of \ToA{}-estimates for both types of impacts at the first impact position $\imp{1}$ is noteworthy.
So, it will be interesting to see how well these results translate to the second impact location, where the excitation occurs very close to sensor $\sensor_1$.
As for the first impact, we first consider the idealized impact $\imp{2}^{\iob}$. 
The scalograms obtained from the four sensor signals are illustrated in Fig.~\ref{fig:I2_wavelet_S0_IOB}.
Note that the time-window is increased to \SI{0.2}{\milli\second}, since the distance between the impact location and the farthermost sensor ($\sensor_4$) is greater than for impact $\imp{1}$. 
The frequency-range, however, remains the same, and so does the threshold, which is chosen as $\sth = \num{1e-2}$. 
For sensors $\sensor_2$--$\sensor_4$, see Figs~\ref{fig:I2_wavelet_S0_IOB}~(b)--(d), we find similar \ToA{}-estimates as for the first impact location $\imp{1}^\iob$, also in terms of their frequency-dependency.
The scalogram for sensor $\sensor_1$, which is closest to the impact, clearly differs from the scalograms of the other sensors, and also from what we observed for the impacts at position $\imp{1}$, cf.~Figs.~\ref{fig:I1_wavelet_S0_IOB} and \ref{fig:I1_wavelet_S0_IOP}.
It does not show distinct local maxima in the given time-frequency window. 
Moreover, the scalogram of $\sensor_1$ crosses the threshold value only for comparatively high frequencies of \SI{80}{\kilo\hertz} and above, and the estimated \ToA{} lies to the left of the \coi{} boundary, where the CWT is affected by the cut-off at the boundaries of the time-domain.
The reasons for the --at first sight-- odd scalogram in Fig.~\ref{fig:I2_wavelet_S0_IOB} (a) are twofold: First of all, recall that scalograms shown above are normalized relative to their respective maximum values, i.e., the colors of the individual contour plots cannot be compared directly. 
Secondly, the scalogram of sensor $\sensor_1$ exceeds the threshold value in almost the entire time-frequency window except for high frequencies at the beginning of the time-interval.
That is why no ``crossing'' occurs for much of the considered frequency-domain.
\begin{figure}[pos=!htbp]
    \centering
    \includegraphics[width=0.8\linewidth]{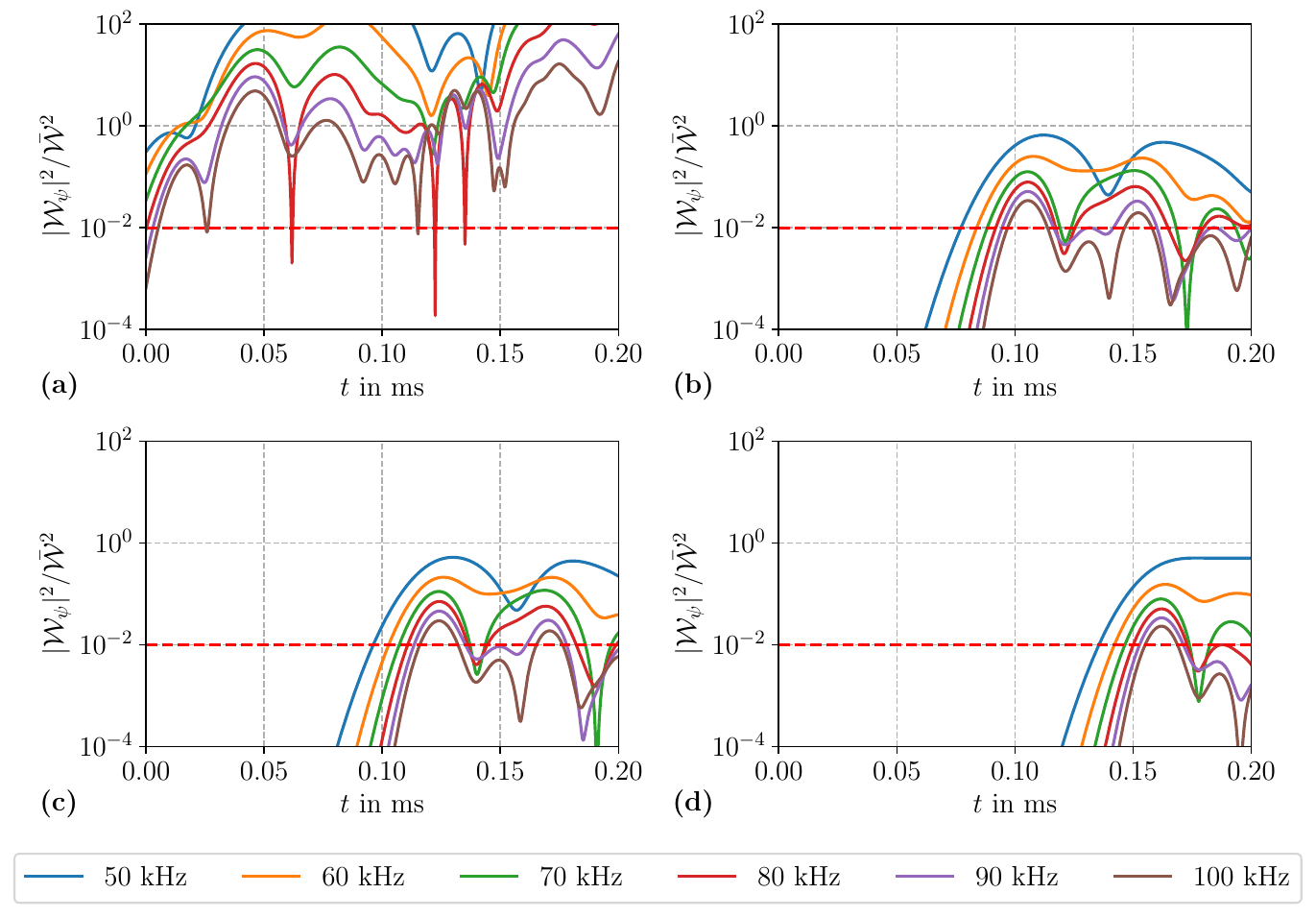}
    \caption{
        Sections of scalograms for noise-free sensor signals at \SIlist{50;60;70;80;90;100}{\kilo\hertz} for impact $\imp{2}^{\iob}$.
        The threshold $\sth = \num{1e-2}$ is indicated by dashed red lines; their (first) intersections with the scalograms' sections determine the \ToA{} for the respective frequencies. 
    } 
    \label{fig:I2_wavelet_freq_lines_IOB}
\end{figure}
This fact is clearly illustrated by Fig.~\ref{fig:I2_wavelet_freq_lines_IOB}, which shows sections through scalograms, which are normalized by a common factor $\bar{\mathcal W}^2$, see Eq.~\eqref{eq:cwt_tc}. 
The threshold $\sth = \num{1e-2}$ is again indicated by dashed red lines.
The sections further reveal a fundamental limitation of the proposed approach to estimate the \ToA{}:
If scalograms differ too much in amplitude, we may not find a threshold level that is crossed in the specified time-frequency for all sensor signals. 
For sensor $\sensor_1$, increasing the threshold to \num{1e-1} approximately doubles the frequency-range, in which a \ToA{} can be estimated.
However, sensors $\sensor_2$--$\sensor_4$ reach this threshold value only for frequencies up to approximately \SI{70}{\kilo\hertz}.
So, the frequency-range, in which all four sensors give a \ToA{}, is effectively smaller for $\sth = \num{1e-1}$. 

\begin{figure}[pos=!htbp]
    \centering
    \includegraphics[width=0.9\linewidth]{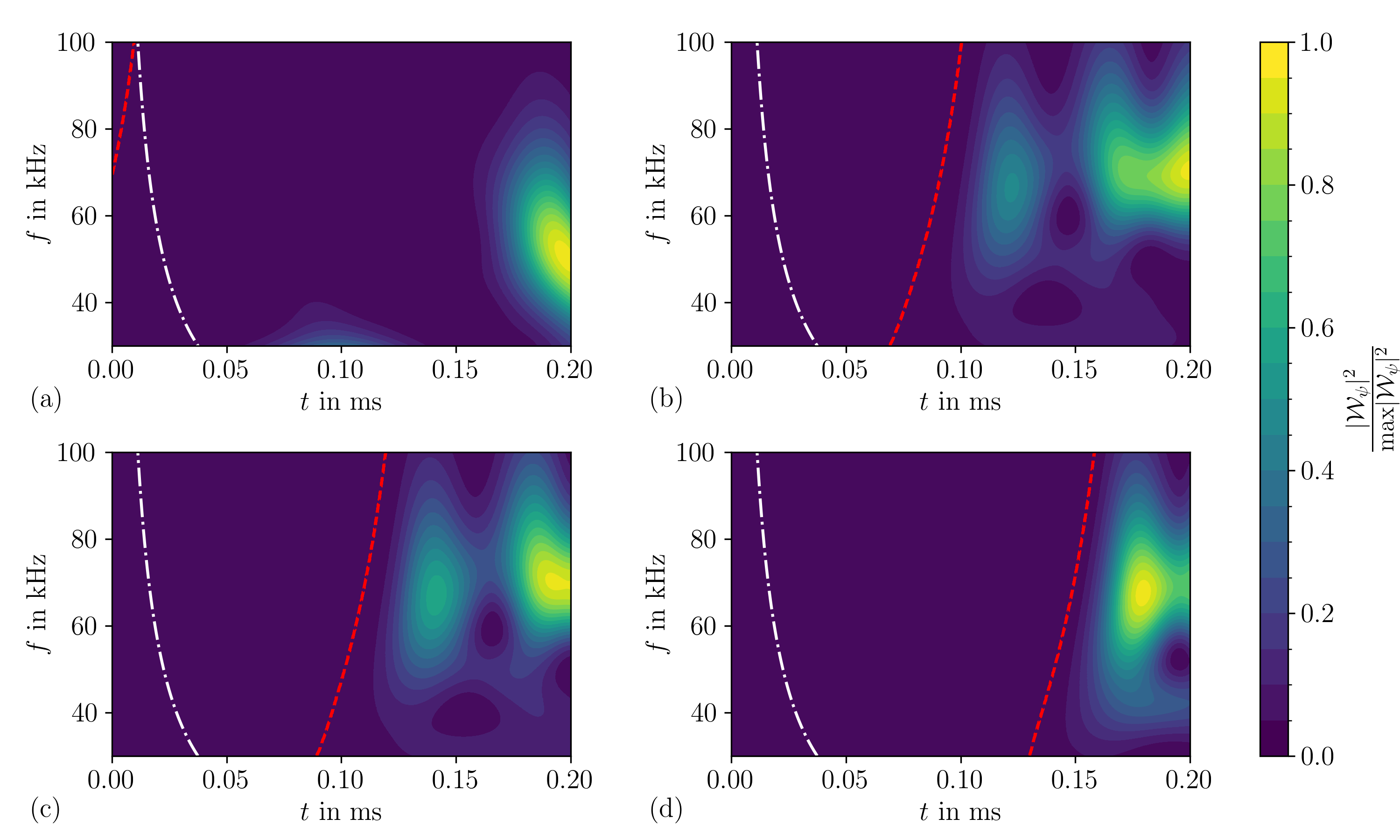}
    \caption{
        Scalograms of noise-free sensor signals $\sensor_1$--$\sensor_4$ for the experiment-based $\imp{2}^{\iop}$: The normalized square of the CWT's absolute value is illustrated in time-spans of \SI{0.2}{\milli\second} and a frequency-range of \SIrange{30}{100}{\kilo \hertz}. 
        Dashed red lines illustrate \ToA{}-estimates as functions of the frequency, i.e., times at which a (non-dimensional) threshold of $\sth = \num{1e-2}$ is reached. Dash-dotted white lines represent the \coi{}. 
    } 
    \label{fig:I2_wavelet_S0_IOP}
\end{figure}

\begin{figure}[pos=!htbp]
    \centering
    \includegraphics[width=0.8\linewidth]{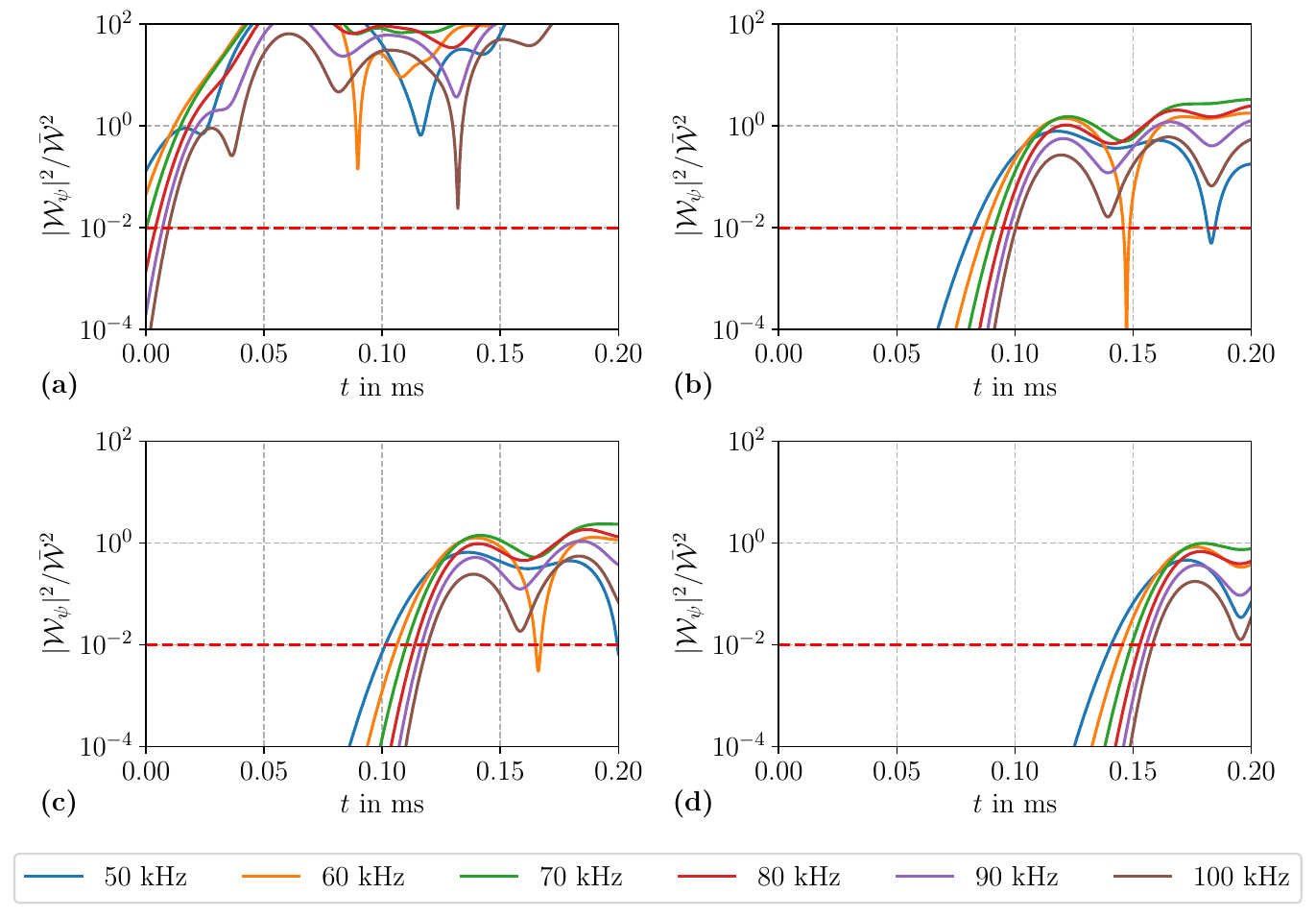}
    \caption{
        Sections of scalograms for noise-free sensor signals at \SIlist{50;60;70;80;90;100}{\kilo\hertz} for impact $\imp{2}^{\iop}$:
        The threshold $\sth = \num{1e-2}$ is indicated by dashed red lines; their (first) intersections with the scalograms' sections determine the \ToA{} for the respective frequencies.  
    } 
    \label{fig:I2_wavelet_freq_lines_IOP}
\end{figure}

Above considerations also hold for the experiment-based impact $\imp{2}^{\iop}$, for which the scalograms are shown in Fig.~\ref{fig:I2_wavelet_S0_IOP}.
The threshold is again set to a value of $\sth = \num{1e-2}$.
As for impact position $\imp{1}$, local maxima in the scalogram occur at higher frequencies for the experiment-based impact $\imp{2}^{\iop}$ than for idealized impact $\imp{2}^{\iob}$.
By eyeball, also the frequency-dependency of \ToA{}-estimates is close to what we have observed for the $\imp{1}^{\iop}$ impact, see Fig.~\ref{fig:I1_wavelet_S0_IOP}. 
The sections through the scalograms shown in Fig.~\ref{fig:I2_wavelet_freq_lines_IOP} again illustrate that no threshold exists such that \ToA{}-estimates can be made for the whole frequency-range of \SIrange{50}{100}{\kilo\hertz} in steps of \SI{10}{\kilo\hertz}.
For sensor $\sensor_1$, the range of frequencies, in which a \ToA{} can be determined by threshold crossing, is slightly larger as compared to the idealized impact $\imp{2}^{\iob}$ (compare Figs. \ref{fig:I2_wavelet_freq_lines_IOB} (a) and \ref{fig:I2_wavelet_freq_lines_IOP} (a)).
For both types of impacts, Tab.~\ref{tab:I2_wavelet_freq_lines}~lists the numerical values for \ToA{}-estimates at the frequencies illustrated in Figs.~\ref{fig:I2_wavelet_freq_lines_IOB} and~\ref{fig:I2_wavelet_freq_lines_IOP}.
Note that \ToA{}-estimates are assumed as zero if scalograms exceed the threshold from the outset.
\begin{table}[pos=!htbp]
    \small
    \centering
    \begin{tabular}[]{r|cccc|cccc}
        \toprule
        Frequency in \SI{}{\kilo\hertz}
        & \multicolumn{4}{l|}{Threshold crossing in \si{\milli\second} – $\imp{2}^{\iob}$} 
        & \multicolumn{4}{l}{Threshold crossing in \si{\milli\second} – $\imp{2}^{\iop}$} \\
        & $\sensor_1$ &  $\sensor_2$ &  $\sensor_3$ &  $\sensor_4$ 
        & $\sensor_1$ &  $\sensor_2$ &  $\sensor_3$ &  $\sensor_4$ 
        \\
        \midrule
        \num{50} & 
        \num{0.0000} & \num{0.0772} & \num{0.0964} & \num{0.1356} &
        \num{0.0000} & \num{0.0824} & \num{0.1014} & \num{0.1408}
        \\
        \num{60} & 
        \num{0.0000} & \num{0.0838} & \num{0.1028} & \num{0.1422} &  
        \num{0.0000} & \num{0.0874} & \num{0.1064} & \num{0.1456}
        \\
        \num{70} & 
        \num{0,0000} & \num{0.0884} & \num{0.1074} & \num{0.1466} & 
        \num{0.0002} & \num{0.0914} & \num{0.1104} & \num{0.1494}
        \\
        \num{80} & 
        \num{0,0002} & \num{0.0920} & \num{0.1108} & \num{0.1502} & 
        \num{0.0042} & \num{0.0950} & \num{0.1138} & \num{0.1528}
        \\
        \num{90} & 
        \num{0,0030} & \num{0.0946} & \num{0.1136} & \num{0.1528} & 
        \num{0.0072} & \num{0.0978} & \num{0.1168} & \num{0.1558}
        \\
        \num{100} & 
        \num{0,0052} & \num{0.0970} & \num{0.1160} & \num{0.1552} &
        \num{0.0096} & \num{0.1004} & \num{0.1194} & \num{0.1584}
        \\
        \bottomrule
    \end{tabular}
    \caption{
        Estimated \ToA{} at frequencies \SIlist{50;60;70;80;90;100}{\kilo\hertz} for a threshold of $\sth = \num{1e-2}$: Idealized impact $\imp{2}^{\iob}$ and experiment-based impact $\imp{2}^{\iop}$ for noise-free sensor signals.
    }
    \label{tab:I2_wavelet_freq_lines}
\end{table}%
In spite of the limited frequency-range (\SIrange{80}{100}{\kilo\hertz}), in which \ToA{}-estimates can be determined for the four sensors, the results for relative times, i.e., differences to the \ToA{} of sensor $\sensor_1$, show excellent consistency, see Tab. \ref{tab:I2_wavelet_freq_lines_diff}.
Across the different frequencies considered, the relative times differ with no more than a single sampling time ($\dt=\SI{0.2}{\micro\second}$) for either kind of impact. 
Comparing the two types of impact, the relative \ToA{} of sensor $\sensor_4$ (c.f. third and last column in Tab. \ref{tab:I2_wavelet_freq_lines_diff}) shows the largest deviation at \qty{80}{\kilo\hertz}, i.e., only $\qty{0.1500}{\milli\second} - \qty{0.1486}{\milli\second} = \SI{1.4}{\micro\second}$.
\begin{table}[pos=!htbp]
    \small
    \centering
    \begin{tabular}[]{r|ccc|ccc}
        \toprule
        Frequency in \SI{}{\kilo\hertz}
        & \multicolumn{3}{l|}{Relative times in \si{\milli\second} – $\imp{2}^{\iob}$} 
        & \multicolumn{3}{l}{Relative times in \si{\milli\second} – $\imp{2}^{\iop}$} \\
        & ~~$\test_2 - \test_1$ ~~& ~~ $\test_3 - \test_1$ ~~& ~~ $\test_4 - \test_1$~~ 
        & ~~$\test_2 - \test_1$ ~~& ~~ $\test_3 - \test_1$ ~~ &~~  $\test_4 - \test_1$ ~~
        \\
        \midrule
        \num{80}& 
        \num{0.0918} & \num{0.1106} & \num{0.1500} & 
        \num{0.0908} & \num{0.1096} & \num{0.1486} 
        \\
        \num{90}& 
        \num{0.0916} & \num{0.1106} & \num{0.1498} & 
        \num{0.0906} & \num{0.1096} & \num{0.1486} 
        \\
        \num{100}& 
        \num{0.0918} & \num{0.1108} & \num{0.1500} & 
        \num{0.0908} & \num{0.1098} & \num{0.1488} 
        \\
        \bottomrule
    \end{tabular}
    \caption{
        Time-differences of \ToA{}-estimates for noise-free sensor signals at frequencies \SIlist{80;90;100}{\kilo\hertz}, relative to the closest sensor to the impact, for the idealized impact $\imp{2}^{\iob}$ and experiment-based impact $\imp{2}^{\iop}$.
    }
    \label{tab:I2_wavelet_freq_lines_diff}
\end{table}%

\begin{figure}[pos=!htbp]
    \centering
    \includegraphics[width=0.9\linewidth]{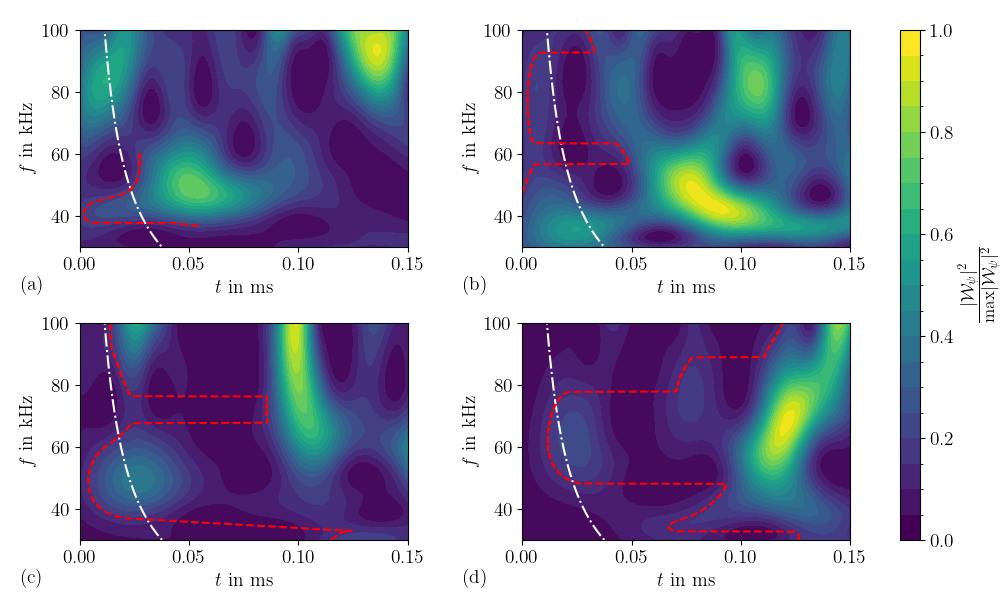}
    \caption{ 
        Scalograms of noise-contaminated sensor signals $\sensor_1$--$\sensor_4$ for the \emph{idealized impact} $\imp{1}^{\iob}$: The normalized square of the CWT's absolute value is illustrated in time-spans of \SI{0.15}{\milli\second} and a frequency-range of \SIrange{30}{100}{\kilo \hertz}.
        Dashed red lines illustrate \ToA{}-estimates as a function of the frequency, i.e., times at which a (non-dimensional) threshold of $\sth = \num{2e-1}$ is reached.
        Dash-dotted white lines represent the \coi{}.   
    } 
    \label{fig:I1_wavelet_S0_IOB_snr50db}
\end{figure}

\begin{figure}[pos=!htbp]
    \centering
    \includegraphics[width=0.8\linewidth]{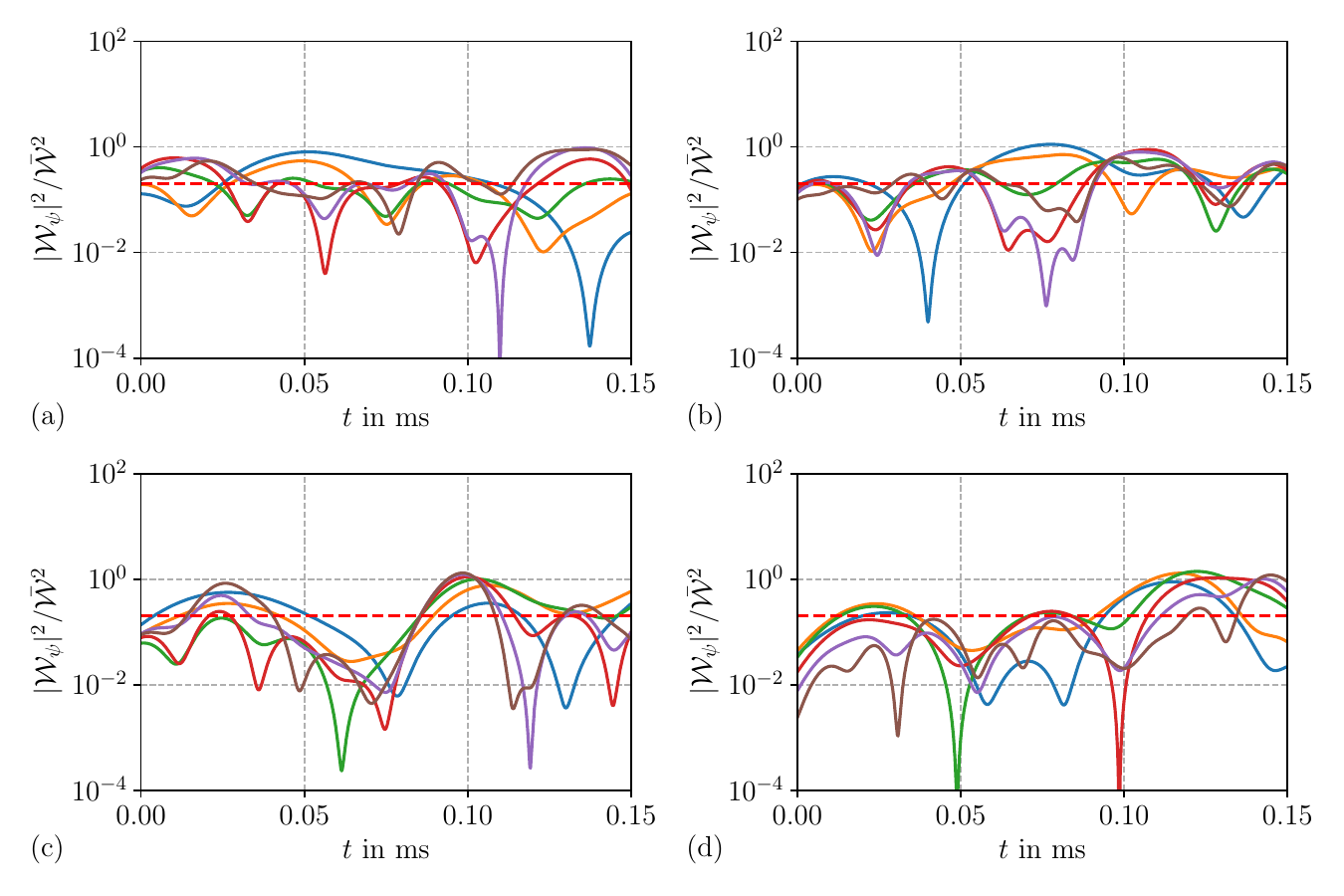}
    \caption{
        Sections of scalograms for noise-contaminated sensor signals at \SIlist{50;60;70;80;90;100}{\kilo\hertz} for impact $\imp{1}^{\iob}$.
        The threshold $\sth = \num{2e-1}$ is indicated by dashed red lines; their (first) intersections with the scalograms' sections determine the \ToA{} for the respective frequencies.
    } 
    \label{fig:I1_wavelet_freq_S0_lines_IOB_snr50db}
\end{figure}

So, even for the challenging case of an impact occuring right next to one of the sensors, we obtain \ToA{}-estimates and relative time-differences accurately and consistently, though possibly only at frequencies that are comparatively high in view of the spectrum of excitation.
As stated above, by restricting the time-domain to intervals, in which sensors begin to measure incident lamb waves, we focus on the detection of the symmetric $S_0$ mode, which propagates faster than the antisymmetric $A_0$ mode in the frequency-range considered.
In more realistic scenarios, however, such as increased damping and/or the presence of measurement noise, the $S_0$ mode may no longer be measurable.
This is primarily due to its comparatively small amplitudes and the limited sensitivity of surface-bonded piezoceramic patch transducers w.r.t. symmetric modes \cite{Ciampa_2010, Merlo_2017, Kundu_2009, Zhu_2017, Grasboeck_smart2023}.
In the present study, this situation is explicitly encountered for the noise-contaminated sensor signals, where the initial $S_0$ arrival is masked by noise and cannot be robustly detected, as demonstrated in Figs. \ref{fig:I1_wavelet_S0_IOB_snr50db} and \ref{fig:I1_wavelet_freq_S0_lines_IOB_snr50db}.
For this analysis, the threshold level is already increased to $\sth = \num{2e-1}$ in order to prevent threshold-crossings caused by the noise floor at the very beginning of the time-window.
Consequently, for noise-contaminated signals the analysis is shifted towards the $A_0$ mode, which exhibits significantly larger amplitudes.
This naturally raises the question of whether threshold-crossing in the frequency domain can be exploited not only for identifying the earliest $S_0$ in ideal signals, but also for extracting meaningful \ToA{}-estimates of the $A_0$ mode in the presence of noise; an issue that is addressed in the following.

\begin{figure}[pos=!htbp]
    \centering
    \includegraphics[width=0.9\linewidth]{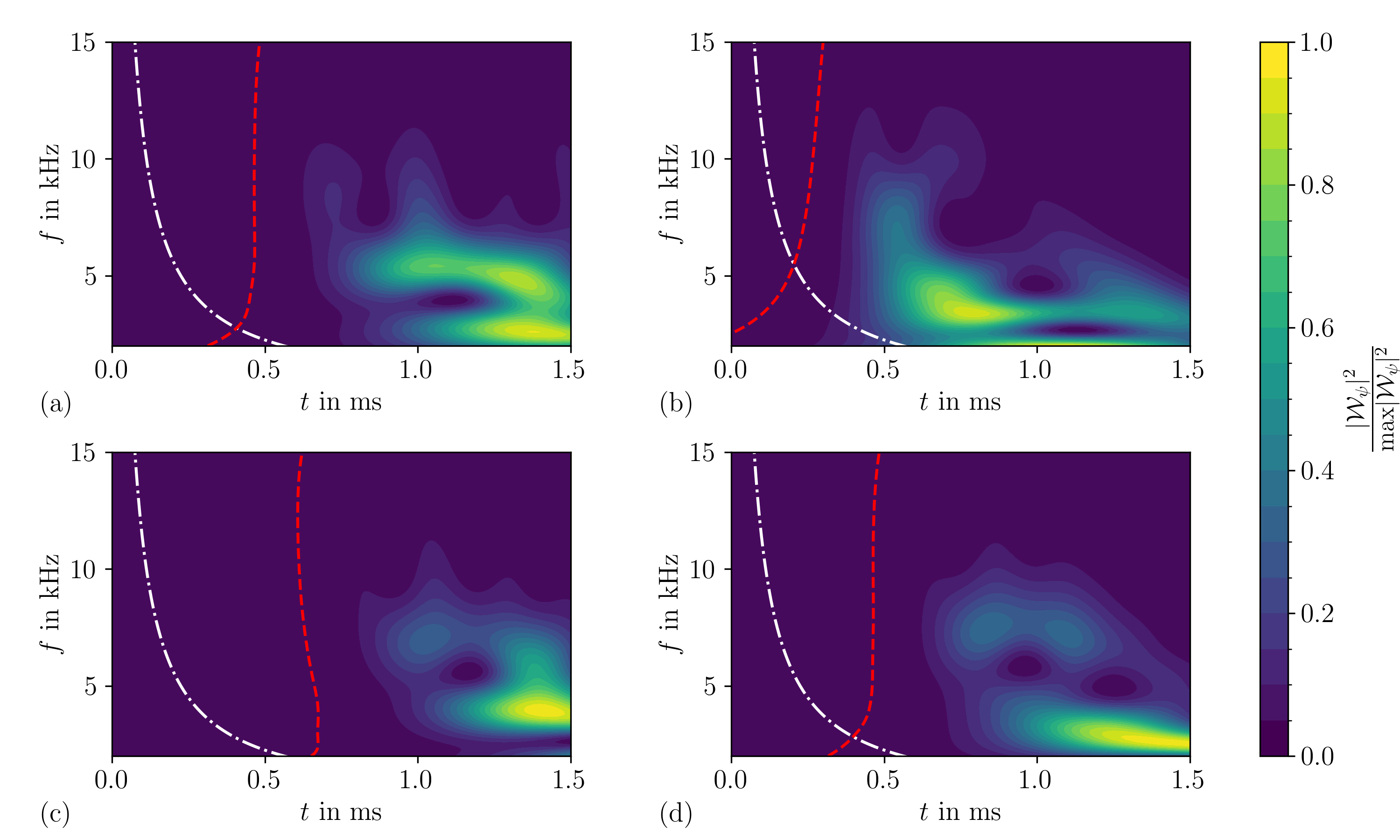}
    \caption{ 
        Scalograms of noise-contaminated sensor signals $\sensor_1$--$\sensor_4$ for the \emph{idealized impact} $\imp{1}^{\iob}$: The normalized square of the CWT's absolute value is illustrated in time-spans of \SI{1.5}{\milli\second} and a frequency-range of \SIrange{2}{15}{\kilo \hertz}.
        Dashed red lines illustrate \ToA{}-estimates as functions of the frequency, i.e., times at which a (non-dimensional) threshold of $\sth = \num{1e-3}$ is reached. Dash-dotted white lines represent the \coi{}.   
    } 
    \label{fig:I1_wavelet_A0_IOB_snr50db}
\end{figure}

\begin{figure}[pos=!htbp]
    \centering
    \includegraphics[width=0.8\linewidth]{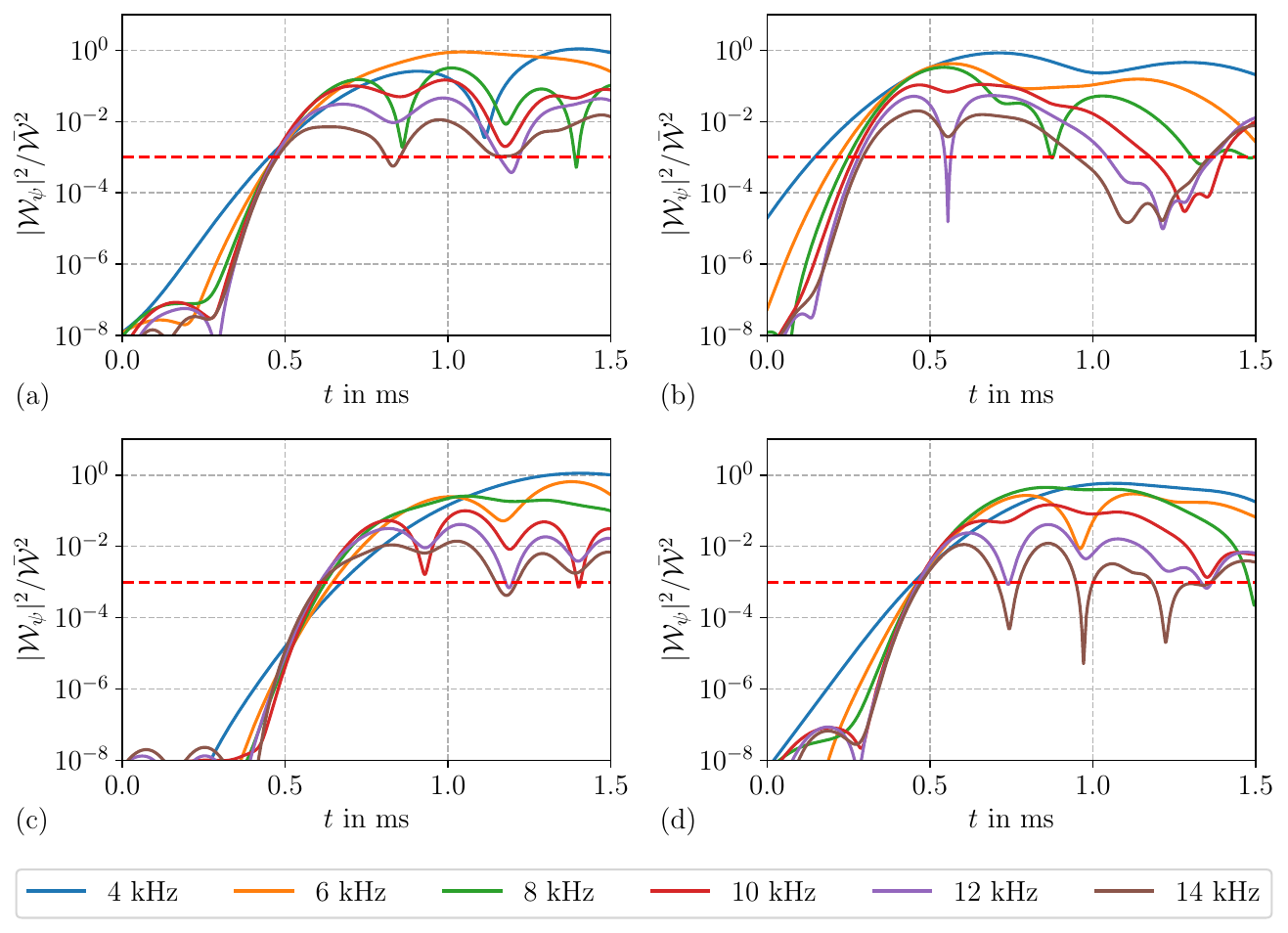}
    \caption{
        Sections of scalograms for noise-contaminated sensor signals at \SIlist{4;6;8;10;12;14}{\kilo\hertz} for impact $\imp{1}^{\iob}$.
        The threshold $\sth = \num{1e-3}$ is indicated by dashed red lines; their (first) intersections with the scalograms' sections determine the \ToA{} for the respective frequencies. 
    } 
    \label{fig:I1_wavelet_freq_A0_lines_IOB_snr50db}
\end{figure}

To adapt the proposed approach to the $A_0$ mode, larger time-windows need to be considered, since the $A_0$ mode propagates with less than half of the velocity of the $S_0$ mode as shown by the dispersion curves in Fig.~\ref{fig::lamb_wave_disp_curve}.
For instance, while the $S_0$ mode propagates with a group wave speed of $\cgroup = \SI{5391}{\metre\per\second}$ at \SI{50}{\kilo\hertz}, the $A_0$ mode propagates at the same frequency with a significantly slower group wave speed of $\cgroup = \SI{1775}{\metre\per\second}$, for the present aluminum plate.
Secondly, we narrow down the frequency-range to \SI{2}{\kilo\hertz}~-~\SI{15}{\kilo\hertz} in order to capture those regions of the scalograms which were already visible in the global view of Fig.~\ref{fig:I1_wavelet_full_IOB}. 
Figure~\ref{fig:I1_wavelet_A0_IOB_snr50db} shows the scalograms of noise-contaminated signals in that time-frequency window for the idealized impact applied at the first position $\imp{1}^{\iob}$.
The (non-dimensional) threshold is chosen as $\sth = \num{1e-3}$, which results in \ToA{}-estimates as indicated by the dashed red lines. 
Note that the threshold is smaller than what we have used for determining the \ToA{} of the $S_0$ mode. 
Due to the much larger amplitudes induced by the $A_0$ mode, which is now contained in the time-frequency window, it is approximately eight orders of magnitude greater in terms of absolute values.
As indicated by dashed red lines, the chosen threshold gives \ToA{}-estimates throughout the entire frequency-range from \SIrange{2}{15}{\kilo\hertz} for all four sensors. 
Unlike the previous results for the $S_0$ mode, the dependency of \ToA{}-estimates on the frequency is less consistent for the $A_0$ mode.
For sensors $\sensor_1$ and $\sensor_4$, the \ToA{} is almost constant for frequencies above \SI{5}{\kilo\hertz}.
Recall that sensors $\sensor_1$ and $\sensor_4$ have approximately the same distance to impact $\imp{1}$; so, similar \ToA{}-estimates are expected.
\ToA{}-estimates aside, the scalograms for the present time-frequency window differ substantially from a qualitative point of view, cf. Figs.~\ref{fig:I1_wavelet_A0_IOB_snr50db}~(a) and (d).
In the time-frequency window relevant for the $S_0$ mode, the scalograms of these two sensors bear a close resemblance to one another, cf. Figs.~\ref{fig:I1_wavelet_S0_IOB}~(a) and (d).  
The \ToA{} obtained from sensor $\sensor_2$ increases with frequency, see Fig.~\ref{fig:I1_wavelet_A0_IOB_snr50db} (b).
This implies a decrease in wave speed, which is contrary to the dispersion curves in Fig.~\ref{fig::lamb_wave_disp_curve}, where the wave speed increases with frequency.
Sensor $\sensor_3$, however, shows the reverse behavior, i.e., the \ToA{} descreases slightly as the frequency increases. 
Figure~\ref{fig:I1_wavelet_freq_A0_lines_IOB_snr50db} shows sections through the scalograms at constant frequencies ranging from \SIrange{4}{14}{\kilo\hertz} in steps of \SI{2}{\kilo\hertz}.
From the intersections of these curves, with the dashed red lines that represent the threshold level, we can identify the \ToA{}-estimates from the four sensor signals, for which numerical values are listed in Tab.~\ref{tab:I1_wavelet_freq_lines_A0_snr50db}.
The \ToA{}-estimates show a much greater variance as compared to the results which we obtained for the $S_0$ mode above, cf. Tab.~\ref{tab:I1_wavelet_freq_lines}.
As illustrated by the scalograms, the dependency of the \ToA{} on the frequency varies both quantitatively and qualitatively among the sensors, even if we disregard low frequencies, for which the \ToA{} of sensor $\sensor_2$ lies to the left of the \coi{} boundary, see Fig. \ref{fig:I1_wavelet_A0_IOB_snr50db} (b).
\begin{table}[pos=!htbp]
    \small
    \centering
    \begin{tabular}[]{r|cccc|cccc}
        \toprule
        Frequency in \SI{}{\kilo\hertz}
        & \multicolumn{4}{l|}{Threshold crossing in \si{\milli\second} – $\imp{1}^{\iob}$} 
        & \multicolumn{4}{l}{Threshold crossing in \si{\milli\second} – $\imp{1}^{\iop}$} \\
        & $\sensor_1$ &  $\sensor_2$ &  $\sensor_3$ &  $\sensor_4$ 
        & $\sensor_1$ &  $\sensor_2$ &  $\sensor_3$ &  $\sensor_4$ 
        \\
        \midrule
        \num{4} & 
        \num{0.4502} & \num{0.1458} & \num{0.6738} & \num{0.4510} &
        \num{0.4488} & \num{0.1426} & \num{0.6728} & \num{0.4490}
        \\
        \num{6} & 
        \num{0.4666} & \num{0.2174} & \num{0.6462} & \num{0.4620} &  
        \num{0.4916} & \num{0.2350} & \num{0.6768} & \num{0.4870}
        \\
        \num{8} & 
        \num{0.4654} & \num{0.2484} & \num{0.6240} & \num{0.4646} & 
        \num{0.4954} & \num{0.2700} & \num{0.6638} & \num{0.4952}
        \\
        \num{10} & 
        \num{0.4654} & \num{0.2664} & \num{0.6120} & \num{0.4638} & 
        \num{0.4832} & \num{0.2786} & \num{0.6324} & \num{0.4818}
        \\
        \num{12} & 
        \num{0.4682} & \num{0.2806} & \num{0.6074} & \num{0.4656} & 
        \num{0.4714} & \num{0.2820} & \num{0.6118} & \num{0.4680}
        \\
        \num{14} & 
        \num{0.4756} & \num{0.2934} & \num{0.6120} & \num{0.4744} &
        \num{0.4650} & \num{0.2846} & \num{0.5990} & \num{0.4662}
        \\
        \bottomrule
    \end{tabular}
    \caption{
        Estimated \ToA{} at frequencies \SIlist{4;6;8;10;12;14}{\kilo\hertz} for a threshold of $\sth = \num{1e-3}$: Idealized impact $\imp{1}^{\iob}$ and experiment-based impact $\imp{1}^{\iop}$ for noise-contaminated sensor signals.
    }
    \label{tab:I1_wavelet_freq_lines_A0_snr50db}
\end{table}%
The consistency improves if we consider time-differences with respect to the sensor closest to the excitation, i.e., $\sensor_2$ in the present case. 
The relative times for the respective frequencies, which are given in Tab.~\ref{tab:I1_wavelet_freq_lines_A0_diff_snr50db}, also show a decrease as the frequency increases. 
In other words, the $A_0$ mode propagates faster for higher frequencies, which is in line with the dispersion curves of Fig.~\ref{fig::lamb_wave_disp_curve}.
As a matter of fact, both the phase velocity and the group velocity show the largest gradients at low frequencies.
\begin{table}[pos=!htbp]
    \small
    \centering
    \begin{tabular}[]{r|ccc|ccc}
        \toprule
        Frequency in \SI{}{\kilo\hertz}
        & \multicolumn{3}{l|}{Relative times in \si{\milli\second} – $\imp{1}^{\iob}$} 
        & \multicolumn{3}{l}{Relative times in \si{\milli\second} – $\imp{1}^{\iop}$} \\
        & ~~$\test_1 - \test_2$ ~~& ~~ $\test_3 - \test_2$ ~~& ~~ $\test_4 - \test_2$~~ 
        & ~~$\test_1 - \test_2$ ~~& ~~ $\test_3 - \test_2$ ~~ &~~  $\test_4 - \test_2$ ~~
        \\
        \midrule
        \num{4} & 
        \num{0.3044} & \num{0.5280} & \num{0.3052} &
        \num{0.3062} & \num{0.5302} & \num{0.3064} 
        \\
        \num{6} & 
        \num{0.2492} & \num{0.4288} & \num{0.2450} &
        \num{0.2566} & \num{0.4418} & \num{0.2520} 
        \\
        \num{8} & 
        \num{0.2170} & \num{0.3756} & \num{0.2162} &
        \num{0.2254} & \num{0.3938} & \num{0.2252} 
        \\
        \num{10} & 
        \num{0.1990} & \num{0.3456} & \num{0.1974} & 
        \num{0.2046} & \num{0.3538} & \num{0.2032} 
        \\
        \num{12} & 
        \num{0.1876} & \num{0.3268} & \num{0.1850} &
        \num{0.1894} & \num{0.3298} & \num{0.1860}  
        \\
        \num{14} & 
        \num{0.1822} & \num{0.3186} & \num{0.1810} & 
        \num{0.1804} & \num{0.3144} & \num{0.1816}  
        \\
        \bottomrule
    \end{tabular}
    \caption{
        Time-differences of \ToA{}-estimates for noise-contaminated signals at frequencies \SIlist{4;6;8;10;12;14}{\kilo\hertz}, relative to the closest sensor to the impact, for both idealized impact $\imp{1}^{\iob}$ and experiment-based impact $\imp{1}^{\iop}$.
    }
    \label{tab:I1_wavelet_freq_lines_A0_diff_snr50db}
\end{table}%

\begin{figure}[pos=!htbp]
    \centering
    \includegraphics[width=0.9\linewidth]{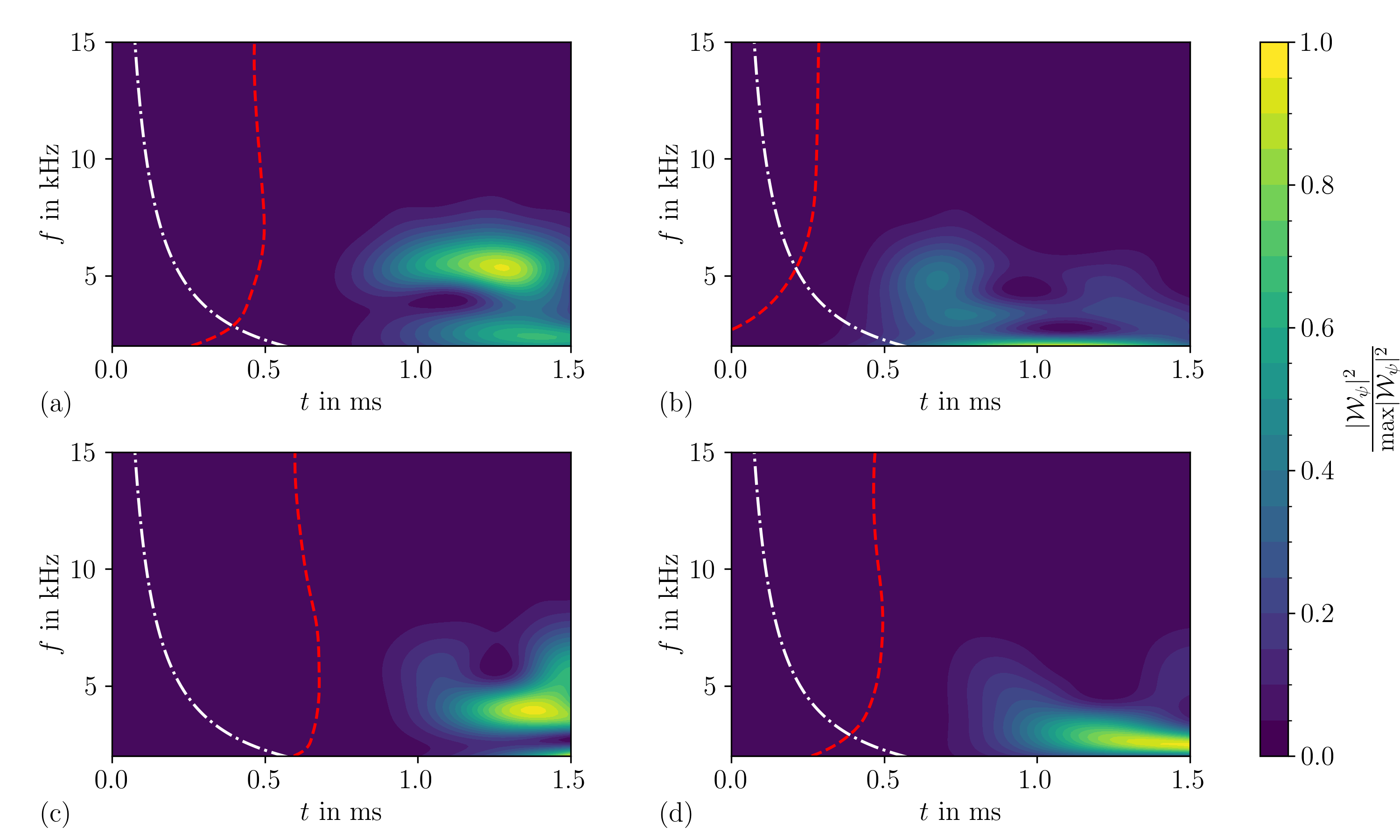}
    \caption{ 
        Scalograms of noise-contaminated sensor signals $\sensor_1$--$\sensor_4$ for the \emph{experiment-based impact} $\imp{1}^{\iop}$: The normalized square of the CWT's absolute value is illustrated in time-spans of \SI{1.5}{\milli\second} and a frequency-range of \SIrange{2}{15}{\kilo \hertz}.
        Dashed red lines illustrate \ToA{}-estimates as functions of the frequency, i.e., times at which a (non-dimensional) threshold of $\sth = \num{1e-3}$ is reached. Dash-dotted white lines represent the \coi{}.  
    } 
    \label{fig:I1_wavelet_A0_IOP_snr50db}
\end{figure}

\begin{figure}[pos=!htbp]
    \centering
    \includegraphics[width=0.8\linewidth]{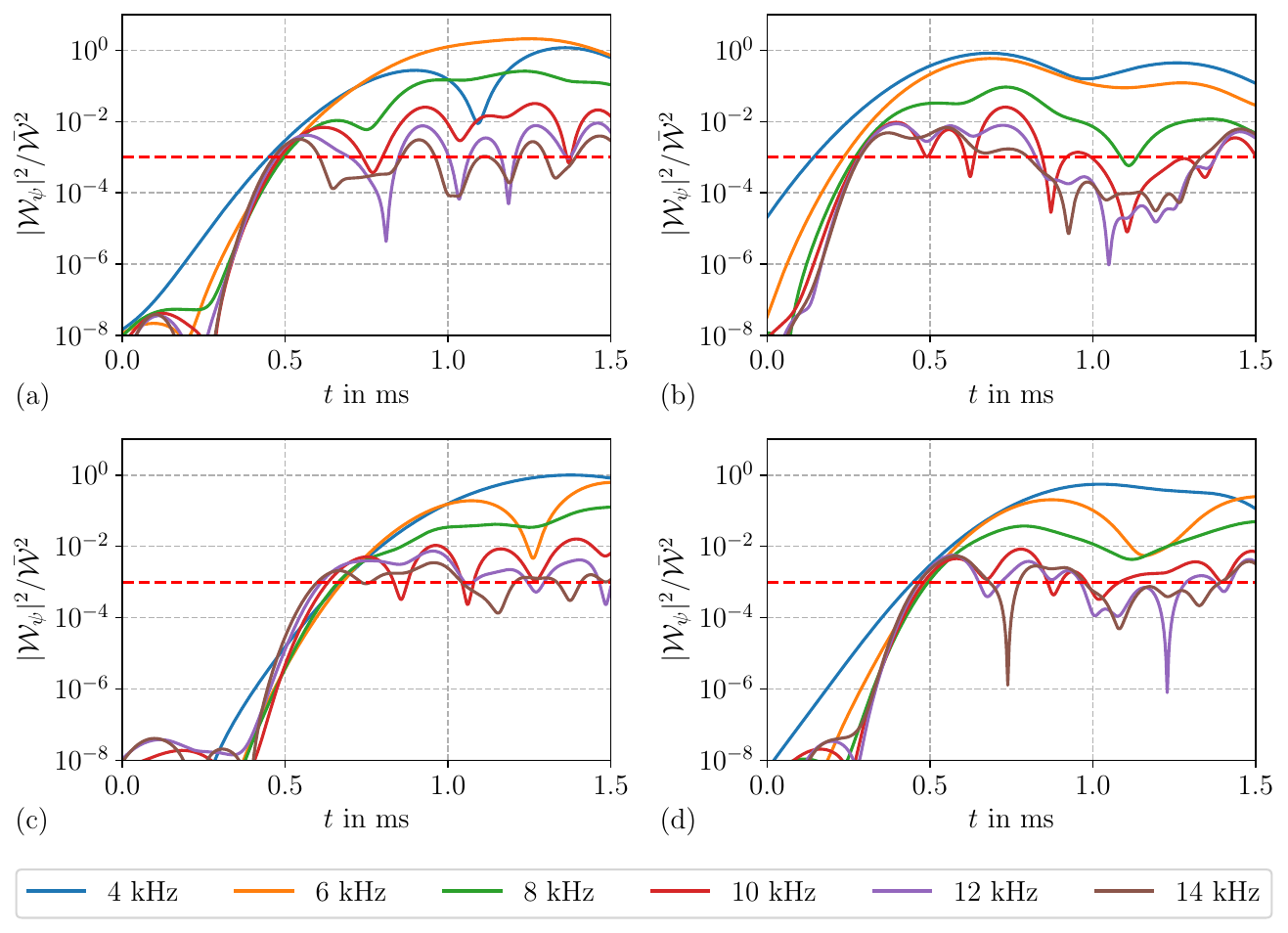}
    \caption{
        Sections of scalograms for noise-contaminated sensor signals at \SIlist{4;6;8;10;12;14}{\kilo\hertz} for impact $\imp{1}^{\iop}$.
        The threshold $\sth = \num{1e-3}$ is indicated by dashed red lines; their (first) intersections with the scalograms' sections determine the \ToA{} for the respective frequencies. 
    } 
    \label{fig:I1_wavelet_freq_A0_lines_IOP_snr50db}
\end{figure}

For the second type of impact $\imp{1}^{\iop}$, we obtain very similar results.
Using the same time-frequency range as for the idealized impact, the scalograms are shown in Fig.~\ref{fig:I1_wavelet_A0_IOP_snr50db}, where the differences between sensor $\sensor_1$ and sensor $\sensor_4$ stand out once again.  
Sections through the scalograms are illustrated in Fig.~\ref{fig:I1_wavelet_freq_A0_lines_IOP_snr50db}; the last four colums of Tab.~\ref{tab:I1_wavelet_freq_lines_A0_snr50db} list the corresponding numerical values for the \ToA{}-estimates.
The relative times listed in the last three columns of Tab.~\ref{tab:I1_wavelet_freq_lines_A0_diff_snr50db} do not only show the same qualitative behavior in terms of frequency-dependency as the idealized impact $\imp{1}^{\iob}$, but the agreement among the two types of impact is also excellent, even more so as the frequency increases.
For frequencies of \SI{10}{\kilo\hertz} and above, the relative times differ by no more than \SI{8.2}{\micro\second}. 
If we relate that difference in relative times to the absolute values $\test_3 - \test_2$, we obtain a relative deviation of \SI{2.37}{\percent}.

\begin{figure}[pos=!htbp]
    \centering
    \includegraphics[width=0.9\linewidth]{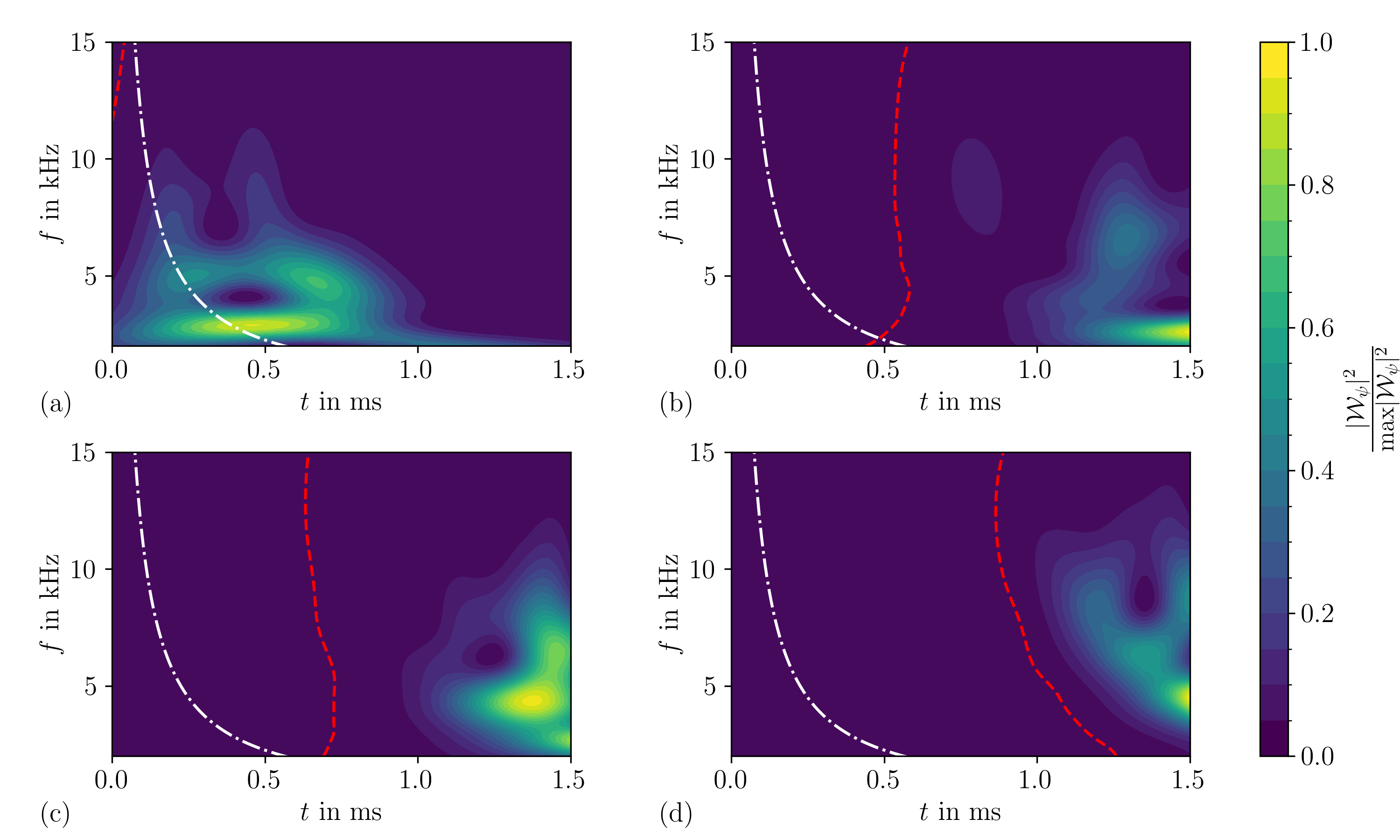}
    \caption{ 
        Scalograms of noise-contaminated sensor signals $\sensor_1$--$\sensor_4$ for the \emph{idealized impact} $\imp{2}^{\iob}$: The normalized square of the CWT's absolute value is illustrated in time-spans of \SI{1.5}{\milli\second} and a frequency-range of \SIrange{2}{15}{\kilo \hertz}.
        Dashed red lines illustrate \ToA{}-estimates as functions of the frequency, i.e., times at which a (non-dimensional) threshold of $\sth = \num{5e-3}$ is reached. Dash-dotted white lines represent the \coi{}.  
    } 
    \label{fig:I2_wavelet_A0_IOB_snr50db}
\end{figure}

\begin{figure}[pos=!htbp]
    \centering
    \includegraphics[width=0.9\linewidth]{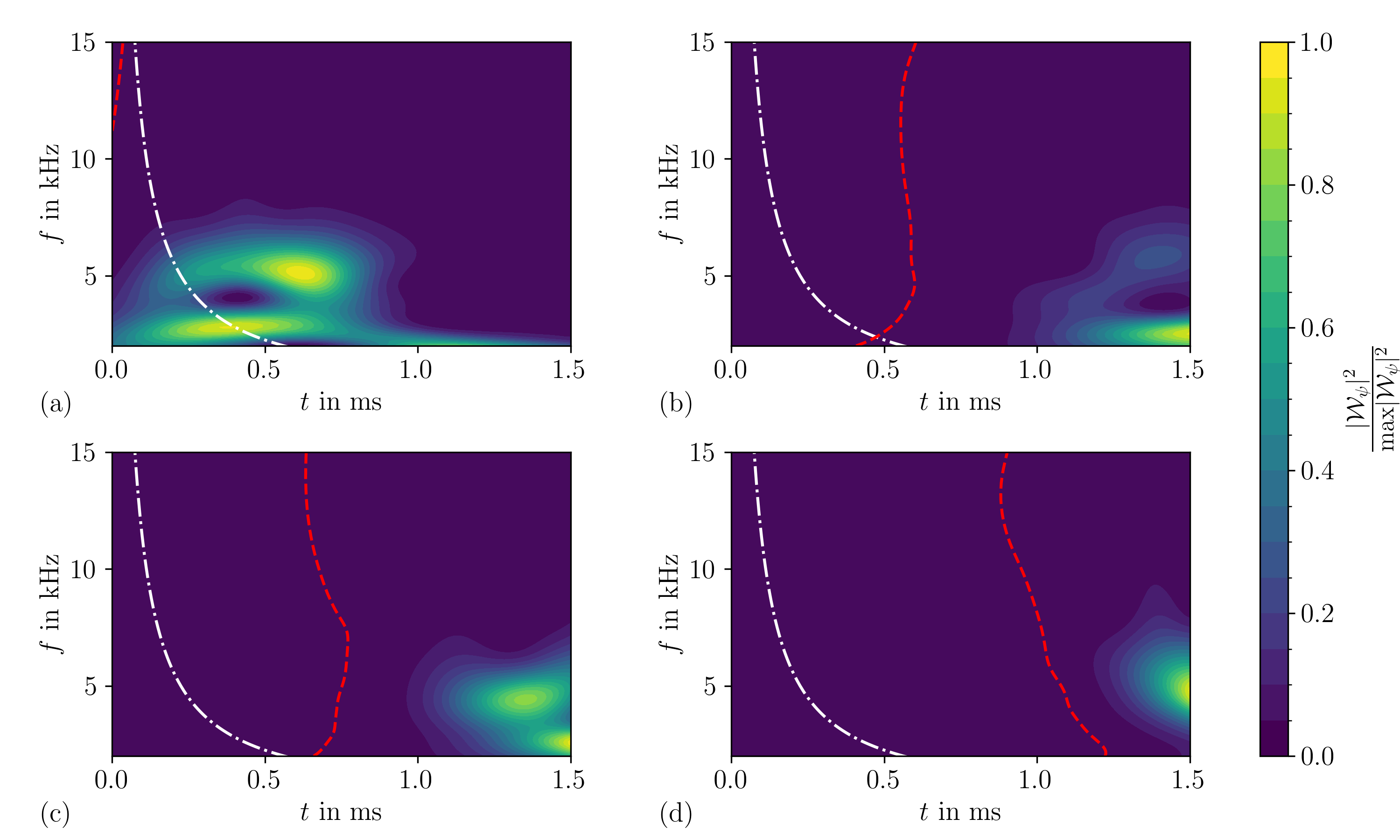}
    \caption{ 
        Scalograms of noise-contaminated sensor signals $\sensor_1$--$\sensor_4$ for the \emph{experiment-based impact} $\imp{2}^{\iop}$: The normalized square of the CWT's absolute value is illustrated in time-spans of \SI{1.5}{\milli\second} and a frequency-range of \SIrange{2}{15}{\kilo \hertz}.
        Dashed red lines illustrate \ToA{}-estimates as functions of the frequency, i.e., times at which a (non-dimensional) threshold of $\sth = \num{5e-3}$ is reached. Dash-dotted white lines represent the \coi{}. 
    } 
    \label{fig:I2_wavelet_A0_IOP_snr50db}
\end{figure}

\begin{figure}[pos=!htbp]
    \centering
    \includegraphics[width=0.8\linewidth]{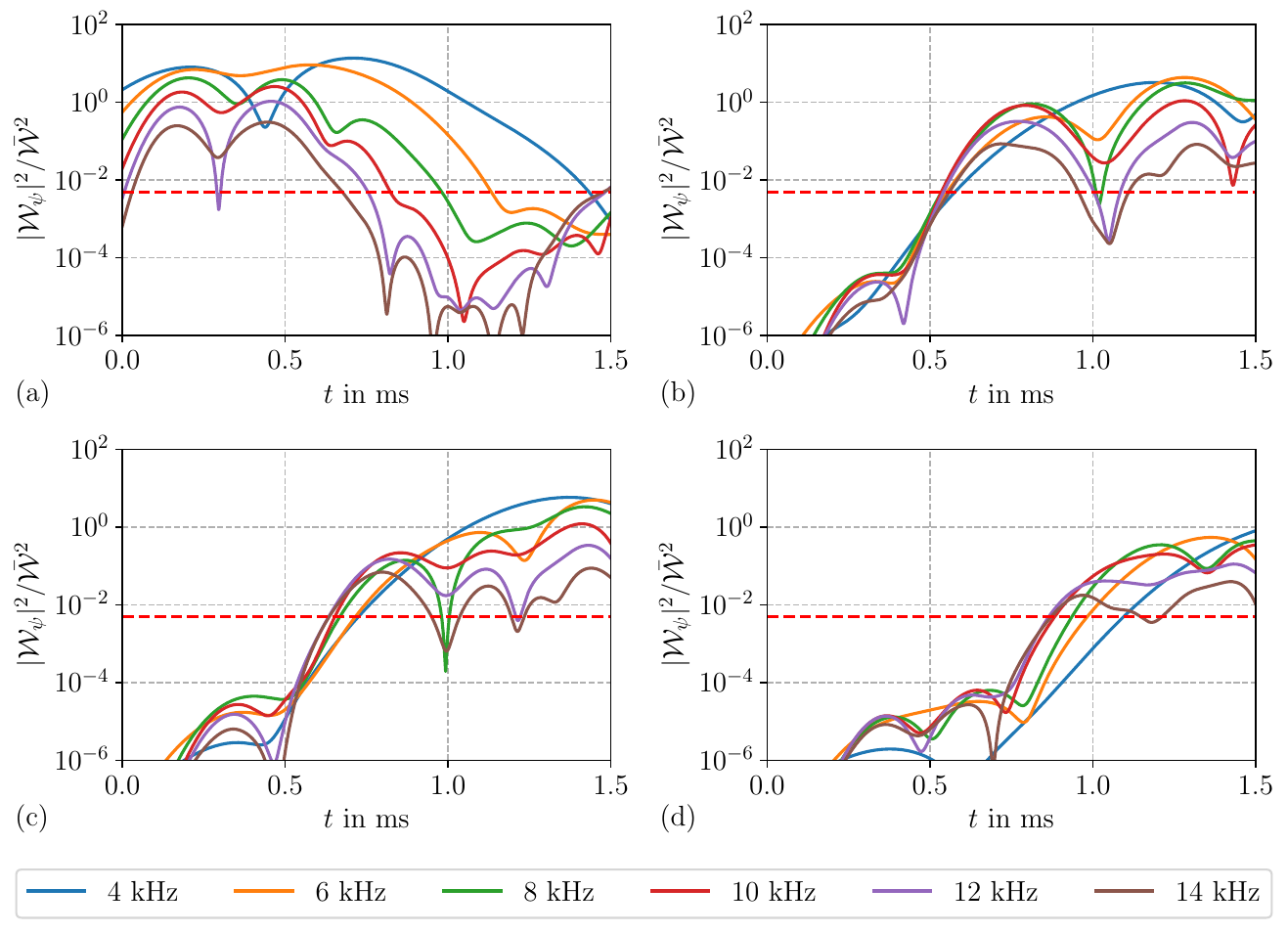}
    \caption{
        Sections of scalograms for noise-contaminated sensor signals at \SIlist{4;6;8;10;12;14}{\kilo\hertz} for impact $\imp{2}^{\iob}$.
        The threshold $\sth = \num{5e-3}$ is indicated by dashed red lines; their (first) intersections with the scalograms' sections determine the \ToA{} for the respective frequencies. 
    } 
    \label{fig:I2_wavelet_freq_A0_lines_IOB_snr50db}
\end{figure}

\begin{figure}[pos=!htbp]
    \centering
    \includegraphics[width=0.8\linewidth]{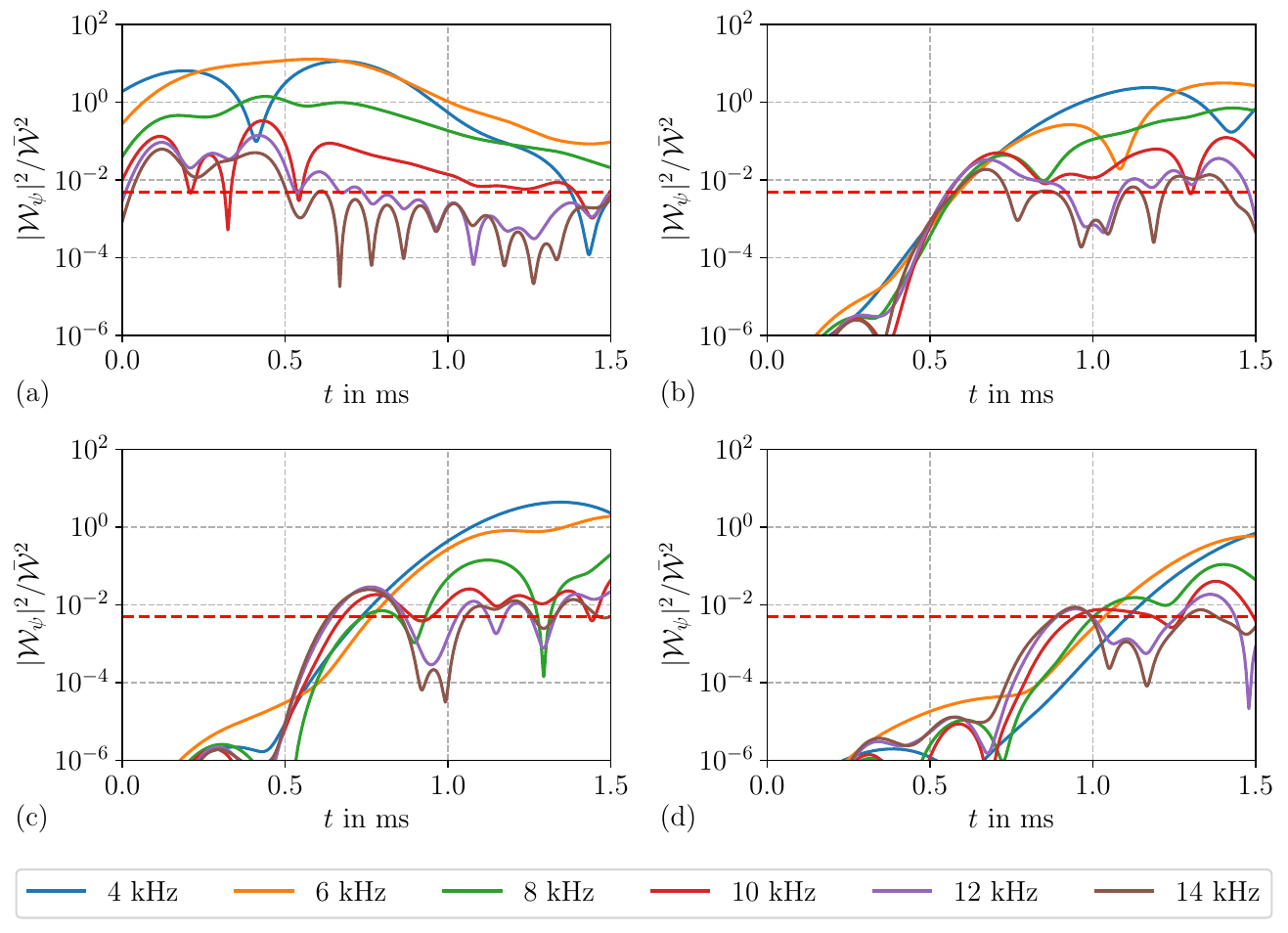}
    \caption{
        Sections of scalograms for noise-contaminated sensor signals at \SIlist{4;6;8;10;12;14}{\kilo\hertz} for impact $\imp{2}^{\iop}$.
        The threshold $\sth = \num{5e-3}$ is indicated by dashed red lines; their (first) intersections with the scalograms' sections determine the \ToA{} for the respective frequencies. 
    } 
    \label{fig:I2_wavelet_freq_A0_lines_IOP_snr50db}
\end{figure}

We conclude our investigations on the detection of $A_0$ mode by means of CWT with the second impact position, which is very close to sensor $\sensor_1$.
Both type of impacts $\imp{2}^{\iob}$ and $\imp{2}^{\iop}$ give very similar results, allowing to discuss them jointly in what follows.
The scalograms shown in Figs.~\ref{fig:I2_wavelet_A0_IOB_snr50db} and \ref{fig:I2_wavelet_A0_IOP_snr50db} reveal that we face the same problems as with the $S_0$ mode, though, in a different frequency-range.
Within the current frequencies ranging from \SIrange{2}{15}{\kilo\hertz}, the scalogram of sensor $\sensor_1$, Figs.~\ref{fig:I2_wavelet_A0_IOB_snr50db}~(a) and \ref{fig:I2_wavelet_A0_IOP_snr50db}~(a), respectively, exceeds the threshold level from the outset.
Only for frequencies above approximately \SI{12}{\kilo\hertz}, \ToA{}-estimates (different from zero) can be obtained by the proposed approach, where a (non-dimensional) threshold of $\sth = \num{5e-3}$ has been used. 
Nonetheless, the \ToA{}-estimates lie entirely to the left of the \coi{} boundary for $\sensor_1$, meaning that cutoff-effects influence the convolution, on which CWT is based. 
For sensors $\sensor_2$--$\sensor_4$, the \ToA{}-estimates for each type of impact agree comparatively well, both qualitatively and quantitatively.
Sections through the scalograms are illustrated in Figs.~\ref{fig:I2_wavelet_freq_A0_lines_IOB_snr50db} and~\ref{fig:I2_wavelet_freq_A0_lines_IOP_snr50db}, respectively. 
They again show the difficulty of choosing a proper threshold value $\sth$. 
If the threshold is too large, we miss \ToA{}-estimates for higher frequencies within the considered range; for small thresholds, the scalogram of sensor $\sensor_1$ lies above the threshold value (in the time-domain of interest).
Numerical values for the \ToA{}-esimates are given in Tab.~\ref{tab:I2_wavelet_freq_lines_A0} for both idealized and experiment-based impacts. 
The corresponding relative times $\test_i - \test_1$, $i=2,3,4$, are listed in Tab.~\ref{tab:I2_wavelet_freq_lines_A0_diff}.
Note that we have only included relative times, for which a \ToA{} different from zero has been obtained for sensor $\sensor_1$. 
The results confirm what we have already observed for the impact at position $\imp{1}$.
The \ToA{}-estimates for the experiment-based impact $\imp{2}^{\iop}$ are slightly delayed as compared to the idealized impact $\imp{2}^{\iob}$. 
Upon considering relative times, this delay is somewhat reduced. 
Nonetheless, we observe a maximum (absolute) deviation between the two impact types of \SI{21.6}{\micro\second} ($\test_4 - \test_1$ at \SI{12}{\kilo\hertz}: $\SI{0.8814}{\milli\second} - \SI{0.8598}{\milli\second} = \SI{0.0216}{\milli\second} = \SI{21.6}{\micro\second}$), which translates into a relative deviation of $\SI{21.6}{\micro\second} / \SI{859.8}{\micro\second} = \SI{2.51}{\percent}$.
However, the maximum relative deviation between the two impacts occurs at \SI{14}{\kilo\hertz} for $\test_2 - \test_1$ and is \SI{3.58}{\percent}: $(\SI{0.5492}{\milli\second} - \SI{0.5302}{\milli\second}) / \SI{0.5302}{\milli\second} = \SI{3.58}{\percent}$.

\begin{table}[pos=!htbp]
    \small
    \centering
    \begin{tabular}[]{r|cccc|cccc}
        \toprule
        Frequency in \SI{}{\kilo\hertz}
        & \multicolumn{4}{l|}{Threshold crossing in \si{\milli\second} – $\imp{2}^{\iob}$} 
        & \multicolumn{4}{l}{Threshold crossing in \si{\milli\second} – $\imp{2}^{\iop}$} \\
        & $\sensor_1$ &  $\sensor_2$ &  $\sensor_3$ &  $\sensor_4$ 
        & $\sensor_1$ &  $\sensor_2$ &  $\sensor_3$ &  $\sensor_4$ 
        \\
        \midrule
        \num{4}& 
        \num{0.0000} & \num{0.5772} & \num{0.7256} & \num{1.0956} &
        \num{0.0000} & \num{0.5864} & \num{0.7350} & \num{1.1086}
        \\
        \num{6}& 
        \num{0.0000} & \num{0.5534} & \num{0.7162} & \num{0.9838} &  
        \num{0.0000} & \num{0.5878} & \num{0.7668} & \num{1.0364}
        \\
        \num{8}& 
        \num{0.0000} & \num{0.5356} & \num{0.6682} & \num{0.9364} & 
        \num{0.0000} & \num{0.5788} & \num{0.7416} & \num{1.0030}
        \\
        \num{10} & 
        \num{0.0000} & \num{0.5356} & \num{0.6528} & \num{0.8836} & 
        \num{0.0000} & \num{0.5588} & \num{0.6768} & \num{0.9486}
        \\
        \num{12} & 
        \num{0.0058} & \num{0.5420} & \num{0.6342} & \num{0.8656} & 
        \num{0.0098} & \num{0.5550} & \num{0.6432} & \num{0.8912}
        \\
        \num{14} & 
        \num{0.0298} & \num{0.5600} & \num{0.6354} & \num{0.8730} &
        \num{0.0282} & \num{0.5774} & \num{0.6332} & \num{0.8860}
        \\
        \bottomrule
    \end{tabular}
    \caption{
        Estimated \ToA{} at frequencies \SIlist{4;6;8;10;12;14}{\kilo\hertz} for a threshold of $\sth = \num{5e-3}$: Idealized impact $\imp{2}^{\iob}$ and experiment-based impact $\imp{2}^{\iop}$ for noise-contaminated sensor signals.
    }
    \label{tab:I2_wavelet_freq_lines_A0}
\end{table}%

\begin{table}[pos=!htbp]
    \small
    \centering
    \begin{tabular}[]{r|ccc|ccc}
        \toprule
        Frequency in \SI{}{\kilo\hertz}
        & \multicolumn{3}{l|}{Relative times in \si{\milli\second} – $\imp{2}^{\iob}$} 
        & \multicolumn{3}{l}{Relative times in \si{\milli\second} – $\imp{2}^{\iop}$} \\
        & ~~$\test_2 - \test_1$ ~~& ~~ $\test_3 - \test_1$ ~~& ~~ $\test_4 - \test_1$~~ 
        & ~~$\test_2 - \test_1$ ~~& ~~ $\test_3 - \test_1$ ~~& ~~ $\test_4 - \test_1$~~
        \\
        \midrule
        \num{12}& 
        \num{0.5362} & \num{0.6284} & \num{0.8598} &
        \num{0.5452} & \num{0.6334} & \num{0.8814}  
        \\
        \num{14}& 
        \num{0.5302} & \num{0.6056} & \num{0.8432} & 
        \num{0.5492} & \num{0.6050} & \num{0.8578}  
        \\
        \bottomrule
    \end{tabular}
    \caption{
        Time-differences of \ToA{}-estimates for noise-contaminated signals at frequencies \SIlist{12;14}{\kilo\hertz}, relative to the closest sensor to the impact, for both the idealized impact $\imp{2}^{\iob}$ and the experiment-based impact $\imp{2}^{\iop}$.
    }
    \label{tab:I2_wavelet_freq_lines_A0_diff}
\end{table}%
The results demonstrate that threshold-crossing in the frequency domain is an effective approach for \ToA{}-estimation of impact-induced Lamb waves.
For noise-free sensor signals, the method enables a robust detection of the earliest arriving $S_0$ mode, yielding consistent relative \ToA{}-differences across sensors, impact types, and impact locations.
In the presence of noise, however, the approach fails for $S_0$ mode detection.
Due to the comparatively small amplitudes of the $S_0$ mode and the limited sensitivity of surface-bonded piezoceramic sensors, the initial $S_0$ arrival is masked by noise and cannot be reliably identified in the scalograms, even for increased threshold levels.
For noise-contaminated signals, frequency-domain threshold-crossing applied to the $A_0$ mode provides physically consistent \ToA{}-estimates and meaningful relative \ToA{} differences between sensors.
Limitations primarily occur for impacts located very close to individual sensors, where threshold-crossings may occur within the COI or from the outset of the time-window.
We have observed this issue with both the $S_0$ and the $A_0$ modes, where a careful choice of threshold values has turned out crucial.
\subsection{Short/Long Term Average}\label{ssec::3d_results_SLA}
The SLA method employs two distinct averaging windows: $\wins$ for short-term averaging and $\winl$ for long-term averaging.
According to Akram and Eaton \cite{Akram_2016}, $\wins$ should span \numrange{2}{3} periods of the dominant frequency, while $\winl$ should be \numrange{5}{10} times greater than $\wins$.
The dominant frequency is determined by the maxima in the scalograms of the sensor signals. 
As shown in Sec. \ref{ssec::3d_results_CWT}, these dominant frequencies vary across sensor signals depending on the impact position and type.
To address this variability and ensure comprehensive frequency coverage, we set $\tdom = \SI{10}{\micro\second}$ corresponding to a frequency of \SI{100}{\kilo\hertz}.
This value represents the upper limit of the frequency-range considered in the scalograms, see e.g. Fig. \ref{fig:I1_wavelet_S0_IOB}.
Following the guidelines by Akram and Eaton \cite{Akram_2016}, we conduct a parameter study in which the lengths of the short-term and long-term windows are varied according to
\begin{align}\label{eq::sla_alpha_beta}
    \wins = \alpha f_s\tdom, \quad \quad \winl = \beta\wins.
\end{align}
The parameters $\alpha$ and $\beta$ are varied systematically as multiples of $\tdom$, in small steps from \numrange{1}{100}.
With the time-discretization $\dt = \SI{0.2}{\micro\second}$ (Sec. \ref{ssec:3D_fe_model}), the sampling frequency is $f_s = \SI{5}{\mega\hertz}$.
At the lowest parameter combination ($\alpha = \beta = \num{1}$), the windows $\wins$ and $\winl$ span time-intervals corresponding to \SI{100}{\kilo\hertz}.
At the highest combination ($\alpha = \beta = \num{100}$), the window lengths are $\wins=\num{5000}$ and $\winl=\num{500000}$, corresponding to \SI{1}{\kilo\hertz} and \SI{10}{\hertz}, respectively.
The literature recommendations mentioned at the beginning of this section are not disregarded; rather, the recommended parameter value range is fully encompassed within this expanded parameter study, ensuring a comprehensive analysis of the entire parameter spectrum of interest.

\begin{figure}[pos=!htbp]
    \centering
    \begin{tikzpicture}				
        \node[anchor=south west,inner sep=0] (image1) at (0, 0) {\includegraphics[width=\textwidth, trim=0 0 0 0, clip, ]{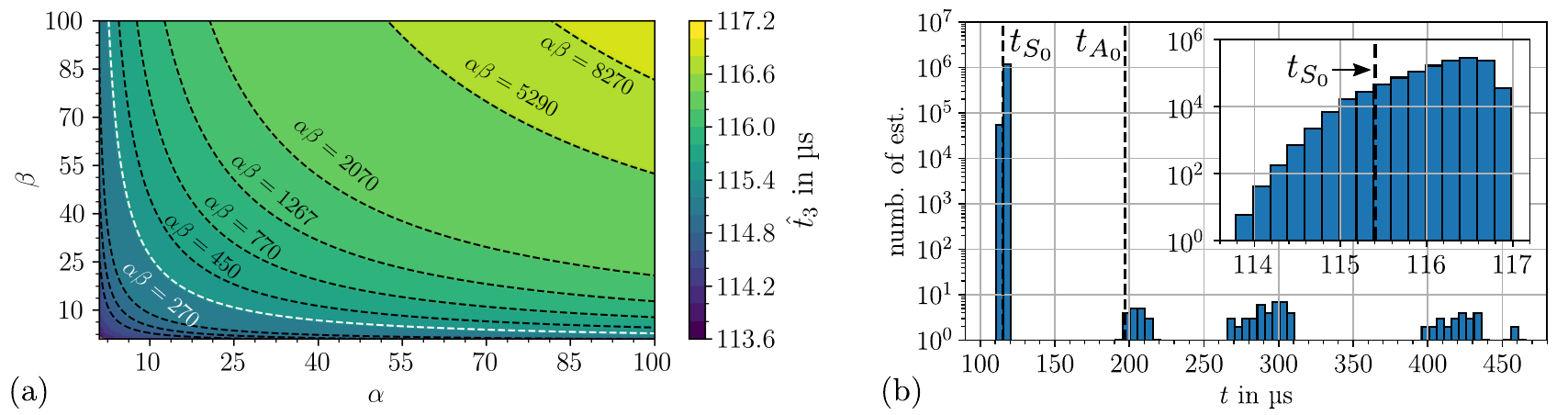}};        
        \begin{scope}[x={(image1.south east)},y={(image1.north west)}]
        \end{scope} 
    \end{tikzpicture}
    \caption{
        Parametric study of the SLA algorithm for signal of sensor $\sensor_3$ (noise-free) upon impact $\imp{1}^\iop$:
        (a) Contour plot of \ToA{}-estimates as a function of $\alpha$ and $\beta$.
        The white dashed line at $\alpha\beta=\num{270}$ corresponds to $\tso=\SI{115.4}{\micro\second}$.
        (b) Histogram of \ToA{}-estimates, including a zoomed view centered on $\tso$. Most estimates cluster near $\tso$, while only few parameter combinations capture $\tao$.}
    \label{fig::iop_560_400_sens3_contour_and_hist}					
\end{figure}

Figure \ref{fig::iop_560_400_sens3_contour_and_hist} presents a parametric study of the SLA method using sensor signal $ \sensor_3 $ (noise-free) for the experiment-based impact $\imp{1}^{\iop}$ as an illustrative example.
The contour plot in Fig. \ref{fig::iop_560_400_sens3_contour_and_hist} (a), shows that the \ToA{}-estimates span a narrow range of $\SIrange{113.6}{117.2}{\micro\second}$ over the entire $(\alpha,\beta)$-parameter space.
A key observation is that the \ToA{}-estimation is primarily governed by the product $\alpha\beta$: dashed lines of constant $\alpha\beta$ align with regions of nearly constant \ToA{}, indicating that $\alpha\beta$, i.e., the combined window-length parameter scaled by $\tdom$, dominates the SLA output.
The dashed line at $\alpha\beta = \num{270}$ highlights the parameter combination yielding the reference $S_0$ arrival time $\tso=\SI{115.4}{\micro\second}$.
The histogram in Fig. \ref{fig::iop_560_400_sens3_contour_and_hist} (b) confirms that most parameter sets produce \ToA{}-estimates near $\tso$, while only few combinations identify the first arrival of the $A_0$ mode $\tao$. 
A zoomed view centered on $\tso$ shows that most \ToA{}-estimates lie within  $\pm\num{8}\dt = \pm\SI{1.6}{\micro\second}$

\begin{figure}[pos=!htbp]
    \centering
    \begin{tikzpicture}				
        \node[anchor=south west,inner sep=0] (image1) at (0, 0) {\includegraphics[width=0.95\textwidth, trim=0 0 0 0, clip, ]{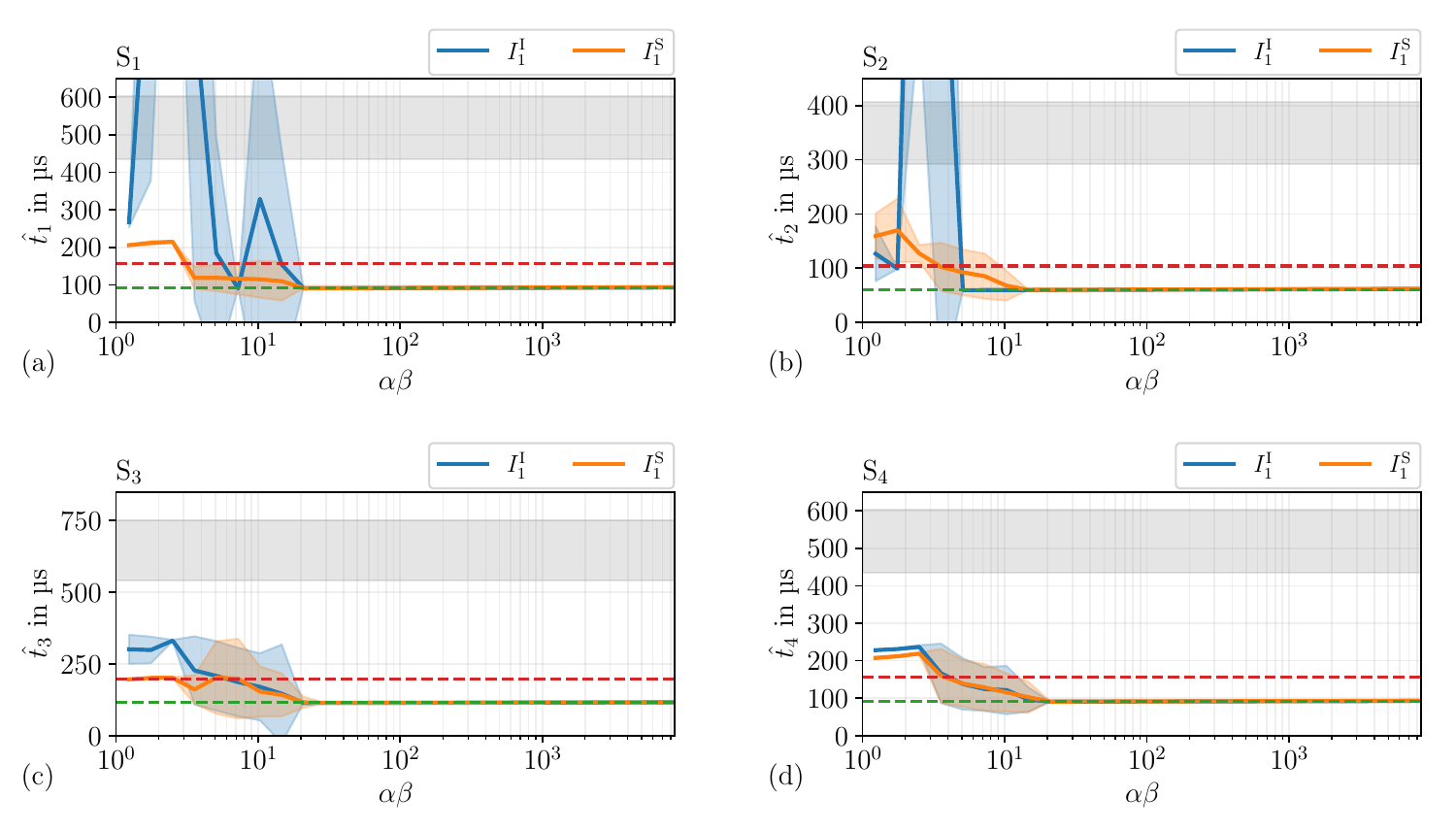}};        
        \begin{scope}[x={(image1.south east)},y={(image1.north west)}]
        \end{scope} 
    \end{tikzpicture}
    \caption{
        Result of \ToA{}-estimation using the SLA method on noise-free sensor signals for impact position $\imp{1}$.
        The solid line represents the mean \ToA{}-estimates, while the color-matched shaded regions indicate the corresponding standard deviations over all $(\alpha, \beta)$-combinations producing that product.
        The red and green dashed lines indicate the first arrival time of the $S_0$ and $A_0$ modes $\tso$ and $\tao$, respectively.
        The grey shaded area marks the earliest possible arrival window of the $A_0$ mode between \qty{10}{\kilo\hertz} and \qty{20}{\kilo\hertz}.
    }
    \label{fig::3d_sla_I1_log}					
\end{figure}

\begin{figure}[pos=!htbp]
    \centering
    \begin{tikzpicture}				
        \node[anchor=south west,inner sep=0] (image1) at (0, 0) {\includegraphics[width=0.95\textwidth, trim=0 0 0 0, clip, ]{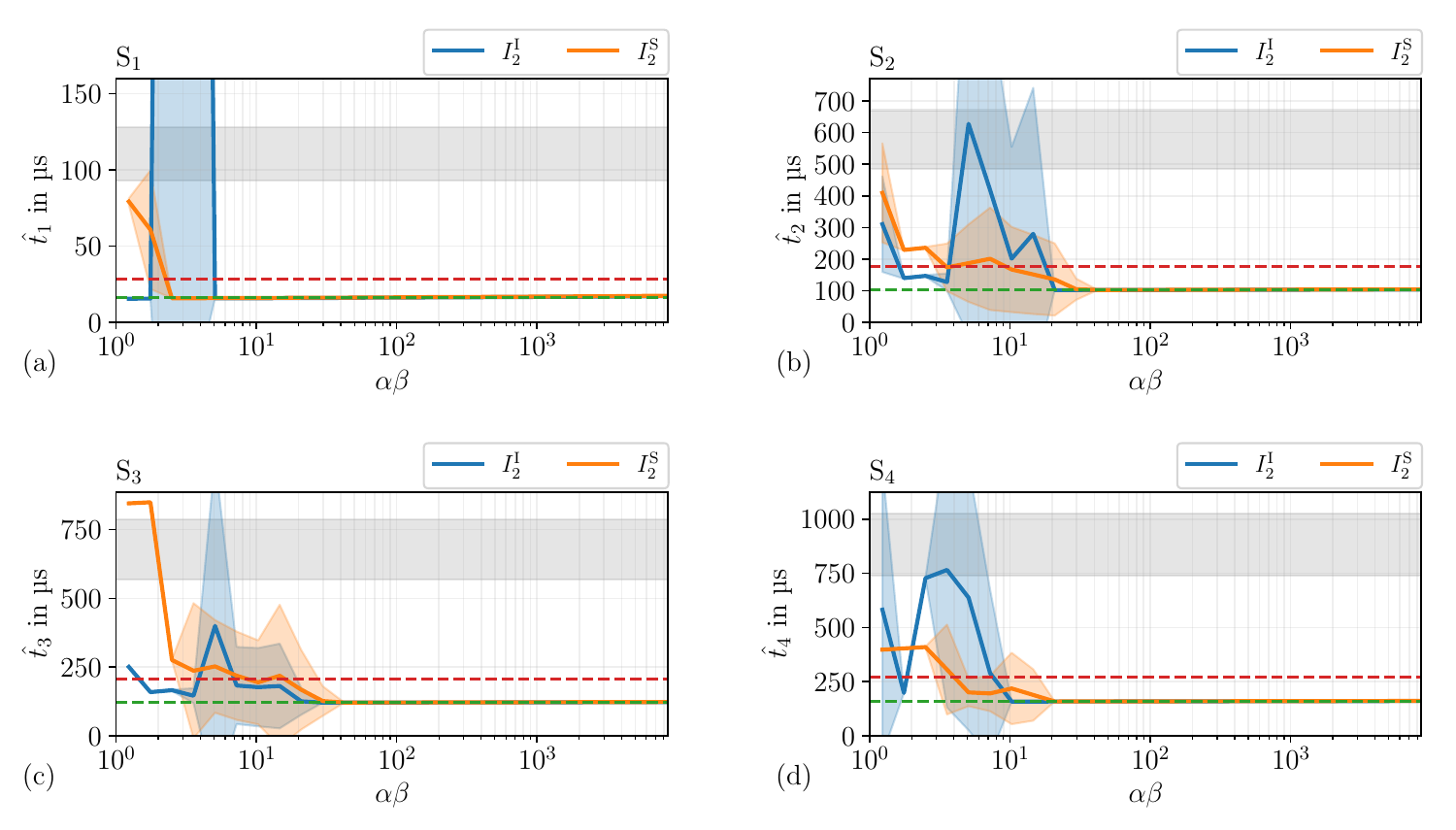}};        
        \begin{scope}[x={(image1.south east)},y={(image1.north west)}]
        \end{scope} 
    \end{tikzpicture}
    \caption{
        Result of \ToA{}-estimation using the SLA method on noise-free sensor signals for impact position $\imp{2}$.
        The solid line represents the mean \ToA{}-estimates, while the color-matched shaded regions indicate the corresponding standard deviations over all $(\alpha, \beta)$-combinations producing that product.
        The red and green dashed lines indicate the first arrival time of the $S_0$ and $A_0$ modes $\tso$ and $\tao$, respectively.
        The grey shaded area marks the earliest possible arrival window of the $A_0$ mode between \qty{10}{\kilo\hertz} and \qty{20}{\kilo\hertz}.
    }
    \label{fig::3d_sla_I2_log}					
\end{figure}

\begin{figure}[pos=!htbp]
    \centering
    \begin{tikzpicture}				
        \node[anchor=south west,inner sep=0] (image1) at (0, 0) {\includegraphics[width=\textwidth, trim=0 0 0 0, clip, ]{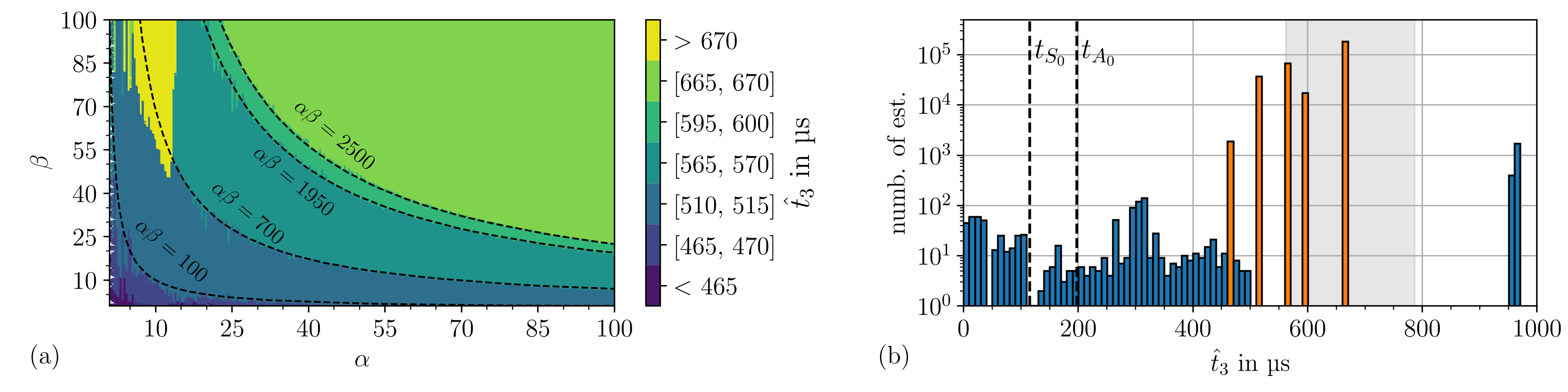}};        
        \begin{scope}[x={(image1.south east)},y={(image1.north west)}]
        \end{scope} 
    \end{tikzpicture}
    \caption{
        Parametric study of the SLA algorithm for signal of sensor $\sensor_3$ (noise-contaminated) upon impact $\imp{1}^\iop$:
        (a) Contour plot of \ToA{}-estimates as a function of $\alpha$ and $\beta$.
        (b) Histogram of \ToA{}-estimates.
        The orange bars in (b) correspond to the \ToA{}-intervals highlighted as discrete colored bands in the contour plot in (a), i.e. [\num{465}, \num{470}] \si{\micro\second}, [\num{510}, \num{515}] \si{\micro\second}, [\num{565}, \num{570}] \si{\micro\second}, [\num{595}, \num{600}] \si{\micro\second}, and [\num{665}, \num{670}] \si{\micro\second}.
        The grey shaded area marks the earliest possible arrival window of the $A_0$ mode between \qty{10}{\kilo\hertz} and \qty{20}{\kilo\hertz}.}
    \label{fig::iop_560_400_sens3_contour_and_hist_snr50db}					
\end{figure}

\begin{figure}[pos=!htbp]
    \centering
    \begin{tikzpicture}				
        \node[anchor=south west,inner sep=0] (image1) at (0, 0) {\includegraphics[width=0.95\textwidth, trim=0 0 0 0, clip, ]{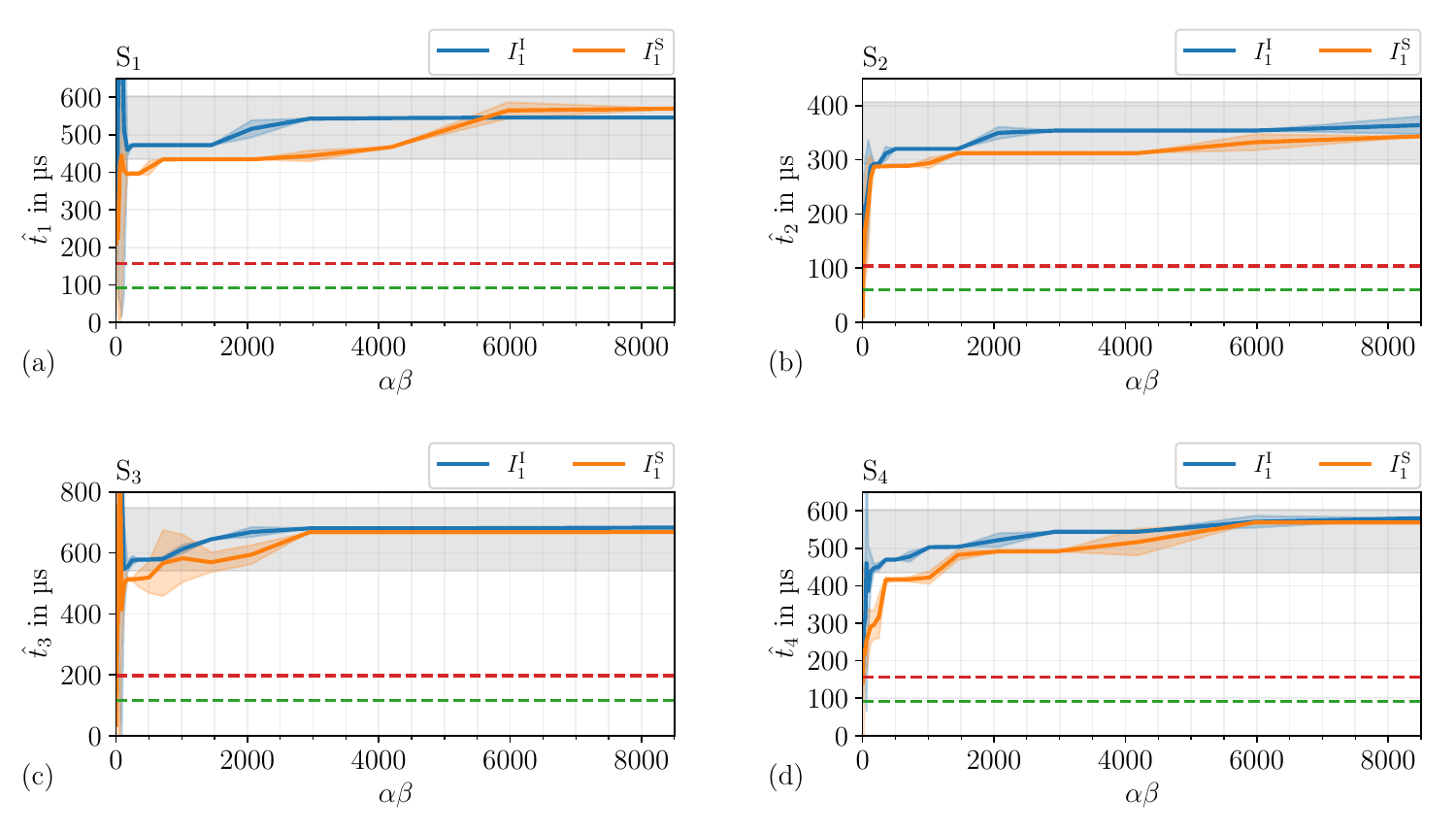}};        
        \begin{scope}[x={(image1.south east)},y={(image1.north west)}]
        \end{scope} 
    \end{tikzpicture}
    \caption{
        Result of \ToA{}-estimation using the SLA method on noise-contaminated sensor signals for impact position $\imp{1}$.
        The solid line represents the mean \ToA{}-estimates, while the color-matched shaded regions indicate the corresponding standard deviations over all $(\alpha, \beta)$-combinations producing that product.
        The red and green dashed lines indicate the first arrival time of the $S_0$ and $A_0$ modes $\tso$ and $\tao$, respectively.
        The grey shaded area marks the earliest possible arrival window of the $A_0$ mode between \qty{10}{\kilo\hertz} and \qty{20}{\kilo\hertz}.
    }
    \label{fig::3d_sla_noise_50snr_I1_lin}
\end{figure}

\begin{figure}[pos=!htbp]
    \centering
    \begin{tikzpicture}				
        \node[anchor=south west,inner sep=0] (image1) at (0, 0) {\includegraphics[width=0.95\textwidth, trim=0 0 0 0, clip, ]{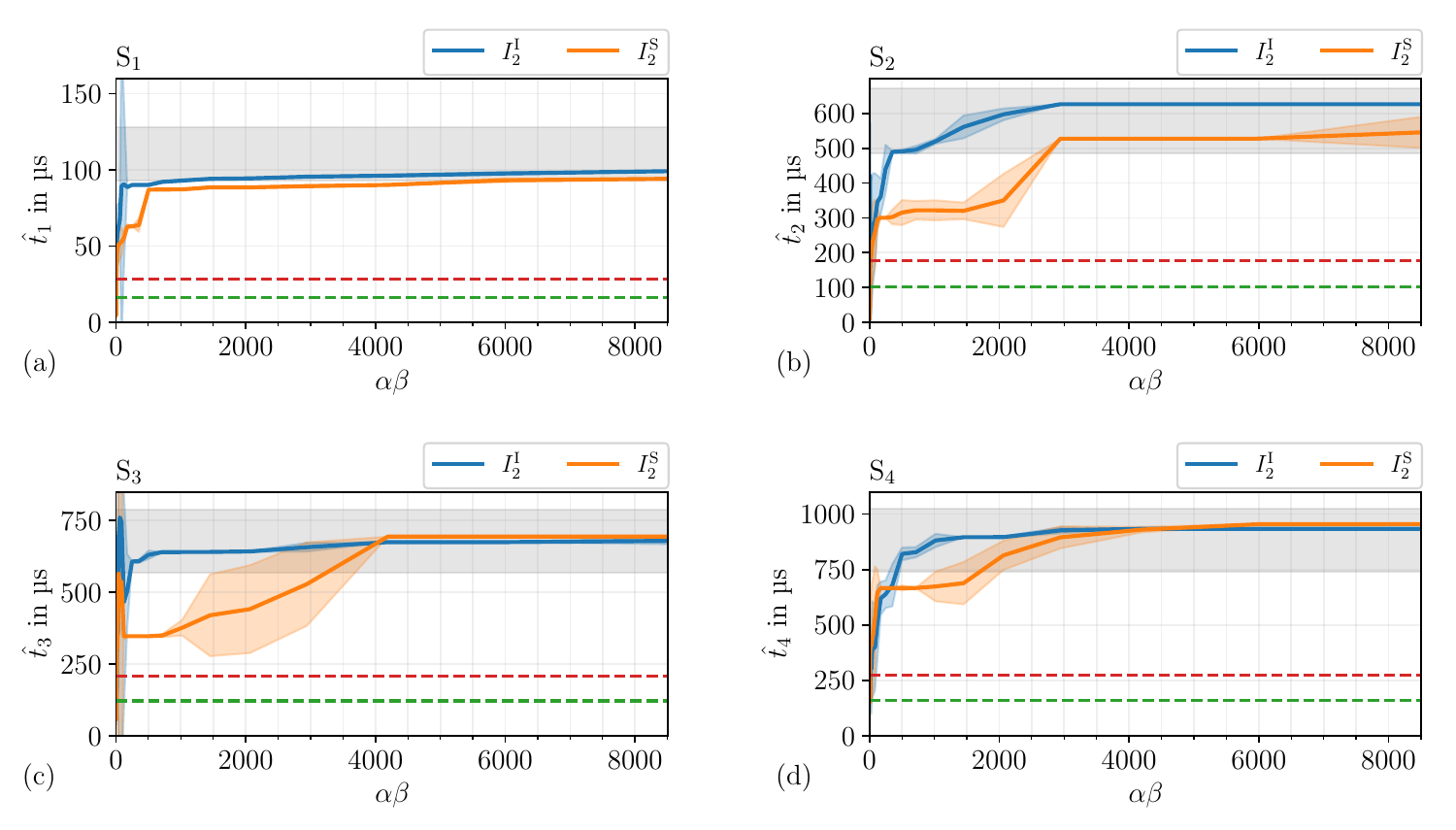}};        
        \begin{scope}[x={(image1.south east)},y={(image1.north west)}]
        \end{scope} 
    \end{tikzpicture}
    \caption{
        Result of \ToA{}-estimation using the SLA method on noise-contaminated sensor signals for impact position $\imp{2}$.
        The solid line represents the mean \ToA{}-estimates, while the color-matched shaded regions indicate the corresponding standard deviations over all $(\alpha, \beta)$-combinations producing that product.
        The red and green dashed lines indicate the first arrival time of the $S_0$ and $A_0$ modes $\tso$ and $\tao$, respectively.
        The grey shaded area marks the earliest possible arrival window of the $A_0$ mode between \qty{10}{\kilo\hertz} and \qty{20}{\kilo\hertz}.
    }
    \label{fig::3d_sla_noise_50snr_I2_lin}					
\end{figure}

To avoid repeating two-dimensional contour plots for all sensors and impact cases, the results are presented in a condensed form by collapsing the $(\alpha, \beta)$-space onto the combined parameter $\alpha\beta$, as illustrated in Fig. \ref{fig::3d_sla_I1_log}.
For each $\alpha\beta$, the solid line denotes the mean \ToA{}-estimate across all $(\alpha, \beta)$-combinations producing that product, while the shaded region indicates the corresponding standard deviation.
In addition, and in analogy to the CWT-based analysis (Sec. \ref{ssec::3d_results_CWT}), we highlight the earliest possible arrival window of the $A_0$ mode between \qty{10}{\kilo\hertz} and \qty{20}{\kilo\hertz} (grey shaded region), inferred from the dispersion characteristics of the plate, see Fig. \ref{fig::lamb_wave_disp_curve}.
Figures \ref{fig::3d_sla_I1_log} and \ref{fig::3d_sla_I2_log} show that, for noise-free sensor signals, the \ToA{}-estimates converge to the first arrival of the $S_0$ mode for all sensors and both impact positions, provided that the combined parameter $\alpha\beta$ is sufficiently large ($\alpha\beta > 20$ for impact position $\imp{1}$ and $\alpha\beta > 40$ for impact position $\imp{2}$).
Above these values, the mean \ToA{}-estimates coincide with $\tso$, and the standard deviations remain small, indicating a high robustness of the SLA output with respect to the averaging window parameters $\alpha$ and $\beta$.
Notably, this behavior is observed even when the impact occurs in close proximity to a sensor, e.g., sensor $\sensor_1$ for impact position $\imp{2}$.
\\
We now investigate how the SLA performs for noise-contaminated sensor signals and whether physically meaningful \ToA{}-values remain detectable.
Figure \ref{fig::iop_560_400_sens3_contour_and_hist_snr50db} repeats the parametric study of Fig. \ref{fig::iop_560_400_sens3_contour_and_hist} for the same sensor/impact case $(\sensor_3, \imp{1}^{\iop})$, but for noise-contaminated sensor signals.
In contrast to the noise-free case, the contour plot in Fig. \ref{fig::iop_560_400_sens3_contour_and_hist_snr50db} (a) exhibits a substantially broader range of \ToA{}-estimates, spanning several hundred microseconds across the 
$(\alpha, \beta)$-space.
Nevertheless, the overall structure remains largely organized along constant $\alpha\beta$-trajectories (dashed reference lines): between these trajectories, the \ToA{}-estimates vary no more than \qty{5}{\micro\second}.
Noise induces additional fringe-like deviations along the $\alpha\beta$-trajectories, which are most pronounced for smaller $\alpha\beta$, i.e., for shorter averaging windows.
The histogram in Fig. \ref{fig::iop_560_400_sens3_contour_and_hist_snr50db} (b) no longer displays a single dominant concentration around $\tso$.
Instead, many parameter combinations produce late-time estimates, including pronounced clusters (orange bars) in the range of approximately \qtyrange{465}{670}{\micro\second}, and additional estimates beyond this range.
This is consistent with the observation in Sec. \ref{ssec::3d_results_CWT} that, in the presence of noise, the initial $S_0$ arrival is masked and cannot be detected reliably.
Importantly, three of the five highlighted bins in Fig. \ref{fig::iop_560_400_sens3_contour_and_hist_snr50db} (b) lie within the time-window associated with the earliest possible $A_0$ arrivals in the \qtyrange{10}{20}{\kilo\hertz} range, whereas two bins occur at earlier times, corresponding to higher-frequency components of the $A_0$ mode.
As in the noise-free case, Figs. \ref{fig::3d_sla_noise_50snr_I1_lin} and \ref{fig::3d_sla_noise_50snr_I2_lin} summarize the results as a function of the combined parameter $\alpha\beta$.
In contrast to Figs. \ref{fig::3d_sla_I1_log} and \ref{fig::3d_sla_I2_log} the \ToA{}-estimates do not converge to the first $S_0$ arrival time. 
Instead, they shift to substantially later times and, for sufficiently large $\alpha\beta$, enter a parameter-insensitive regime (a plateau) in which the mean \ToA{} varies only weakly.
These plateaus predominantly fall within the reference window associated with the earliest possible $A_0$ arrivals in the \qtyrange{10}{20}{\kilo\hertz} range.
However, the plateau regime should not be interpreted as a sensor-consistent identification of a unique $A_0$-related \ToA{}.
While differences in absolute times of the plateau across sensors are expected due to different impact-to-sensor distances, the relevant observation here is that the plateau levels are not aligned consistently within the \qtyrange{10}{20}{\kilo\hertz} reference window: depending on the sensor, the plateau occurs closer to the lower (earlier-time) boundary of the grey band (suggesting a bias towards the faster, higher-frequency parts), whereas for other sensors it is located in the upper part of the same band (suggesting a bias towards later, lower-frequency contributions within that range); cf. Fig. \ref{fig::3d_sla_noise_50snr_I2_lin} (a) versus Fig. \ref{fig::3d_sla_noise_50snr_I2_lin} (d).
Consequently, for the same impact, SLA may estimate \ToA{}-values associated with different effective $A_0$ frequency components across sensors, i.e., different portions of the dispersive $A_0$ wave propagation.
This lack of cross-sensor alignment complicates a robust \ToA{}-estimation under noise.

Compared with the noise-free case, a pronounced dependence on the impact type becomes visible.
This behavior is physically plausible, as the two impact types exhibit different excitation spectra, c.f., Figs. \ref{fig::3d_shaker_signal} (b) and \ref{fig::3d_hammer_signal} (b), and the $A_0$ mode is strongly dispersive in the relevant frequency range, such that the time-domain response is sensitive to which spectral components remain dominant under noise.

The question at hand is whether the sensitivity of the SLA method to later $A_0$ mode arrivals can be enhanced by suppressing higher-frequency content.
From a conceptual point of view, this is closely related to the CWT-based analysis (Sec. \ref{ssec::3d_results_CWT}), where focusing on sections of the scalogram at low frequencies is used to detect the $A_0$ mode.
Motivated by this analogy, low-pass filtering was applied prior to the SLA algorithm.
However, within the investigated settings, low-pass filtering does not enable enhanced or reliable $A_0$ mode detection using the SLA method, whereas it proves effective for the MER method described in Sec. \ref{ssec::3d_results_MER}.

In summary, the SLA result is primarily governed by the combined parameter $\alpha\beta$.
For noise-free signals, sufficiently large $\alpha\beta$ yields stable convergence to the first $S_0$ arrival across sensors and impact positions.
In the presence of noise, the first $S_0$ arrival is masked and the \ToA{}-estimates shift to later times; for large $\alpha\beta$ \ToA{}-estimates approach a plateau that largely falls into the $A_0$ arrival window in the \qtyrange{10}{20}{\kilo\hertz} range, with a noticeable dependence on the impact type.
Low-pass filtering does not systematically improve $A_0$ detection with SLA under the conditions studied.

\subsection{Modified Energy Ratio}\label{ssec::3d_results_MER}
Based on the guidance provided by Akram and Eaton \cite{Akram_2016}, the window-size $\wine$ of the MER method should cover \numrange{2}{3} periods of the signal's dominant frequency, analogous to the choice of the short-term averaging window in the SLA method.
Accordingly, $\wine$ is parameterized via Eq. \eqref{eq::sla_alpha_beta} as $\wine=\alpha f_s \tdom$.
Since the dominant frequency depends on both impact location and impact type, cf. Sec. \ref{ssec::3d_results_CWT}, we set $\tdom=\SI{10}{\micro\second}$.
Varying $\alpha$ from \numrange{1}{100} then corresponds to window lengths associated with frequencies between \SI{100}{\kilo\hertz} and \SI{1}{\kilo\hertz}, thus covering the range recommended in the literature.

\begin{figure}[pos=!htbp]
    \centering
    \begin{tikzpicture}				
        \node[anchor=south west,inner sep=0] (image1) at (0, 0) {\includegraphics[width=0.9\textwidth, trim=0 0 0 0, clip, ]{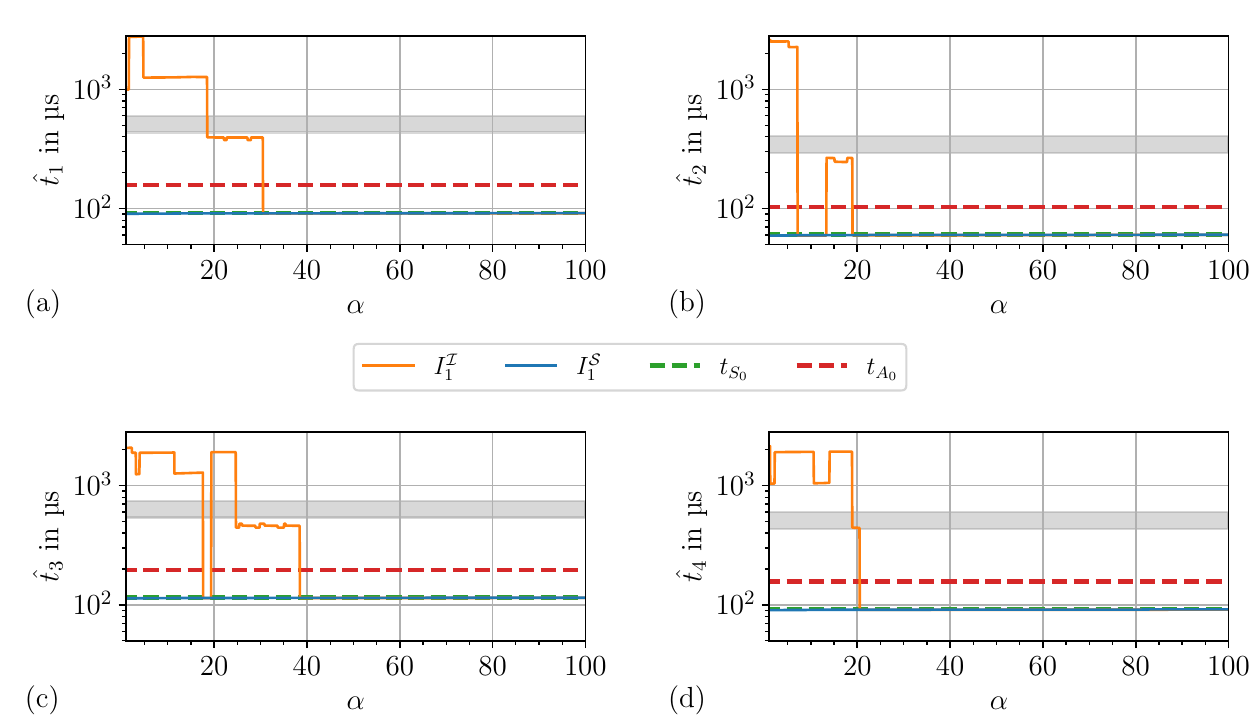}};        
        \begin{scope}[x={(image1.south east)},y={(image1.north west)}]
        \end{scope} 
    \end{tikzpicture}
    \caption{
        Results of the parametric study of MER algorithm applied to noise-free signals from sensors $\sensor_1$--$\sensor_4$ in subfigures (a)--(d), for the idealized and experiment-based impacts at location $\imp{1}$. 
        The dashed reference lines indicate the first arrival time of the $S_0$ and $A_0$ modes, $\tso$ and $\tao$.
        The grey shaded area marks the earliest possible arrival window of the $A_0$ mode between \qtyrange{10}{20}{\kilo\hertz}.
        }
    \label{fig::3d_para_var_mer_I1}					
\end{figure}

\begin{figure}[pos=!htbp]
    \centering
    \begin{tikzpicture}				
        \node[anchor=south west,inner sep=0] (image1) at (0, 0) {\includegraphics[width=0.9\textwidth, trim=0 0 0 0, clip, ]{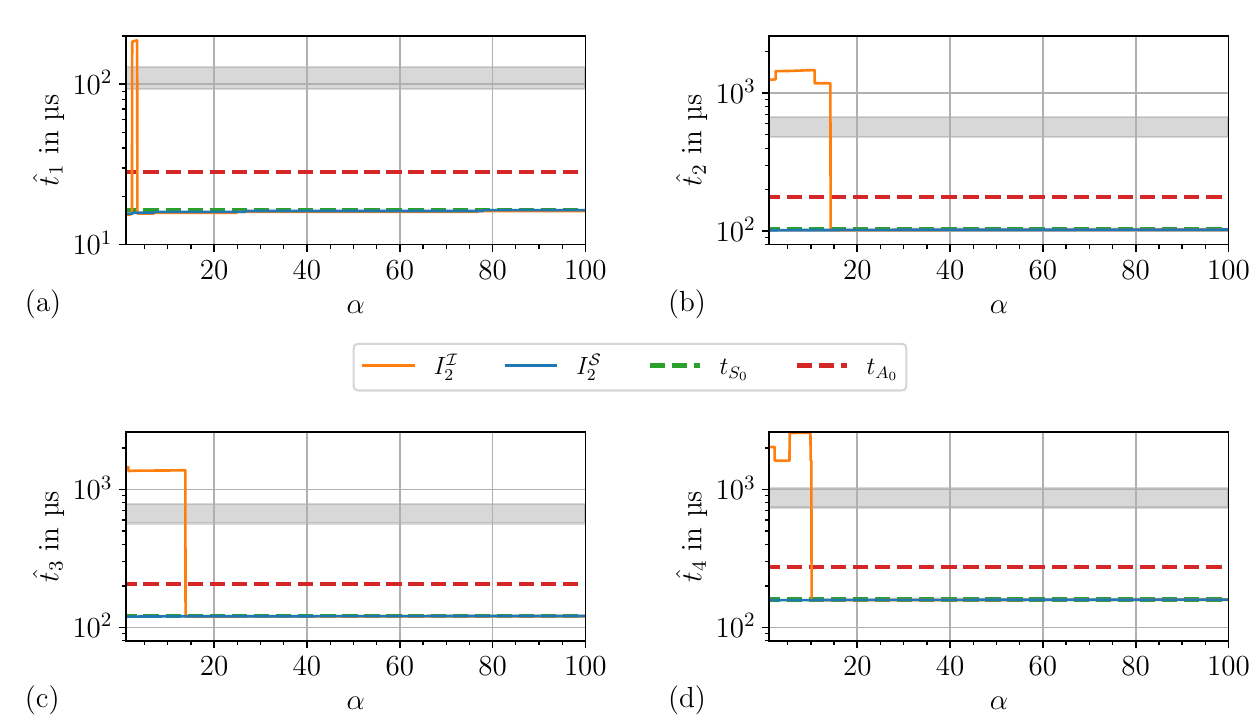}};        
        \begin{scope}[x={(image1.south east)},y={(image1.north west)}]
        \end{scope} 
    \end{tikzpicture}
    \caption{
        Results of parametric study of MER algorithm applied to noise-free signals of sensors $\sensor_1$--$\sensor_4$ in subfigures (a)--(d) for idealized and experiment-based impact at location $\imp{2}$. 
        The dashed reference lines indicate the first arrival time of the $S_0$ and $A_0$ modes, $\tso$ and $\tao$.
        The grey shaded area marks the earliest possible arrival window of the $A_0$ mode between \qtyrange{10}{20}{\kilo\hertz}.
        }
    \label{fig::3d_para_var_mer_I2}					
\end{figure}

We first consider the application of MER to the noise-free sensor signals.
Figures \ref{fig::3d_para_var_mer_I1} and \ref{fig::3d_para_var_mer_I2} show the corresponding parameter studies for impacts at locations $\imp{1}$ and $\imp{2}$, respectively.
For both impact locations, the method is capable of detecting the earliest arrival associated with the $S_0$ mode, i.e., $\tso$.
For the idealized impact (orange lines), the estimated \ToA{}-values show significant fluctuations for small values of $\alpha$, particularly for $\alpha<40$ at $\imp{1}$ and $\alpha<15$ at $\imp{2}$, where they are substantially delayed relative to the reference events ($\tso$ and $\tao$).
To better visualize these variations, the ordinate is plotted on a logarithmic scale.
The delayed \ToA{}-estimates correlate with local maxima of the sensor signals, indicating that the method is initially sensitive to secondary wave features.
With increasing $\alpha$, these delayed \ToA{}-estimates disappear and the results converge to $\tso$, which suggests that larger window lengths suppress the influence of local maxima and improve robustness.
For the experiment-based impacts (blue lines), the behavior is even more stable: the \ToA{}-estimates remain nearly constant over the full parameter range and coincide with $\tso$.
Overall, these observations show that MER reliably identifies the first $S_0$-arrival in noise-free signals, provided that the window length is chosen sufficiently large.

\begin{figure}[pos=!htbp]
    \centering
    \begin{tikzpicture}				
        \node[anchor=south west,inner sep=0] (image1) at (0, 0) {\includegraphics[width=0.9\textwidth, trim=0 0 0 0, clip, ]{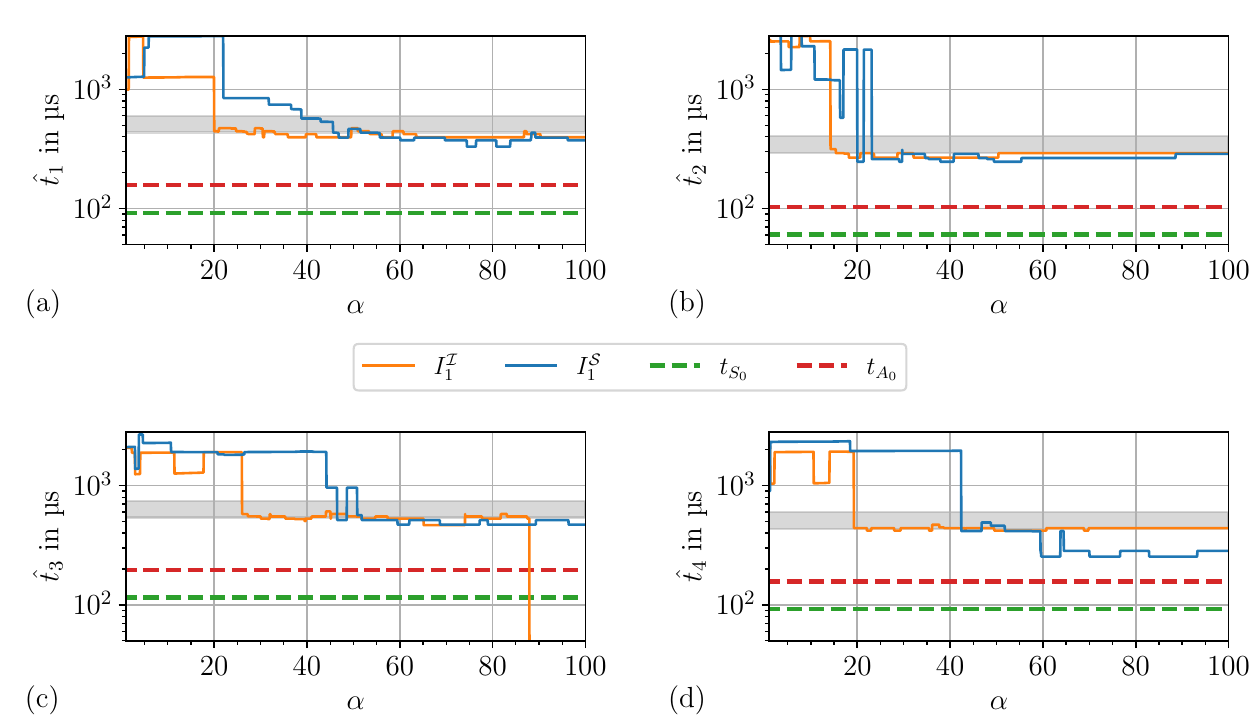}};        
        \begin{scope}[x={(image1.south east)},y={(image1.north west)}]
        \end{scope} 
    \end{tikzpicture}
    \caption{
        Results of the parametric study of the MER algorithm applied to noise-contaminated signals of sensors $\sensor_1$--$\sensor_4$, shown in subfigures (a)--(d), for the idealized and experiment-based impacts at location $\imp{1}$. 
        The dashed reference lines indicate the first arrival time of the $S_0$ and $A_0$ modes, $\tso$ and $\tao$.
        The grey shaded area marks the earliest possible $A_0$-arrival window for frequencies between \qtyrange{10}{20}{\kilo\hertz}.
        }
    \label{fig::3d_para_var_mer_I1_snr50db}					
\end{figure}

\begin{figure}[pos=!htbp]
    \centering
    \begin{tikzpicture}				
        \node[anchor=south west,inner sep=0] (image1) at (0, 0) {\includegraphics[width=0.9\textwidth, trim=0 0 0 0, clip, ]{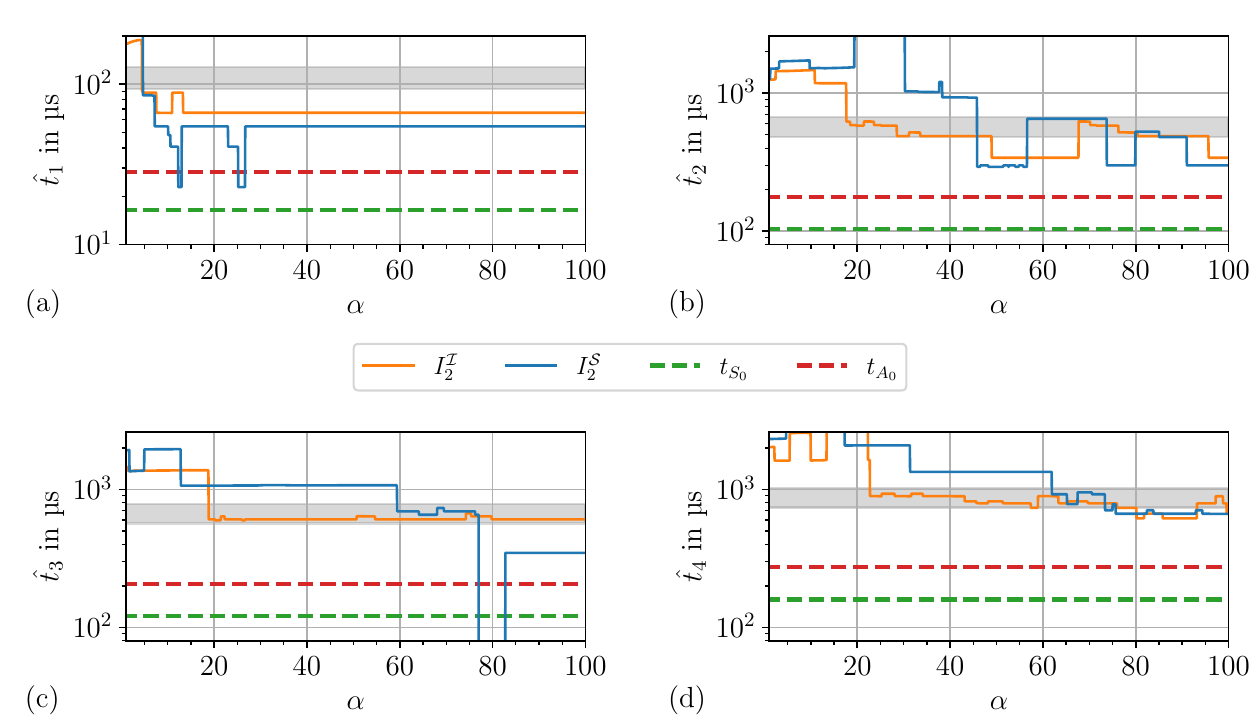}};        
        \begin{scope}[x={(image1.south east)},y={(image1.north west)}]
        \end{scope} 
    \end{tikzpicture}
    \caption{
        Results of parametric study of MER algorithm applied to noise-contaminated signals of sensors $\sensor_1$--$\sensor_4$ in subfigures (a)--(d) for idealized and experiment-based impact at location $\imp{2}$. 
        The dashed reference lines indicate the first arrival time of the $S_0$ and $A_0$ modes, $\tso$ and $\tao$.
        The grey shaded area marks the earliest possible arrival window of the $A_0$ mode between \qtyrange{10}{20}{\kilo\hertz}.
        }
    \label{fig::3d_para_var_mer_I2_snr50db}					
\end{figure}

The situation changes markedly when noise-contaminated sensor signals are considered.
The corresponding parameter studies are shown in Figs. \ref{fig::3d_para_var_mer_I1_snr50db} and \ref{fig::3d_para_var_mer_I2_snr50db}.
The gray shaded regions denote the earliest possible arrival window of the $A_0$ mode in the \qtyrange{10}{20}{\kilo\hertz} at the respective sensors, analogous to Sec. \ref{ssec::3d_results_SLA}.
In contrast to the noise-free case, the \ToA{}-estimates no longer converge clearly and robustly to the reference event $\tso$, but instead exhibit a fragmented dependence on $\alpha$.
For individual sensors, piecewise nearly constant regions can still be identified, and some of these \ToA{}-estimates fall into the $A_0$ arrival window in the \qtyrange{10}{20}{\kilo\hertz} range.
However, these regions are neither aligned across all sensors nor persistent over comparable parameter ranges.
Instead, they are accompanied by jumps between different time levels whose occurrence and extent differ across sensors, reflecting differences in the respective signal time-histories.
Consequently, MER applied directly to noise-contaminated raw signals does not provide a sufficiently robust basis for consistent \ToA{}-estimation across sensors.

To investigate whether physically meaningful \ToA{}-estimates can nevertheless be recovered, low-pass filtering is applied to the noise-contaminated signals, following the approach already introduced in Sec. \ref{ssec::3d_results_SLA}. 
The underlying idea is that suppressing higher-frequency contributions may facilitate the detection of low-frequency wave features associated with the $A_0$ mode.
Before turning to the noise-contaminated filtered signals, the same procedure is first applied to noise-free signals in order to assess under controlled coniditions whether MER can isolate $A_0$-related wave arrivals.

\begin{figure}[pos=!htbp]
    \centering
    \begin{tikzpicture}				
        \node[anchor=south west,inner sep=0] (image1) at (0, 0) {\includegraphics[width=\textwidth, trim=0 0 0 0, clip, ]{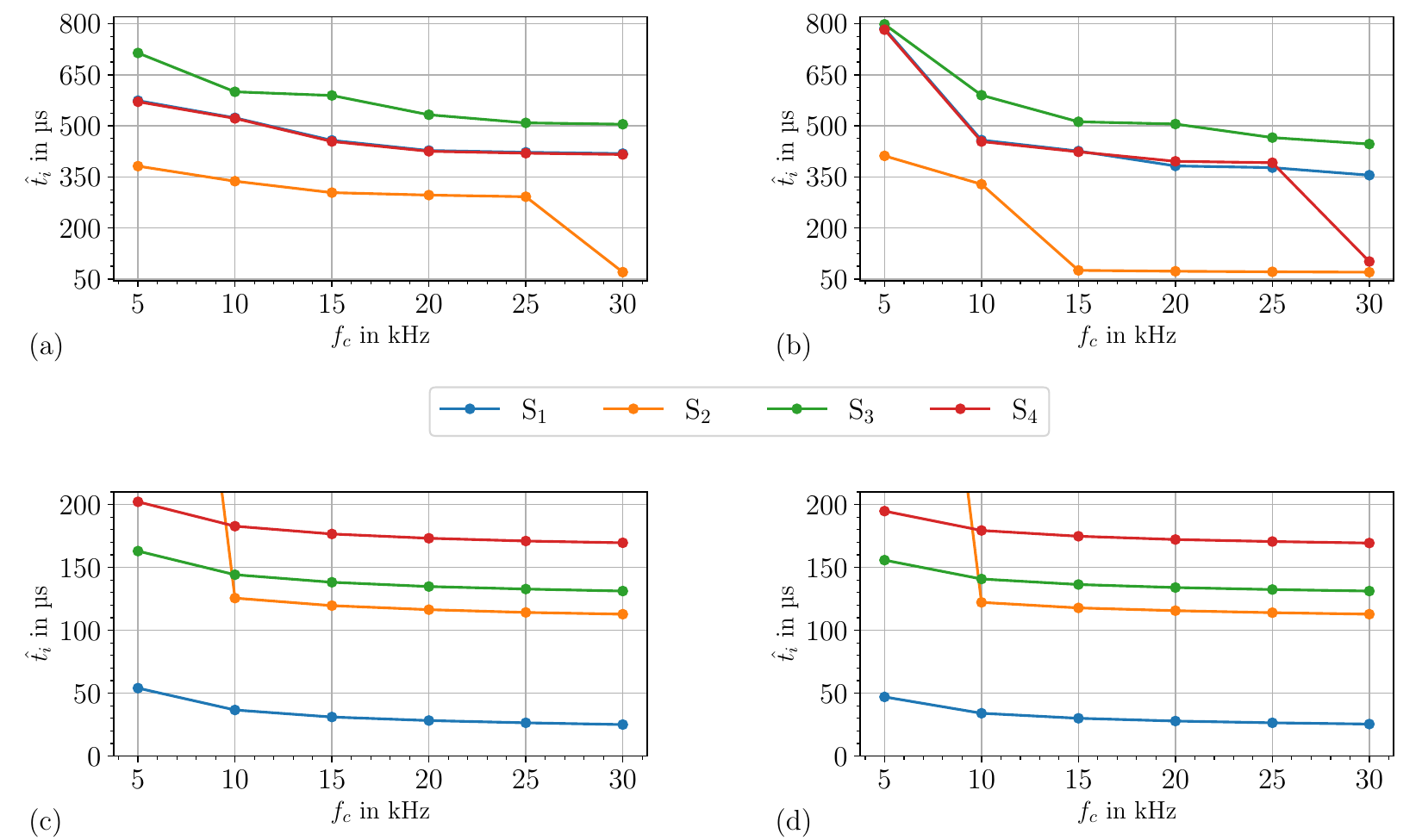}};        
        \begin{scope}[x={(image1.south east)},y={(image1.north west)}]
        \end{scope} 
    \end{tikzpicture}
    \caption{
        \ToA{}-estimation results obtained by MER with a window size of $\wine=1250$ ($\alpha=25$) after low-pass filtering of the noise-free sensor signals with varying cutoff frequency $f_c$, shown for all impacts under consideration: (a) $\imp{1}^{\iob}$, (b) $\imp{1}^{\iop}$, (c) $\imp{2}^{\iob}$ and (d) $\imp{2}^{\iop}$.
    }
    \label{fig::3D_mer_all_imp_filter}					
\end{figure}

For illustration, we focus on the case $\alpha=\num{25}$, shown in Fig. \ref{fig::3D_mer_all_imp_filter}, where the \ToA{}-estimates obained after low-pass filtering are plotted as functions of the cutoff frequency $f_c$.
To assess whether these \ToA{}-estimates correspond to a common physical wave mode, the associated group wave speeds are evaluated next.
From the distances between the impact location and the sensors, c.f., Tab. \ref{tab::3d_fe_model_sens_imp_pos}, together with the corresponding \ToA{}-estimates, an estimated group wave speed $\hat{c}_{g,i}^{j}$ is computed for each sensor $\sensor_i;~i=1\mhyphen4$ and impact $\imp{j};~j=1\mhyphen4$:
\begin{equation}\label{eq::def_est_cg}
    \hat{c}_{g,i}^{j} = \frac{\Vert \xb_{s_i} - \xb_{\imp{j}} \Vert}{\test[i]}.
\end{equation}
These velocities indicate whether the identified \ToA{}-estimates correspond to the $S_0$ or $A_0$ mode.
This additional step is essential, since a \ToA{}-estimate is only physically meaningful if the associated wave speeds are both plausible and sufficiently consistent across the sensors.
For the noise-free filtered signals, the \ToA{}-estimates shown in Fig. \ref{fig::3D_mer_all_imp_filter} and the corresponding group wave speeds listed in Tab. \ref{tab:cgest_mer_filter} indicate that this combined evaluation is successful for impacts at location $\imp{1}$.
Over a range of cutoff frequencies, the estimated \ToA{}-values can be associated with the $A_0$ mode, and the corresponding group wave speeds at $f_c=\SI{10}{\kilo\hertz}$ are tightly clustered. 
For the idealized impact $\imp{1}^{\iob}$, the inferred wave speeds range from \SIrange{991}{1074}{\metre\per\second}, while for the experiment-based impact $\imp{1}^{\iop}$ they range from \SIrange{1065}{1140}{\metre\per\second}, c.f., first row in Tab. \ref{tab:cgest_mer_filter}. 
This agreement across sensors strongly suggests that the identified \ToA{}-values correspond to a common physical wave phenomenon, namely the $A_0$ mode.
For impacts at location $\imp{2}$, however, the same procedure does not yield $A_0$-related estimates at any cutoff frequency.
Thus, even under noise-free conditions, the recovery of $A_0$-arrivals by MER in combination with filtering is impact-dependent and succeeds only for $\imp{1}$.

\begin{figure}[pos=!htbp]
    \centering
    \begin{tikzpicture}				
        \node[anchor=south west,inner sep=0] (image1) at (0, 0) {\includegraphics[width=\textwidth, trim=0 0 0 0, clip, ]{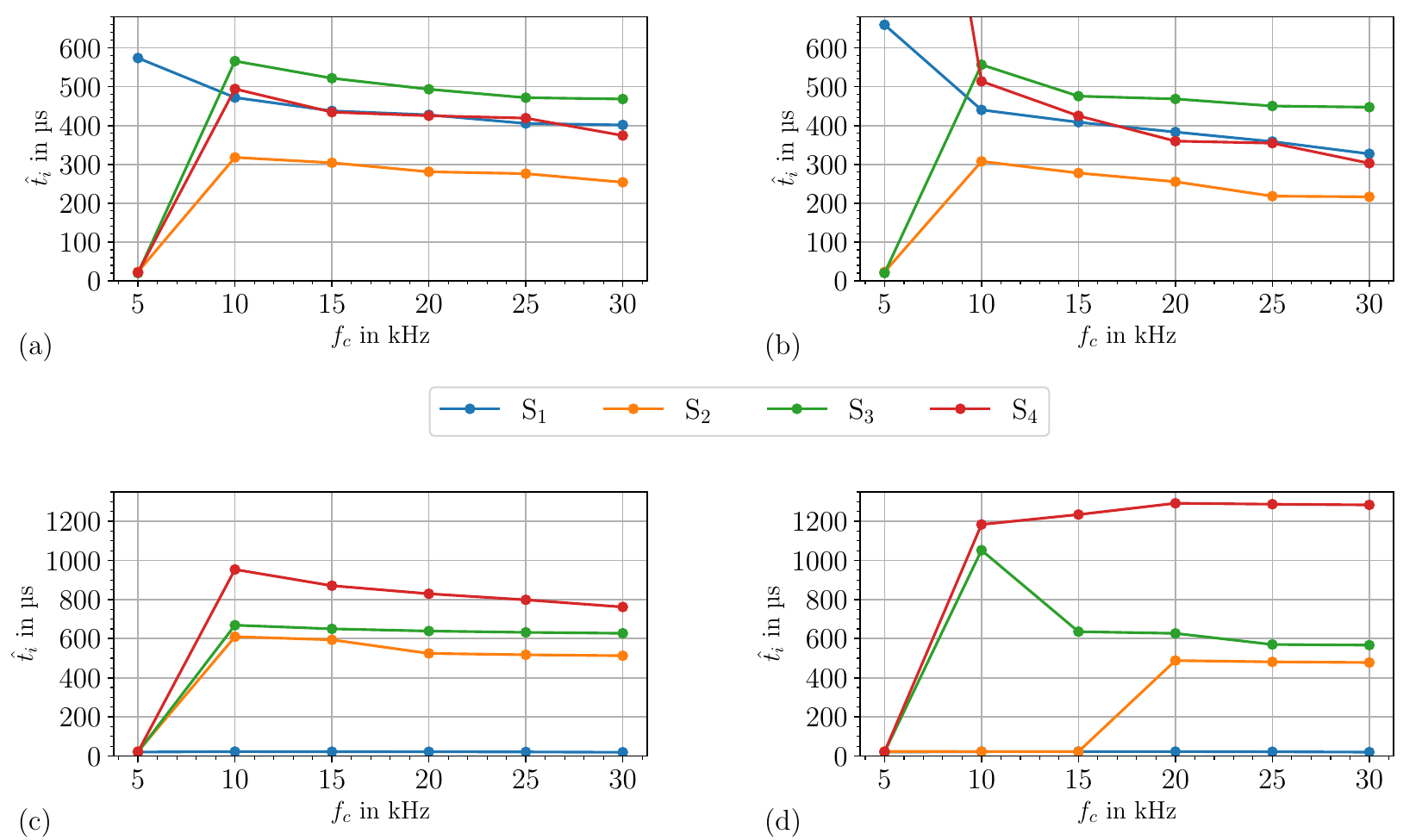}};        
        \begin{scope}[x={(image1.south east)},y={(image1.north west)}]
        \end{scope} 
    \end{tikzpicture}
    \caption{
        \ToA{}-estimation results obtained by MER with a window size of $\wine=1250$ ($\alpha=25$) after low-pass filtering of the noise-contaminated sensor signals with varying cutoff frequency $f_c$, shown for all impacts under consideration: (a) $\imp{1}^{\iob}$, (b) $\imp{1}^{\iop}$, (c) $\imp{2}^{\iob}$ and (d) $\imp{2}^{\iop}$.
    }
    \label{fig::3D_mer_all_imp_filter_snr50db}					
\end{figure}

We next consider the corresponding noise-contaminated filtered signals shown in Fig. \ref{fig::3D_mer_all_imp_filter_snr50db} and summarized in Tab. \ref{tab:cgest_mer_filter_snr50db} for a cutoff frequency of $f_c=\SI{10}{\kilo\hertz}$, in order to assess whether physically meaningful and cross-sensor-consistent \ToA{}-estimates can still be recovered in the presence of noise.
For impact $\imp{1}$, low-pass filtering again leads to a set of \ToA{}-estimates whose inferred group wave speeds are mutually consistent and lie in a range that can reasonably be assigned to the $A_0$ mode. 
At $f_c=\SI{10}{\kilo\hertz}$, the estimated wave speeds range from \SIrange{1047}{1137}{\metre\per\second} for $\imp{1}^{\iob}$, and from \SIrange{1107}{1179}{\metre\per\second} for $\imp{1}^{\iop}$.
Hence, despite the presence of noise, filtering enables MER to recover physically meaningful and cross-sensor-consistent \ToA{}-estimates for this impact location.
For impacts at location $\imp{2}$, by contrast, the filtered noise-contaminated signals do not lead to a comparably coherent picture.
The inferred wave speeds vary widely across the sensors.
This is particularly evident for the experiment-based impact $\imp{2}^{\iop}$, where the values listed in Tab. \ref{tab:cgest_mer_filter_snr50db} include \qty{643}{\metre\per\second}, \qty{745}{\metre\per\second}, \qty{4925}{\metre\per\second}, alongside one value of \qty{5320}{\metre\per\second}.
Such a set of wave speeds does not support a consistent assignment of the estimated \ToA{}-values across sensors. Instead, it indicates that at different sensors different features of the broadband impact response are identified, rather than a common arrival of the same wave mode.
The same conclusion follows qualitatively for the idealized impact $\imp{2}^{\iob}$, for which the estimated wave speeds are likewise inconsistent with an $A_0$-based interpretation.
Therefore, although filtering improves the results for $\imp{1}$, it does not provide a universally reliable approach for noise-contaminated sensor signals.

\begin{table}[pos=!htbp]
    \small
    \centering
    \begin{tabularx}{0.8\textwidth}{>{\raggedright\arraybackslash}p{0.5cm}|>{\raggedleft\arraybackslash}X>{\raggedleft\arraybackslash}X>{\raggedleft\arraybackslash}X>{\raggedleft\arraybackslash}X|>{\raggedleft\arraybackslash}X>{\raggedleft\arraybackslash}X>{\raggedleft\arraybackslash}X>{\raggedleft\arraybackslash}X} 
        \toprule        
        & \multicolumn{4}{c|}{$\hat{c}_g$ in $\SI{}{\meter\per\second}$: idealized impact}
        & \multicolumn{4}{c}{$\hat{c}_g$ in $\SI{}{\meter\per\second}$: experiment-based impact}
        \\
        & $\sensor_1$ 
        & $\sensor_2$ 
        & $\sensor_3$ 
        & $\sensor_4$ 
        & $\sensor_1$ 
        & $\sensor_2$ 
        & $\sensor_3$ 
        & $\sensor_4$ 
        \\        
        \midrule
        $\imp{1}$
        & \num{991} 
        & \num{1036}
        & \num{1074}
        & \num{992}
        & \num{1133}
        & \num{1065}
        & \num{1092}
        & \num{1140}
        \\     
        $\imp{2}$
        & \num{3014}
        & \num{4596}
        & \num{4692}
        & \num{4828}
        & \num{3244}
        & \num{4724}
        & \num{4805}
        & \num{4919}
        \\        
        \bottomrule
    \end{tabularx}
    \caption{
        Estimated values of group wave speed $\hat{c}_{g}$ with the associated \ToA{} values (obtained by MER)  when applying a low-pass filter to noise-free sensor signals with a cutoff frequency of $f_c=\SI{10}{\kilo\hertz}$.
    }
    \label{tab:cgest_mer_filter}
\end{table}

\begin{table}[pos=!htbp]
    \small
    \centering
    \begin{tabularx}{0.8\textwidth}{>{\raggedright\arraybackslash}p{0.5cm}|>{\raggedleft\arraybackslash}X>{\raggedleft\arraybackslash}X>{\raggedleft\arraybackslash}X>{\raggedleft\arraybackslash}X|>{\raggedleft\arraybackslash}X>{\raggedleft\arraybackslash}X>{\raggedleft\arraybackslash}X>{\raggedleft\arraybackslash}X} 
        \toprule        
        & \multicolumn{4}{c|}{$\hat{c}_g$ in $\SI{}{\meter\per\second}$: idealized impact}
        & \multicolumn{4}{c}{$\hat{c}_g$ in $\SI{}{\meter\per\second}$: experiment-based impact}
        \\
        & $\sensor_1$ 
        & $\sensor_2$ 
        & $\sensor_3$ 
        & $\sensor_4$ 
        & $\sensor_1$ 
        & $\sensor_2$ 
        & $\sensor_3$ 
        & $\sensor_4$ 
        \\        
        \midrule
        $\imp{1}$
        & \num{1099}
        & \num{1100}
        & \num{1137}
        & \num{1047}
        & \num{1179}
        & \num{1137}
        & \num{1157}
        & \num{1107}
        \\     
        $\imp{2}$
        & \num{4925}
        & \num{946}
        & \num{1012}
        & \num{925}
        & \num{4925}
        & \num{5320}
        & \num{643}
        & \num{745}
        \\        
        \bottomrule
    \end{tabularx}
    \caption{
        Estimated values of group wave speed $\hat{c}_{g}$ with the associated \ToA{} values (obtained by MER) when applying a low-pass filter to noise-contaminated sensor signals with a cutoff frequency of $f_c=\SI{10}{\kilo\hertz}$.
    }
    \label{tab:cgest_mer_filter_snr50db}
\end{table}

Overall, the MER method can be assessed as follows:
For noise-free sensor signals, it provides reliable \ToA{}-estimates of the earliest $S_0$-arrival and thus works well as a first-arrival detector.
For noise-contaminated raw signals, however, this robustness is lost, and no sufficiently consistent \ToA{}-estimation across sensors can be achieved.
Low-pass filtering can partially restore physically meaningful behavior by shifting the analysis towards the $A_0$ mode.
In the present study, this succeeds for impacts at location $\imp{1}$, for which the estimated \ToA{}-values yield consistent group wave speeds that can be associated with $A_0$.
For impacts at location $\imp{2}$, on the other hand, the resulting \ToA{}-estimates show substantial scatter across sensors, indicating that at different sensors different features of the broadband impact response are identified, rather than a common arrival that can be interpreted consistently across all sensors.
Hence, MER combined with filtering constitutes a viable strategy only under favorable conditions and cannot be regarded as a generally reliable method for \ToA{}-estimation on noise-contaminated sensor signals.

\subsection{Akaike Information Criterion}\label{ssec::3D_results_AIC}
In real-world measurements, poor SNR are a common challenge, and high-quality signals are often a rare luxury.
In our simulated case, however, the opposite occurs: the signal has an effectively perfect SNR, with values near zero before the wave arrival except for negligible numerical noise.
This poses a problem for the AIC, as the variance becomes so small that its logarithm in Eq. \eqref{eq::def_aic} approaches $-\infty$, disrupting the analysis.
To address this issue, Gaussian noise with a mean of $\mu_\mathrm{G} = \qty{0}{\volt}$ and a standard deviation of $\sigma_\mathrm{G} = \qty{1e-4}{\volt}$ was added to the sensor signals, ensuring a finite variance that allows the AIC to effectively identify the transition point.

\begin{figure}[pos=!htbp]
        \centering
        \begin{tikzpicture}				
            \node[anchor=south west,inner sep=0] (image1) at (0, 0) {\includegraphics[width=\textwidth, trim=0 0 0 0, clip, ]{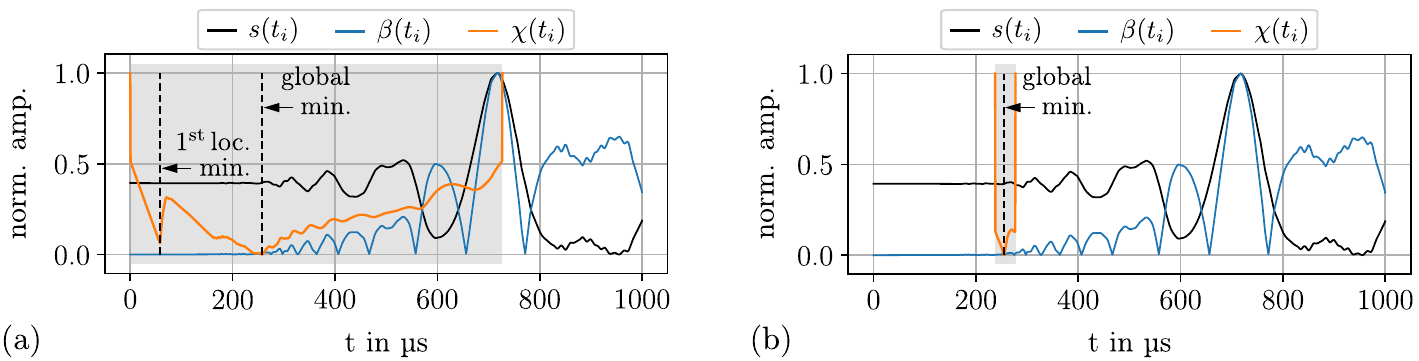}};        
            \begin{scope}[x={(image1.south east)},y={(image1.north west)}]
            \end{scope} 
        \end{tikzpicture}
        \caption{Demonstration of the two-step AIC procedure on a representative noise-free sensor signal ($\sensor_2$ for $\imp{1}^{\iop}$).
        The grey shaded area indicates the window, where the AIC, i.e., $\aic(t_i)$, is applied to the signal.
        In the first AIC step (a), the AIC window is determined by the AIC picker via the characteristic function $\beta(t_i)$ (Allen's formula \eqref{eq::allens_formula}) and in the second AIC step (b) the \ToA{}-estimation is repeated in a small time-domain around the first estimate.}
        \label{fig::aic_illustrate_aic_steps_3D}					
    \end{figure}

\begin{figure}[pos=!htbp]
    \centering
    \begin{tikzpicture}				
        \node[anchor=south west,inner sep=0] (image1) at (0, 0) {\includegraphics[width=\textwidth, trim=0 0 0 0, clip, ]{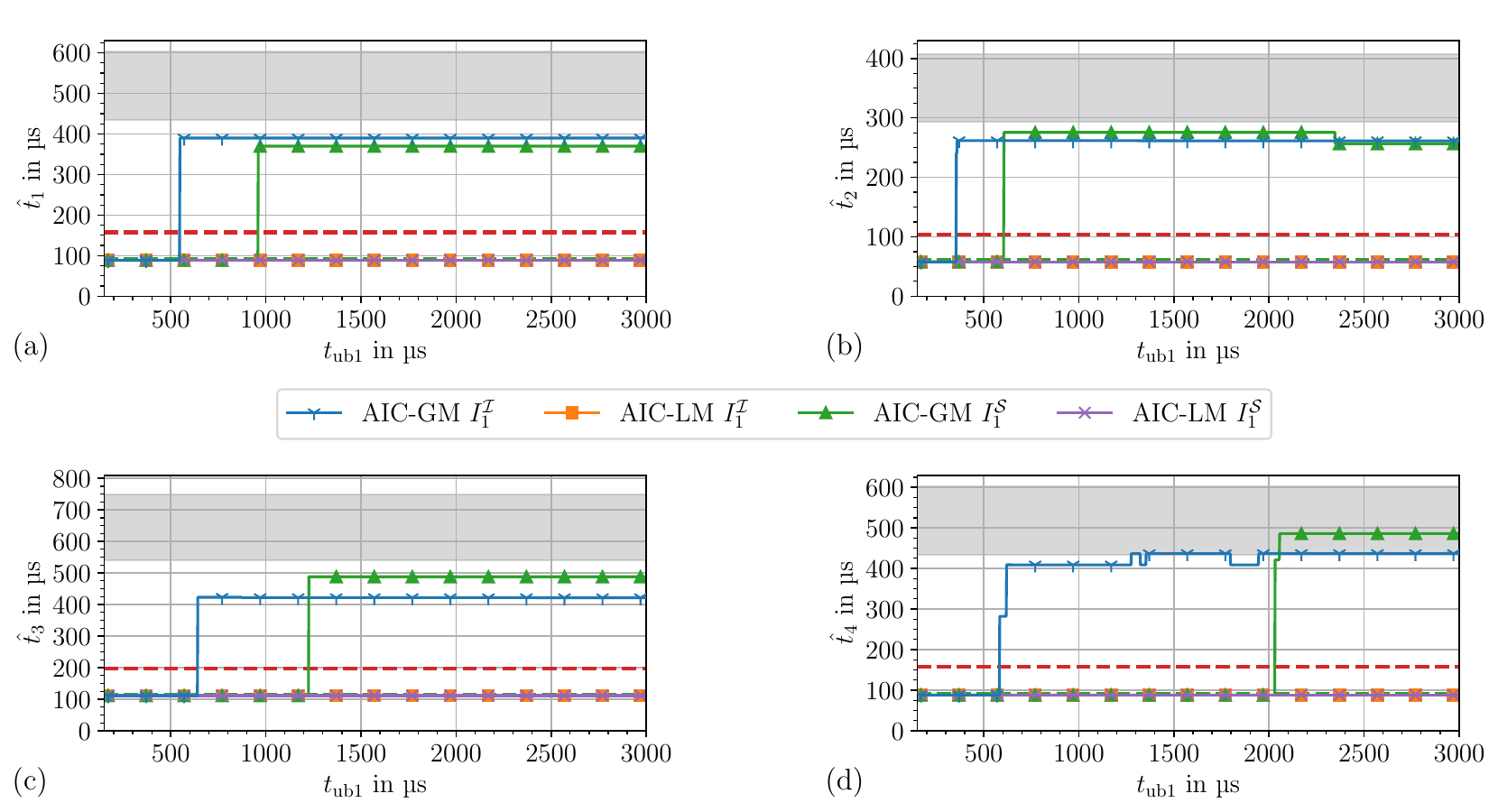}};        
        \begin{scope}[x={(image1.south east)},y={(image1.north west)}]
        \end{scope} 
    \end{tikzpicture}
    \caption{
    Influence of varying the upper bound $\tfirstub$ of the first AIC window on the estimated \ToA{} for sensors (a) $\sensor_1$, (b) $\sensor_2$, (c) $\sensor_3$ and (d) $\sensor_4$ at impact location $\imp{1}$.
    Shown in each case is the comparison between \aicglob{} and \aicloc{} for idealized and experiment-based impacts.
    The green and red dashed lines indicate the reference events $\tso$ and $\tao$, respectively.
    The grey shaded area marks the earliest possible arrival window of the $A_0$ mode between \qtyrange{10}{20}{\kilo\hertz}.}
    \label{fig::aic_vary_win_len_and_1st_loc_min_I1}					
\end{figure}

Although the AIC depends on several parameters, the present work shows that not all of them are equally relevant for the resulting \ToA{}-estimates.
In particular $\Ra$, $\tam$, and $\tfirstlb$ were found to have no noticeable influence on the noise-free sensor signals within the ranges investigated here.
This includes values of $\Ra$ from \numrange{0}{10} and of $\tam$ from \qtyrange{0}{100}{\micro\second}, which clearly extends beyond the recommendations by Sedlak et al. \cite{Sedlak_2009} discussed in Sec. \ref{subsec::aic}.
Likewise, setting the lower bound of the first AIC window to $\tfirstlb=\qty{0}{\micro\second}$, i.e. starting the analysis at the beginning of the signal, does not change the \ToA{}-estimates.
Hence, for the present signals, the first AIC window is taken to extend from the beginning of the signal up to its maximum absolute value, so that $\Ra$, $\tam$, and $\tfirstlb$ can be omitted from the subsequent parametric study.
We also examined the influence of the second AIC window parameters $\tfb$ and $\tfa$, by varying them from \SIrange{0.2}{600}{\micro\second}, corresponding to \SIrange{0.01}{20}{\percent} of the total time-range considered.
Over this wide range, the resulting \ToA{}-estimates changed only marginally, with deviations of at most $\pm\SI{1}{\micro\second}$.
In view of this minor influence, we set $\tfb=\tfa=\SI{40}{\micro\second}$ in all subsequent analyses and omit a separate parametric discussion of these quantities.
Figure \ref{fig::aic_illustrate_aic_steps_3D} illustrates the resulting two-step AIC procedure for a representative noise-free sensor signal ($\sensor_2$ for $\imp{1}^{\iop}$) in the time interval from \SIrange{0}{1000}{\micro\second}.
In Fig. \ref{fig::aic_illustrate_aic_steps_3D} (a), the shaded region denotes the time window in which the characteristic function $\allen(t_i)$ (blue curve), see Eq. \eqref{eq::allens_formula}, is used to define the first AIC step, while the corresponding AIC function $\aic(t_i)$ (orange curve) is evaluated from the sensor signal $s(t_i)$ (black curve).
The global minimum of $\aic(t_i)$ in this first step provides the first estimate $\tfe$ of the \ToA{}.
In Fig. \ref{fig::aic_illustrate_aic_steps_3D} (b), the minimization is repeated in a narrower time interval around $\tfe$ in order to refine the \ToA{}-estimate.
Notably, in this example, the {\em first local minimum} of $\aic(t_i)$ coincides with the first arrival of the $S_0$ mode, $\tso$.
Motivated by this observation, we now investigate how the size of the {\em first} AIC window affects the \ToA{}-estimate defined by the global minimum of $\aic(t_i)$.
At the same time, we examine whether the first local minimum of $\aic(t_i)$, which in Fig. \ref{fig::aic_illustrate_aic_steps_3D} coincides with $\tso$, can serve as a useful additional indicator for the earliest $S_0$-related arrival.
For this purpose, $\tfirstlb = \SI{0}{\micro\second}$ is kept fixed, while $\tfirstub$ is varied from \SIrange{160}{3000}{\micro\second} in increments of $\dt=\SI{0.2}{\micro\second}$.
The lower bound of this range is chosen just above the largest reference value of $\tso$ listed in Tab. \ref{tab:ts0_and_ta0_markers}, whereas the upper bound corresponds to the full simulated time-range $T$.
Thus, instead of determining the upper bound of the first AIC window by the picker, a broad range of possible first-window sizes is prescribed explicitly.
In the following, two evaluations are compared:
\begin{enumerate}
    \item {\em \aicglob{}}, in which the \ToA{} is taken as the {\em global} minimum (GM) of $\aic(t_i)$, consistent with the AIC definition used in this work, and
    \item {\em \aicloc{}}, in which the \ToA{} is taken as the {\em first local} minimum (LM) of $\aic(t_i)$, in order to assess its suitability as an indicator of $\tso$.
\end{enumerate}
To begin with, we consider the noise-free sensor signals, and the corresponding results for impact location $\imp{1}$ are shown in Fig. \ref{fig::aic_vary_win_len_and_1st_loc_min_I1}.
Each subfigure in Fig. \ref{fig::aic_vary_win_len_and_1st_loc_min_I1} (a)--(d) corresponds to one of the four sensor signals $\sensor_1-\sensor_4$, while the green and red dashed reference lines indicate $\tso$ and $\tao$, respectively.
It can be seen that the \ToA{}-estimates obtained from \aicglob{} coincide with $\tso$ only up to certain values of $\tfirstub$.
For larger values of $\tfirstub$, the global minimum of $\aic(t_i)$ shifts to a later minimum, which is reflected by the abrupt jumps in the estimated \ToA{}.
By contrast, the first local minimum remains at $\tso$ over the entire investigated range of $\tfirstub$.
Thus, for the present noise-free signals at impact location $\imp{1}$, the first local minimum \aicloc{} provides a stable indicator of the earliest $S_0$-related arrival, whereas the global minimum \aicglob{} becomes sensitive to the size of the first AIC window.

\begin{figure}[pos=!htbp]
    \centering
    \begin{tikzpicture}				
        \node[anchor=south west,inner sep=0] (image1) at (0, 0) {\includegraphics[width=\textwidth, trim=0 0 0 0, clip, ]{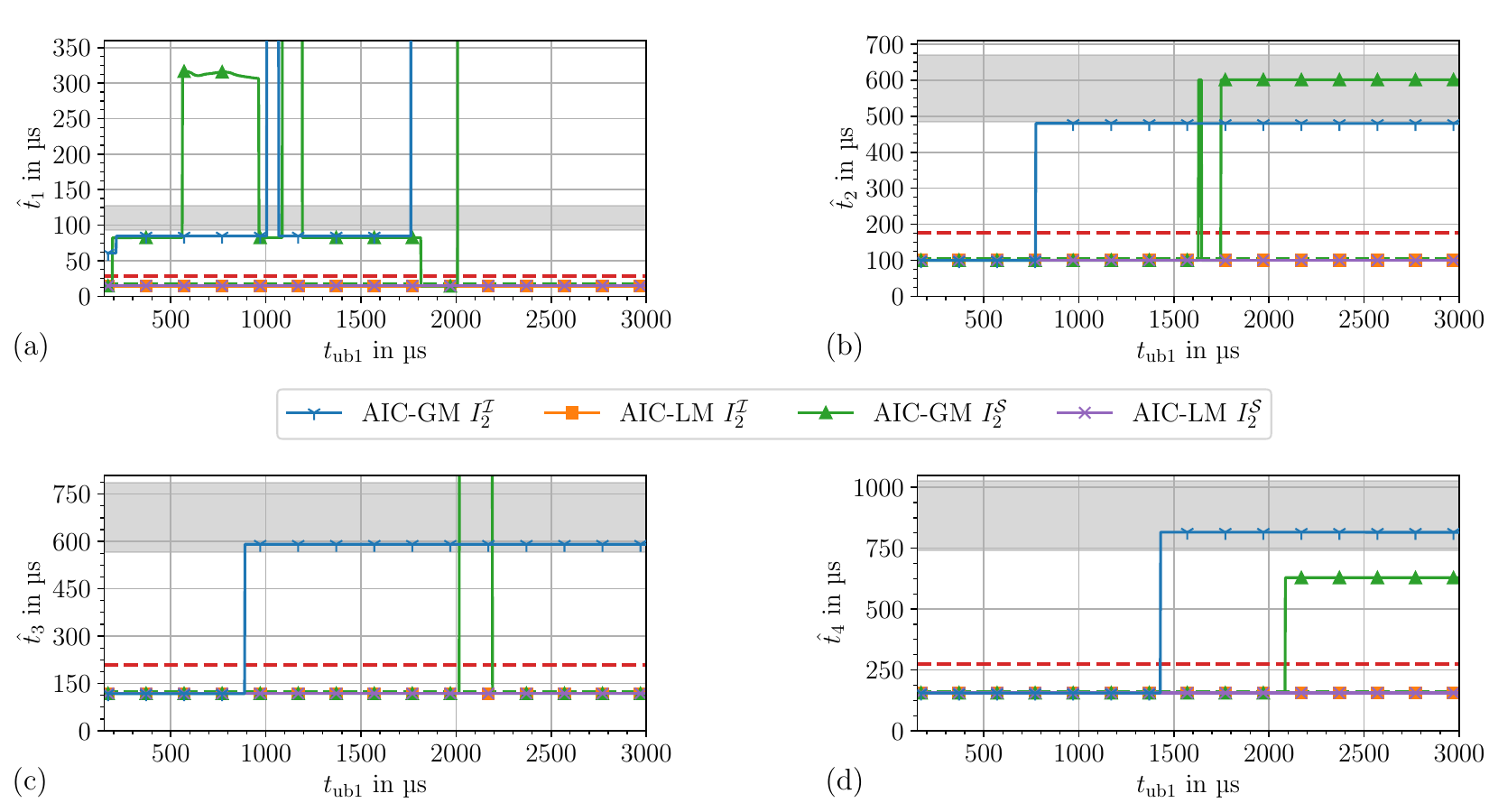}};        
        \begin{scope}[x={(image1.south east)},y={(image1.north west)}]
        \end{scope} 
    \end{tikzpicture}
    \caption{
    Same analysis as in Fig. \ref{fig::aic_vary_win_len_and_1st_loc_min_I1}, but for impact location $\imp{2}$.
    Shown in each case is the comparison between \aicglob{} and \aicloc{} for idealized and experiment-based impacts.
    The green and red dashed lines indicate the reference events $\tso$ and $\tao$, respectively.
    The grey shaded area marks the earliest possible arrival window of the $A_0$ mode between \qtyrange{10}{20}{\kilo\hertz}.}
    \label{fig::aic_vary_win_len_and_1st_loc_min_I2}					
\end{figure}
\begin{table}[pos=!htbp]
    \small
    \centering
    \begin{tabular}[]{r|r|r|r|r}
        \toprule
        & $\test[1]$ in \SI{}{\micro\second}
        & $\test[2]$ in \SI{}{\micro\second}
        & $\test[3]$ in \SI{}{\micro\second}
        & $\test[4]$ in \SI{}{\micro\second}
        \\        
        \midrule
        $\imp{1}^{\iob}$
        & \num{88.2}        
        & \num{57.6}        
        & \num{111.6}        
        & \num{88.2}        
        \\
        $\imp{1}^{\iop}$
        & \num{89.0}        
        & \num{58.0}        
        & \num{111.6}        
        & \num{88.6}        
        \\
        $\imp{2}^{\iob}$
        & \num{14.2}        
        & \num{99.8}        
        & \num{117.8}        
        & \num{155.4}        
        \\
        $\imp{2}^{\iop}$
        & \num{14.6}        
        & \num{99.8}        
        & \num{118.2}        
        & \num{155.6}        
        \\        
        \bottomrule
    \end{tabular}
    \caption{
        Numerical \ToA{}-estimates for sensors $\sensor_1$--$\sensor_4$ for all impact positions and types when \aicloc{} is used.
    }
    \label{tab:aic_results_mod_AIC}
\end{table}
An even clearer difference is observed for impact location $\imp{2}$.
As shown in Fig. \ref{fig::aic_vary_win_len_and_1st_loc_min_I2}, the sensor closest to the impact, $\sensor_1$, does not detect $\tso$ at all when the \ToA{} is evaluated from the global minimum of $\aic(t_i)$.
By contrast, the first local minimum remains associated with $\tso$ over the entire investigated range of $\tfirstub$.
The same qualitative behavior is also found for the other sensors, although it is most evident for $\sensor_1$.
Table \ref{tab:aic_results_mod_AIC} lists the corresponding numerical \ToA{}-estimates obtained from \aicloc{} for all sensors, impact locations, and impact types.
A comparison with the reference values of $\tso$ in Tab. \ref{tab:ts0_and_ta0_markers} shows that these estimates agree closely for all sensors and impact cases, with only a slight tendency towards earlier values by a few microseconds.
Hence, the results for $\imp{2}$ confirm even more clearly that, for the present noise-free signals, the first local minimum provides a stable indicator of the earliest $S_0$-related arrival, whereas the global minimum can shift to later minima depending on the size of the first AIC window.

\begin{figure}[pos=!htbp]
    \centering
    \begin{tikzpicture}				
        \node[anchor=south west,inner sep=0] (image1) at (0, 0) {\includegraphics[width=\textwidth, trim=0 0 0 0, clip, ]{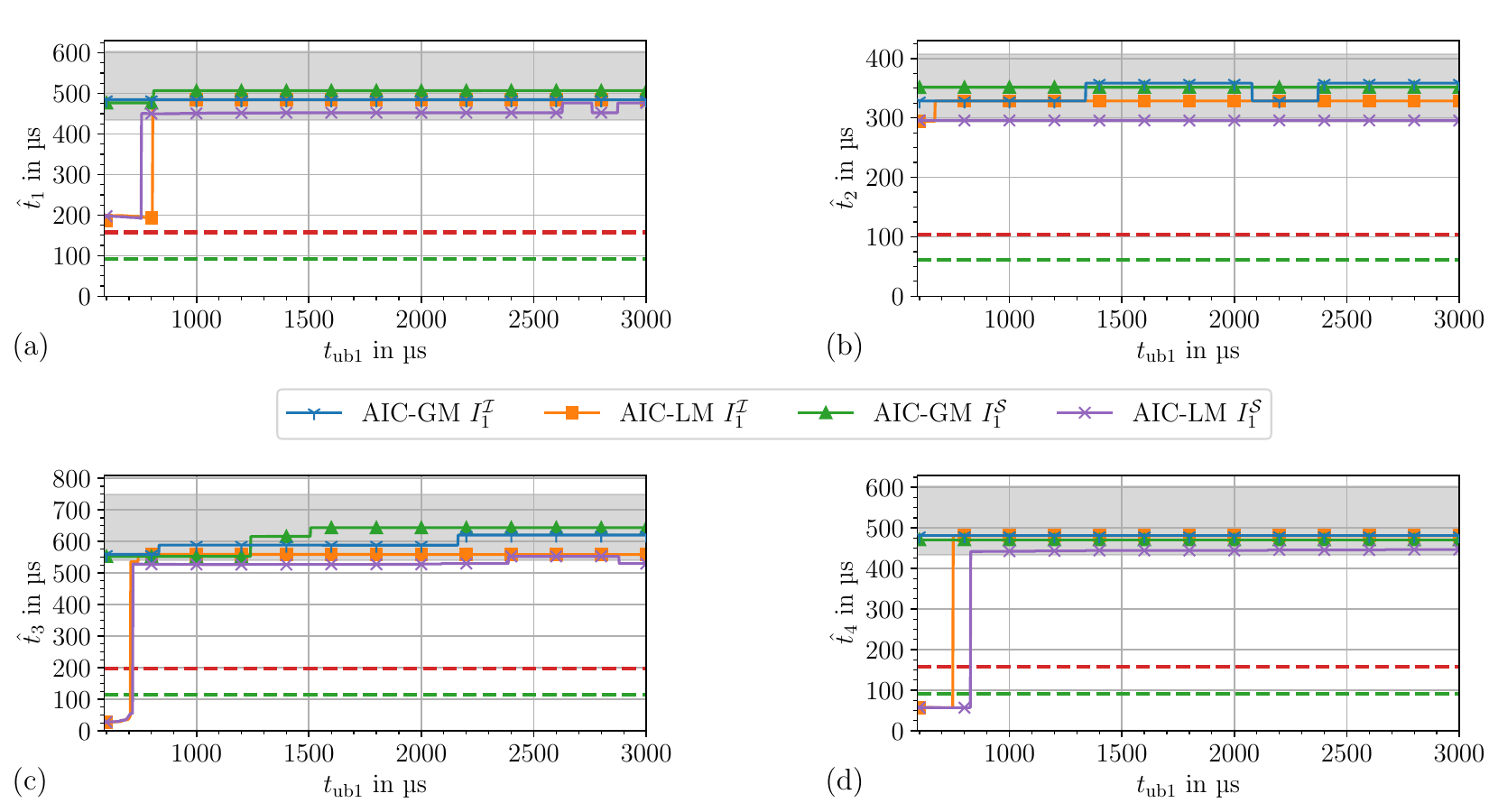}};        
        \begin{scope}[x={(image1.south east)},y={(image1.north west)}]
        \end{scope} 
    \end{tikzpicture}
    \caption{
    Influence of varying the upper bound $\tfirstub$ of the first AIC window on the estimated \ToA{} for sensors (a) $\sensor_1$, (b) $\sensor_2$, (c) $\sensor_3$ and (d) $\sensor_4$ at impact location $\imp{1}$.
    Shown in each case is the comparison between \aicglob{} and \aicloc{} for idealized and experiment-based impacts.
    The green and red dashed lines indicate the reference events $\tso$ and $\tao$, respectively.
    The grey shaded area marks the earliest possible arrival window of the $A_0$ mode between \qtyrange{10}{20}{\kilo\hertz}.}
    \label{fig::aic_vary_win_len_and_1st_loc_min_I1_noise50dB}					
\end{figure}

\begin{figure}[pos=!htbp]
    \centering
    \begin{tikzpicture}				
        \node[anchor=south west,inner sep=0] (image1) at (0, 0) {\includegraphics[width=\textwidth, trim=0 0 0 0, clip, ]{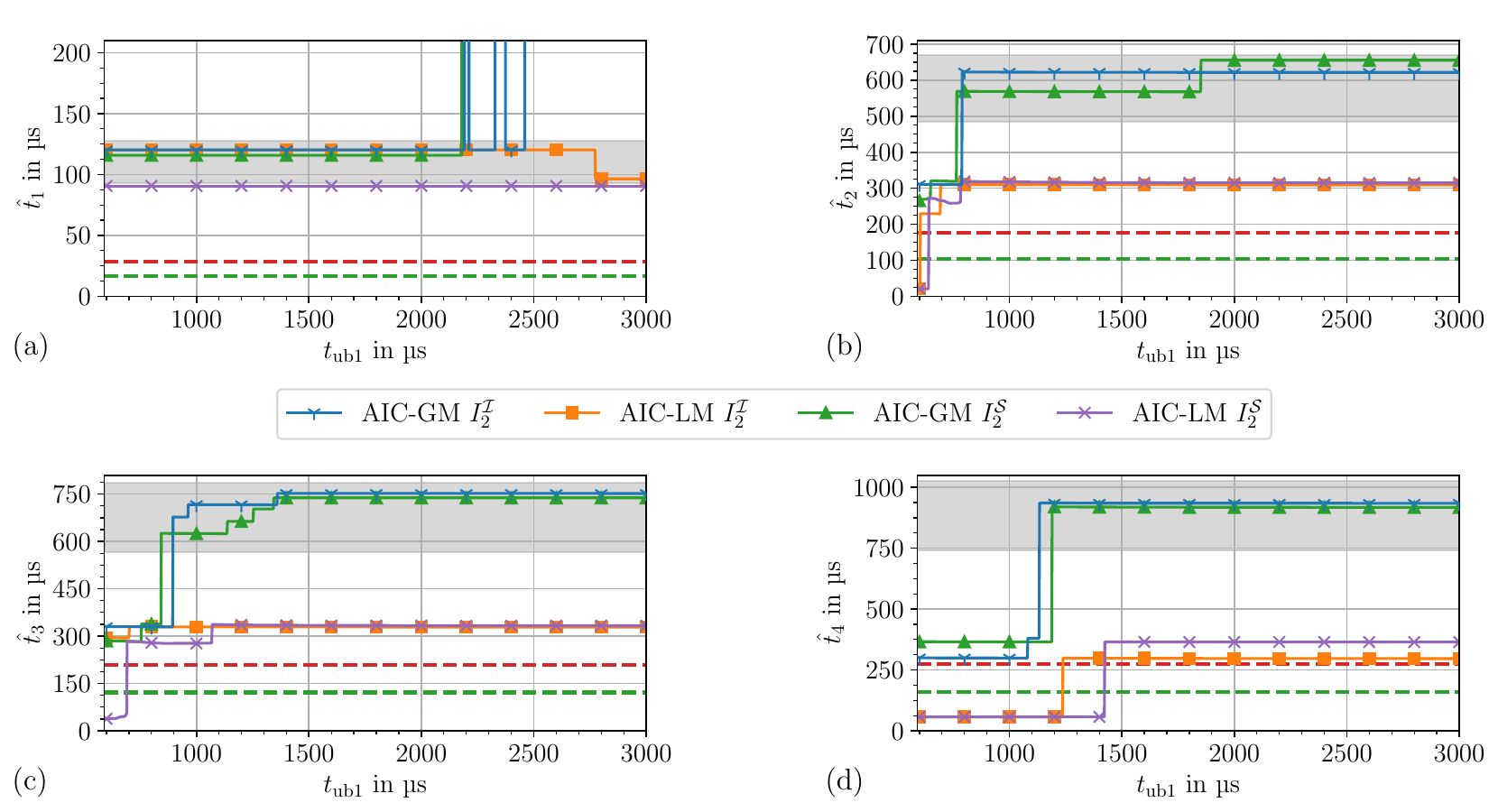}};        
        \begin{scope}[x={(image1.south east)},y={(image1.north west)}]
        \end{scope} 
    \end{tikzpicture}
    \caption{
    Same analysis as in Fig. \ref{fig::aic_vary_win_len_and_1st_loc_min_I1_noise50dB}, but for impact location $\imp{2}$.
    Shown in each case is the comparison between \aicglob{} and \aicloc{} for idealized and experiment-based impacts.
    The green and red dashed lines indicate the reference events $\tso$ and $\tao$, respectively.
    The grey shaded area marks the earliest possible arrival window of the $A_0$ mode between \qtyrange{10}{20}{\kilo\hertz}.}
    \label{fig::aic_vary_win_len_and_1st_loc_min_I2_noise50dB}					
\end{figure}

We now investigate how the AIC performs for noise-contaminated sensor signals and whether physically meaningful \ToA{}-values remain detectable.
The corresponding results are shown in Figs. \ref{fig::aic_vary_win_len_and_1st_loc_min_I1_noise50dB} and \ref{fig::aic_vary_win_len_and_1st_loc_min_I2_noise50dB} for impact locations $\imp{1}$ and $\imp{2}$, respectively.
In contrast to the noise-free case, neither \aicglob{} nor \aicloc{} now provides a uniformly reliable detection of $\tso$ across all sensors and both impact types at both impact positions.
Instead, many parameter combinations produce later \ToA{}-estimates.
This is consistent with the observations made in the previous sections, namely that in the presence of noise, the initial $S_0$ arrival is masked and can no longer be detected reliably in the sensor signals.
For impact location $\imp{1}$ in Fig. \ref{fig::aic_vary_win_len_and_1st_loc_min_I1_noise50dB}, both AIC variants predominantly settle at clearly later times, which in most cases lie within the reference window associated with $A_0$-related arrivals in the frequency range from \qtyrange{10}{20}{\kilo\hertz}.
Thus, although the robust detection of $\tso$ is lost, the TOA-estimates do not become arbitrary.
Instead, \ToA{}-estimates exhibit parameter-insensitive regimes that point to later wave features with a plausible physical interpretation.
A similar, though more differentiated, behavior is observed for impact location $\imp{2}$, see Fig. \ref{fig::aic_vary_win_len_and_1st_loc_min_I2_noise50dB}.
Here, \aicglob{} again tends to lock onto comparatively late and largely constant \ToA{}-estimates, which are mostly located within the arrival-time window associated with the $A_0$ mode between \qtyrange{10}{20}{\kilo\hertz}.
\aicloc{}, by contrast, often yields earlier \ToA{}-estimates that remain nearly unchanged, particularly for sensors $\sensor_1$--$\sensor_3$; when changes occur, they usually take the form of jumps to another level, which again remains nearly constant.
However, these levels differ significantly from sensor to sensor.
Thus, for the noise-contaminated signals, a stable dependence on the window size does not automatically imply cross-sensor consistency or detection of the same wave portion.
Against this background, we now examine whether suppressing higher-frequency signal contributions by low-pass filtering, following the same approach as in Sec. \ref{ssec::3d_results_MER}, shifts the AIC output towards later low-frequency wave features associated with the $A_0$ mode.
More specifically, the question is whether this preprocessing step enables the detection of a consistent portion of the $A_0$ mode across all sensors, rather than merely producing parameter-insensitive but sensor-dependent late-time estimates.


\begin{figure}[pos=!htbp]
    \centering
    \begin{tikzpicture}				
        \node[anchor=south west,inner sep=0] (image1) at (0, 0) {\includegraphics[width=0.9\textwidth, trim=0 0 0 0, clip, ]{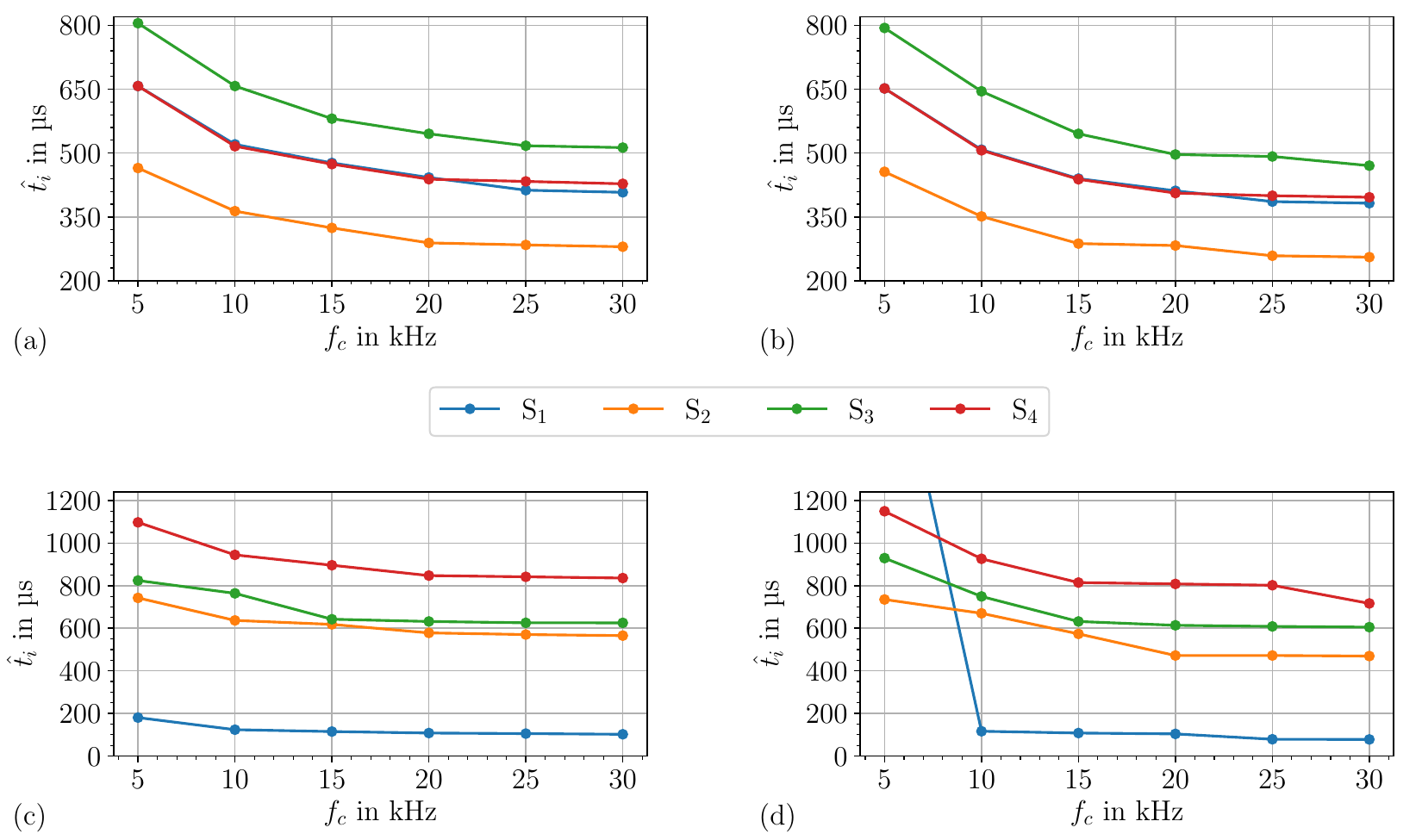}};        
        \begin{scope}[x={(image1.south east)},y={(image1.north west)}]
        \end{scope} 
    \end{tikzpicture}
    \caption{
        \ToA{}-estimation results (obtained by AIC) by filtering the sensor signals with different cutoff frequencies for all impacts under consideration: (a) $\imp{1}^{\iob}$, (b) $\imp{1}^{\iop}$, (c) $\imp{2}^{\iob}$ and (d) $\imp{2}^{\iop}$.
    }
    \label{fig::3D_aic_all_imp_filter}					
\end{figure}

The corresponding results for the noise-free sensor signals are shown in Fig. \ref{fig::3D_aic_all_imp_filter}, while Tab. \ref{tab:cgest_aic_filter} lists the associated group wave speeds $\hat{c}_{g}$ for a cutoff frequency of $f_c = \qty{10}{\kilo\hertz}$.
These results were obtained using \aicglob{}.
For all four impact cases, the \ToA{}-estimates shift to clearly later times and can be associated with the arrival of the $A_0$ mode.
For impact location $\imp{1}$, sensors $\sensor_1$ and $\sensor_4$ yield very similar \ToA{}-estimates, as expected from their approximately equal distance to the impact.
Across all sensors, the associated group wave speeds show only minor variation, see Tab. \ref{tab:cgest_aic_filter}.
For the idealized impact at $\imp{1}$, they range from \qtyrange{962}{1003}{\metre\per\second}, while for the experiment-based impact they lie between \qtyrange{996}{1022}{\metre\per\second}.
For impact location $\imp{2}$, the estimated wave speeds are somewhat lower, namely \qtyrange{886}{934}{\metre\per\second} for the idealized impact and \qtyrange{861}{953}{\metre\per\second} for the experiment-based impact, but they again remain closely grouped for all sensors.
Hence, in the noise-free case, low-pass filtering enables the AIC to identify a consistent portion of the $A_0$ mode across all sensors and for both impact locations.

\begin{figure}[pos=!htbp]
    \centering
    \begin{tikzpicture}				
        \node[anchor=south west,inner sep=0] (image1) at (0, 0) {\includegraphics[width=0.9\textwidth, trim=0 0 0 0, clip, ]{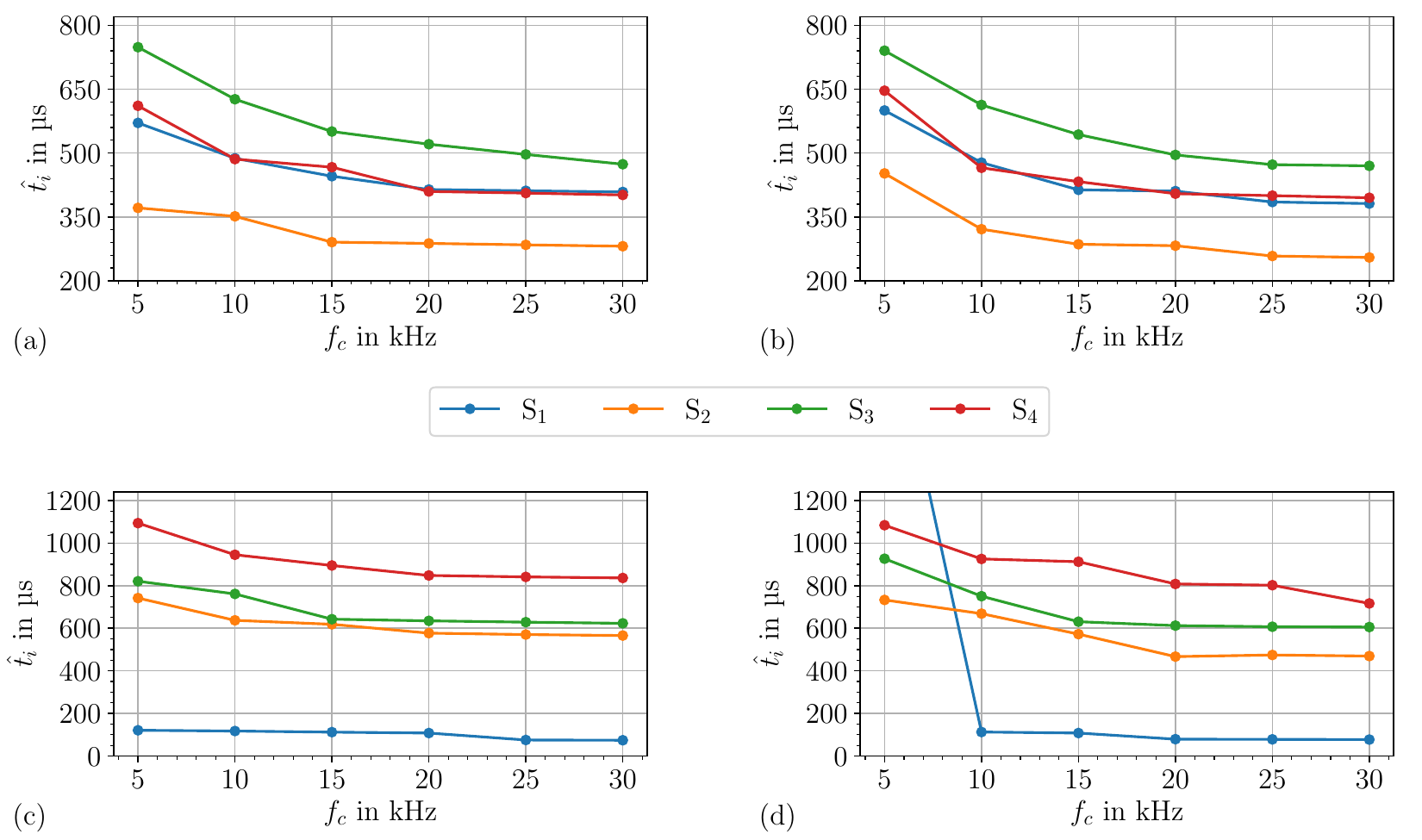}};        
        \begin{scope}[x={(image1.south east)},y={(image1.north west)}]
        \end{scope} 
    \end{tikzpicture}
    \caption{
        \ToA{}-estimation results (obtained by AIC) by filtering the noise-contaminated sensor signals with different cutoff frequencies for all impacts under consideration: (a) $\imp{1}^{\iob}$, (b) $\imp{1}^{\iop}$, (c) $\imp{2}^{\iob}$ and (d) $\imp{2}^{\iop}$.
    }
    \label{fig::3D_aic_all_imp_filter_snr50db}					
\end{figure}

Figure \ref{fig::3D_aic_all_imp_filter_snr50db} shows that low-pass filtering also shifts the \ToA{}-estimates of the noise-contaminated sensor signals to later times that can be associated with the arrival of the $A_0$ mode.
The associated group wave speeds listed in Tab. \ref{tab:cgest_aic_filter_snr50db} remain comparatively close across the sensors for both impact locations and both impact types, indicating that the cross-sensor consistency observed for the filtered noise-free case is largely retained even in the presence of noise. 
Thus, in contrast to the unfiltered noise-contaminated results, low-pass filtering enables the AIC to identify a consistent portion of the $A_0$ mode also under noisy signal conditions.


\begin{table}[pos=!htbp]
    \small
    \centering
    \begin{tabularx}{0.8\textwidth}{>{\raggedright\arraybackslash}p{0.5cm}|>{\raggedleft\arraybackslash}X>{\raggedleft\arraybackslash}X>{\raggedleft\arraybackslash}X>{\raggedleft\arraybackslash}X|>{\raggedleft\arraybackslash}X>{\raggedleft\arraybackslash}X>{\raggedleft\arraybackslash}X>{\raggedleft\arraybackslash}X}
        \toprule        
        & \multicolumn{4}{c|}{$\hat{c}_g$ in $\SI{}{\meter\per\second}$: idealized impact}
        & \multicolumn{4}{c}{$\hat{c}_g$ in $\SI{}{\meter\per\second}$: experiment-based impact}
        \\
        & $\sensor_1$ 
        & $\sensor_2$ 
        & $\sensor_3$ 
        & $\sensor_4$ 
        & $\sensor_1$ 
        & $\sensor_2$ 
        & $\sensor_3$ 
        & $\sensor_4$ 
        \\        
        \midrule
        $\imp{1}$
        & \num{997}
        & \num{962}
        & \num{980}
        & \num{1003}
        & \num{1021}
        & \num{996}
        & \num{998}
        & \num{1022}
        \\     
        $\imp{2}$
        & \num{893}
        & \num{907}
        & \num{886}
        & \num{934}
        & \num{949}
        & \num{861}
        & \num{903}
        & \num{953}
        \\        
        \bottomrule
    \end{tabularx}
    \caption{
        Estimated values of group wave speed $\hat{c}_{g}$ with the associated \ToA{} values (obtained by \aicglob{}) when applying a low-pass filter to noise-free sensor signals with a cutoff frequency of $f_c=\SI{10}{\kilo\hertz}$.
    }
    \label{tab:cgest_aic_filter}
\end{table}

\begin{table}[pos=!htbp]
    \small
    \centering
    \begin{tabularx}{0.8\textwidth}{>{\raggedright\arraybackslash}p{0.5cm}|>{\raggedleft\arraybackslash}X>{\raggedleft\arraybackslash}X>{\raggedleft\arraybackslash}X>{\raggedleft\arraybackslash}X|>{\raggedleft\arraybackslash}X>{\raggedleft\arraybackslash}X>{\raggedleft\arraybackslash}X>{\raggedleft\arraybackslash}X}
        \toprule        
        & \multicolumn{4}{c|}{$\hat{c}_g$ in $\SI{}{\meter\per\second}$: idealized impact}
        & \multicolumn{4}{c}{$\hat{c}_g$ in $\SI{}{\meter\per\second}$: experiment-based impact}
        \\
        & $\sensor_1$ 
        & $\sensor_2$ 
        & $\sensor_3$ 
        & $\sensor_4$ 
        & $\sensor_1$ 
        & $\sensor_2$ 
        & $\sensor_3$ 
        & $\sensor_4$ 
        \\        
        \midrule
        $\imp{1}$
        & \num{1064}
        & \num{995}
        & \num{1028}
        & \num{1065}
        & \num{1086}
        & \num{1089}
        & \num{1050}
        & \num{1112}
        \\     
        $\imp{2}$
        & \num{941}
        & \num{906}
        & \num{888}
        & \num{934}
        & \num{980}
        & \num{863}
        & \num{901}
        & \num{953}
        \\        
        \bottomrule
    \end{tabularx}
    \caption{
        Estimated values of group wave speed $\hat{c}_{g}$ with the associated \ToA{} values (obtained by \aicglob{}) when applying a low-pass filter to noise-contaminated sensor signals with a cutoff frequency of $f_c=\SI{10}{\kilo\hertz}$.
    }
    \label{tab:cgest_aic_filter_snr50db}
\end{table}
Overall, the AIC results reveal two complementary strengths of the method.
For unfiltered noise-free sensor signals, the first local minimum \aicloc{} provides a robust estimator of the earliest $S_0$-related arrival $\tso$.
In the presence of noise, this earliest arrival is masked and can no longer be identified consistently.
However, after low-pass filtering, the AIC yields $A_0$-related \ToA{}-estimates whose corresponding group wave speeds remain comparatively close across the sensors for both impact locations and both impact types.
Thus, the method combines reliable $\tso$-detection under favorable signal conditions with a successful and cross-sensor-consistent detection of $A_0$-related arrivals after suitable preprocessing, even in the presence of noise.
\section{Summary and Conclusions}
The main objective of this work is to provide a good understanding of the behavior of different \ToA{}-estimation methods and to derive guidelines for obtaining the most accurate and reliable results possible.
We have examined different methods (TC, CWT, \sla{}, MER, and AIC) based on fundamental considerations and applied them to sensor signals obtained from a 3D FE model of an aluminum plate, simulating Lamb wave propagation induced by different impact types and locations.
In our case, \ToA{}-estimation of Lamb waves in 3D isotropic elastic continua differs, due to the characteristics of the impact, e.g., broadness in frequency-domain, from typical SHM applications.
In the latter case, as mentioned in Sec. \ref{sec::results}, the excitation is ``well-defined'' and also narrow-banded and in our case there is no such thing as a single \ToA{}.
Instead, the challenge is to identify parameter ranges for each method where estimates are consistent and physically meaningful.
To assess robustness under more realistic conditions, the methods were applied not only to idealized noise-free signals, but also to noise-contaminated signals with an SNR of \qty{45}{\decibel}, derived from experimental observations.
Under these conditions, the earliest arrivals are partially obscured by noise, which fundamentally affects the detectability of the $S_0$ mode.
This difficulty is further increased by the limited sensitivity of surface-bonded piezoceramic sensors to symmetric modes.

The novelty of this work lies in extending and refining existing methods: we introduced a frequency-domain threshold crossing within the CWT framework that enhances both robustness and accuracy of \ToA{}-estimation, and we demonstrated that using local minima in the AIC improves the reliable detection of the $S_0$ mode.
Beyond these methodological contributions, the study provides practical guidelines and insights into the specific characteristics of each assessed method, supporting accurate and reliable \ToA{}-estimation for applications such as impact localization.
Based on the obained findings, we summarize the results of this work:
\begin{itemize}    
    \item TC:
    This method is attractive for its simplicity.
    For noise-free signals, very small thresholds can detect the earliest $S_0$-related arrival $\tso$.
    In the presence of noise, however, this advantage is lost, since the first $S_0$ arrival is masked and the method becomes even more sensitive to the threshold choice.
    For very low thresholds, the noise floor itself triggers the estimation; for higher thresholds, the estimated \ToA{} shifts monotonically to later times.
    Hence, TC is of limited robustness under realistic conditions and is best regarded as a supporting tool rather than a stand-alone method.
    \item CWT:    
    Although this method requires considerable a priori knowledge such as wavelet type, frequency resolution, and preprocessing, it offers a powerful combination of temporal and spectral resolutions, enabling detailed analysis of how signal frequency content evolves over time.
    For noise-free signals, slicing the scalogram at selected frequencies and applying threshold crossing in the frequency domain enables robust detection of the earliest $S_0$ mode and yields remarkably consistent relative \ToA{}-differences across sensors, impact types, and impact locations.    
    In the presence of noise, however, the initial $S_0$ arrival can no longer be robustly identified, because it is masked by noise.
    In that case, the same framework remains useful for the $A_0$ mode: by shifting the analysis to lower frequencies, physically consistent \ToA{}-estimates and meaningful relative \ToA{}-differences can still be obtained, though limitations remain for impacts very close to individual sensors and near the COI.    
    \item SLA:
    By introducing parameters $\alpha$ and $\beta$, which represent multiples of the dominant period of the sensor signal, we provided a systematic way to investigate the influence of the short- and long-term averaging windows.
    For noise-free signals, the combined parameter $\alpha\beta$ largely governs the result, and sufficiently large values lead to stable convergence to the earliest $S_0$ arrival across sensors and impact positions.
    In the presence of noise, the first $S_0$ arrival is masked and the estimates shift to later times; for large $\alpha\beta$ they approach plateaus that mostly fall into the $A_0$-related arrival window in the \qtyrange{10}{20}{\kilo\hertz} range, but with noticeable dependence on the impact type.
    Low-pass filtering does not systematically improve $A_0$ detection for SLA under the conditions studied.
    \item MER:
    Similar to SLA, MER relies on window-based averaging, but with two windows of equal length, so that only one governing parameter remains.
    For noise-free signals, MER provides reliable \ToA{}-estimates of the earliest $S_0$ arrival and works well as a first-arrival detector, especially for the experiment-based impacts, while larger window lengths are required for the idealized impacts.
    For noise-contaminated signals, this robustness is lost and no sufficiently consistent \ToA{}-estimation across sensors can be achieved.
    Low-pass filtering can partially restore physically meaningful behavior by shifting the analysis towards the $A_0$ mode; this works well for impacts at location $\imp{1}$, but fails to provide a universally reliable solution for impacts at location $\imp{2}$.
    Hence, MER combined with filtering is viable only under favorable conditions.
    \item AIC:
    The AIC method reveals two complementary strengths in the present assessment.
    For unfiltered noise-free sensor signals, it reliably estimates the first $S_0$-related arrival $\tso$ over a wide range of parameter settings, even when the impact occurs very close to a sensor.
    In particular, using the proposed first local minimum (\aicloc{}) consistently yields $\tso$-detection, whereas the commonly used global minimum (\aicglob{}) tends to produce later estimates that cannot be associated with the reference events and may shift depending on the first AIC window.
    This highlights the benefit of the local-minimum-based approach as a major contribution of the present work.
    In the presence of noise, the earliest $S_0$ arrival is masked and can no longer be identified consistently. Nevertheless, the AIC estimates do not become arbitrary, but tend to settle in parameter-insensitive regimes associated with later wave features.
    Most importantly, after low-pass filtering, the AIC framework also proves capable of detecting $A_0$-related arrivals.
    The corresponding \ToA{}-estimates yield group wave speeds that remain comparatively close across sensors for both impact locations and both impact types, indicating a high degree of cross-sensor consistency even under noise-contaminated conditions.
    Overall, the AIC method emerges as a stable and versatile technique: it enables reliable $S_0$-detection under favorable signal conditions, while also providing meaningful and physically plausible $A_0$-related \ToA{}-estimates after suitable preprocessing. 
\end{itemize}
This assessment makes clear that \ToA{}-estimation is not about identifying a universally superior method, but about understanding the specific strengths and limitations of each approach, including their sensitivity to realistic noise levels.
In particular, the results show that noise primarily compromises the detection of the earliest $S_0$-related arrivals, while meaningful $A_0$-related \ToA{}-estimates can still be recovered by suitable time-frequency analysis or preprocessing.
Building on the results obtained in the present work, the application of selected \ToA{}-estimation methods on an anisotropic plate-like structure made of carbon fibre reinforced polymer has already been conducted (see Grasboeck et al. \cite{Grasboeck_2025}).
The next step is the investigation of impact localization methods using proven \ToA{}-estimation methods, employing them on both isotropic and anisotropic plate-like structures.

\appendix
\section{PIC255 Piezoceramic Material Parameters}\label{app:a}
\begin{table}[pos=H]
    \centering
    \small
    \caption{Material parameters of modeled piezoelectric patches of PIC255 \label{tab::param_piezo}}
    \begin{tabular}{lcl}
        \toprule
        \textbf{Parameter} & \textbf{Value} & \textbf{Description}\\
        \midrule
        $ d_p $                   & $ \SI{0.5}{\milli\metre} $              & Thickness                                \\
        $ w_p $                   & $ \SI{30}{\milli\metre} $               & Length and width                         \\
        $ \rho_p $                & $ \SI{7800}{\kilogram\per\metre^3} $    & Volume mass density                      \\
        \hline
        $ \epsoneone^S $          & $ \SI{930}{} $                          & Relative permittivity at constant strain \\
        $ \epsthrthr^S $          & $ \SI{857}{} $                          & -                                        \\
        \hline
        $ \ethrthr $              & $ \SI{13.75}{\newton\metre\per\volt} $  & Stress Piezoelectric constant            \\
        $ \ethrone $              & $ \SI{-7.15}{\newton\metre\per\volt} $  & -                                        \\
        $ \eonefiv $              & $ \SI{11.91}{\newton\metre\per\volt} $  & -                                        \\
        \hline
        $ C_{11}^E $              & $ \SI{1.23e11}{\newton\per\metre^2} $   & Stiffness at constant electrical field   \\
        $ C_{12}^E $              & $ \SI{7.67e10}{\newton\per\metre^2} $   & -                                        \\
        $ C_{13}^E $              & $ \SI{7.025e10}{\newton\per\metre^2} $  & -                                        \\
        $ C_{33}^E $              & $ \SI{9.711e10}{\newton\per\metre^2} $  & -                                        \\
        $ C_{44}^E = C_{55}^E $   & $ \SI{2.226e10}{\newton\per\metre^2} $  & -                                        \\
        $ C_{66}^E $              & $ \SI{2.315e10}{\newton\per\metre^2} $  & -                                        \\
        \bottomrule
    \end{tabular}
\end{table}

\section{Experimental Setup: Free Aluminum Plate}\label{app:b}
\begin{figure}[pos=!htbp]
    \centering
    \begin{tikzpicture}
        \node[anchor=south west,inner sep=0] (image1) at (0, 0) {\includegraphics[width=0.7\textwidth, trim=0 0 0 0, clip, ]{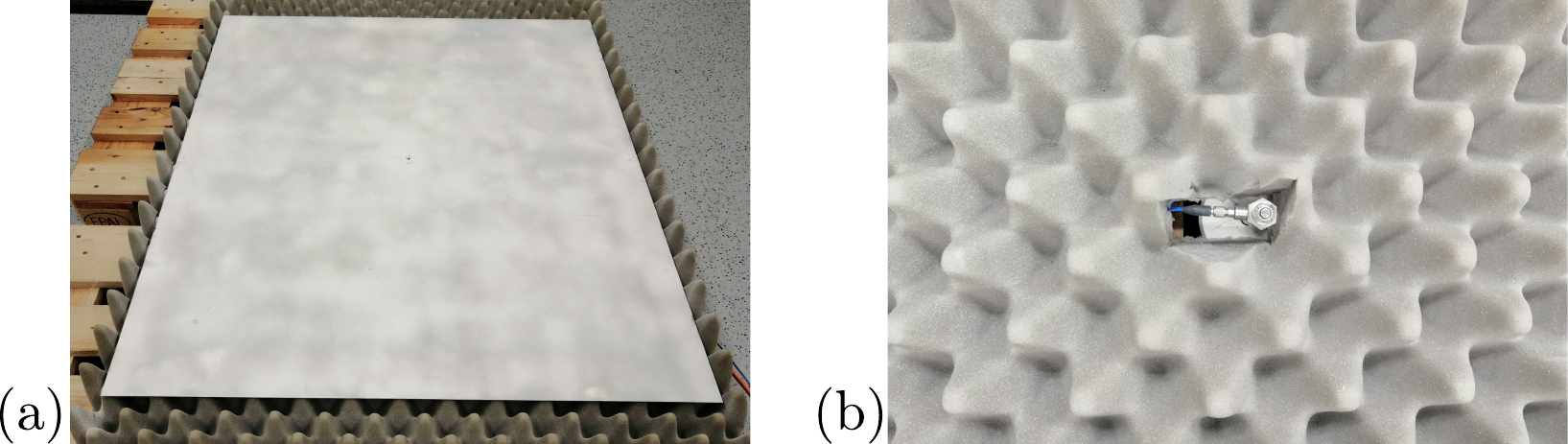}};
    \end{tikzpicture}
    \caption{Experimental setup of impacting the aluminum plate with the shaker tip. (a) Plate on foam mat with wooden pallet underneath and (b) shaker tip with the force sensor through a hole in the foam mat.}
    \label{fig::al_plate_meas_setup_shaker}
\end{figure}

Using 1D laser vibrometry allows point-by-point measurement of the deflection at the specimen surface.
Since it is an optical method, feedback on the structure can be excluded.
In this work, we used the \psvfull (PSV) \cite{polytec}.
Although it is a 3D scanning vibrometer, we decided to use only one laser head and operate it as a 1D vibrometer.
This has the advantage that the effort for setting up a measurement is considerably reduced, but only the displacement parallel to the laser beam can be measured \cite{Raddatz_04112016}.
To ensure the reproducibility of our experiments, we have included a detailed list of all used PSV components in Tab.~\ref{tab::app_psv_comp}.
The reflective surface of the aluminum plate is coated with a chalk layer to turn the specular reflection into a diffuse reflection.
As a result, a larger portion of the emitted laser is reflected back to the source, which increases the signal quality.
Since the measuring system can only measure one point at a time, a separate measurement must be performed for each measuring point.
Thus, repeatable impact excitation is required.
In order to impact the plate with the shaker, the plate is placed on a foam mat with a distance of $ d_s = \SI{1935}{\milli\metre}$ to the scanning head.
Under the foam mat a wooden pallet is used, such that the shaker can be placed beneath the aluminum plate.
To allow the shaker tip to reach the aluminum plate, a small hole is cut into the foam mat.
This experimental setup is depicted in Fig.~\ref{fig::al_plate_meas_setup_shaker}.
The individual measuring points are arranged on the entire plate in the form of a regular grid with a grid size of $ \dx=\dy=\SI{12.5}{\milli\meter} $.
\begin{table}[pos=H]
    \centering
    \small
    \caption{Used components of the \psvfull system \label{tab::app_psv_comp}}
    \begin{tabular}{lc}
        \toprule
        \textbf{Description} & \textbf{Model name} \\
        \midrule
        Controller       & OFV-5000                         \\
        Junction Box     & PSV-E-401-3D \& PSV-E-408        \\
        Scanning Head    & PSV-I-400                        \\
        Acquisition Card & PCI-6110                         \\
        Mode             & ``PSV-3D as PSV-1D''             \\
        Velocity Decoder & VD-09                            \\
        Path Decoder     & DD-900                           \\
        \bottomrule
    \end{tabular}
\end{table}

\bibliographystyle{cas-model2-names}
\bibliography{literature}

@article{newmark1959method,
	title={A Method of Computation for Structural Dynamics},
	author = {Newmark, N. M.},
	journal={Journal of the Engineering Mechanics Division},
	volume={Vol. 85},
	number={Issue 3},
	pages={67--94},
	year={1959},
	publisher={ASCE}
}

@book{Giurgiutiu_2014,
 author = {Giurgiutiu, V.},
 year = {2014},
 title = {Structural health monitoring with piezoelectric wafer active sensors}, 
 address = {Amsterdam},
 edition = {2. ed.},
 publisher = {{Academic Press/Elsevier}},
 isbn = {9780124186910}
}

@article{Cheng_2021,
 author = {Cheng, L. and Xin, H. and Groves, R. M. and Veljkovic, M.},
 year = {2021},
 title = {Acoustic emission source location using Lamb wave propagation simulation and artificial neural network for {I}-shaped steel girder},
 pages = {121706},
 volume = {273},
 issn = {09500618},
 journal = {Construction and Building Materials},
 doi = {10.1016/j.conbuildmat.2020.121706}
}

@article{Mallardo_2013,
 author = {Mallardo, V. and Aliabadi, M. H. and {Sharif Khodaei}, Z.},
 year = {2013},
 title = {Optimal sensor positioning for impact localization in smart composite panels},
 pages = {559--573},
 volume = {24},
 number = {5},
 issn = {1045389X},
 journal = {Journal of Intelligent Material Systems and Structures},
 doi = {10.1177/1045389X12464280}
}

@article{Kundu_2009,
 author = {Kundu, T. and Das, S. and Jata, K. V.},
 year = {2009},
 title = {Detection of the point of impact on a stiffened plate by the acoustic emission technique},
 pages = {035006},
 volume = {18},
 number = {3},
 issn = {0957-4484},
 journal = {Smart Materials and Structures},
 doi = {10.1088/0964-1726/18/3/035006}
}

@article{Sanchez_2016,
    author = {Sanchez, N. and Meruane, V. and Ortiz-Bernardin, A.},
    year = {2016},
    title = {A novel impact identification algorithm based on a linear approximation with maximum entropy},
    pages = {095050},
    volume = {25},
    number = {9},
    issn = {0957-4484},
    journal = {Smart Materials and Structures},
    doi = {10.1088/0964-1726/25/9/095050}
}

@article{SharifKhodaei_2012b,
    author = {{Sharif Khodaei}, Z. and Ghajari, M. and Aliabadi, M. H.},
    year = {2012},
    title = {Determination of impact location on composite stiffened panels},
    pages = {105026},
    volume = {21},
    number = {10},
    issn = {0957-4484},
    journal = {Smart Materials and Structures},
    doi = {10.1088/0964-1726/21/10/105026}
}

@article{Gaul_1998,    
    author = {Gaul, L. and Hurlebaus, S.},
    year = {1998},
    title = {IDENTIFICATION OF THE IMPACT LOCATION ON A PLATE USING WAVELETS},
    pages = {783--795},
    volume = {12},
    number = {6},
    issn = {08883270},
    journal = {Mechanical Systems and Signal Processing},
    doi = {10.1006/mssp.1998.0163}
}

@book{Cohen_1995,
    author = {Cohen, L.},
    year = {1995},
    title = {Time-frequency analysis},
    address = {Upper Saddle River, NJ},
    publisher = {{Prentice Hall PTR}},
    isbn = {0-13-594532-1},
    series = {Prentice Hall signal processing series}
}

@book{Mallat_2009,
    author = {Mallat, S. G.},
    year = {2009},
    title = {A wavelet tour of signal processing: The Sparse way},
    address = {Amsterdam},
    edition = {3.ed.},
    publisher = {{Elsevier /Academic Press}},
    isbn = {9780123743701},
    doi = {10.1016/B978-0-12-374370-1.00002-1}
}

@article{Dris_2020,
    author = {Dris, E. Y. and Drai, R. and Bentahar, M. and Berkani, D.},
    year = {2020},
    title = {Adaptive Algorithm for Estimating and Tracking the Location of Multiple Impacts on a Plate-Like Structure},
    pages = {1--23},
    volume = {31},
    number = {1},
    issn = {0934-9847},
    journal = {Research in Nondestructive Evaluation},
    doi = {10.1080/09349847.2019.1617913}    
}

@article{Li_2009,
    author = {Li, F. and Meng, G. and Kageyama, K. and Su, Z. and Ye, L.},
    year = {2009},
    title = {{Optimal Mother Wavelet Selection for Lamb Wave Analyses}},
    pages = {1147--1161},
    volume = {20},
    number = {10},
    issn = {1045389X},
    journal = {Journal of Intelligent Material Systems and Structures},
    doi = {10.1177/1045389X09102562}    
}

@article{Haase_2003,
    author = {Haase, M. and Widjajakusuma, J.},
    year = {2003},
    title = {Damage identification based on ridges and maxima lines of the wavelet transform},
    pages = {1423--1443},
    volume = {41},
    number = {13-14},
    issn = {00207225},
    journal = {International Journal of Engineering Science},
    doi = {10.1016/S0020-7225(03)00026-0}
}

@article{Ciampa_2010,
    author = {Ciampa, F. and Meo, M.},
    year = {2010},
    title = {{Acoustic emission source localization and velocity determination of the fundamental mode A0 using wavelet analysis and a Newton-based optimization technique}},
    pages = {045027},
    volume = {19},
    number = {4},
    issn = {0957-4484},
    journal = {Smart Materials and Structures},
    doi = {10.1088/0964-1726/19/4/045027}
}

@article{Ciampa_2012b,
 author = {Ciampa, F. and Meo, M. and Barbieri, E.},
 year = {2012},
 title = {Impact localization in composite structures of arbitrary cross section},
 pages = {643--655},
 volume = {11},
 number = {6},
 issn = {1475-9217},
 journal = {Structural Health Monitoring: An International Journal},
 doi = {10.1177/1475921712451951},
}

@article{Ciampa_2012,
 author = {Ciampa, F. and Meo, M.},
 year = {2012},
 title = {Impact detection in anisotropic materials using a time reversal approach},
 pages = {43--49},
 volume = {11},
 number = {1},
 issn = {1475-9217},
 journal = {Structural Health Monitoring: An International Journal},
 doi = {10.1177/1475921710395815},
}

@article{Akram_2016,    
    author = {Akram, J. and Eaton, D. W.},
    year = {2016},
    title = {A review and appraisal of arrival-time picking methods for downhole microseismic data},    
    pages = {KS71-KS91},
    volume = {81},
    number = {2},
    issn = {0016-8033},
    journal = {Geophysics},
    doi = {10.1190/geo2014-0500.1}    
}

@article{J.Wong_2009,    
    author = {Wong, J. and Han, L. and Bancroft, J. C. and Stewart, R. R.},
    year = {2009},
    title = {Automatic time picking of first arrivals on noisy microseismic data},
    journal = {CSEG Conference Abstracts}    
}

@article{Han_2009,
    author = {Han, L. and Wong, J. and Bancroft, J. C.},
    year = {2009},
    title = {Time picking and random noise reduction on microseismic data},
    pages = {1--13},
    number = {21},
    journal = {CREWES Research Report}    
}

@article{Sedlak_2013,
    author = {Sedlak, P. and Hirose, Y. and Enoki, M.},
    year = {2013},
    title = {Acoustic emission localization in thin multi-layer plates using first-arrival determination},
    pages = {636--649},
    volume = {36},
    number = {2},
    issn = {08883270},
    journal = {Mechanical Systems and Signal Processing},
    doi = {10.1016/j.ymssp.2012.11.008}
}

@article{Maeda_1985,
    author = {Maeda, N.},
    year = {1985},
    title = {A Method for Reading and Checking Phase Time in Auto-Processing System of Seismic Wave Data},
    pages = {365--379},
    volume = {38},
    number = {3},
    issn = {0037-1114},
    journal = {Zisin (Journal of the Seismological Society of Japan. 2nd ser.)},
    doi = {10.4294/zisin1948.38.3_365}
}

@article{Simone_2017,    
    author = {de Simone, M. E. and Ciampa, F. and Boccardi, S. and Meo, M.},
    year = {2017},
    title = {Impact source localisation in aerospace composite structures},
    pages = {125026},
    volume = {26},
    number = {12},
    issn = {0957-4484},
    journal = {Smart Materials and Structures},
    doi = {10.1088/1361-665X/aa973e}    
}

@misc{pi_ceramic,
    author = {{PI Ceramic GmbH}},
    title = {},
    note = {\url{https://web.archive.org/web/20211021221548/https://www.piceramic.de/de/}},
    urldate = {21.10.21}
}

@misc{ansys_help,
    author = {{Ansys Help}},
    title = {},
    note = {\url{https://ansyshelp.ansys.com/}},
    urldate = {09.10.21}
}

@phdthesis{Raddatz_04112016,
    author = {Raddatz, F. F.},
    year = {04/11/2016},
    title = {Lokalisierung der Interaktionsorte von Lambwellen in komplexen Faserverbundstrukturen},
    address = {Braunschweig},
    school = {{Technische Universit{\"a}t Carolo-Wilhelmina zu Braunschweig}},
    type = {Dissertation}    
}

@misc{polytec,
    author = {{Polytec GmbH}},
    title = {},
    note = {\url{https://web.archive.org/web/20210930074250/https://www.polytec.com/de}},
    urldate = {30.09.21}
}

@article{Coverley_2003,
    author = {Coverley, P. T. and Staszewski, W. J.},
    year = {2003},
    title = {Impact damage location in composite structures using optimized sensor triangulation procedure},
    pages = {795--803},
    volume = {12},
    number = {5},
    issn = {0957-4484},
    journal = {Smart Materials and Structures},
    doi = {10.1088/0964-1726/12/5/017}
}

@article{Staszewski_2009,
 author = {Staszewski, W. J. and Mahzan, S. and Traynor, R.},
 year = {2009},
 title = {Health monitoring of aerospace composite structures -- Active and passive approach},
 pages = {1678--1685},
 volume = {69},
 number = {11-12},
 issn = {02663538},
 journal = {Composites Science and Technology},
 doi = {10.1016/j.compscitech.2008.09.034}
}

@article{Merlo_2017,
    author = {Merlo, E. M. and Bulletti, A. and Giannelli, P. and Calzolai, M. and Capineri, L.},
    year = {2017},
    title = {A Novel Differential Time-of-Arrival Estimation Technique for Impact Localization on Carbon Fiber Laminate Sheets},
    volume = {17},
    number = {10},
    journal = {Sensors},
    doi = {10.3390/s17102270}
    }

@article{Tabian_2019b,
    author = {Tabian, I. and Fu, H. and {Sharif Khodaei}, Z.},
    year = {2019},
    title = {A Convolutional Neural Network for Impact Detection and Characterization of Complex Composite Structures},
    volume = {19},
    number = {22},
    journal = {Sensors},
    doi = {10.3390/s19224933}
    }

@misc{hbm,
    author = {{Hottinger Br{\"u}el \& Kjaer GmbH (HBM)}},
    title = {},
    note = {\url{https://web.archive.org/web/20210116055950/https://www.hbm.com/en/2961/x60-2-component-fast-curing-adhesive/}},
    urldate = {16.01.21}
}

@misc{piezomech,
    author = {{Piezomechanik GmbH}},
    title = {},
    note = {\url{https://web.archive.org/web/20210728034527/https://www.piezomechanik.com/}},
    urldate = {28.07.21}
}

@article{Allen_1982,
    author = {Allen, R.},
    year = {1982},
    title = {Automatic phase pickers: Their present use and future prospects},
    pages = {S225-S242},
    volume = {72},
    number = {6B},
    issn = {0037-1106},
    journal = {Bulletin of the Seismological Society of America},
    doi = {10.1785/BSSA07206B0225}
}

@book{Altenbach_2015,
    author = {Altenbach, H.},
    year = {2015},
    title = {Kontinuumsmechanik},
    address = {Berlin, Heidelberg},
    publisher = {{Springer Berlin Heidelberg}},
    isbn = {978-3-662-47069-5},
    doi = {10.1007/978-3-662-47070-1}
}

@article{Rioul_1991,
    author = {Rioul, O. and Vetterli, M.},
    year = {1991},
    title = {Wavelets and signal processing},
    pages = {14--38},
    volume = {8},
    number = {4},
    issn = {1053-5888},
    journal = {IEEE Signal Processing Magazine},
    doi = {10.1109/79.91217}    
}

@book{Shearer_2009,
    author = {Shearer, P. M.},
    year = {2011},
    title = {Introduction to seismology},
    address = {Cambridge},
    edition = {2. ed., repr. with corr},
    publisher = {{Cambridge Univ. Press}},
    isbn = {9780521882101}    
}

@article{Trnkoczy_,
    author = {Trnkoczy, A.},
    title = {Understanding and parameter setting of STA/LTA trigger algorithm~},
    doi = {10.2312/GFZ.NMSOP-2_IS_8.1},
}

@article{Bennett_1948,
    author = {Bennett, W. R.},
    year = {1948},
    title = {Spectra of Quantized Signals},
    pages = {446--472},
    volume = {27},
    number = {3},
    issn = {00058580},
    journal = {Bell System Technical Journal},
    doi = {10.1002/j.1538-7305.1948.tb01340.x}
}

@article{Gabor_1946,
    author = {Gabor, D.},
    year = {1946},
    title = {Theory of communication. Part 1: The analysis of information},
    pages = {429--441},
    volume = {93},
    number = {26},
    journal = {Journal of the Institution of Electrical Engineers - Part III: Radio and Communication Engineering},
    doi = {10.1049/ji-3-2.1946.0074}
}

@article{LIGHTHILL_1965,
    author = {Lighthill, M. J.},
    year = {1965},
    title = {Group Velocity},
    pages = {1--28},
    volume = {1},
    number = {1},
    issn = {0272-4960},
    journal = {IMA Journal of Applied Mathematics},
    doi = {10.1093/imamat/1.1.1},    
}

@book{Tipler_2012,
    author = {Tipler, P. A. and Llewellyn, R.},
    year = {2012},
    title = {Modern physics},
    address = {New York},
    edition = {6. ed.},
    publisher = {Freeman},
    isbn = {9781429250788},   
}

@book{Crawford_1968,
    author = {Crawford, Jr., F. S.},
    year = {1968},
    title = {Waves, vol. 3},
    publisher = {McGraw-Hill},
    isbn = {9780070048607}    
}

@book{Hellier_2001,    
    author = {Hellier, C. J.},
    year = {2001},
    title = {Handbook of nondestructive evaluation: Rodrick Rules},    
    address = {New York},
    volume = {2},
    publisher = {McGraw-Hill},
    isbn = {9780070281219},
    series = {McGraw-Hill Handbooks}    
}

@ARTICLE{2020SciPy-NMeth,
    author  = {Virtanen, P. and Gommers, R. and Oliphant, T. E. and
    Haberland, M. and Reddy, T. and Cournapeau, D. and
    Burovski, E. and Peterson, P. and Weckesser, W. and
    Bright, J. and {van der Walt}, S. J. and
    Brett, M. and Wilson, J. and Millman, K. J. and
    Mayorov, N. and Nelson, A. R. J. and Jones, E. and
    Kern, R. and Larson, E. and Carey, C. J. and
    Polat, I. and Feng, Y. and Moore, E. W. and
    {VanderPlas}, J. and Laxalde, D. and Perktold, J. and
    Cimrman, R. and Henriksen, I. and Quintero, E. A. and
    Harris, C. R. and Archibald, A. M. and
    Ribeiro, A. H. and Pedregosa, F. and
    {van Mulbregt}, P. and {SciPy 1.0 Contributors}},
    title   = {{{SciPy} 1.0: Fundamental Algorithms for Scientific
    Computing in Python}},
    journal = {Nature Methods},
    year    = {2020},
    volume  = {17},
    pages   = {261--272},
    adsurl  = {https://rdcu.be/b08Wh},
    doi     = {10.1038/s41592-019-0686-2},
}

@article{Kundu_2014,
    author = {Kundu, T.},
    year = {2014},
    title = {Acoustic source localization},
    pages = {25--38},
    volume = {54},
    number = {1},
    issn = {0041624X},
    journal = {Ultrasonics},
    doi = {10.1016/j.ultras.2013.06.009}
}

@article{Meo_2005,
    author = {Meo, M. and Zumpano, G. and Piggott, M. and Marengo, G.},
    year = {2005},
    title = {Impact identification on a sandwich plate from wave propagation responses},
    pages = {302--306},
    volume = {71},
    number = {3-4},
    issn = {02638223},
    journal = {Composite Structures},
    doi = {10.1016/j.compstruct.2005.09.028}
}

@article{Torrence_1998,
    author = {Torrence, C. and Compo, G. P.},
    year = {1998},
    title = {A Practical Guide to Wavelet Analysis},
    pages = {61--78},
    volume = {79},
    number = {1},
    issn = {0003-0007},
    journal = {Bulletin of the American Meteorological Society},
    doi = {10.1175/1520-0477(1998)079<0061:APGTWA>2.0.CO;2}
}

@book{Achenbach_1999,
 author = {Achenbach, J. D.},
 year = {1999},
 title = {Wave propagation in elastic solids},
 address = {Amsterdam},
 edition = {7. impr},
 volume = {16},
 publisher = {Elsevier},
 isbn = {0720403251},
 series = {North Holland series in applied mathematics and mechanics},
 doi = {10.1016/C2009-0-08707-8},
}

@article{Lamb_1917,
 author = {Lamb, H.},
 year = {1917},
 title = {On waves in an elastic plate},
 pages = {114--128},
 volume = {93},
 number = {648},
 issn = {0950-1207},
 journal = {Proceedings of the Royal Society of London. Series A, Containing Papers of a Mathematical and Physical Character},
 doi = {10.1098/rspa.1917.0008},
}

@article{Guyomar_2009,
 author = {Guyomar, D. and Lallart, M. and Monnier, T. and Wang, X. and Petit, L.},
 year = {2009},
 title = {Passive Impact Location Estimation Using Piezoelectric Sensors},
 pages = {357--367},
 volume = {8},
 number = {5},
 issn = {1475-9217},
 journal = {Structural Health Monitoring: An International Journal},
 doi = {10.1177/1475921709102090}
}

@article{Guyomar_2011b,
 author = {Guyomar, D. and Lallart, M. and Petit, L. and Wang, X.-J.},
 year = {2011},
 title = {Impact localization and energy quantification based on the power flow: A low-power requirement approach},
 pages = {3270--3283},
 volume = {330},
 number = {13},
 issn = {10958568},
 journal = {Journal of Sound and Vibration},
 doi = {10.1016/j.jsv.2011.01.013}
}

@article{Haywood_2005,
 author = {Haywood, J. and Coverley, P. T. and Staszewski, W. J. and Worden, K.},
 year = {2005},
 title = {An automatic impact monitor for a composite panel employing smart sensor technology},
 pages = {265--271},
 volume = {14},
 number = {1},
 issn = {0957-4484},
 journal = {Smart Materials and Structures},
 doi = {10.1088/0964-1726/14/1/027}
}

@article{LeClerc_2007,
 author = {LeClerc, J. R. and Worden, K. and Staszewski, W. J. and Haywood, J.},
 year = {2007},
 title = {Impact detection in an aircraft composite panel---A neural-network approach},
 pages = {672--682},
 volume = {299},
 number = {3},
 issn = {10958568},
 journal = {Journal of Sound and Vibration},
 doi = {10.1016/j.jsv.2006.07.019}
}

@article{Jang_2015,
 author = {Jang, B.-W. and Lee, Y.-G. and Kim, C.-G. and Park, C.-Y.},
 year = {2015},
 title = {Impact source localization for composite structures under external dynamic loading condition},
 pages = {359--374},
 volume = {24},
 number = {4},
 issn = {0924-3046},
 journal = {Advanced Composite Materials},
 doi = {10.1080/09243046.2014.917239}
}

@article{LanzaDiScalea_2007,
 author = {{Di Lanza Scalea}, F. and Matt, H. and Bartoli, I.},
 year = {2007},
 title = {{The response of rectangular piezoelectric sensors to Rayleigh and Lamb ultrasonic waves}},
 pages = {175--187},
 volume = {121},
 number = {1},
 issn = {0001-4966},
 journal = {The Journal of the Acoustical Society of America},
 doi = {10.1121/1.2400668}
}

@article{Park_2012,
 author = {Park, B. and Sohn, H. and Olson, S. E. and DeSimio, M. P. and Brown, K. S. and Derriso, M. M.},
 year = {2012},
 title = {Impact localization in complex structures using laser-based time reversal},
 pages = {577--588},
 volume = {11},
 number = {5},
 issn = {1475-9217},
 journal = {Structural Health Monitoring: An International Journal},
 doi = {10.1177/1475921712449508}
}

@article{Qiu_2011,
 author = {Qiu, L. and Yuan, S. and Zhang, X. and Wang, Y.},
 year = {2011},
 title = {A time reversal focusing based impact imaging method and its evaluation on complex composite structures},
 pages = {105014},
 volume = {20},
 number = {10},
 issn = {0957-4484},
 journal = {Smart Materials and Structures},
 doi = {10.1088/0964-1726/20/10/105014}
}

@article{Seno_2019,
 author = {Seno, A. H. and {Sharif Khodaei}, Z. and Aliabadi, M. F. H.},
 year = {2019},
 title = {Passive sensing method for impact localisation in composite plates under simulated environmental and operational conditions},
 pages = {20--36},
 volume = {129},
 issn = {08883270},
 journal = {Mechanical Systems and Signal Processing},
 doi = {10.1016/j.ymssp.2019.04.023}
}

@article{McLaskey_2010,
 author = {McLaskey, G. C. and Glaser, S. D. and Grosse, C. U.},
 year = {2010},
 title = {Beamforming array techniques for acoustic emission monitoring of large concrete structures},
 pages = {2384--2394},
 volume = {329},
 number = {12},
 issn = {10958568},
 journal = {Journal of Sound and Vibration},
 doi = {10.1016/j.jsv.2009.08.037} 
}

@article{He_2012,
 author = {He, T. and Pan, Q. and Liu, Y. and Liu, X. and Hu, D.},
 year = {2012},
 title = {Near-field beamforming analysis for acoustic emission source localization},
 pages = {587--592},
 volume = {52},
 number = {5},
 issn = {0041624X},
 journal = {Ultrasonics},
 doi = {10.1016/j.ultras.2011.12.003}
}

@article{vandecar1990determination,
  title={Determination of teleseismic relative phase arrival times using multi-channel cross-correlation and least squares},
  author={VanDecar, J. C. and Crosson, R. S.},
  journal={Bulletin of the Seismological Society of America},
  volume={80},
  number={1},
  pages={150--169},
  year={1990},
  publisher={The Seismological Society of America}
}

@article{DeMeersman1998,
author = {De Meersman, K. and Kendall, J.-M. and van der Baan, M.},
title = {The 1998 Valhall microseismic data set: An integrated study of relocated sources, seismic multiplets, and S-wave splitting},
journal = {GEOPHYSICS},
volume = {74},
number = {5},
pages = {B183-B195},
year = {2009},
doi = {10.1190/1.3205028}
}

@article{Molyneux_1999,
author = {Molyneux, J. B. and Schmitt, D. R.},
title = {First‐break timing: Arrival onset times by direct correlation},
journal = {GEOPHYSICS},
volume = {64},
number = {5},
pages = {1492-1501},
year = {1999},
doi = {10.1190/1.1444653}
}

@article{McCormack_1993,
author = {McCormack, M. D. and Zaucha, D. E. and Dushek, D. W.},
title = {First‐break refraction event picking and seismic data trace editing using neural networks},
journal = {GEOPHYSICS},
volume = {58},
number = {1},
pages = {67-78},
year = {1993},
doi = {10.1190/1.1443352},
}

@article{Dai_1995,
 author = {Dai, H. and MacBeth, C.},
 year = {1995},
 title = {Automatic picking of seismic arrivals in local earthquake data using an artificial neural network},
 pages = {758--774},
 volume = {120},
 number = {3},
 issn = {0956540X},
 journal = {Geophysical Journal International},
 doi = {10.1111/j.1365-246X.1995.tb01851.x}
}

@article{Zonzini_2022,
 author = {Zonzini, F. and Bogomolov, D. and Dhamija, T. and Testoni, N. and de Marchi, L. and Marzani, A.},
 year = {2022},
 title = {Deep Learning Approaches for Robust Time of Arrival Estimation in Acoustic Emission Monitoring},
 volume = {22},
 number = {3},
 journal = {Sensors (Basel, Switzerland)},
 doi = {10.3390/s22031091}
}

@article{Saragiotis_2004,
 author = {Saragiotis, C. D. and Hadjileontiadis, L. J. and Rekanos, I. T. and Panas, S. M.},
 year = {2004},
 title = {Automatic P Phase Picking Using Maximum Kurtosis and {$\kappa$}-Statistics Criteria},
 pages = {147--151},
 volume = {1},
 number = {3},
 issn = {1545-598X},
 journal = {IEEE Geoscience and Remote Sensing Letters},
 doi = {10.1109/LGRS.2004.828915}
}

@article{Lokajicek_2006,
 author = {Lokaj\'i\'cek, T. and Kl\'ima, K.},
 year = {2006},
 title = {{A first arrival identification system of acoustic emission (AE) signals by means of a high-order statistics approach}},
 pages = {2461--2466},
 volume = {17},
 number = {9},
 issn = {0957-0233},
 journal = {Measurement Science and Technology},
 doi = {10.1088/0957-0233/17/9/013}
}

@article{Liu_2021,
 author = {Liu, S. and Li, Z. and Wu, T. and Zhang, W.},
 year = {2021},
 title = {Determining Ultrasound Arrival Time by HHT and Kurtosis in Wind Speed Measurement},
 pages = {93},
 volume = {10},
 number = {1},
 journal = {Electronics},
 doi = {10.3390/electronics10010093}
}

@article{Stoica_2004,
 author = {Stoica, P. and Selen, Y.},
 year = {2004},
 title = {Model-order selection},
 pages = {36--47},
 volume = {21},
 number = {4},
 issn = {1053-5888},
 journal = {IEEE Signal Processing Magazine},
 doi = {10.1109/MSP.2004.1311138}
}

@article{Kitagawa_1978,
 author = {Kitagawa, G. and Akaike, H.},
 year = {1978},
 title = {A procedure for the modeling of non-stationary time series},
 pages = {351--363},
 volume = {30},
 number = {2},
 issn = {0020-3157},
 journal = {Annals of the Institute of Statistical Mathematics},
 doi = {10.1007/BF02480225}
}

@book{Billingsley_1995,
 author = {Billingsley, P.},
 year = {1995},
 title = {Probability and measure},
 address = {New York, NY},
 edition = {3. ed.},
 publisher = {Wiley},
 isbn = {9780471007104},
 series = {A Wiley-Interscience publication}
}

@article{More_2011, 
 author = {Mor\'e, J. J. and Wild, S. M.},
 year = {2011},
 title = {Estimating Computational Noise},
 pages = {1292--1314},
 volume = {33},
 number = {3},
 issn = {1064-8275},
 journal = {SIAM Journal on Scientific Computing},
 doi = {10.1137/100786125} 
}

@article{Arts_2022,
 author = {Arts, L. P. A. and {van den Broek}, E. L.},
 year = {2022},
 title = {{The fast continuous wavelet transformation (fCWT) for real-time, high-quality, noise-resistant time--frequency analysis}},
 pages = {47--58},
 volume = {2},
 number = {1},
 journal = {Nature Computational Science},
 doi = {10.1038/s43588-021-00183-z}
}

@article{Zhu_2017, 
 author = {Zhu, J. and Parvasi, S. M. and Ho, S. C. M. and Patil, D. and Ge, M. and Li, H. and Song, G.},
 year = {2017},
 title = {An innovative method for automatic determination of time of arrival for Lamb waves excited by impact events}, 
 volume = {26},
 number = {5},
 journal = {Smart Materials and Structures},
 doi = {10.1088/1361-665X/aa63e1}
}

@inproceedings{Grasboeck_smart2023,
 author = {Grasboeck, L. and Humer, A. and Benjeddou, A.},
 title = {Revisiting Multidisciplinary Time-of-Arrival Estimation Methods for Impact Detection on Plates},
 pages = {107--118},
 publisher = {{Dept. of Mechanical Engineering {\&} Aeronautics, University of Patras}},
 isbn = {978-960-88104-6-4},
 editor = {Saravanos, Dimitris A. and Benjeddou, Ayech and Chrysochoidis, Nikolaos A. and Theodosiou, Theodosis C.},
 booktitle = {10th ECCOMAS Thematic Conference on Smart Structures and Materials},
 year = {03.07.2023 - 05.07.2023},
 doi = {10.7712/150123.9769.451409}
}

@article{Sleeman_1999,
 author = {Sleeman, R. and {van Eck}, T.},
 year = {1999},
 title = {Robust automatic P-phase picking: an on-line implementation in the analysis of broadband seismogram recordings},
 pages = {265--275},
 volume = {113},
 number = {1-4},
 issn = {00319201},
 journal = {Physics of the Earth and Planetary Interiors},
 doi = {10.1016/S0031-9201(99)00007-2}
}

@article{Leonard_2000,
 author = {Leonard, M.},
 year = {2000},
 title = {Comparison of Manual and Automatic Onset Time Picking},
 pages = {1384--1390},
 volume = {90},
 number = {6},
 issn = {0037-1106},
 journal = {Bulletin of the Seismological Society of America},
 doi = {10.1785/0120000026}
}

@book{Kobayashi_2012,
 author = {Kobayashi, H. and Mark, B. L. and Turin, W.},
 year = {2012},
 title = {Probability, random processes, and statistical analysis},
 address = {Cambridge and New York},
 publisher = {{Cambridge University Press}},
 isbn = {9780521895446}
}

@article{Akaike_1974,
 author = {Akaike, H.},
 year = {1974},
 title = {A new look at the statistical model identification},
 pages = {716--723},
 volume = {19},
 number = {6},
 issn = {0018-9286},
 journal = {IEEE Transactions on Automatic Control},
 doi = {10.1109/TAC.1974.1100705}
}

@article{Sedlak_2009,
 author = {Sedlak, P. and Hirose, Y. and Khan, S. A. and Enoki, M. and Sikula, J.},
 year = {2009},
 title = {New automatic localization technique of acoustic emission signals in thin metal plates},
 pages = {254--262},
 volume = {49},
 number = {2},
 issn = {0041624X},
 journal = {Ultrasonics},
 doi = {10.1016/j.ultras.2008.09.005}
}

@article{Grasboeck_2025,
author = {Grasboeck, L. and Humer, A. and Benjeddou, A.},
title ={Detection and time-of-arrival estimation of impact-induced waves in composite laminates},
journal = {Structural Health Monitoring},
volume = {24},
number = {4},
pages = {2428-2446},
year = {2025},
doi = {10.1177/14759217241260848}
}

@misc{humer2025,
      title={Localization of Impacts on Thin-Walled Structures by Recurrent Neural Networks: End-to-end Learning from Real-World Data}, 
      author={Humer, A. and Grasboeck, L. and Benjeddou, A.},
      year={2025},
      eprint={2505.08362},
      archivePrefix={arXiv},
      primaryClass={cs.LG},
      url={https://arxiv.org/abs/2505.08362}, 
}

@article{Barile_2025,
title = {Investigation of an improved time of arrival detection method for acoustic Emission signals and its applications to damage characterisation in composite materials},
journal = {Mechanical Systems and Signal Processing},
volume = {223},
pages = {111906},
year = {2025},
issn = {0888-3270},
doi = {10.1016/j.ymssp.2024.111906},
author = {Claudia Barile and Caterina Casavola and Giovanni Pappalettera and Vimalathithan {Paramsamy Kannan}}
}

@article{Hou_2025,
title = {Arrival time detection of AE signals using multi-threshold wavelet denoising and improved AIC for composites damage localization},
journal = {Mechanical Systems and Signal Processing},
volume = {237},
pages = {113080},
year = {2025},
issn = {0888-3270},
doi = {10.1016/j.ymssp.2025.113080},
author = {Dongming Hou and Kuihui Huang and Xinxin Qi and Wentao Wang and Xiaoyang Bi and Jinzhen Kong and Yang Chang and Libin Zhao and Ning Hu}
}

@article{Jana_2026,
title = {Fast and accurate decomposition of overlapping Lamb wave modes in metallic and composite structures},
journal = {Mechanical Systems and Signal Processing},
volume = {250},
pages = {114105},
year = {2026},
issn = {0888-3270},
doi = {10.1016/j.ymssp.2026.114105},
author = {Souvik Jana and Santosh Kapuria}
}

@article{Hamstad_2005,
  author = {Marvin Hamstad and Agnes O'Gallagher},
  title = {Effects of Noise on Lamb-Mode Acoustic-Emission Arrival Times Determined by Wavelet Transform},
  year = {2005},
  number = {23},
  month = {2005-12-01 00:12:00},
  publisher = {J. Acoust. Emiss.},
  url = {https://tsapps.nist.gov/publication/get_pdf.cfm?pub_id=50139},
  language = {en},
}

\section*{CRediT authorship contribution statement}
\textbf{Lukas Grasboeck:} Conceptualization, Methodology, Software, Investigation, Writing - Original Draft, Visualization
\textbf{Alexander Humer:} Supervision, Conceptualization, Methodology, Software, Validation, Writing - Review \& Editing
\textbf{Ayech Benjeddou:} Co-Supervision, Validation, Resources, Writing - Review \& Editing

\section*{Acknowledgements}
This work has been supported by the COMET-K2 Symbiotic Mechatronics Center of the Linz Center of Mechatronics (LCM) funded by the Austrian federal government and the federal state of Upper Austria.
The authors would like to thank the Fraunhofer Institute for Structural Durability and System Reliability LBF for the support during the first author research stay, particularly for the opportunity to conduct the experimental part of the work at LBF.

\section*{Declaration of conflicting interests}
The authors declare that they have no known competing financial interests or personal relationships that could have appeared to influence the work reported in this paper.

\section*{Declaration of generative AI and AI-assisted technologies in the writing process}
During the preparation of this work, the authors used ChatGPT by OpenAI to assist with language editing, readability improvement, and structural refinement of selected manuscript sections.
The authors reviewed and edited all AI-assisted output and take full responsibility for the content of the published article.

\section*{ORCiD}
\noindent\textbf{Lukas Grasboeck:} \href{https://orcid.org/0009-0007-0662-4913}{0009-0007-0662-4913} \\
\textbf{Alexander Humer:} \href{https://orcid.org/0000-0001-8510-5423}{0000-0001-8510-5423} \\
\textbf{Ayech Benjeddou:} \href{https://orcid.org/0000-0002-4760-4800}{0000-0002-4760-4800}

\end{document}